\documentclass[11pt,reqno]{amsart}

\usepackage{amssymb}
\usepackage{booktabs}
\usepackage[expansion=false]{microtype}
\usepackage{xcolor}
\usepackage{xurl}
\usepackage{pgfplots}
\usepackage{listings}
\pgfplotsset{compat=1.18}
\usepackage[colorlinks=true,linkcolor=blue!50!black,citecolor=blue!50!black,urlcolor=blue!50!black]{hyperref}

\newtheorem{theorem}{Theorem}[section]
\newtheorem{lemma}[theorem]{Lemma}
\newtheorem{proposition}[theorem]{Proposition}
\newtheorem{corollary}[theorem]{Corollary}
\newtheorem{conjecture}[theorem]{Conjecture}
\newtheorem*{theoremA}{Theorem A}
\newtheorem*{theoremB}{Theorem B}
\newtheorem*{claimC}{Claim C}
\theoremstyle{definition}
\newtheorem{definition}[theorem]{Definition}
\theoremstyle{remark}
\newtheorem{remark}[theorem]{Remark}
\newtheorem{example}[theorem]{Example}

\newcommand{\erfc}{\operatorname{erfc}}
\newcommand{\erfi}{\operatorname{erfi}}
\newcommand{\erf}{\operatorname{erf}}
\newcommand{\Ci}{\operatorname{Ci}}
\newcommand{\sinc}{\operatorname{sinc}}
\newcommand{\Ein}{\operatorname{Ein}}
\newcommand{\Eip}{\mathcal{E}}
\newcommand{\Li}{\operatorname{Li}}
\newcommand{\arctanh}{\operatorname{arctanh}}
\newcommand{\dd}{\,\mathrm{d}}
\newcommand{\QQ}{\mathbb{Q}}
\newcommand{\Qbar}{\overline{\mathbb{Q}}}
\newcommand{\Reg}{\mathcal{R}}
\newcommand{\RegQ}{\mathcal{R}_{\Qbar}}
\newcommand{\ZZ}{\mathbb{Z}}
\newcommand{\CC}{\mathbb{C}}
\newcommand{\RR}{\mathbb{R}}
\newcommand{\Mul}{\mathsf{M}}
\newcommand{\CT}{\operatorname*{CT}}

\title[Computing infinite sums of holonomic functions]{Computing infinite sums of holonomic functions}
\author{Fredrik Johansson}
\address{INRIA, Universit\'{e} de Bordeaux, CNRS, CANARI, 33400 Talence, France}
\email{fredrik.johansson@gmail.com}
\date{Version 1, November 2026}

\pgfplotstableread[row sep=\\]{
n gm z\\
1 1.000000 -0.9878\\
2 1.414214 -0.9069\\
3 2.154435 -0.3816\\
4 1.778279 -0.8244\\
5 1.584893 -1.1792\\
6 1.647549 -1.1967\\
7 1.534127 -1.4813\\
8 1.454215 -1.7349\\
9 1.576059 -1.5988\\
10 1.895899 -1.1010\\
11 1.905134 -1.1386\\
12 1.805504 -1.3753\\
13 1.877421 -1.2906\\
14 1.885922 -1.3224\\
15 2.234445 -0.7121\\
16 2.124939 -0.9364\\
17 2.168487 -0.8816\\
18 2.158765 -0.9262\\
19 2.196480 -0.8761\\
20 2.111742 -1.0748\\
21 2.037895 -1.2645\\
22 1.973004 -1.4460\\
23 2.329504 -0.6819\\
24 2.314748 -0.7277\\
25 2.238327 -0.9106\\
26 2.387657 -0.5993\\
27 2.311921 -0.7782\\
28 2.243747 -0.9509\\
29 2.182084 -1.1178\\
30 2.126061 -1.2794\\
31 2.185527 -1.1469\\
32 2.179476 -1.1809\\
33 2.247395 -1.0230\\
34 2.194502 -1.1772\\
35 2.258481 -1.0244\\
36 2.276363 -0.9916\\
37 2.226313 -1.1405\\
38 2.179914 -1.2857\\
39 2.175104 -1.3163\\
40 2.220840 -1.2014\\
41 2.215174 -1.2327\\
42 2.173621 -1.3704\\
43 2.169417 -1.3993\\
44 2.131567 -1.5322\\
45 2.096016 -1.6624\\
46 2.125671 -1.5854\\
47 2.091838 -1.7126\\
48 2.089882 -1.7372\\
49 2.088008 -1.7615\\
50 2.103197 -1.7281\\
51 2.101123 -1.7523\\
52 2.071336 -1.8724\\
53 2.043071 -1.9903\\
54 2.068649 -1.9176\\
55 2.041489 -2.0333\\
56 2.015637 -2.1471\\
57 2.015361 -2.1672\\
58 2.047183 -2.0668\\
59 2.046374 -2.0876\\
60 2.022097 -2.1976\\
61 1.998889 -2.3060\\
62 2.012023 -2.2733\\
63 2.099065 -1.9554\\
64 2.143409 -1.8036\\
65 2.154525 -1.7759\\
66 2.129613 -1.8840\\
67 2.149744 -1.8212\\
68 2.160305 -1.7944\\
69 2.136324 -1.9002\\
70 2.183953 -1.7295\\
71 2.282413 -1.3702\\
72 2.256403 -1.4771\\
73 2.252677 -1.5014\\
74 2.249059 -1.5255\\
75 2.283365 -1.4047\\
76 2.312577 -1.3032\\
77 2.308219 -1.3283\\
78 2.303982 -1.3531\\
79 2.361200 -1.1437\\
80 2.376809 -1.0920\\
81 2.392133 -1.0410\\
82 2.386915 -1.0672\\
83 2.381835 -1.0931\\
84 2.388387 -1.0744\\
85 2.364048 -1.1752\\
86 2.340515 -1.2749\\
87 2.370174 -1.1649\\
88 2.347045 -1.2635\\
89 2.367074 -1.1905\\
90 2.344520 -1.2880\\
91 2.340429 -1.3118\\
92 2.318896 -1.4076\\
93 2.338134 -1.3356\\
94 2.344342 -1.3171\\
95 2.323411 -1.4115\\
96 2.303097 -1.5049\\
97 2.283374 -1.5974\\
98 2.280289 -1.6190\\
99 2.277270 -1.6405\\
101 2.255888 -1.7517\\
102 2.294957 -1.5870\\
103 2.330143 -1.4403\\
104 2.463796 -0.8785\\
105 2.442728 -0.9707\\
107 2.449431 -0.9516\\
108 2.429197 -1.0422\\
109 2.468547 -0.8793\\
110 2.448351 -0.9695\\
112 2.482279 -0.8326\\
113 2.462387 -0.9218\\
114 2.481700 -0.8425\\
115 2.462162 -0.9309\\
117 2.424530 -1.1056\\
118 2.451629 -0.9896\\
120 2.510969 -0.7359\\
121 2.491936 -0.8227\\
122 2.487448 -0.8460\\
124 2.464896 -0.9543\\
125 2.490567 -0.8423\\
127 2.521669 -0.7092\\
128 2.517107 -0.7324\\
130 2.494880 -0.8393\\
131 2.477529 -0.9224\\
133 2.499630 -0.8270\\
134 2.482599 -0.9092\\
136 2.586627 -0.4373\\
137 2.568746 -0.5201\\
139 2.559512 -0.5663\\
140 2.562417 -0.5549\\
142 2.601714 -0.3775\\
144 2.608774 -0.3476\\
145 2.643757 -0.1884\\
147 2.704019 0.0835\\
149 2.713025 0.1247\\
150 2.695033 0.0436\\
152 2.700708 0.0698\\
154 2.697286 0.0546\\
156 2.723019 0.1735\\
158 2.798577 0.5187\\
159 2.780522 0.4387\\
161 2.764221 0.3668\\
163 2.768788 0.3902\\
165 2.746332 0.2880\\
167 2.731213 0.2184\\
169 2.710022 0.1184\\
171 2.689485 0.0196\\
173 2.669574 -0.0780\\
175 2.671338 -0.0697\\
177 2.668722 -0.0831\\
179 2.660135 -0.1267\\
181 2.641629 -0.2214\\
183 2.653634 -0.1612\\
185 2.645534 -0.2037\\
187 2.627872 -0.2964\\
189 2.634353 -0.2641\\
192 2.594783 -0.4759\\
194 2.605519 -0.4209\\
196 2.607702 -0.4113\\
198 2.615193 -0.3730\\
201 2.586843 -0.5304\\
203 2.600175 -0.4598\\
205 2.622166 -0.3415\\
208 2.594592 -0.4964\\
210 2.610345 -0.4111\\
212 2.643631 -0.2285\\
215 2.616434 -0.3818\\
217 2.601639 -0.4671\\
220 2.603145 -0.4617\\
222 2.588876 -0.5457\\
225 2.609562 -0.4300\\
228 2.584680 -0.5775\\
230 2.588031 -0.5604\\
233 2.608432 -0.4442\\
236 2.599640 -0.4989\\
238 2.640315 -0.2615\\
241 2.608597 -0.4508\\
244 2.592714 -0.5490\\
247 2.606238 -0.4706\\
250 2.630172 -0.3289\\
252 2.631689 -0.3210\\
255 2.627379 -0.3491\\
258 2.605022 -0.4884\\
261 2.615010 -0.4294\\
264 2.629250 -0.3437\\
267 2.600846 -0.5231\\
270 2.655743 -0.1828\\
274 2.687740 0.0141\\
277 2.679151 -0.0391\\
280 2.664168 -0.1331\\
283 2.646893 -0.2433\\
286 2.630088 -0.3523\\
290 2.630843 -0.3499\\
293 2.624523 -0.3928\\
297 2.621045 -0.4184\\
300 2.640200 -0.2944\\
303 2.646052 -0.2573\\
307 2.618622 -0.4416\\
310 2.594340 -0.6077\\
314 2.613040 -0.4844\\
318 2.611781 -0.4960\\
321 2.599655 -0.5818\\
325 2.584696 -0.6894\\
329 2.609849 -0.5180\\
333 2.610116 -0.5192\\
337 2.609342 -0.5278\\
340 2.625766 -0.4144\\
344 2.621165 -0.4494\\
348 2.630744 -0.3840\\
352 2.619063 -0.4697\\
356 2.609030 -0.5447\\
361 2.628179 -0.4096\\
365 2.616952 -0.4936\\
369 2.638299 -0.3403\\
373 2.625590 -0.4354\\
378 2.626137 -0.4342\\
382 2.616687 -0.5070\\
386 2.607465 -0.5790\\
391 2.589439 -0.7199\\
395 2.576270 -0.8249\\
400 2.601687 -0.6338\\
404 2.600813 -0.6437\\
409 2.623196 -0.4744\\
414 2.611958 -0.5646\\
419 2.609351 -0.5885\\
424 2.610423 -0.5835\\
428 2.611566 -0.5772\\
433 2.615366 -0.5503\\
438 2.600454 -0.6731\\
443 2.587026 -0.7859\\
449 2.576493 -0.8777\\
454 2.568951 -0.9450\\
459 2.574697 -0.9023\\
464 2.565099 -0.9877\\
470 2.561378 -1.0255\\
475 2.562184 -1.0241\\
481 2.563467 -1.0196\\
486 2.571217 -0.9583\\
492 2.556139 -1.0947\\
498 2.555367 -1.1081\\
503 2.551803 -1.1449\\
509 2.553762 -1.1344\\
515 2.558064 -1.1029\\
521 2.547432 -1.2043\\
527 2.558635 -1.1105\\
533 2.546859 -1.2233\\
539 2.528885 -1.3946\\
545 2.519155 -1.4924\\
552 2.512329 -1.5657\\
558 2.531073 -1.3986\\
565 2.537335 -1.3486\\
571 2.537674 -1.3525\\
578 2.515240 -1.5743\\
584 2.508750 -1.6449\\
591 2.512196 -1.6213\\
598 2.529565 -1.4624\\
605 2.519811 -1.5660\\
612 2.521649 -1.5569\\
619 2.551131 -1.2766\\
626 2.550055 -1.2944\\
633 2.532990 -1.4705\\
641 2.535696 -1.4528\\
648 2.533434 -1.4834\\
656 2.516342 -1.6659\\
663 2.515704 -1.6813\\
671 2.506001 -1.7915\\
679 2.513006 -1.7294\\
687 2.510108 -1.7698\\
694 2.500973 -1.8749\\
702 2.489689 -2.0055\\
711 2.482203 -2.0986\\
719 2.467480 -2.2699\\
727 2.472691 -2.2256\\
736 2.487404 -2.0784\\
744 2.500824 -1.9429\\
753 2.500757 -1.9553\\
761 2.498881 -1.9864\\
770 2.489710 -2.1001\\
779 2.474970 -2.2781\\
788 2.471865 -2.3264\\
797 2.468470 -2.3785\\
806 2.473970 -2.3287\\
816 2.480104 -2.2724\\
825 2.480146 -2.2844\\
835 2.465911 -2.4645\\
844 2.469315 -2.4377\\
854 2.460400 -2.5578\\
864 2.465798 -2.5083\\
874 2.484625 -2.2979\\
884 2.480434 -2.3612\\
894 2.476315 -2.4242\\
905 2.479184 -2.4042\\
915 2.478177 -2.4298\\
926 2.472270 -2.5170\\
936 2.467982 -2.5836\\
947 2.454419 -2.7683\\
958 2.444609 -2.9083\\
969 2.447400 -2.8895\\
981 2.470228 -2.6165\\
992 2.469755 -2.6372\\
1003 2.466747 -2.6903\\
1015 2.471976 -2.6389\\
1027 2.488798 -2.4371\\
1039 2.471919 -2.6707\\
1051 2.467923 -2.7385\\
1063 2.460432 -2.8532\\
1075 2.470317 -2.7378\\
1087 2.490099 -2.4901\\
1100 2.501357 -2.3553\\
1113 2.482288 -2.6245\\
1126 2.483607 -2.6220\\
1139 2.481201 -2.6698\\
1152 2.469222 -2.8492\\
1165 2.470105 -2.8530\\
1179 2.473531 -2.8225\\
1192 2.458437 -3.0494\\
1206 2.453287 -3.1401\\
1220 2.451986 -3.1768\\
1234 2.437525 -3.4027\\
1248 2.435229 -3.4553\\
1263 2.431161 -3.5354\\
1277 2.436442 -3.4774\\
1292 2.439333 -3.4551\\
1307 2.439983 -3.4655\\
1322 2.453109 -3.2903\\
1338 2.459730 -3.2115\\
1353 2.477644 -2.9625\\
1369 2.485177 -2.8677\\
1384 2.487754 -2.8448\\
1400 2.487359 -2.8671\\
1417 2.486794 -2.8930\\
1433 2.493538 -2.8068\\
1450 2.498867 -2.7421\\
1466 2.494233 -2.8283\\
1483 2.496120 -2.8155\\
1501 2.488763 -2.9469\\
1518 2.504433 -2.7190\\
1535 2.485824 -3.0264\\
1553 2.488122 -3.0077\\
1571 2.488787 -3.0145\\
1589 2.493577 -2.9550\\
1608 2.496756 -2.9216\\
1626 2.500967 -2.8699\\
1645 2.499657 -2.9079\\
1664 2.515487 -2.6671\\
1683 2.526199 -2.5079\\
1703 2.526938 -2.5107\\
1723 2.523780 -2.5774\\
1742 2.529216 -2.5017\\
1763 2.531161 -2.4845\\
1783 2.540521 -2.3427\\
1804 2.545183 -2.2786\\
1824 2.554345 -2.1377\\
1846 2.557737 -2.0935\\
1867 2.551545 -2.2101\\
1889 2.549506 -2.2579\\
1910 2.545163 -2.3449\\
1932 2.548992 -2.2923\\
1955 2.558635 -2.1389\\
1977 2.566305 -2.0178\\
2000 2.580927 -1.7755\\
2023 2.580993 -1.7845\\
2047 2.586093 -1.7057\\
2071 2.581040 -1.8047\\
2094 2.591281 -1.6335\\
2119 2.594871 -1.5795\\
2143 2.592451 -1.6316\\
2168 2.597940 -1.5426\\
2193 2.614296 -1.2576\\
2218 2.619058 -1.1790\\
2244 2.621721 -1.1378\\
2270 2.618795 -1.1975\\
2296 2.626026 -1.0722\\
2323 2.622127 -1.1502\\
2350 2.623918 -1.1237\\
2377 2.630170 -1.0141\\
2404 2.625213 -1.1124\\
2432 2.621474 -1.1891\\
2460 2.633351 -0.9717\\
2489 2.641561 -0.8221\\
2518 2.646811 -0.7273\\
2547 2.637512 -0.9091\\
2576 2.639077 -0.8841\\
2606 2.631274 -1.0404\\
2636 2.628430 -1.1019\\
2667 2.626882 -1.1388\\
2697 2.629676 -1.0900\\
2729 2.625526 -1.1789\\
2760 2.623599 -1.2242\\
2792 2.625129 -1.2005\\
2824 2.619095 -1.3296\\
2857 2.620986 -1.2988\\
2890 2.616336 -1.4017\\
2923 2.616125 -1.4141\\
2957 2.614404 -1.4580\\
2991 2.626630 -1.2112\\
3026 2.619302 -1.3720\\
3061 2.615123 -1.4682\\
3096 2.611417 -1.5555\\
3132 2.614230 -1.5043\\
3169 2.610948 -1.5839\\
3205 2.603130 -1.7626\\
3242 2.591926 -2.0184\\
3280 2.586949 -2.1402\\
3318 2.593680 -2.0029\\
3356 2.596662 -1.9478\\
3395 2.600432 -1.8745\\
3434 2.603893 -1.8073\\
3474 2.615396 -1.5580\\
3514 2.627053 -1.3033\\
3555 2.621185 -1.4442\\
3596 2.622623 -1.4197\\
3637 2.628287 -1.2976\\
3679 2.630355 -1.2574\\
3722 2.637069 -1.1092\\
3765 2.641468 -1.0133\\
3809 2.645678 -0.9209\\
3853 2.646586 -0.9049\\
3897 2.652757 -0.7647\\
3942 2.651730 -0.7934\\
3988 2.646806 -0.9154\\
4034 2.654375 -0.7393\\
4081 2.651297 -0.8177\\
4128 2.653172 -0.7770\\
4176 2.648209 -0.9025\\
4224 2.654124 -0.7626\\
4273 2.659278 -0.6402\\
4322 2.662784 -0.5573\\
4372 2.666343 -0.4722\\
4423 2.664788 -0.5137\\
4474 2.658009 -0.6871\\
4526 2.654421 -0.7819\\
4578 2.654569 -0.7826\\
4631 2.658663 -0.6823\\
4684 2.654375 -0.7966\\
4739 2.652399 -0.8525\\
4793 2.651715 -0.8753\\
4849 2.659002 -0.6893\\
4905 2.657793 -0.7251\\
4962 2.658288 -0.7162\\
5019 2.660634 -0.6578\\
5077 2.663134 -0.5946\\
5136 2.669790 -0.4192\\
5195 2.665903 -0.5266\\
5255 2.666865 -0.5035\\
5316 2.672986 -0.3392\\
5378 2.673084 -0.3385\\
5440 2.670688 -0.4066\\
5503 2.669728 -0.4356\\
5566 2.671234 -0.3960\\
5631 2.675687 -0.2734\\
5696 2.671579 -0.3909\\
5762 2.680315 -0.1453\\
5828 2.682615 -0.0807\\
5896 2.682734 -0.0778\\
5964 2.683446 -0.0577\\
6033 2.679762 -0.1648\\
6103 2.680056 -0.1571\\
6173 2.676590 -0.2597\\
6245 2.672514 -0.3816\\
6317 2.667667 -0.5281\\
6390 2.666000 -0.5811\\
6464 2.666463 -0.5705\\
6539 2.665746 -0.5956\\
6614 2.663423 -0.6699\\
6691 2.663935 -0.6580\\
6768 2.666007 -0.5979\\
6846 2.663947 -0.6653\\
6926 2.668215 -0.5359\\
7006 2.665652 -0.6194\\
7087 2.661514 -0.7538\\
7169 2.658374 -0.8581\\
7252 2.665746 -0.6272\\
7335 2.663647 -0.6982\\
7420 2.668561 -0.5435\\
7506 2.669268 -0.5237\\
7593 2.675914 -0.3100\\
7681 2.671985 -0.4406\\
7770 2.675592 -0.3242\\
7859 2.677776 -0.2538\\
7950 2.677585 -0.2616\\
8042 2.674710 -0.3594\\
8135 2.674680 -0.3625\\
8229 2.675903 -0.3231\\
8325 2.682694 -0.0938\\
8421 2.689656 0.1435\\
8518 2.688944 0.1200\\
8617 2.682937 -0.0870\\
8716 2.684537 -0.0318\\
8817 2.686832 0.0483\\
8919 2.681510 -0.1387\\
9022 2.684745 -0.0250\\
9127 2.684417 -0.0368\\
9232 2.680505 -0.1772\\
9339 2.687780 0.0837\\
9447 2.690919 0.1977\\
9556 2.691056 0.2038\\
9667 2.694684 0.3374\\
9779 2.692509 0.2595\\
9892 2.685592 0.0052\\
10006 2.689218 0.1402\\
10122 2.689721 0.1598\\
10239 2.683669 -0.0672\\
10357 2.686000 0.0208\\
10477 2.683169 -0.0871\\
10598 2.684670 -0.0300\\
10721 2.687850 0.0924\\
10845 2.683844 -0.0624\\
10970 2.682794 -0.1037\\
11097 2.681194 -0.1672\\
11226 2.680821 -0.1829\\
11355 2.683460 -0.0791\\
11487 2.684648 -0.0321\\
11620 2.681330 -0.1656\\
11754 2.681271 -0.1689\\
11890 2.677199 -0.3356\\
12027 2.669344 -0.6598\\
12167 2.664705 -0.8555\\
12307 2.664808 -0.8561\\
12450 2.667439 -0.7510\\
12594 2.670047 -0.6456\\
12739 2.669661 -0.6656\\
12887 2.675939 -0.4029\\
13036 2.676344 -0.3879\\
13186 2.674585 -0.4656\\
13339 2.673205 -0.5279\\
13493 2.675360 -0.4374\\
13649 2.676775 -0.3781\\
13807 2.674742 -0.4696\\
13967 2.680462 -0.2198\\
14128 2.681717 -0.1654\\
14292 2.681302 -0.1849\\
14457 2.682337 -0.1396\\
14624 2.682972 -0.1117\\
14793 2.682285 -0.1435\\
14964 2.678556 -0.3145\\
15137 2.679778 -0.2602\\
15313 2.677326 -0.3750\\
15490 2.678464 -0.3243\\
15669 2.679571 -0.2744\\
15850 2.677011 -0.3963\\
16033 2.681453 -0.1887\\
16219 2.678824 -0.3147\\
16406 2.674220 -0.5369\\
16596 2.673047 -0.5965\\
16788 2.674167 -0.5456\\
16982 2.673716 -0.5708\\
17179 2.673656 -0.5770\\
17377 2.676669 -0.4318\\
17578 2.678608 -0.3383\\
17782 2.680233 -0.2594\\
17987 2.680985 -0.2233\\
18195 2.679092 -0.3199\\
18406 2.678326 -0.3605\\
18619 2.676598 -0.4506\\
18834 2.675581 -0.5054\\
19052 2.677991 -0.3840\\
19272 2.678267 -0.3719\\
19495 2.677324 -0.4232\\
19721 2.679150 -0.3300\\
19949 2.681962 -0.1837\\
20179 2.682199 -0.1722\\
20413 2.683515 -0.1031\\
20649 2.682410 -0.1629\\
20888 2.684529 -0.0497\\
21129 2.685128 -0.0175\\
21374 2.689532 0.2220\\
21621 2.688322 0.1570\\
21871 2.684789 -0.0365\\
22124 2.684828 -0.0346\\
22380 2.683245 -0.1230\\
22638 2.683470 -0.1111\\
22900 2.682776 -0.1509\\
23165 2.682618 -0.1607\\
23433 2.683951 -0.0856\\
23704 2.681904 -0.2035\\
23978 2.684427 -0.0591\\
24256 2.684054 -0.0811\\
24536 2.679734 -0.3339\\
24820 2.679170 -0.3690\\
25107 2.680809 -0.2742\\
25397 2.681887 -0.2117\\
25691 2.678779 -0.3988\\
25988 2.677873 -0.4556\\
26289 2.679798 -0.3417\\
26593 2.679150 -0.3832\\
26900 2.680424 -0.3073\\
27212 2.681880 -0.2196\\
27526 2.678089 -0.4555\\
27845 2.680105 -0.3326\\
28167 2.679378 -0.3800\\
28492 2.680008 -0.3425\\
28822 2.680860 -0.2905\\
29155 2.679288 -0.3924\\
29492 2.677638 -0.5004\\
29834 2.677952 -0.4831\\
30179 2.677800 -0.4957\\
30528 2.676221 -0.6016\\
30881 2.678642 -0.4462\\
31238 2.680646 -0.3166\\
31599 2.680915 -0.3006\\
31965 2.680324 -0.3417\\
32334 2.679060 -0.4285\\
32708 2.676105 -0.6306\\
33087 2.675961 -0.6440\\
33469 2.677075 -0.5716\\
33856 2.677831 -0.5229\\
34248 2.678672 -0.4678\\
34644 2.676942 -0.5908\\
35045 2.677436 -0.5597\\
35450 2.673988 -0.8055\\
35860 2.672567 -0.9108\\
36275 2.673311 -0.8630\\
36694 2.675998 -0.6755\\
37119 2.676373 -0.6525\\
37548 2.673600 -0.8571\\
37982 2.671376 -1.0242\\
38422 2.672709 -0.9323\\
38866 2.672403 -0.9603\\
39315 2.671024 -1.0681\\
39770 2.669712 -1.1723\\
40230 2.668870 -1.2423\\
40695 2.669982 -1.1655\\
41166 2.671621 -1.0477\\
41642 2.670721 -1.1224\\
42124 2.670327 -1.1592\\
42611 2.670399 -1.1603\\
43104 2.669078 -1.2698\\
43602 2.668792 -1.2994\\
44107 2.669493 -1.2518\\
44617 2.667380 -1.4263\\
45133 2.668146 -1.3735\\
45655 2.668093 -1.3857\\
46183 2.667033 -1.4790\\
46717 2.665205 -1.6358\\
47257 2.665174 -1.6477\\
47804 2.663788 -1.7710\\
48357 2.661023 -2.0095\\
48916 2.661487 -1.9826\\
49482 2.662872 -1.8783\\
50054 2.663643 -1.8244\\
50633 2.665803 -1.6525\\
51219 2.665576 -1.6813\\
51811 2.665004 -1.7398\\
52410 2.664648 -1.7804\\
53016 2.667050 -1.5832\\
53630 2.665743 -1.7059\\
54250 2.667165 -1.5915\\
54877 2.666270 -1.6793\\
55512 2.668251 -1.5140\\
56154 2.668000 -1.5450\\
56803 2.668870 -1.4762\\
57460 2.668082 -1.5555\\
58125 2.670218 -1.3715\\
58797 2.667232 -1.6508\\
59477 2.668652 -1.5305\\
60165 2.669382 -1.4722\\
60861 2.670157 -1.4091\\
61565 2.670599 -1.3762\\
62277 2.671418 -1.3075\\
62997 2.674040 -1.0689\\
63726 2.675842 -0.9049\\
64463 2.676561 -0.8420\\
65208 2.676916 -0.8130\\
65963 2.675098 -0.9921\\
66726 2.676138 -0.8975\\
67497 2.676159 -0.9006\\
68278 2.675704 -0.9503\\
69068 2.675833 -0.9430\\
69866 2.677002 -0.8331\\
70674 2.677792 -0.7594\\
71492 2.679234 -0.6199\\
72319 2.680266 -0.5198\\
73155 2.683458 -0.2009\\
74001 2.681891 -0.3610\\
74857 2.682505 -0.3004\\
75723 2.684898 -0.0567\\
76599 2.684755 -0.0719\\
77485 2.684043 -0.1461\\
78381 2.681550 -0.4070\\
79287 2.682895 -0.2682\\
80204 2.682803 -0.2795\\
81132 2.682347 -0.3295\\
82070 2.683332 -0.2262\\
83019 2.683851 -0.1718\\
83980 2.685224 -0.0246\\
84951 2.685678 0.0245\\
85933 2.687225 0.1935\\
86927 2.686029 0.0633\\
87933 2.688973 0.3886\\
88950 2.687947 0.2769\\
89978 2.685732 0.0313\\
91019 2.686155 0.0790\\
92072 2.685233 -0.0247\\
93137 2.686392 0.1068\\
94214 2.686721 0.1451\\
95303 2.688264 0.3231\\
96406 2.689490 0.4665\\
97521 2.690451 0.5808\\
98649 2.690261 0.5620\\
99790 2.689816 0.5129\\
100944 2.689660 0.4974\\
102111 2.689840 0.5217\\
103292 2.689882 0.5297\\
104487 2.691358 0.7101\\
105695 2.691683 0.7534\\
106918 2.690883 0.6606\\
108154 2.692393 0.8490\\
109405 2.690699 0.6456\\
110671 2.691163 0.7067\\
111950 2.690395 0.6153\\
113245 2.691933 0.8112\\
114555 2.691422 0.7516\\
115880 2.691510 0.7671\\
117220 2.691997 0.8335\\
118576 2.691381 0.7595\\
119947 2.691850 0.8242\\
121335 2.691996 0.8478\\
122738 2.691971 0.8495\\
124157 2.691538 0.7976\\
125593 2.691849 0.8432\\
127046 2.690615 0.6846\\
128515 2.691715 0.8351\\
130002 2.691530 0.8151\\
131505 2.690807 0.7224\\
133026 2.691103 0.7667\\
134565 2.690587 0.7008\\
136121 2.690716 0.7225\\
137695 2.690905 0.7527\\
139288 2.690182 0.6568\\
140899 2.689437 0.5566\\
142529 2.690997 0.7787\\
144177 2.690624 0.7306\\
145845 2.691938 0.9213\\
147531 2.692775 1.0459\\
149238 2.691674 0.8941\\
150964 2.691145 0.8229\\
152710 2.690902 0.7923\\
154476 2.690342 0.7150\\
156262 2.689893 0.6532\\
158070 2.689581 0.6108\\
159898 2.689067 0.5379\\
161747 2.687799 0.3513\\
163618 2.686275 0.1240\\
165510 2.686315 0.1307\\
167425 2.684604 -0.1292\\
169361 2.685794 0.0524\\
171320 2.684839 -0.0946\\
173301 2.684648 -0.1247\\
175305 2.684420 -0.1610\\
177333 2.684500 -0.1494\\
179384 2.685248 -0.0321\\
181459 2.685188 -0.0418\\
183557 2.686071 0.0987\\
185680 2.686584 0.1815\\
187828 2.685023 -0.0693\\
190000 2.684108 -0.2182\\
192198 2.683777 -0.2736\\
194421 2.683496 -0.3213\\
196669 2.683898 -0.2567\\
198944 2.684137 -0.2185\\
201245 2.684019 -0.2395\\
203572 2.684192 -0.2118\\
205927 2.682674 -0.4696\\
208308 2.683344 -0.3583\\
210718 2.682272 -0.5440\\
213155 2.682419 -0.5217\\
215620 2.683630 -0.3151\\
218114 2.682872 -0.4489\\
220636 2.682323 -0.5477\\
223188 2.682589 -0.5039\\
225770 2.684207 -0.2203\\
228381 2.685016 -0.0776\\
231022 2.684401 -0.1882\\
233694 2.684994 -0.0824\\
236397 2.685205 -0.0448\\
239131 2.684424 -0.1872\\
241897 2.684734 -0.1314\\
244694 2.685199 -0.0466\\
247525 2.685498 0.0085\\
250387 2.685053 -0.0743\\
253283 2.685031 -0.0789\\
256213 2.685489 0.0069\\
259176 2.686152 0.1326\\
262173 2.686115 0.1264\\
265206 2.686173 0.1383\\
268273 2.685628 0.0339\\
271376 2.684690 -0.1479\\
274514 2.684669 -0.1528\\
277689 2.685085 -0.0720\\
280901 2.685791 0.0668\\
284150 2.685879 0.0848\\
287436 2.685977 0.1048\\
290761 2.686329 0.1760\\
294123 2.686985 0.3096\\
297525 2.687231 0.3612\\
300966 2.686669 0.2486\\
304447 2.686102 0.1336\\
307968 2.686803 0.2792\\
311530 2.686521 0.2222\\
315133 2.687624 0.4538\\
318778 2.687576 0.4463\\
322465 2.688204 0.5817\\
326194 2.687532 0.4422\\
329967 2.687420 0.4207\\
333783 2.686889 0.3091\\
337644 2.686728 0.2760\\
341549 2.686307 0.1860\\
345499 2.685463 0.0024\\
349495 2.686276 0.1814\\
353537 2.685214 -0.0528\\
357626 2.684292 -0.2583\\
361762 2.684348 -0.2472\\
365946 2.684634 -0.1842\\
370179 2.685382 -0.0158\\
374460 2.686144 0.1576\\
378791 2.685735 0.0648\\
383172 2.684878 -0.1322\\
387604 2.685284 -0.0390\\
392087 2.685025 -0.0995\\
396621 2.684808 -0.1510\\
401208 2.684397 -0.2489\\
405849 2.683985 -0.3481\\
410543 2.684074 -0.3288\\
415291 2.684263 -0.2854\\
420094 2.684529 -0.2229\\
424953 2.684073 -0.3349\\
429868 2.683834 -0.3951\\
434839 2.682980 -0.6073\\
439868 2.682825 -0.6490\\
444956 2.683397 -0.5107\\
450102 2.683407 -0.5112\\
455308 2.682970 -0.6240\\
460574 2.684033 -0.3587\\
465901 2.684587 -0.2199\\
471289 2.684614 -0.2141\\
476740 2.684989 -0.1190\\
482254 2.684787 -0.1721\\
487831 2.684060 -0.3622\\
493473 2.684575 -0.2295\\
499181 2.684035 -0.3728\\
504954 2.684122 -0.3519\\
510794 2.684941 -0.1359\\
516702 2.685129 -0.0865\\
522678 2.685021 -0.1160\\
528723 2.685278 -0.0472\\
534838 2.685497 0.0122\\
541024 2.685629 0.0486\\
547281 2.686314 0.2375\\
553611 2.686496 0.2892\\
560014 2.686168 0.1995\\
566491 2.685764 0.0875\\
573043 2.685832 0.1071\\
579670 2.686036 0.1655\\
586374 2.686111 0.1879\\
593156 2.686346 0.2563\\
600017 2.686427 0.2811\\
606956 2.686970 0.4402\\
613976 2.686601 0.3353\\
621077 2.687533 0.6106\\
628260 2.687334 0.5553\\
635527 2.687886 0.7222\\
642877 2.688247 0.8341\\
650312 2.688023 0.7717\\
657833 2.688203 0.8303\\
665442 2.687697 0.6817\\
673138 2.687895 0.7460\\
680923 2.688631 0.9761\\
688799 2.688180 0.8427\\
696765 2.688045 0.8055\\
704824 2.688018 0.8017\\
712976 2.687786 0.7335\\
721222 2.687571 0.6698\\
729563 2.687828 0.7555\\
738001 2.687945 0.7970\\
746536 2.687196 0.5610\\
755171 2.687223 0.5729\\
763905 2.687023 0.5113\\
772740 2.687248 0.5875\\
781677 2.687409 0.6441\\
790718 2.686920 0.4861\\
799863 2.687418 0.6546\\
809114 2.686967 0.5074\\
818472 2.687245 0.6037\\
827938 2.687244 0.6071\\
837514 2.687380 0.6568\\
847200 2.686851 0.4792\\
856999 2.687057 0.5529\\
866910 2.687315 0.6459\\
876937 2.686890 0.5014\\
887079 2.687256 0.6324\\
897339 2.687360 0.6727\\
907717 2.688277 1.0019\\
918216 2.687561 0.7522\\
928835 2.688054 0.9335\\
939578 2.687455 0.7228\\
950445 2.687768 0.8406\\
961437 2.688385 1.0704\\
972557 2.687964 0.9220\\
983805 2.687580 0.7858\\
995184 2.687326 0.6958\\
1006694 2.687258 0.6746\\
1018337 2.687797 0.8806\\
1030115 2.687721 0.8572\\
1042029 2.687042 0.6043\\
1054080 2.687085 0.6242\\
1066272 2.687441 0.7644\\
1078604 2.687186 0.6704\\
1091078 2.686733 0.4981\\
1103698 2.686310 0.3354\\
1116463 2.686897 0.5683\\
1129375 2.686740 0.5094\\
1142437 2.686699 0.4962\\
1155650 2.686686 0.4937\\
1169016 2.686525 0.4319\\
1182537 2.686880 0.5782\\
1196214 2.687149 0.6908\\
1210049 2.686755 0.5337\\
1224044 2.686593 0.4698\\
1238201 2.686768 0.5451\\
1252521 2.686818 0.5691\\
1267007 2.686877 0.5971\\
1281661 2.686702 0.5269\\
1296485 2.686921 0.6225\\
1311479 2.686690 0.5279\\
1326647 2.686685 0.5289\\
1341991 2.686743 0.5566\\
1357512 2.686356 0.3919\\
1373213 2.686679 0.5353\\
1389095 2.686965 0.6637\\
1405161 2.686933 0.6535\\
1421412 2.686606 0.5120\\
1437852 2.686349 0.4006\\
1454482 2.686492 0.4668\\
1471304 2.686187 0.3319\\
1488320 2.686035 0.2647\\
1505534 2.685851 0.1821\\
1522946 2.686123 0.3083\\
1540560 2.686057 0.2795\\
1558378 2.686115 0.3079\\
1576402 2.686272 0.3832\\
1594634 2.686496 0.4909\\
1613077 2.686403 0.4497\\
1631733 2.686178 0.3451\\
1650605 2.685955 0.2406\\
1669696 2.686136 0.3290\\
1689007 2.686468 0.4914\\
1708542 2.686506 0.5129\\
1728302 2.686357 0.4431\\
1748291 2.686186 0.3614\\
1768511 2.685966 0.2545\\
1788965 2.686080 0.3130\\
1809656 2.686373 0.4615\\
1830586 2.686366 0.4604\\
1851758 2.686457 0.5091\\
1873175 2.686083 0.3217\\
1894839 2.685677 0.1154\\
1916754 2.685750 0.1536\\
1938923 2.685976 0.2716\\
1961348 2.686369 0.4780\\
1984032 2.686500 0.5496\\
2006979 2.686394 0.4967\\
2030191 2.686341 0.4714\\
2053672 2.686373 0.4914\\
2077424 2.686642 0.6385\\
2101451 2.686946 0.8060\\
2125755 2.687063 0.8744\\
2150341 2.687090 0.8944\\
2175211 2.687155 0.9348\\
2200369 2.687271 1.0046\\
2225818 2.686955 0.8350\\
2251561 2.686785 0.7446\\
2277602 2.687061 0.9041\\
2303944 2.687127 0.9466\\
2330591 2.687097 0.9347\\
2357545 2.687082 0.9318\\
2384812 2.686811 0.7813\\
2412394 2.686659 0.6982\\
2440295 2.686468 0.5909\\
2468519 2.686721 0.7420\\
2497069 2.686943 0.8773\\
2525949 2.687146 1.0021\\
2555164 2.687221 1.0525\\
2584716 2.687292 1.1011\\
2614610 2.687666 1.3324\\
2644850 2.687652 1.3320\\
2675439 2.687419 1.1974\\
2706382 2.687562 1.2921\\
2737684 2.687565 1.3016\\
2769347 2.687531 1.2880\\
2801376 2.687674 1.3843\\
2833776 2.687878 1.5204\\
2866551 2.687956 1.5779\\
2899704 2.688215 1.7511\\
2933241 2.688120 1.7010\\
2967166 2.687878 1.5556\\
3001484 2.688042 1.6704\\
3036198 2.688053 1.6870\\
3071314 2.688172 1.7741\\
3106836 2.688286 1.8589\\
3142768 2.688344 1.9080\\
3179117 2.688265 1.8669\\
3215885 2.688147 1.7988\\
3253079 2.688253 1.8804\\
3290703 2.688260 1.8957\\
3328763 2.688102 1.7992\\
3367262 2.687940 1.6991\\
3406207 2.688012 1.7583\\
3445602 2.688065 1.8053\\
3485453 2.687978 1.7553\\
3525764 2.687780 1.6272\\
3566542 2.687973 1.7721\\
3607792 2.688318 2.0258\\
3649518 2.688103 1.8847\\
3691727 2.687881 1.7372\\
3734425 2.687875 1.7430\\
3777616 2.688263 2.0331\\
3821307 2.687998 1.8525\\
3865503 2.688141 1.9674\\
3910210 2.688237 2.0495\\
3955434 2.688502 2.2579\\
4001182 2.688533 2.2935\\
4047458 2.688773 2.4862\\
4094270 2.689107 2.7521\\
4141623 2.689298 2.9127\\
4189524 2.689274 2.9110\\
4237978 2.689134 2.8205\\
4286993 2.689233 2.9129\\
4336575 2.689323 2.9999\\
4386731 2.689560 3.2019\\
4437466 2.689266 2.9898\\
4488789 2.688965 2.7698\\
4540705 2.688951 2.7745\\
4593221 2.688841 2.7030\\
4646345 2.688880 2.7496\\
4700083 2.688642 2.5740\\
4754443 2.688512 2.4829\\
4809431 2.688591 2.5618\\
4865056 2.688366 2.3918\\
4921323 2.688207 2.2751\\
4978242 2.688211 2.2915\\
5035819 2.688283 2.3641\\
5094061 2.688052 2.1842\\
5152978 2.687884 2.0549\\
5212576 2.687982 2.1502\\
5272863 2.687954 2.1386\\
5333847 2.687852 2.0634\\
5395536 2.687750 1.9872\\
5457940 2.687460 1.7461\\
5521064 2.687421 1.7219\\
5584919 2.687098 1.4483\\
5649513 2.686935 1.3118\\
5714853 2.686890 1.2797\\
5780950 2.686934 1.3264\\
5847810 2.686904 1.3072\\
5915444 2.686915 1.3249\\
5983860 2.686859 1.2816\\
6053068 2.686853 1.2829\\
6123076 2.686847 1.2849\\
6193893 2.686872 1.3156\\
6265530 2.686566 1.0378\\
6337995 2.686788 1.2518\\
6411299 2.686856 1.3233\\
6485450 2.687032 1.4979\\
6560458 2.686945 1.4236\\
6636335 2.687020 1.5038\\
6713088 2.687076 1.5661\\
6790730 2.687209 1.7044\\
6869269 2.687317 1.8191\\
6948717 2.687364 1.8765\\
7029084 2.687421 1.9437\\
7110380 2.687366 1.9001\\
7192617 2.687402 1.9469\\
7275804 2.687234 1.7890\\
7359954 2.687076 1.6398\\
7445077 2.687143 1.7175\\
7531184 2.687114 1.6984\\
7618288 2.687117 1.7110\\
7706398 2.687111 1.7143\\
7795528 2.687078 1.6896\\
7885689 2.687120 1.7440\\
7976892 2.686957 1.5823\\
8069151 2.686895 1.5255\\
8162476 2.686596 1.2168\\
8256881 2.686539 1.1632\\
8352377 2.686536 1.1659\\
8448978 2.686455 1.0855\\
8546696 2.686610 1.2601\\
8645545 2.686564 1.2171\\
8745536 2.686670 1.3412\\
8846685 2.686531 1.1950\\
8949003 2.686688 1.3762\\
9052504 2.686492 1.1653\\
9157202 2.686507 1.1884\\
9263112 2.686710 1.4256\\
9370246 2.686652 1.3678\\
9478620 2.686697 1.4265\\
9588246 2.686720 1.4620\\
9699141 2.686720 1.4701\\
}\khcfdata
\begin{document}

\begin{abstract}
For a holonomic function $F$ that is regular at infinity we study the zeta function
$\zeta_F(s) = \sum_{k=1}^{\infty} F(k)k^{-s}$.
We show that when the expansion of $F$ at infinity has algebraic coefficients, the
values and derivatives of $\zeta_F$ at algebraic $s$ can be computed
to $p$ bits in $p^{3/2+o(1)}$ bit operations
using a version of Riemann's theta splitting
combined with bit-burst evaluation of holonomic functions.
The same holds for polylogarithm-type generalizations of $\zeta_F$.
This gives subquadratic algorithms for a large class of numbers,
including every sum $\sum_{k=1}^{\infty} A(k)$ or $\sum_{k=1}^{\infty} \log A(k)$ where $A$ is an
algebraic function with algebraic coefficients, most of which had no previously known algorithm with better than
quadratic complexity.
For example, we obtain the first subquadratic algorithm for
Khinchin's constant, which we compute to $10^7$ digits.
We demonstrate the method and its speedup on many other examples taken from analysis,
geometry, number theory and physics.
\end{abstract}


\maketitle

\section{Introduction}

\subsection{Two constants}

Two strikingly similar series
\begin{equation}
\label{eq:twoconstants}
\begin{aligned}
  \gamma &= \sum_{k=1}^{\infty}\Bigl(\frac1k - \log\Bigl(1+\frac1k\Bigr)\Bigr), \\
  \log K\,\log 2 &= -\sum_{k=2}^{\infty}\log\Bigl(1-\frac1k\Bigr)\log\Bigl(1+\frac1k\Bigr)
\end{aligned}
\end{equation}
express two seemingly unrelated numbers: Euler's constant, the regularized limit of the harmonic numbers
\[
\gamma = \lim_{n \to \infty} ( ( \tfrac{1}{1} + \tfrac{1}{2} + \cdots + \tfrac{1}{n} ) - \log n) = 0.5772156649\ldots,
\]
and Khinchin's constant, the limit
\[
  K = \lim_{n\to\infty}(a_1a_2\cdots a_n)^{1/n} = 2.6854520010\ldots
\]
of the geometric means of the partial quotients of the continued fraction $x = [a_0; a_1, a_2, \dots]$ of almost
every real number $x$ \cite{Khi35,Khi64}.

Euler's constant is known to be rapidly computable: along with many other classical constants
like $e$, $\pi$, $\log 2$, $\zeta(3)$, Catalan's constant and $\Gamma(\frac14)$,
it can be expressed finitely in terms of holonomic functions (solutions
of linear ordinary differential equations with polynomial
coefficients)
which can be evaluated in time quasilinear in the precision~\cite{BB87,HP98,BM80,BZ10,CC90,vdH99}.
Khinchin's constant is not known to have this property,
and all known algorithms for $K$ until the present take at least quadratic time.
In the words of Weisstein \cite{MWKdigits}, ``the numerical value of $K$ is
notoriously difficult to calculate to high precision''.

In this work, we show that both sums \eqref{eq:twoconstants} and many others
belong to a common class which can be computed in subquadratic time.
We demonstrate with several examples that the method is practical
and readily scales to millions of digits.

\subsection{Zeta functions of holonomic functions}

The starting point is the familiar device of viewing any infinite sum
as a Dirichlet series.
In general, let $\Reg$ be the class of sequences $F$ that for large indices
agree with a holonomic function regular at infinity (Definition~\ref{def:regular}), and let $\RegQ \subset \Reg$ be
the subclass in which the differential equation and the expansion of $F$ at infinity have algebraic coefficients.
This is a condition on the data describing $F$, not on $F$ itself, which need not be an algebraic function:
$\log(1+\frac1k)$ is in $\RegQ$, but $\log(1+\frac{\pi}{k})$ is not. To $F \in \Reg$ we attach the following
functions:
\begin{equation}\label{eq:functions}
\begin{aligned}
  \zeta_F(s) &= \sum_{k=1}^{\infty}F(k)\,k^{-s} &&\text{(zeta function)}, \\
  L_F(s,\chi) &= \sum_{k=1}^{\infty}\chi(k)F(k)\,k^{-s} &&\text{(twisted zeta function)}, \\
  \Li_F(s,z) &= \sum_{k=1}^{\infty}F(k)\,z^kk^{-s} &&\text{(polylogarithm)}, \\
  \Phi_F(z,s,a) &= \sum_{k=0}^{\infty}F(k)\,z^k(k+a)^{-s} &&\text{(Lerch transcendent)}, \\
  \Lambda_F(s;\tau,r) &= \sum_{k=1}^{\infty}e^{-\tau k^r}F(k)\,k^{-s} &&\text{(damped zeta function)},
\end{aligned}
\end{equation}
with $\chi$ periodic, $|z| < 1$, $\tau > 0$ and $0 < r < 2$ in the last.
For $F = 1$ these are the classical
Riemann zeta function $\zeta(s)$, Dirichlet $L$-functions (when $\chi$ is a Dirichlet character), polylogarithm $\Li_s(z)$, Lerch
transcendent $\Phi(z,s,a)$, and sums $\sum_ke^{-\tau k^r}k^{-s}$ of stretched exponentials, with
$\Lambda(s;\tau,1) = \Li_s(e^{-\tau})$, where $\Lambda = \Lambda_1$.
Rational $F$ give the Dirichlet series attached to rational functions
studied since Mellin and Mahler \cite{Mah28,CN81,Sar84,Pet98,Ess97}.

Sums of holonomic functions are values and derivatives of these functions (Section~\ref{sec:sums}). We compute them by an extension of Riemann's theta splitting of $\zeta(s)$, the formula behind his second proof
of the functional equation.

\subsection{Main results}
Let $F$ be analytic at infinity, $F(k) = f(1/k)$ for $k \ge k_0$, where
$f(x) = \sum_{n=0}^{\infty} a_nx^n$ has radius of convergence $\rho > 1/k_0$. The following formula needs only this;
holonomy of $f$ makes its pieces computable (Theorem~B).

\begin{theoremA}[exact approximate functional equation; Theorem~\ref{thm:split}]
For every $u_0 > 0$ and every $s \in \CC$ except the poles,
\[
  \zeta_F(s) = \sum_{k=1}^{k_0-1}F(k)k^{-s} + \sum_{k=k_0}^{\infty}I_k(s)
  + \sqrt\pi\sum_{m=1}^{\infty}J_m(s) + M(s),
\]
where the \emph{$k$-terms} and the \emph{dual terms}
\[
  I_k(s) = \int_{u_0}^\infty e^{-k^2u}h_s(u)\dd u, \qquad
  J_m(s) = \int_0^{u_0}u^{-1/2}e^{-\pi^2m^2/u}h_s(u)\dd u
\]
are entire in $s$ (and, when moreover $F \in \Reg$, values at $\kappa = k^2$ and $c = \pi^2m^2$ of holonomic functions of
$\kappa$ and $c$), the \emph{kernel}
\[
h_s(u) = \sum_{n=0}^{\infty} a_n \frac{u^{(s+n)/2-1}}{\Gamma(\frac{s+n}{2})}
\]
is a Borel-type transform of $f$, and the \emph{main term} $M$ is an
explicit sum of elementary functions and incomplete gamma functions carrying all the poles.
\end{theoremA}

For $F = 1$, $k_0 = 1$ and $u_0 = \pi$ this is Riemann's formula, and the two infinite sums
are symmetric in $s$ and $1-s$, which gives the
functional equation. The two sums differ in general
but both are rapidly convergent: the $k$-terms decay
like $e^{-k^2u_0}$ and the dual terms like $e^{-\pi^2m^2/u_0}$.
Formally the dual side is the Poisson dual
$\sum_{m\in\ZZ} \widehat G_s(m)$ of $G_s(x) = |x|^{-s}F(|x|)$, and the \emph{split point} $u_0$ is a smoothing parameter
(Remark~\ref{rem:poisson}). The other functions have splittings of the same shape. For $\Li_F(s,z)$ and $\Phi_F$ the
factor $z^k = e^{2\pi i\alpha k}$, with complex $\alpha$, is treated like a periodic weight: the dual points become
$\pi^2(n-\alpha)^2$, complex numbers, and for $F = 1$ the formula is Crandall's \cite{Cra12}
(Theorem~\ref{thm:geometric}). For $\Lambda_F$ the factor $e^{-\tau k^r}$ moves into the kernel $h_s$, which
becomes an entire function of $u^{-r/2}$ (Propositions~\ref{prop:expsplit} and~\ref{prop:prodsplit}).

\begin{theoremB}[Theorems~\ref{thm:complexity} and~\ref{thm:geometric}, Section~\ref{sec:dampedcost}]
Let $F \in \RegQ$, let $s_0$ be algebraic, and let
$j \ge 0$ be fixed. Then the $j$-th Taylor (or Laurent) coefficient at $s_0$ of each of
\begin{enumerate}
\item $\zeta_F(s)$, and $L_F(s,\chi)$ for $\chi$ periodic with algebraic values,
\item $\Li_F(s,z)$ and $\Phi_F(z,s,a)$, for algebraic $z$ with $0 < |z| < 1$ and algebraic $a > 0$,
\end{enumerate}
can be computed to $p$ bits in $O(p^{1/2}\Mul(p)\log^2 p)$ bit operations.
\end{theoremB}

For the damped zeta functions the splitting identities are proved (Propositions~\ref{prop:expsplit}
and~\ref{prop:prodsplit}), but we only sketch the cost analysis, so we state that bound separately.

\begin{claimC}[Section~\ref{sec:dampedcost}, Appendix~\ref{app:stretched}]
Let $\tau > 0$ and $s_0$ be algebraic and $j \ge 0$ fixed. When $F = 1$ and $r \in \QQ$, $0 < r \le 1$, or when
$F \in \RegQ$ is analytic at infinity and $r = 1/q$ with an integer $q \ge 2$, the $j$-th Taylor coefficient of
$\Lambda_F(s;\tau,r)$ at $s_0$ can be computed to $p$ bits in $p^{3/2+o(1)}$ bit operations.
\end{claimC}

Here $\Mul(p)$ is the cost of $p$-bit integer multiplication, so the bound is $p^{3/2+o(1)}$, against
$p^{2+o(1)}$ for the classical methods (Euler--Maclaurin summation, the expansion in zeta values, and, for the
functions with exponential decay, direct summation, which needs $p^{1+1/r+o(1)}$ for $\Lambda_F$). We write
$\widetilde O(X)$ for $X(\log X)^{O(1)}$; all bounds of the form $p^{c+o(1)}$ in this paper are in fact
$\widetilde O(p^c)$. We use the exponent form when comparing methods, and $\widetilde O$ for quasilinear costs and for
costs measured against the size of the output, such as $\widetilde O(p^2)$ for $\Theta(p)$ numbers of $p$ bits.

The expansion $\zeta_F(s) = \sum_{k<k_0}F(k)k^{-s} + \sum_{n=0}^{\infty}a_n\zeta(s+n,k_0)$ in Hurwitz zeta values
\cite{FV96}, the \emph{zeta-value expansion}, gives the meromorphic continuation, with simple poles at $s = 1-n$ of residue $a_n$ and the
zeta-regularized values of divergent sums $\sum_kk^mF(k)$ at $s = -m$ \cite{Eli95,Kir02}
(Proposition~\ref{prop:cont}). Other analytic properties of $\zeta_F$, such as its growth and zeros, play no role in the
algorithms and are treated in a companion note \cite{Joh26}.

\subsection{Sums in the class}\label{sec:sums}

Euler's constant is the constant term of $\zeta(s)$ at its pole $s = 1$, and by \eqref{eq:twoconstants} also
$\zeta_{F_1}(0)$ with $F_1(k) = \frac1k - \log(1+\frac1k)$; we will similarly express Khinchin's constant, and many other
constants, as values of $\zeta_F$ for various $F$. The functions \eqref{eq:functions} cover familiar ways in which sums of holonomic functions arise:
\begin{itemize}
\item a sum $\sum_{k=1}^{\infty} F(k)$ is the value $\zeta_F(0)$;
\item an alternating sum $\sum_{k=1}^{\infty} (-1)^k F(k)$ is the value $L_F(0, \chi)$ where $\chi(k) = (-1)^k$;
\item a sum $\sum_{k=1}^{\infty}(\log k)^jF(k)$ is $(-1)^j\zeta_F^{(j)}(0)$;
\item a product $\prod_k \exp(F(k))$ is $\exp(\zeta_F(0))$;
\item a product $\prod_k k^{F(k)}$ is $\exp(-\zeta_F'(0))$;
\item a divergent sum $\sum_{k=1}^{\infty} k^m F(k)$ has the regularized value $\zeta_F(-m)$;
\item the Dirichlet generating function of $F$ is $\zeta_F(s)$, and its ordinary generating
function $\sum_kF(k)z^k$ is $\Li_F(0,z)$; weights $(\log k)^j$ and powers $k^{-s}$ in a generating function are
derivatives and values in $s$;
\item shifts of the power, $\sum_kF(k)(k+a)^{-s}$, are Lerch transcendents, and at fixed $s$ reduce to the other
functions (Section~\ref{sec:twistshift});
\item summands with stretched-exponential decay are values of damped zeta functions.
\end{itemize}

The class $\Reg$ of admissible sequences is large. It is closed under sums and products, and
contains every algebraic function $A(k)$,
the compositions $\log A(k)$ and $\arctan A(k)$,
and more generally $G(A(k))$ where $G$ is any holonomic function that is
analytic or regular singular at the limit of $A$ at infinity
such as the generalized hypergeometric function $G(z) = {}_pF_{p-1}(\ldots; z)$.
The underlying class of holonomic functions is
also closed under integration, and contains the periods of
families of algebraic curves depending algebraically on $k$ such as complete elliptic, hyperelliptic and abelian integrals.
Bose--Einstein integrals lead to such sums (Section~\ref{sec:physics}).

Table~\ref{tab:constants} lists constants from classical analysis, number theory, mathematical physics and probability expressed in terms of $\zeta_F$ and its relatives, including examples from the tables of Flajolet and Vardi~\cite{FV96} and of Finch~\cite{Fin03},
Sloane's OEIS~\cite{OEIS}, and other sources. By Theorem~B and Claim~C, all of them can be computed in
$p^{3/2+o(1)}$ bit operations at algebraic values of their parameters.

\begin{table}[p]
\centering\footnotesize
\renewcommand{\arraystretch}{1.03}
\setlength{\aboverulesep}{0.22ex}\setlength{\belowrulesep}{0.33ex}
\caption{Constants and families of constants as special values of the functions \eqref{eq:functions}.}
\label{tab:constants}
\begin{tabular*}{\textwidth}{@{\extracolsep{\fill}}lll@{}}
\toprule
constant & $F(k)$ & expression \\
\midrule
\multicolumn{3}{@{}p{\textwidth}@{}}{\emph{Limiting constants of partial sums and products} (Section~\ref{sec:stieltjes}), with
$F_1(k) = \frac1k - \log(1+\frac1k)$, $F_2(k) = \log(1+\frac1k)$, $F_3(k) = 3k + \frac12 - k(3k+2)\log(1+\frac1k)$ and
$F_4(k) = (k+\frac12)\log(1+\frac1k) - 1$.} \\
\midrule
Euler's $\gamma$ & $1$; or $F_1$ & $\CT_{s=1}\zeta(s)$; or $\zeta_{F_1}(0)$ \\
Stieltjes $\gamma_1$ & $F_1$; $F_2^2$ & $-\zeta_{F_1}'(0) - \frac12\zeta_{F_2^2}(0)$ \\
Stirling's $\sqrt{2\pi}$ & $1$; or $F_4$ & $e^{-\zeta'(0)}$; or $e^{1-\zeta_{F_4}(0)}$ \\
Wallis's $\frac\pi2 = \prod_k\frac{4k^2}{4k^2-1}$ & $-\log(1-\frac1{4k^2})$ & $e^{\zeta_F(0)}$ \\
Glaisher's $A$ & $1$; or $F_3$ & $e^{1/12-\zeta'(-1)}$; or $(2\pi e)^{\frac1{12}}e^{\zeta_{F_3}(0)/6}$ \\
\midrule
\multicolumn{3}{@{}p{\textwidth}@{}}{\emph{Continued fractions and L\"uroth expansions} (Sections~\ref{sec:khinchin}, \ref{sec:AG}, \ref{sec:KH}):
Gauss--Kuzmin probabilities $\mu_k$, Khinchin zeta function $Z$; Alladi--Grinstead and Kalpazidou constants
$e^{\mathcal C-1}$, $e^{\mathcal C}$.} \\
\midrule
Khinchin's $K$ & $\mu_k = \log_2(1+\frac{1}{k(k+2)})$ & $\exp(-Z'(0))$ \\
H\"older means $K_q$, $q<1$ & $\mu_k$ & $Z(-q)^{1/q}$ \\
$\mathcal A = \sum_{k=2}^{\infty}\frac{\log k}{k^2-1}$ & $\frac1{k^2-1}$ ($k\ge2$) & $-\zeta_F'(0)$ \\
$\mathcal C = \sum_{k=1}^{\infty}\frac{\log k}{k(k+1)}$ & $\frac{1}{k(k+1)}$ & $-\zeta_F'(0)$ \\
\midrule
\multicolumn{3}{@{}p{\textwidth}@{}}{\emph{Analysis and geometry} (Sections~\ref{sec:theodorus}, \ref{sec:HL}, \ref{sec:somos},
Remark~\ref{rem:puiseux}): the spiral of Theodorus, with Hlawka's constant
$\lim_{n\to\infty}(\sum_{k<n}\arctan k^{-1/2} - 2\sqrt n)$; an algebraic example; $t > 1$.} \\
\midrule
Theodorus $\sum_{k=1}^{\infty}\frac{1}{(k+1)\sqrt k}$ & $\frac1{k+1}$ & $\zeta_F(\frac12)$ \\
Hlawka's constant & $\arctan\frac1{\sqrt k}$ & $\zeta_F(0)$ (regularized) \\
Hardy--Littlewood $H(x)$ & $\sin(x/k)$ & $\zeta_F(1)$ \\
$\sum_{k=1}^{\infty}\frac{1}{k^{3/2}+1}$ & $\frac1{k^3-1}$ ($k\ge2$) & $\frac12+\zeta_F(-\frac32)-\zeta_F(0)$ \\
$\sum_{k=1}^{\infty}(k\arcsin\frac1k - 1)$ & $k\arcsin\frac1k - 1$ & $\zeta_F(0)$ (A323737) \\
$\sum_{k=2}^{\infty}\zeta''(k)$ \cite{Fin18} & $\frac1{k(k-1)}$ ($k\ge2$) & $\zeta_F''(0)$ \\
Boyadzhiev's constant \cite{Boy19} & $\frac1k\log(1+\frac1k)$ & $\zeta_F(0)$ \\
Somos's $\sigma_2$; $\sigma_t = \prod_kk^{t^{-k}}$ & $1$; or $F_2$ ($t = 2$) & $e^{-\partial_s\Li_s(\frac1t)|_{s=0}}$; or $e^{\Li_{F_2}(0,\frac12)}$ \\
Sondow--Hadjicostas $\gamma(z)$ & $F_1$ & $z^{-1}\Li_{F_1}(0,z)$ \\
\midrule
\multicolumn{3}{@{}p{\textwidth}@{}}{\emph{Fields and spectra} (Sections~\ref{sec:vortex}, \ref{sec:physics}):
coaxial rings of unit radius spaced $\lambda$ apart ($\theta/2\pi \in \QQ$); Laplacian $\Delta$ on the unit sphere $S^2$
\cite{Eli95,Kir02}; Kaluza--Klein tower of the top quark.} \\
\midrule
vortex rings $U(\lambda)$ & ${}_2F_1(\frac32,\frac32;2;-\frac{4}{\lambda^2k^2})$ & $\lambda^{-3}\zeta_F(3)$ \\
ring resonators $S_\lambda(\theta)$ & ${}_2F_1(\frac32,\frac32;3;-\frac{4}{\lambda^2k^2})$ & $\frac{\pi}{2\lambda^3}L_F(3,\cos k\theta)$ \\
Epstein--Hurwitz $E_a(t)$, $t > \frac12$ & $(1+\frac{a^2}{k^2})^{-t}$ & $\zeta_F(2t)$ \\
$\operatorname{Tr}\Delta^{-3/2}$ & $(2+\frac1k)(1+\frac1k)^{-3/2}$ & $\zeta_F(2)$ \\
Casimir energy $\frac12\operatorname{Tr}\Delta^{1/2}$ & $(2+\frac1k)(1+\frac1k)^{1/2}$ & $\frac12\zeta_F(-2) + \frac1{48}$ \\
KK top tower \cite[(27)]{Pet02} & $(\tau_k - 1)y_k^2 - 1$ & $\zeta_F(0)$, $y_k = \arcsin\tau_k^{-1/2}$, \\
 & & $\tau_k = 4(m_t^2 + k^2/R^2)/m_H^2$ \\
\midrule
\multicolumn{3}{@{}p{\textwidth}@{}}{\emph{Statistical mechanics and probability} (Sections~\ref{sec:beyond}, \ref{sec:physics},
\ref{sec:expsqrt}): hydrogen \cite{EKR11}, $\xi = I/(k_BT)$; $j \in \frac12 + \ZZ$ for Bose--Einstein and Fermi--Dirac;
Landau levels of graphene, $\tau = E_1/(k_BT)$; Fisher's log-series distribution; wrapped stable density $p_t$.} \\
\midrule
Planck--Larkin $\sigma(T)$ & $e^{\xi/k^2} - 1 - \xi/k^2$ & $2\zeta_F(-2)$ \\
Bose--Einstein $g_{j}(z)$ & $1$ & $\Li_j(z)$ \\
Fermi--Dirac $F_{j-1}(\eta)$, $\eta < 0$ & $1$ & $-\Li_j(-e^\eta)$ \\
graphene $\sum_{n=1}^{\infty}e^{-\tau\sqrt n}$ & $1$ & $\Lambda(0;\tau,\frac12)$ \\
log-series entropy $H(z)$ & $1$ & $\partial_s\Li_s(z)|_{s=1}$ \\
$2\pi p_t(0) - 1$, $0 < \alpha < 2$ & $1$ & $2\Lambda(0;t,\alpha)$ \\
\bottomrule
\end{tabular*}
\end{table}

\subsubsection{Special cases with faster algorithms}

Special cases of sums in this class were already known to be efficiently computable.
Among the constants of Table~\ref{tab:constants},
Euler's constant $\gamma$ and the two closed forms involving $\pi$
can be computed in quasilinear time.
More generally, so can every number that is the value at an algebraic point of a holonomic function
with algebraic coefficients and initial values \cite{CC90,vdH99}, such as the integer zeta values $\zeta(3), \zeta(5),
\dots$ \cite{Kar95,Kar98}, ordinary polylogarithms at algebraic points and Catalan's constant; this includes many periods and
exponential periods in the sense of Kontsevich and Zagier \cite{KZ01}.
The Weierstrass products of the gamma function give sums of the same kind as the series for $\gamma$,
\[
  \log\Gamma(1+x) = -\gamma x + \sum_{k=1}^{\infty}\Bigl(\frac xk - \log\Bigl(1+\frac xk\Bigr)\Bigr),
\]
\[
  \psi(1+x) = -\gamma + \sum_{k=1}^{\infty}\frac{x}{k(k+x)},
\]
and for algebraic $x$ these too can be computed in quasilinear time, from the incomplete gamma integral
(\cite{BZ10}; Section~\ref{sec:rational}).
For rational $A$, the sum $\sum_k A(k)$
reduces by partial fractions to polygamma values, and sums of logarithms or inverse trigonometric or hyperbolic
functions of a rational function of $k$ reduce to logarithms of gamma values at algebraic points, as in
$\sum_{k=1}^{\infty} (\frac1k - \arctan(\frac1k)) = \gamma + \arg\Gamma(1+i)$ and
$\sum_{k=2}^{\infty}\arctanh(1/k^2) = \tfrac{1}{2} \log(\sinh(\pi)/\pi)$
(OEIS A352619, A393668), all computable in quasilinear time. When the coefficients of the rational function are arbitrary complex numbers,
the same reductions still give $p^{3/2+o(1)}$ (Section~\ref{sec:rational}).

On the level of summands, the structure leading to such quasilinear-time algorithms is fragile:
an extra factor $\frac1k$, as in $\sum_k\frac1k\arctan\frac1k$, destroys it.

For constants of Table~\ref{tab:constants} other than the above exceptions, and for sums of
algebraic functions that are not rational, we know of no quasilinear
algorithm, and no earlier subquadratic algorithm
except the restriction of this paper's method to $F = 1$
which is classical (see comments in Section~\ref{sec:related}).

\subsection{Further results}

We compare the theta splitting with the classical methods in the setting of $\zeta_F$ (Section~\ref{sec:complexity}): the
zeta-value expansion is Theorem~A with the order of two summations exchanged, and Euler--Maclaurin summation is Poisson
summation with a sharp instead of a smoothed cutoff. At large height and moderate precision, $\zeta_F(\sigma + i t)$ can be computed
in $O(t^{1/2})$ operations by combining the zeta-value expansion with the Riemann--Siegel formula.

The algorithms of Theorem~B(1) need only $O(p)$ bits of memory, and their cost is uniform in the data: it stays
$O(p^{1/2}\Mul(p)\log^2p)$ when $s_0$ varies with $p$ in a fixed number field with height $p^{O(1)}$ and
$|s_0| = O(p/\log p)$, and when the coordinates of $F$ in a fixed basis of solutions of a fixed equation have
logarithmic height $O(p)$ (Theorem~\ref{thm:uniform}). At negative integers $s_0$ this gives exact algorithms for many classical
integer sequences in $n^{3/2+o(1)}$ bit operations and $O(n\log n)$ bits of memory. These include the Bernoulli,
Euler, tangent, Genocchi and Springer numbers, the Fubini numbers, and in general the coefficients of every
exponential generating function that is rational in $e^{\lambda t}$, subject to a mild condition on its poles
(Corollaries~\ref{cor:bernoulli} and~\ref{cor:egf}, Table~\ref{tab:sequences}). For the Bernoulli numbers this matches
Harvey's algorithm \cite{Har14} in its quasilinear-memory setting. For the others we know of no earlier subquadratic
algorithm, apart from \cite{Joh21} for the Euler numbers.

For Khinchin's constant we develop three complete algorithms, K1, K2 and K3 (Section~\ref{sec:khinchin}): from
the derivative of what we call the Khinchin zeta function,
from the logarithm series for $\log K\log 2$ at the beginning of this introduction, and from a
dilogarithm series of Bailey, Borwein
and Crandall, a value of the twisted zeta function $L_G$ of the non-elementary function $G(k) = \Li_2(4/k^2)$. With the
same machinery, the third is about $5$ times faster than the second and $19$ times faster than the first,
and we discuss to what extent such rewritings generalize. With an implementation of
the last series in FLINT we computed $K$ to $10^7$ digits, and with a generic program several other constants of
Table~\ref{tab:constants} to $10^6$ digits (Section~\ref{sec:other}).

\subsection{Related work}
\label{sec:related}
The meromorphic continuation of a Dirichlet series $\sum_{k=1}^{\infty} F(k)k^{-s}$ whose coefficients have an asymptotic
expansion at infinity is classical; it follows from Euler--Maclaurin summation or the Mellin transform method
\cite{Zag06,FGD95}, and is used routinely for zeta regularization in physics \cite{Eli95,Kir02}. Dirichlet series
attached to polynomials and rational functions, in one and several variables, have been studied by Mahler
\cite{Mah28}, Cassou-Nogu\`es \cite{CN81}, Sargos \cite{Sar84}, Peter \cite{Pet98}, Essouabri \cite{Ess97}
and others; Apostol and Vu \cite{AV84} treat series such as $\sum_{k=1}^{\infty} k^{-s}(k+1)^{-1}$, whose value at $s = \frac12$
is the constant of Theodorus; Brink \cite{Bri12} expresses it and related constants as series of zeta values at
half-integers.

For summands with an expansion in powers of $k^{-1/q}$, which include the algebraic functions, a
general-purpose numerical method is Monien summation~\cite{Mon10} as implemented in PARI/GP \cite{BC21}, which evaluates the
summand at $O(p/\log p)$ points; like Euler--Maclaurin summation it costs $p^{2+o(1)}$. Mellin--Barnes integrals
give the continuation of hypergeometric cases like $\sum_{k=1}^{\infty} k^{-s}(k+a)^{-t}$ and their asymptotic
expansions in parameters \cite{Kat97}.

Approximate functional equations go back to Riemann \cite{Rie59} and
Hardy and Littlewood \cite{HL23}; the theta-function (``smoothed'') form for general $L$-functions is due to
Lavrik \cite{Lav68} and Chandrasekharan and Narasimhan \cite{CN62}, and is the basis of the algorithms of Rubinstein
\cite{Rub05} and Dokchitser \cite{Dok04}; for the Lerch zeta function see \cite{GLS03}. In mathematical physics the
same device is Ewald summation of lattice sums \cite{Ewa21}, the Chowla--Selberg formula for Epstein zeta functions
\cite{CS67} (see \cite{BGMWZ13}), and the proper-time and heat-kernel representations of sums $\sum_kF(k^2)$ in quantum
field theory \cite{Sch51,Eli95,Kir02}. Crandall used such splittings systematically as computational tools for $\zeta$
\cite{BBC00} and for polylogarithms, Lerch transcendents and related functions \cite{Cra12} (see also
\cite{Cof12,BB15}), without a complexity analysis.

The $L$-function algorithms and the present results overlap only in part. Both contain the Dirichlet $L$-functions and
the Hurwitz and Lerch zeta functions; there Theorem~\ref{thm:split} with $F = 1$, and its twisted form
(Remark~\ref{rem:twist}), is the smoothed approximate functional equation. Beyond this intersection neither contains the other. The $L$-function algorithms allow arbitrary arithmetic
coefficients and gamma factors of any degree, but they need a functional equation. Theta splitting needs no functional
equation and allows any coefficients $F \in \Reg$, but its theta function is that of the integers, with a gamma factor
of degree~$1$. For a completed $L$-function of degree $d$ and conductor $N$, the weights of the smoothed
approximate functional equation are incomplete Mellin transforms of a Meijer $G$-function, and are therefore
holonomic: incomplete gamma functions in degree $1$, and Bessel and Meijer $G$-functions in general. These methods are
usually analysed by the number of terms, about $\sqrt N$ at fixed precision. As a function of $p$, about
$\sqrt N\,p^{d/2}$ terms are needed, so only degree~$1$ gives $p^{3/2+o(1)}$ at algebraic $s$ (see below and
Section~\ref{sec:limits}); products of functions of degree~$1$, such as Dedekind zeta functions of abelian fields, can
be computed factor by factor.

For $F = 1$ every term of the theta splitting is a single incomplete gamma value, and the method then
gives $p^{3/2+o(1)}$ at algebraic $s$ by fast evaluation of the incomplete gamma function \cite{CC90,vdH99}.
This bound is folklore, but it has rarely been stated precisely.\footnote{Table~1 of Borwein and Borwein \cite{BB88} lists
the bit complexity $p^{1/2+o(1)}\Mul(p)$ for the gamma and zeta functions on a compact region, and
$O(\Mul(p)\log^2p)$ for their values at rational points. As justification it cites only Riemann's integral
representation of $\pi^{-s/2}\Gamma(\frac s2)\zeta(s)$, which leads to \eqref{eq:riemann}, with the remark ``We
truncate both the integral and the sum''; the algorithm is not described further. For $\Gamma$ both bounds are
correct. For~$\zeta$, both bounds miss a factor $p^{1/2}$ by every method known to the author; see also Section~\ref{sec:openexp}.}
To our knowledge the bound $p^{3/2+o(1)}$ at algebraic $s$, with a complete algorithm, was first stated, for
Dirichlet $L$-functions and with applications to
Stieltjes constants and Bernoulli and Euler numbers, in the author's preprint \cite{Joh21}. That preprint is
preliminary work, subsumed by the present paper, which generalizes it substantially: Theorem~B extends it from
$F = 1$ to all $F \in \RegQ$ and to polylogarithms and Lerch transcendents (Claim~C adds the damped zeta functions), and Theorem~\ref{thm:uniform} proves the uniformity in $s$
that \cite{Joh21} argued only briefly.

To our knowledge Dirichlet series for holonomic $F$ have not been studied as such. What holonomy adds is that every piece of the splitting is again holonomic, which is
what makes the fast evaluation possible.

\subsection{Outline}
Section~\ref{sec:defs} defines the classes $\Reg$ and $\RegQ$ of summands and the functions
\eqref{eq:functions}, and recalls the fast evaluation of holonomic functions. Section~\ref{sec:splitting} works out one
example in full, proves Theorem~A and shows that both sides are holonomic. Section~\ref{sec:beyond} gives the splittings of the polylogarithm, the Lerch transcendent and the damped zeta
function, and discusses the limits of the method. Section~\ref{sec:complexity} proves Theorem~B, states the
uniform bounds and their consequences for integer sequences, and compares the theta splitting with classical
methods. Section~\ref{sec:khinchin} defines the Khinchin zeta function and gives three algorithms for Khinchin's
constant, with an implementation and a computation to $10^7$ digits. Section~\ref{sec:other} treats other
constants of Table~\ref{tab:constants}, and Section~\ref{sec:open} lists open problems. Appendix~\ref{app:cost} describes a
Lean formalization of much of the paper and contains the cost analysis;
Appendix~\ref{sec:khpseudo} gives an annotated reference implementation for Khinchin's constant.

The git repository containing this paper and associated code
and data is \url{https://github.com/fredrik-johansson/holonomic-sums/}.

\section{Preliminaries}\label{sec:defs}
This section defines the class of summands and the functions attached to them, recalls how holonomic functions are
evaluated quickly, and fixes notation.

\subsection{Holonomic functions and the regular class}
A function $F$ of one variable, analytic on some open set, is \emph{holonomic} (or \emph{D-finite}) if it satisfies a
linear differential equation $LF = 0$ with
\[
  L = \sum_{i=0}^r p_i(x)\,\partial_x^i, \qquad p_i \in \CC[x], \quad p_r \ne 0;
\]
equivalently, its derivatives span a finite-dimensional vector space over $\CC(x)$. A sequence $(a_n)$ is
\emph{holonomic} (or \emph{P-recursive}) if it satisfies a linear recurrence $\sum_{i=0}^{d}c_i(n)a_{n+i} = 0$ with
polynomial coefficients, $c_d \ne 0$. Holonomic functions are closed under sums, products, differentiation and
integration and under composition with algebraic functions, and the coefficients of a power series are a holonomic
sequence if and only if the series is a holonomic function \cite{Sta80}. Elementary functions such as $\exp$, $\log$
and algebraic functions, the hypergeometric functions ${}_pF_q$, Bessel functions, the error function and the
incomplete gamma functions are holonomic; $\Gamma(x)$, $\zeta(x)$ and $\tan x$ are not. The summands $F$ of this paper
are holonomic functions with a restriction on their behaviour at $x = \infty$, which we now define.

Let $L$ be as above, and suppose that $x = \infty$ is an ordinary point or a regular singular point of $L$, i.e.\ that $t = 1/x$ is such a point of the transformed
operator. Let $R$ exceed the absolute values of the finite singular points of $L$. By Frobenius's method
\cite[\S16]{Inc26}, every solution of $L$ on $(R, \infty)$ can be written as
\begin{equation}\label{eq:frob}
  F(x) = \sum_{\sigma\in\Sigma}\sum_{j=0}^{J} x^{-\sigma}(\log x)^j f_{\sigma,j}(1/x),
\end{equation}
with a finite set $\Sigma \subset \CC$ of exponents, pairwise incongruent modulo $\ZZ$, and power series $f_{\sigma,j}$
convergent for $|t| < 1/R$. The exponents are the roots of the indicial equation and need not be integers, so $F$
may be ramified at infinity (Remark~\ref{rem:puiseux}). In particular every algebraic function
is regular at infinity in the sense of the following definition.

\begin{definition}\label{def:regular}
Let $F$ be a function on $(R,\infty)$.
\begin{enumerate}
\item $F$ is \emph{analytic at infinity} if $F(x) = f(1/x)$ with $f$ analytic for $|t| < 1/R$. No differential
equation is assumed.
\item $F$ is \emph{regular at infinity} if it is annihilated by an operator $L$ as above; then $F$ has the form
\eqref{eq:frob}.
\end{enumerate}
A sequence $(F(k))_{k\ge1}$ is \emph{analytic}, respectively \emph{regular}, if for some integer $k_1 \ge 1$ it
vanishes for $k < k_1$ and agrees for $k \ge k_1$ with a function $G$ analytic, respectively regular, at infinity
that continues analytically to the real interval $[k_1,\infty)$. We write this as $F(k) = G(k)$ $(k \ge k_1)$, or
simply $F(k) = G(k)$ when $k_1 = 1$, and where it does not matter we do not distinguish $F$ from $G$.
We write $\Reg$ for the set of regular sequences, and $\RegQ \subset \Reg$ for the set of those for which $L$ can be
chosen with $p_i \in \Qbar[x]$ and the coefficients of the power series $f_{\sigma,j}$ in \eqref{eq:frob} are
algebraic numbers.
\end{definition}

For convenience, the definition allows prepending a finite run of zeros
to skip poles or branch cuts of the underlying holonomic function
while starting the summation at $k = 1$.
For example, the function $-\log(1-\frac1k)\log(1+\frac1k)$ of \eqref{eq:twoconstants} is not
defined at $k = 1$; we write the sequence as $F(k) = -\log(1-\frac1k)\log(1+\frac1k)$ $(k \ge 2)$ with
the understanding that $F(1) = 0$.

Membership in $\RegQ$ concerns the operator and the expansion at infinity, and hence also the exponents $\sigma$,
which are roots of an indicial polynomial with algebraic coefficients: ${}_2F_1(\frac32,\frac32;2;-\frac4{k^2})$ is in $\RegQ$, and so is $\sin(x/k)$
for algebraic $x$, but not for $x = \pi$, when its expansion has the coefficients $\pm\pi^{2n+1}/(2n+1)!$. Both
conditions are needed: $k^{-\pi}$ has the single expansion coefficient $1$, but its exponent is transcendental.
Algebraic data are what allow binary splitting, and hence the bound of Theorem~B. Without such a condition, a
complexity bound needs some other model of the data, since $F$ could be a hard-to-compute constant times a simple
function. Section~\ref{sec:complexity} also treats generic data, given by $p$-bit approximations, and obtains weaker
bounds for them (Table~\ref{tab:methods}).

The two conditions of Definition~\ref{def:regular} are independent. A regular $F$ need not be analytic at infinity
(it may involve $x^{-\sigma}$ and $\log x$), and a holonomic $F$ that is analytic at infinity need not be regular
there. Regularity is a property of the operator, and an operator that annihilates $F$ also annihilates all analytic
continuations of $F$. For example, $F(x) = e^{x-\sqrt{x^2+1}} = e^{-1/(2x) + O(x^{-3})}$ is holonomic and analytic at
infinity, but continuing it around the branch point $x = i$ gives $e^{x+\sqrt{x^2+1}}$, which grows like $e^{2x}$.
Every operator annihilating $F$ annihilates this function too, so $x = \infty$ is an irregular singular point of all
of them, and $F$ is not regular at infinity. The functions that are both regular and analytic at infinity are the
$F(x) = f(1/x)$ with $f$ analytic at $0$ and annihilated by an operator with an ordinary or regular singular point
at $0$.

Besides the closure properties listed in Section~\ref{sec:sums}, regularity at infinity is preserved by
$F(x) \mapsto F(qx+a)$, by multiplication by $x^{-\sigma}$ and $(\log x)^j$, and by differentiation. It holds for all G-functions of $1/x$ (algebraic functions,
logarithms, polylogarithms, ${}_{q+1}F_q$ with rational parameters) and for entire functions of $1/x$ such as
$e^{c/x^2}$, whose irregular singularity is at $x = 0$ and is never reached. What it excludes is exponential
behaviour at infinity, such as $z^k$ with $|z| \ne 1$, $e^{-\tau k^r}$ or $e^{ik^2}$, and nonconstant periodic
functions, which have an irregular singularity there. This is why the periodic weight $\chi$ and the factors $z^k$
and $e^{-\tau k^r}$ are kept separate from $F$ in \eqref{eq:functions}.

\begin{definition}\label{def:functions}
For a sequence $F$ that is analytic or in $\Reg$, a periodic $\chi\colon\ZZ \to \CC$, $0 < |z| < 1$, $a > 0$,
$\tau > 0$ and $0 < r < 2$, the \emph{zeta function} $\zeta_F(s)$, the \emph{twisted zeta function} $L_F(s,\chi)$, the
\emph{polylogarithm} $\Li_F(s,z)$, the \emph{Lerch transcendent} $\Phi_F(z,s,a)$ and the \emph{damped zeta function}
$\Lambda_F(s;\tau,r)$ of $F$ are the series \eqref{eq:functions}, summed from $k = 1$ (from $k = 0$ for
$\Phi_F$); we write $\Lambda = \Lambda_1$. For internal use we also write the tails as
\[
  \zeta_F^{\ge k_0}(s) = \sum_{k=k_0}^{\infty}F(k)\,k^{-s},
\]
and similarly $L_F^{\ge k_0}$, $\Li_F^{\ge k_0}$ and $\Lambda_F^{\ge k_0}$.
\end{definition}

The series for $\zeta_F$ and $L_F$ converge absolutely for $\operatorname{Re}s > 1 - \beta$, where
$\beta = \min\operatorname{Re}\sigma$ over the exponents that occur; those for $\Li_F$, $\Phi_F$ and $\Lambda_F$
converge for all $s$ and define entire functions. Finitely many terms are entire functions of $s$, so all
analytic questions concern the tails; the splittings below are applied to a tail $\zeta_F^{\ge k_0}$ with $k_0$ large
enough for the expansion of $F$ at infinity to converge at $k_0$, and the terms with $k < k_0$ are added directly. We call $\Li_F$ and $\Lambda_F$ \emph{damped}, as opposed to
$\zeta_F$ and $L_F$, since the factors $z^k$ and $e^{-\tau k^r}$ make the series converge; for $r = 1$ the two
coincide, $\Lambda_F(s;\tau,1) = \Li_F(s,e^{-\tau})$. The name \emph{damped zeta function} is ours.

\begin{lemma}\label{lem:reduction}
If $F$ is given for $x \ge k_0$ by \eqref{eq:frob} (with any power series $f_{\sigma,j}$ convergent for
$|t| < 1/R$), and $F_{\sigma,j}(x) = f_{\sigma,j}(1/x)$, then
\[
  \zeta_F^{\ge k_0}(s) = \sum_{\sigma\in\Sigma}\sum_{j=0}^{J}(-\partial_s)^j\,\zeta_{F_{\sigma,j}}^{\ge k_0}(s+\sigma).
\]
\end{lemma}

\begin{proof}
Termwise, since $x^{-\sigma}(\log x)^jx^{-s} = (-\partial_s)^jx^{-s-\sigma}$, for $\operatorname{Re}s$ large; then by
continuation.
\end{proof}

The same reduction, with the same proof, holds for $L_F$, $\Li_F$ and $\Lambda_F$, since the factors $\chi(k)$, $z^k$
and $e^{-\tau k^r}$ are not affected.

\begin{remark}[Puiseux expansions]\label{rem:puiseux}
Since regularity at infinity is a condition on the operator, $F$ need not be single-valued or analytic at
infinity. In particular, a convergent Puiseux series $\sum_{n=0}^{\infty}a_nx^{-n/q}$
annihilated by such an operator is of the form \eqref{eq:frob}, with $\Sigma \subseteq \{0, \frac1q, \dots,
\frac{q-1}q\}$, by grouping the terms according to $n \bmod q$, and Lemma~\ref{lem:reduction} reduces it to $q$
functions that are analytic at infinity. For example, $F(k) = 1/(k^{3/2}+1) = \sum_{n=0}^{\infty}(-1)^nk^{-3(n+1)/2}$
involves both classes of exponents modulo $\ZZ$. Multiplying by the conjugate $k^{3/2}-1$ separates them,
$F(k) = (k^{3/2}-1)G(k)$ for $k \ge 2$, with $G(k) = 1/(k^3-1)$ $(k \ge 2)$ rational and analytic at infinity, so that
\[
  \sum_{k=1}^{\infty}\frac{1}{k^{3/2}+1} = \frac12 + \zeta_G\bigl(-\tfrac32\bigr) - \zeta_G(0) = 1.9520200662\ldots,
\]
where neither $-\frac32$ nor $0$ is a pole of $\zeta_G$, as the poles are at $s = -2, -5, -8, \dots$. In general, a summand
$R(\sqrt k)$ with $R$ rational, such as the summand of the constant of Theodorus in Table~\ref{tab:constants}, splits as $A(k) + \sqrt k\,B(k)$, where $A(y^2)$ and $yB(y^2)$ are the even and odd
parts of $R(y)$; this is the decomposition into monodromy eigenspaces used in the proof of
Theorem~\ref{thm:complexity} (Appendix~\ref{app:cost}).
\end{remark}

\subsection{Evaluating holonomic functions}\label{sec:evalhol}
Two classical techniques evaluate holonomic functions with algebraic data in quasilinear time, and we use them
throughout; Appendix~\ref{app:tools} states them precisely. \emph{Binary splitting} sums a series whose terms satisfy a
linear recurrence with polynomial coefficients, such as a hypergeometric series or the Taylor series of a holonomic
function at a point with few bits: the terms are products of matrices with small entries, and the products are formed
in a balanced tree, so that the numbers being multiplied have similar sizes. For a geometrically convergent series
with algebraic coefficients this costs $O(\Mul(p)\log^2p)$ bit operations for $p$ bits \cite{HP98,CC90,CHTZZ07}. The
\emph{bit-burst algorithm} extends this to a point $x$ given to $p$ bits: it continues the solution of the
differential equation from a point with few bits through truncations of $x$ to $2^j$ bits, $j = 1, 2, \dots$, summing
one Taylor series by binary splitting at each step, again in $O(\Mul(p)\log^2p)$ \cite{CC90,vdH99}; see
\cite{Mez10,Mez16} for rigorous implementations.

We use the error functions $\operatorname{erf}$, $\erfc = 1 - \operatorname{erf}$ and
$\erfi(x) = -i\operatorname{erf}(ix)$, the incomplete gamma functions $\Gamma(a,x)$ and $\gamma(a,x)$ with their
regularized forms $Q(a,x) = \Gamma(a,x)/\Gamma(a)$ and $P(a,x) = 1 - Q(a,x)$, the exponential integrals
$E_1(x) = \Gamma(0,x)$ and $\operatorname{Ei}$, the entire functions
\[
  \Ein(y) = \sum_{n=1}^{\infty}\frac{(-1)^{n+1}y^n}{n\cdot n!}, \qquad \Eip(y) = \sum_{n=1}^{\infty}\frac{y^n}{n\cdot n!}
  = \operatorname{Ei}(y) - \gamma - \log y,
\]
and the cosine integral $\Ci$. All are holonomic.

\subsection{The analytic case}\label{sec:scope}
The constructions of Sections~\ref{sec:splitting}--\ref{sec:complexity} are carried out for summands that are
analytic at infinity. Where this is assumed, we write
\begin{equation}\label{eq:f}
  F(k) = f(1/k) \quad (k \ge k_0), \qquad f(x) = \sum_{n=0}^{\infty}a_nx^n, \qquad |a_n| \le C R^n,
\end{equation}
with $R < k_0$, and let $n_0 = \min\{n : a_n \ne 0\}$ (we exclude $f = 0$). General $F \in \Reg$, of the form \eqref{eq:frob},
reduce to this case by Lemma~\ref{lem:reduction}, at the price of shifts and derivatives in $s$; this is how
Theorem~\ref{thm:complexity} is proved in that generality. For $F$ as in \eqref{eq:f}, the splitting identity
(Theorem~\ref{thm:split}) needs nothing more. That its two sides are holonomic (Lemma~\ref{lem:kernelhol}, Proposition~\ref{prop:dual}), and hence
Theorem~\ref{thm:complexity} and the algorithms of Section~\ref{sec:khinchin}, need $F$ regular and analytic at
infinity; condition~(H) of Section~\ref{sec:splitting} is what gives the dual side a first-order system with
polynomial coefficients. The cost bounds for the classical methods in Section~\ref{sec:complexity} need only that
the $a_n$, and the Taylor coefficients of $F$, are P-recursive.

\subsection{Operations on summands}\label{sec:twists}
A constant can often be written in several ways as a value of one of the functions \eqref{eq:functions}, and the
cost of computing it depends on the choice (Sections~\ref{sec:khinchin} and~\ref{sec:generalize}). We use five
operations that rewrite a summand. Three of them, dilations, twists and functional equations, give identities between
functions of $s$. The other two, shifts and telescoping, preserve only the value at a fixed point.

\emph{Dilations.} For an integer $m \ge 1$ let $F_m(k) = F(mk)$, again analytic, respectively regular. Then
\[
  \sum_{k=1,\ m\mid k}^{\infty}\chi(k)F(k)k^{-s} = m^{-s}L_{F_m}(s,\chi_m), \qquad \chi_m(k) = \chi(mk).
\]

\emph{Twists.} Every function $\chi$ of period $q$ is a combination of the functions $k \mapsto [d \mid k]\,\psi(k/d)$
with $d \mid q$ and $\psi$ a Dirichlet character modulo $q/d$ (group the $k$ by $\gcd(k,q) = d$ and expand $\chi(dj)$,
as a function on $(\ZZ/(q/d))^\times$, in characters). Hence
\[
  L_F(s,\chi) = \sum_{d\mid q}\sum_{\psi \bmod q/d}c_{d,\psi}\,d^{-s}L_{F_d}(s,\psi).
\]
Periodic weights thus reduce to dilations and twists by Dirichlet characters, but the characters cannot be removed
as long as $s$ is a variable.

\emph{Functional equations.} Identities of the special functions in $F$, such as the duplication and Landen identities
of the dilogarithm, rewrite the summand without changing its values.

\emph{Shifts.} Summing over residue classes removes a periodic weight altogether: for $1 \le a \le q$,
\[
  \sum_{k=1,\ k\equiv a \bmod q}^{\infty}F(k)k^{-s} = q^{-s}\sum_{j=0}^{\infty}F(qj+a)\Bigl(j + \frac aq\Bigr)^{-s},
\]
and $F(qx+a)$ is regular at infinity. Now $s$ enters the summand through
$(j+a/q)^{-s} = j^{-s}(1+\frac{a}{qj})^{-s}$, as in the Hurwitz zeta function. At a fixed point $s_0$ the left-hand side
is, up to the term $j = 0$, the value $\zeta_{F'}(s_0)$ with
\[
  F'(j) = q^{-s_0}F(qj+a)\Bigl(1+\frac a{qj}\Bigr)^{-s_0},
\]
which suffices for computing it; but as a function of $s$ it is not of the form $\zeta_{F'}$, since the summand
depends on $s$. When only the value at a fixed point matters, as for the sums of Table~\ref{tab:constants}, periodic
weights can therefore be removed. We keep them nevertheless, since Theorem~\ref{thm:split} handles them directly
(Remark~\ref{rem:twist}).

\emph{Telescoping.} For a constant $\sum_{k=k_0}^{\infty}F(k) = \zeta_F(0)$, the summands $F(k)$ and
$F(k) + H(k+1) - H(k)$ have the same sum, up to $H(k_0)$ and the limit of $H$ at infinity, for any $H$ regular at
infinity. This holds only at $s = 0$, since $k^{-s}$ does not commute with the shift. Summation by parts is of this
kind.

\subsection{Notation}\label{sec:notation}
Table~\ref{tab:notation} lists the notation used throughout. Symbols that occur only in one example (in
Sections~\ref{sec:khinchin} and~\ref{sec:other}) are defined where they occur.

\begin{table}[ht]
\centering\small
\caption{Principal notation.}
\label{tab:notation}
\begin{tabular}{@{}lp{0.62\textwidth}@{}}
\toprule
$F$, $f$, $a_n$ & summand, $F(k) = f(1/k)$ for $k \ge k_0$, $f(x) = \sum_na_nx^n$ \\
$R$, $\rho$ & $|a_n| \le CR^n$; radius of convergence of $f$ \\
$k_0$, $k_1$ & start of the convergent expansion; $F(k) = 0$ for $k < k_1$ (Definition~\ref{def:regular}) \\
$\Reg$, $\RegQ$ & admissible summands; those with algebraic data (Definition~\ref{def:regular}) \\
$\zeta_F$, $L_F$, $\Li_F$, $\Phi_F$, $\Lambda_F$ & the functions \eqref{eq:functions} \\
$\chi$, $q$ & periodic weight and its period \\
$s$, $s_0$ & complex variable; evaluation point \\
$p$, $\Mul(p)$ & precision in bits; cost of $p$-bit multiplication \\
$\widetilde O(X)$ & $X(\log X)^{O(1)}$ \\
$u_0$ & split point of the theta splitting \\
$h_s$, $g_0$, $g_1$, $G_0$, $G_1$ & kernel and its even and odd parts \eqref{eq:kernel}, \eqref{eq:kernelparts} \\
$I_k$, $J_m$, $M$ & $k$-terms, dual terms and main term (Theorem~\ref{thm:split}) \\
$Q(a,x)$, $P(a,x)$ & regularized upper and lower incomplete gamma functions \\
$\vartheta$ & Euler operator $x\,\mathrm d/\mathrm dx$ or $u\,\mathrm d/\mathrm du$ \\
$\mathcal L[g](\kappa)$, $Z_t[g](c)$ & truncated Laplace transform and dual moments (Section~\ref{sec:twosides}) \\
$E$, $D_a$ & $E = e^{-c/u_0}$, starting values \eqref{eq:startvals} \\
$\kappa$, $x$ & $\kappa = k^2$ and $x = \kappa u_0$ in the $k$-terms \\
$k_{\max}$, $m_{\max}$ & numbers of $k$-terms and dual terms \eqref{eq:counts} \\
$K$, $\mu_k$, $Z$ & Khinchin's constant, Gauss--Kuzmin probabilities, Khinchin zeta function \\
\bottomrule
\end{tabular}
\end{table}

\section{Theta splitting}\label{sec:splitting}

The model is Riemann's second proof of the functional equation \cite{Rie59}. Inserting
\[
k^{-s} = \frac{1}{\Gamma(\tfrac s2)} \int_0^\infty u^{s/2-1}e^{-k^2u}\dd u
\]
into $\zeta(s) = \sum_{k=1}^{\infty}k^{-s}$
allows writing $\zeta(s)$ as the Mellin transform of the theta series $\Theta(u) = \sum_{k=1}^{\infty}e^{-k^2u}$. Splitting the
integral at $u = \pi$, integrating termwise on $[\pi,\infty)$ and using the theta transformation
\[
\Theta(u) = \frac12\sqrt{\pi/u} - \frac12 + \sqrt{\pi/u}\,\Theta(\pi^2/u)
\]
on $(0,\pi]$ gives
\begin{multline}\label{eq:riemann}
  \zeta(s) = \sum_{k=1}^{\infty}k^{-s}Q\Bigl(\frac s2,\pi k^2\Bigr)
  + \frac{\pi^{s-1/2}\Gamma(\frac{1-s}2)}{\Gamma(\frac s2)}\sum_{m=1}^{\infty}m^{s-1}Q\Bigl(\frac{1-s}2,\pi m^2\Bigr) \\
  + \frac{\pi^{s/2}}{\Gamma(\frac s2)\,s(s-1)},
\end{multline}
with $Q$ as in Section~\ref{sec:evalhol}. Both sums converge like $\sum_ke^{-\pi k^2}$, for all $s$, and their symmetry
under $s \mapsto 1-s$ is the functional equation. For general $F$, the factor $f(1/k)$ is absorbed by replacing
$u^{s/2-1}/\Gamma(\frac s2)$ with a kernel $h_s(u)$ (Section~\ref{sec:kernel}); the splitting point $u_0$ becomes a
free parameter that balances the two sides (Theorem~\ref{thm:split}); and when $F$ is holonomic, so are both sides,
as functions of $k^2$ and $m^2$ (Section~\ref{sec:twosides}). 

\subsection{A first example}\label{sec:workedexample}
Before the general construction, here is Theorem~A for a sum with $F \ne 1$ in which every piece is explicit. Let
\[
  S_A = \sum_{k=2}^{\infty}\frac2k\arctanh\frac1k = 1.35313027229593328165\ldots,
\]
which gives the constant $\mathcal A = \frac12\log2 + \frac12S_A$ of Table~\ref{tab:constants}
(Section~\ref{sec:AG}). Here $S_A = \zeta_F(0)$ with $F(k) = f(1/k)$ $(k \ge 2)$ and
$f(x) = 2x\arctanh x = \sum_{m=1}^{\infty}\frac{2}{2m-1}x^{2m}$, so $R = 1$ and $k_0 = 2$. At $s = 0$ the kernel of
Theorem~A is an error function,
\[
  h(u) = \sum_{m=1}^{\infty}\frac{2}{2m-1}\,\frac{u^{m-1}}{\Gamma(m)} = \sqrt{\frac\pi u}\,\erfi\sqrt u, \qquad
  u\,h'(u) = e^u - \tfrac12h(u),
\]
and termwise integration gives $\int_0^\infty e^{-k^2u}h(u)\dd u = \frac2k\arctanh\frac1k$ for $k \ge 2$. For any
$u_0 > 0$, Theorem~A reads
\[
  S_A = \sum_{k=2}^{\infty}I_k + \sqrt\pi\sum_{m=1}^{\infty}J_m + M,
\]
with
\begin{align*}
  I_k &= \int_{u_0}^\infty e^{-k^2u}h(u)\dd u, &
  J_m &= \int_0^{u_0}u^{-1/2}e^{-\pi^2m^2/u}h(u)\dd u, \\
  M &= \int_0^{u_0}\Bigl(\frac12\sqrt{\frac\pi u} - \frac12 - e^{-u}\Bigr)h(u)\dd u .
\end{align*}

\emph{The $k$-terms.} Expanding $h$ termwise, with $x = k^2u_0$ and $Q(m,x) = e^{-x}\sum_{i<m}x^i/i!$,
\[
  I_k = \sum_{m=1}^{\infty}\frac{2}{2m-1}\,k^{-2m}Q(m,x) = e^{-x}\sum_{m=1}^{\infty}\frac{2}{(2m-1)k^{2m}}\sum_{i=0}^{m-1}\frac{x^i}{i!},
\]
a double sum whose terms satisfy first-order recurrences with small rational coefficients, so binary splitting
computes $I_k$ to $p$ bits in quasilinear time. Equivalently, $I_k = \ell(k^2)$, where the truncated Laplace transform
$\ell(\kappa) = \int_{u_0}^\infty e^{-\kappa u}h(u)\dd u$ satisfies the first-order equation
$\kappa\ell' = -\frac12\ell - u_0e^{-\kappa u_0}h(u_0) - e^{-(\kappa-1)u_0}/(\kappa-1)$.

\emph{The dual terms.} Let
\[
  \nu_t(c) = \int_0^{u_0}u^{-t}e^{-c/u}h(u)\dd u, \qquad Y_t(c) = \int_0^{u_0}u^{-t}e^{-c/u}e^u\dd u,
\]
so that $J_m = \nu_{1/2}(\pi^2m^2)$. Integrating by parts and using the equation of $h$ gives, with
$E = e^{-c/u_0}$,
\begin{align*}
  c\,\nu_{t+1} &= Eu_0^{1-t}h(u_0) - (\tfrac12 - t)\nu_t - Y_t, \\
  c\,Y_{t+1} &= Eu_0^{1-t}e^{u_0} - (1-t)Y_t - Y_{t-1},
\end{align*}
and $\partial_c\nu_t = -\nu_{t+1}$, $\partial_cY_t = -Y_{t+1}$. Hence $(\nu_{1/2}, Y_{1/2}, Y_{-1/2})$ satisfies a linear
differential system in $c$ with coefficients linear in $c$, forced by $E$, and $J_m$ is the value of a holonomic
function at $c = \pi^2m^2$. The bit-burst algorithm computes it from starting values at a nearby rational point,
which are series in $\erfc$ and $E$.

\emph{Cost.} The $k$-terms decay like $e^{-k^2u_0}$ and the dual terms like $e^{-\pi^2m^2/u_0}$. With $u_0 = 1$, the
eight $k$-terms $k \le 9$ and three dual terms give more than $30$ digits; for $p$ bits one needs
$O(\sqrt p)$ terms on each side, each costing $\widetilde O(p)$, and $M$ is a single hypergeometric series in $u_0$.
This gives $p^{3/2+o(1)}$, against $\Theta(p)$ zeta values for the classical expansion
$S_A = \sum_{m=1}^{\infty}2(\zeta(2m)-1)/(2m-1)$. The rest of this section carries out the same steps for general
$F$, and Section~\ref{sec:complexity} the cost analysis.

\subsection{The kernel}\label{sec:kernel}
The kernel is the function whose Laplace transform in $\kappa = k^2$ is $k^{-s}f(1/k)$. It is obtained by applying the
formula for $k^{-s}$ used in Riemann's proof termwise to $f$. Under \eqref{eq:f} define, for $s \in \CC$ and $u > 0$,
\begin{equation}\label{eq:kernel}
  h_s(u) = \sum_{n=0}^{\infty} a_n\frac{u^{(s+n)/2-1}}{\Gamma(\frac{s+n}{2})} = u^{s/2-1}E_s(u^{1/2}), \qquad
  E_s(v) = \sum_{n=0}^{\infty}\frac{a_nv^n}{\Gamma(\frac{s+n}2)}.
\end{equation}
Since $1/\Gamma$ is entire and $|a_n| \le CR^n$, $E_s(v)$ is entire in $(s,v)$, and for $s$ in a compact set $\mathcal K$,
$|E_s(v)| \le C(1+|v|)^{c}e^{R^2|v|^2}$. If $\operatorname{Re}(s+n) > 0$ for all $n$ with $a_n \ne 0$, then for
$k > R$
\begin{equation}\label{eq:laplace}
  k^{-s}f(1/k) = \int_0^\infty e^{-k^2u}h_s(u)\dd u,
\end{equation}
by $k^{-w} = \Gamma(w/2)^{-1}\int_0^\infty u^{w/2-1}e^{-k^2u}\dd u$ and Tonelli's theorem. Writing $u = v^2$,
$h_s$ is $u^{s/2}$ times an even and an odd entire function of $v$:
\begin{equation}\label{eq:kernelparts}
\begin{gathered}
  h_s(u) = u^{s/2}\bigl(g_0(u) + u^{-1/2}g_1(u)\bigr), \\
  g_0(u) = \sum_{m=0}^{\infty}\frac{a_{2m}u^{m-1}}{\Gamma(m+\frac s2)}, \qquad
  g_1(u) = \sum_{m=0}^{\infty}\frac{a_{2m+1}u^{m}}{\Gamma(m+\frac{s+1}2)}.
\end{gathered}
\end{equation}
The reason for this split is that the transform halves exponents: $a_nk^{-s-n}$ corresponds to $u^{(s+n)/2-1}$, so
even and odd $n$ give two classes of exponents at $u = 0$, differing by $\frac12$ modulo $\ZZ$. Equivalently, with
$f(x) = A_0(x^2) + xA_1(x^2)$,
\[
  F(k)k^{-s} = k^{-s}A_0(k^{-2}) + k^{-s-1}A_1(k^{-2}),
\]
a sum of two functions of $k^2$ times powers of $k$, which is the form the Gaussians $e^{-k^2u}$ can handle.
Within one class the factors $1/\Gamma(m+\sigma)$ change by the rational factor $m+\sigma$ from one term to the next,
whereas $\Gamma(w)/\Gamma(w+\frac12)$ is not rational in $w$. Each part therefore has a power series with P-recursive
coefficients (Lemma~\ref{lem:kernelhol}), and each needs its own incomplete gamma values below, since the recurrences
for $Q(a,x)$ and for the moments $D_a$ only relate exponents $a$ in the same class modulo $\ZZ$. For $f = 1$ only the
even part occurs.

The rest of this subsection, and Section~\ref{sec:twosides}, assume moreover that $F$ is regular at infinity. Then
$f$ is annihilated by an operator that is regular singular (or ordinary) at $x = 0$. Written in terms of $\vartheta = x\frac{\mathrm d}{\mathrm dx}$ it has the form
\begin{equation}\label{eq:thetaform}
  \sum_{j=0}^{r}x^jq_j(\vartheta), \qquad \deg q_0 \ge \deg q_j \quad (0 \le j \le r),
\end{equation}
by Fuchs's criterion \cite[\S15.3]{Inc26}; we call this condition (H). In terms of the coefficients it says that
$\sum_{j=0}^{r}q_j(n-j)a_{n-j} = 0$ for large $n$, with the coefficient $q_0$ of the newest term of maximal degree.

\begin{lemma}\label{lem:kernelhol}
Let $f$ be annihilated by an operator satisfying (H), with coefficients in a field $\mathbb K \subset \CC$, and fix
$s$. Then each of $G_0 = u\,g_0(u)$ and $G_1 = g_1(u)$ is annihilated by an operator with coefficients in
$\mathbb K(s)[u]$, and for $G = G_0, G_1$ there are $\ell \ge 1$, $A \in \mathbb K(s)[u]^{\ell\times\ell}$ and
$b \in \CC[u]^\ell$ such that $\mathbf G = (G, \vartheta G, \dots, \vartheta^{\ell-1}G)^T$ satisfies
\begin{equation}\label{eq:Gsys}
  \vartheta\mathbf G = A(u)\,\mathbf G + b(u).
\end{equation}
If moreover the coefficients $a_n$ of $f$ lie in $\mathbb K$, and $m_2 \in \ZZ$ is such that $\sigma_0 + m_2$ is not
a pole of $\Gamma$, where $\sigma_0 = \frac s2$ for $G_0$ and $\sigma_0 = \frac{s+1}2$ for $G_1$, then
$\Gamma(\sigma_0+m_2)\,\mathbf G$ satisfies \eqref{eq:Gsys} with the same $A$ and with $b \in \mathbb K(s)[u]^\ell$.
\end{lemma}

\begin{proof}
First, $A_0$ and $A_1$ are annihilated by operators satisfying (H). Let $L$ be the
operator of $f$ and $L^-$ the operator obtained by $x \mapsto -x$, which also satisfies (H). The least common left
multiple of $L$ and $L^-$ annihilates $f(x) \pm f(-x)$; its solution space is the sum of those of $L$ and $L^-$, so by
Fuchs's criterion it is regular singular at $0$ and satisfies (H). In particular, the even part $A_0(x^2)$ is annihilated by an
operator $\sum_jx^jq_j(\vartheta)$ with (H). Split it as $L_{\mathrm{ev}} + L_{\mathrm{od}}$, where $L_{\mathrm{ev}}$
collects the even powers $x^j$; since $\vartheta$ preserves parity, $L_{\mathrm{ev}}A_0(x^2)$ is even and
$L_{\mathrm{od}}A_0(x^2)$ is odd, so both vanish. With $y = x^2$ and $\vartheta_x = 2\vartheta_y$ this reads
$\sum_iy^iq_{2i}(2\vartheta_y)A_0 = 0$, which satisfies (H) because $q_0$ is among the $q_{2i}$. For $A_1$ apply the
same argument to the odd part $xA_1(x^2)$, using $L\,x = x\,L|_{\vartheta\mapsto\vartheta+1}$.

Next, let $B(y) = \sum_{m=0}^{\infty}\beta_my^m$ satisfy an equation with (H),
\[
  \sum_{j=0}^{r}y^jQ_j(\vartheta)B = 0, \qquad\text{that is,}\qquad
  \sum_{j=0}^{r}Q_j(m-j)\beta_{m-j} = 0 \quad (m \ge m_1),
\]
with $\deg Q_0 = d \ge \deg Q_j$. Let $\sigma_0 \in \CC$ and $G(u) = \sum_m\gamma_mu^m$ with
$\gamma_m = \beta_m/\Gamma(m+\sigma_0)$ (with $1/\Gamma = 0$ at the poles). Multiplying the recurrence by
$1/\Gamma(m-r+\sigma_0)$ gives, for $m \ge m_1$,
\[
  \sum_{j=0}^{r}\widetilde Q_j(m-j)\,\gamma_{m-j} = 0, \qquad
  \widetilde Q_j(m-j) = Q_j(m-j)\,(m-r+\sigma_0)_{r-j},
\]
with $\deg\widetilde Q_0 = d+r$ and $\deg\widetilde Q_j \le d+r-j$. Hence $\sum_ju^j\widetilde Q_j(\vartheta)G$ is a
polynomial, and in this operator the coefficient of the highest power $\vartheta^{d+r}$ is the constant leading
coefficient of $\widetilde Q_0$. Dividing by it expresses $\vartheta^{d+r}G$ as a combination of
$G, \dots, \vartheta^{d+r-1}G$ with polynomial coefficients, plus a polynomial: this is \eqref{eq:Gsys}, with a
companion matrix of size $\ell = d + r$, where $r$ is the degree in $y$ of the operator of $B$ and $d$ its order. Taking $B = A_0$, $\sigma_0 = \frac s2$ gives $G_0$, and $B = A_1$, $\sigma_0 = \frac{s+1}2$ gives
$G_1$.

The matrix $A$ depends only on the $\widetilde Q_j$, whose coefficients lie in $\mathbb K(\sigma_0) = \mathbb K(s)$.
The polynomial $b$, on the other hand, is a linear combination of the finitely many initial coefficients
$\gamma_m$, $m < m_1 + r$, which involve the values $1/\Gamma(m+\sigma_0)$ and in general do not lie in
$\mathbb K(s)$. Multiplied by $\Gamma(\sigma_0+m_2)$ they become $\beta_m\,\Gamma(\sigma_0+m_2)/\Gamma(\sigma_0+m)$, and
$\Gamma(x+m_2)/\Gamma(x+m)$ is a rational function of $x$ (a Pochhammer symbol or its reciprocal); this gives the
last statement. This normalization is the one used in Appendix~\ref{app:dual}.
\end{proof}

Condition (H) is what the dual side needs. It makes the coefficient of the highest power of $\vartheta$ in the
operator of $G$ a constant, so that \eqref{eq:Gsys} has polynomial coefficients in $u$, without denominators.
Multiplication by $u^i$ only shifts the index of the moments $Z_t$ in Section~\ref{sec:twosides}, so finitely many
moments form a closed system (Proposition~\ref{prop:dual}).

\subsection{The splitting}
As in Riemann's proof, we insert \eqref{eq:laplace} into the sum and split the integral over $u$ at $u_0$. For
$u \ge u_0$ the Gaussians $e^{-k^2u}$ decay rapidly in $k$, and for $u \le u_0$ the theta transformation turns them into
Gaussians $e^{-\pi^2m^2/u}$ that decay rapidly in $m$. Since \eqref{eq:laplace} holds only for $k > R$, the sum starts
at $k_0 > R$; the theta transformation needs the full lattice, so the terms $k < k_0$ are subtracted again in $M$.

For fixed $x > 0$ the regularized incomplete gamma functions $Q(a,x)$ and $P(a,x) = \gamma(a,x)/\Gamma(a)$ are
entire in $a$.

\begin{theorem}\label{thm:split}
Assume \eqref{eq:f} ($F$ need not be holonomic) and let $u_0 > 0$. For $s \in \CC$ not a pole,
\[
  \zeta_F^{\ge k_0}(s) = \sum_{k=k_0}^{\infty}I_k(s) + \sqrt\pi\sum_{m=1}^{\infty}J_m(s) + M(s),
\]
where
\begin{align*}
  I_k(s) &= \int_{u_0}^\infty e^{-k^2u}h_s(u)\dd u = \sum_{n=0}^{\infty}a_n\,k^{-s-n}\,Q\Bigl(\frac{s+n}2, k^2u_0\Bigr), \\
  J_m(s) &= \int_0^{u_0}u^{-1/2}e^{-\pi^2m^2/u}h_s(u)\dd u, \\
  M(s) &= \sum_{n=0}^{\infty}a_n\biggl(\frac{\sqrt\pi\,u_0^{(s+n-1)/2}}{(s+n-1)\,\Gamma(\frac{s+n}{2})}
     - \frac{u_0^{(s+n)/2}}{2\,\Gamma(\frac{s+n}2+1)} \\
     &\qquad\qquad - \sum_{k=1}^{k_0-1}k^{-s-n}P\Bigl(\frac{s+n}2, k^2u_0\Bigr)\biggr).
\end{align*}
The functions $I_k$ and $J_m$ are entire, and for $s$ in a compact set $\mathcal K$,
$I_k(s) \ll e^{-(k^2-R^2)u_0}$ and $J_m(s) \ll e^{-\pi^2m^2/u_0}$ uniformly; all series converge locally uniformly,
and the only singularities of $M$ are the simple poles of its first term at $s = 1-n$, with residue $a_n$.
\end{theorem}

\begin{proof}
First let $\operatorname{Re}s > 1$. By \eqref{eq:laplace}, $\zeta_F^{\ge k_0}(s) = \int_0^\infty h_s(u)\Theta(u)\dd u$
with $\Theta(u) = \sum_{k=k_0}^{\infty}e^{-k^2u}$, the interchange being justified by Tonelli on $[u_0,\infty)$ and,
on $(0,u_0]$, by $\Theta(u) \le \sqrt{\pi/u}$ and $|h_s(u)| \ll u^{\operatorname{Re}s/2-1}$. Split the integral at $u_0$.
On $[u_0,\infty)$ integrate termwise in $k$, which gives the $I_k$, and termwise in $n$, which gives the series for
$I_k$ (with $\int_{u_0}^\infty e^{-\kappa u}u^{w-1}\dd u = \kappa^{-w}\Gamma(w)Q(w,\kappa u_0)$). On $(0,u_0]$ use the theta
transformation
\[
  \Theta(u) = \frac12\sqrt{\frac\pi u} - \frac12 - \sum_{k=1}^{k_0-1}e^{-k^2u} + \sqrt{\frac\pi u}\sum_{m=1}^{\infty}e^{-\pi^2m^2/u}.
\]
The last sum gives the $J_m$. The first three terms give $M$, by termwise integration in $n$ of
$\int_0^{u_0}u^{w-3/2}\dd u = u_0^{w-1/2}/(w-\frac12)$, $\int_0^{u_0}u^{w-1}\dd u = u_0^w/w$ and
$\int_0^{u_0}e^{-\kappa u}u^{w-1}\dd u = \kappa^{-w}\Gamma(w)P(w,\kappa u_0)$ with $w = (s+n)/2$.

It remains to continue both sides analytically. $I_k$ is entire since $u \ge u_0$, and $|h_s(u)| \ll e^{R^2u}$ there gives the bound.
$J_m$ is entire since $e^{-\pi^2m^2/u}$ kills any power of $u$ at $0$, and
$e^{-\pi^2m^2/u} \le e^{-\pi^2(m^2-1)/u_0}e^{-\pi^2/u}$ gives the bound. In $M$, $\Gamma(w)^{-1}$ cancels the
pole of $u_0^w/w$, and $k^{-s-n}P(\frac{s+n}2,k^2u_0) \ll (k^2u_0)^{n/2}k^{-n}/\Gamma(\frac n2) = u_0^{n/2}/\Gamma(\frac n2)$
for large $n$, so the third term converges for all $s$ even though $k < k_0$ may be less than $R$. The terms of the
first series are $O(R^n u_0^{n/2}/\Gamma(n/2))$ away from the poles. Hence both sides are meromorphic in $\CC$
and agree for $\operatorname{Re}s > 1$. The residue of the first term of $M$ at $s = 1-n$ is
$a_n\sqrt\pi/\Gamma(\frac12) = a_n$.
\end{proof}

\begin{example}[Riemann]\label{ex:riemann}
For $f = 1$, $k_0 = 1$ and $u_0 = \pi$, the substitution $u = \pi^2m^2/v$ gives
$\sqrt\pi J_m(s) = \pi^{s-1/2}m^{s-1}\Gamma(\frac{1-s}2,\pi m^2)/\Gamma(\frac s2)$, and Theorem~\ref{thm:split}
becomes Riemann's formula \eqref{eq:riemann}.
\end{example}

\begin{remark}[Poisson summation]\label{rem:poisson}
Theorem~\ref{thm:split} is a smoothed form of the Poisson summation formula
\[
  \sum_{k\in\ZZ}G_s(k) = \sum_{m\in\ZZ}\widehat G_s(m), \qquad G_s(x) = |x|^{-s}F(|x|).
\]
Formally, for $u_0 = \infty$ the dual terms become
\[
  \sqrt\pi\int_0^\infty u^{-1/2}e^{-\pi^2m^2/u}h_s(u)\dd u = \widehat G_s(m);
\]
for $F = 1$ this is $\pi^{s-1/2}\Gamma(\frac{1-s}{2})\Gamma(\frac s2)^{-1}m^{s-1}$. The functional equation of $\zeta$ expresses that
$\widehat G_s$ is again a power. For a general $F$, $\widehat G_s$ is not of the form $|\xi|^{s'}F'(|\xi|)$ with $F'$
regular, and there is no functional equation; but the split point $u_0$ makes both sides converge rapidly.
\end{remark}

\begin{remark}[Why $k^2$]\label{rem:whyk2}
One can also start from the Laplace representation in $k$ instead of $k^2$: for $k > R$,
\[
  F(k) = \int_0^\infty e^{-kt}B(t)\dd t, \qquad B(t) = \sum_{n=1}^{\infty}a_n\frac{t^{n-1}}{(n-1)!},
\]
with the Borel transform $B$, which is entire of exponential type $R$. Then $\sum_{k\ge k_0}F(k) =
\int_0^\infty B(t)e^{-(k_0-1)t}(e^t-1)^{-1}\dd t$ is a Bose--Einstein integral, and the analogue of the theta
transformation is the partial fraction expansion
\[
  \frac{1}{e^t-1} = \frac1t - \frac12 + \sum_{m=1}^{\infty}\frac{2t}{t^2+4\pi^2m^2},
\]
whose terms decay only like $m^{-2}$, because $e^{-t|x|}$ is not smooth at $x = 0$. Expanding $t/(e^t-1)$ in powers of
$t$ instead gives the Bernoulli numbers and Euler--Maclaurin summation (Section~\ref{sec:em}). Among the weights
$e^{-u|x|^\alpha}$, analyticity at $x = 0$ requires $\alpha$ to be an even integer, and $\alpha = 2$ is the smallest choice
and the only one with an elementary, self-dual transform. For $\alpha = 4$ the dual weight is still holonomic, but the
$k$-terms decay like $e^{-k^4u_0}$ and the dual terms like $e^{-c(m^4/u_0)^{1/3}}$, so that again $k_{\max}m_{\max} \asymp p$
and nothing is gained.
\end{remark}

\subsection{The two sides are holonomic}\label{sec:twosides}
In this subsection $F$ is regular and analytic at infinity, so that $f$ satisfies (H). Holonomy is what makes each term
cheap: a value of a holonomic function with small parameters, at a point with few bits, costs $\widetilde O(p)$ by
binary splitting or the bit-burst algorithm (Section~\ref{sec:evalhol}). Both sides of Theorem~\ref{thm:split} are built from two integral transforms of the kernel, the truncated Laplace
transform and the dual moments
\[
  \mathcal L[g](\kappa) = \int_{u_0}^\infty e^{-\kappa u}g(u)\dd u, \qquad Z_t[g](c) = \int_0^{u_0}u^{-t}e^{-c/u}g(u)\dd u.
\]
For fixed $s$, $I_k(s) = \mathcal L[h_s](k^2)$ and $J_m(s) = Z_{1/2}[h_s](\pi^2m^2)$, and with
\eqref{eq:kernelparts},
\[
  Z_{1/2}[h_s](c) = Z_{(3-s)/2}[G_0](c) + Z_{1-s/2}[G_1](c).
\]
On the dual side, differentiation in $c$ raises the index $t$, while integration by parts together with the equation
of the kernel lowers it again; so a finite window of indices is closed under $\partial_c$. For a $C^1$ function $g$ on $[0,u_0]$, integrating
$\frac{\mathrm d}{\mathrm du}(u^{1-t}e^{-c/u}g) = u^{-t}e^{-c/u}((1-t)g + \frac cug + \vartheta g)$ gives, with $E = e^{-c/u_0}$,
\begin{equation}\label{eq:ibp}
  c\,Z_{t+1}[g] = E\,u_0^{1-t}g(u_0) - (1-t)\,Z_t[g] - Z_t[\vartheta g],
\end{equation}
and differentiation under the integral sign gives $-\partial_cZ_t[g] = Z_{t+1}[g]$. Also $Z_t[u^jg] = Z_{t-j}[g]$.
Write $C_t = Z_t[1]$; by \eqref{eq:ibp}, $c\,C_{t+1} = Eu_0^{1-t} - (1-t)C_t$.

\begin{proposition}\label{prop:dual}
Let $\mathbf G$ satisfy \eqref{eq:Gsys} with $A = \sum_{i=0}^{d}A_iu^i$ and $b = \sum_{i=0}^{e}b_iu^i$, and fix $t$.
Let $Y(c)$ be the vector formed by $W_{t-i} = Z_{t-i}[\mathbf G](c)$ for $0 \le i \le d$, by $C_{t-i}(c)$ for
$0 \le i \le e$, by $E(c)$, and by the $\ell$ numbers $E(c)\,\mathbf G(u_0)_j$. Then
\begin{equation}\label{eq:fuchsys}
  c\,Y'(c) = (B_0 + cB_1)\,Y(c),
\end{equation}
where the entries of $B_0, B_1$ lie in $\mathbb K(s, u_0, u_0^{t})$ and depend only on $A$, $b$, $t$ and $u_0$, and $Y$
is analytic in $\operatorname{Re}c > 0$. In particular $Z_{1/2}[h_s]$ is holonomic, with $c = 0$ as its only finite
singular point.
\end{proposition}

\begin{proof}
By \eqref{eq:ibp} applied componentwise and \eqref{eq:Gsys},
\[
  c\,W_{t+1} = E\,u_0^{1-t}\mathbf G(u_0) - (1-t)W_t - \sum_{i=0}^{d}A_iW_{t-i} - \sum_{i=0}^{e}b_iC_{t-i}.
\]
Since $-\partial_cW_t = W_{t+1}$, $c\,\partial_cW_t = -c\,W_{t+1}$ is the right-hand side above with the opposite sign, a
combination of components of $Y$ with constant coefficients, and $c\,\partial_cW_{t-i} = -c\,W_{t-i+1}$ for $i \ge 1$ is
$c$ times a component of $Y$. The same holds for the $C_{t-i}$, and for the last block, since
$c\,\partial_cE = -(c/u_0)E$. The integrals converge locally uniformly for $\operatorname{Re}c > 0$.
\end{proof}

Similarly $\mathcal L[h_s](\kappa)$ is holonomic in $\kappa$, and on the $k$-side the series in $n$
for $I_k(s)$ has terms satisfying a linear recurrence in $n$ with coefficients in $\QQ(s,k,u_0)$, since
$Q(a+1,x) = Q(a,x) + x^ae^{-x}/\Gamma(a+1)$. The algorithms sum this series for each $k$ (Section~\ref{app:kterms}).

\emph{Starting values.} The point $c = \pi^2m^2$ is transcendental, so binary splitting does not apply there
directly. Instead we compute $Y$ at a nearby rational point $c_0$ by expanding $g$ termwise, and the bit-burst algorithm
carries it to $c$ (Theorem~\ref{thm:complexity}). Let
\[
  D_a(c) = Z_{1-a}[1](c) = \int_0^{u_0}u^{a-1}e^{-c/u}\dd u = c^a\,\Gamma\Bigl(-a, \frac{c}{u_0}\Bigr), \qquad E = e^{-c/u_0},
\]
so that $D_a = C_{1-a}$. For $g = \sum_jg_ju^j$ and a rational $c_0$,
\begin{equation}\label{eq:startvals}
  Z_t[g](c_0) = \sum_{j=0}^{\infty}g_jD_{j-t+1}(c_0),
\end{equation}
and integration by parts with $g = u^a$ gives
\begin{equation}\label{eq:Drec}
  a\,D_a + c\,D_{a-1} = E\,u_0^a .
\end{equation}
For $a \in \frac12 + \ZZ$ this expresses every $D_a$ as $\alpha_aD_{-1/2} + \beta_aE$ with
$D_{-1/2} = \sqrt{\pi/c}\erfc\sqrt{c/u_0}$, and for $a \in \ZZ$ in terms of $E_1(c/u_0)$ and $E$; in general one
incomplete gamma value per class of $a$ modulo $\ZZ$ is needed.

\begin{remark}[Periodic weights and twists]\label{rem:twist}
Let $\chi$ have period $q$. Splitting $\chi$ into its even and odd parts, it suffices to treat $\chi$ of parity
$\delta \in \{0,1\}$, i.e.\ $\chi(-k) = (-1)^\delta\chi(k)$. Then $w(k) = \chi(k)k^\delta$ is even and
$\chi(k)F(k)k^{-s} = w(k)F(k)k^{-s-\delta}$. Poisson summation applied to $x^\delta e^{-ux^2}$ on each residue class
modulo $q$ gives the theta transformation
\[
  \theta_\chi(u) = \sum_{k\in\ZZ}\chi(k)k^\delta e^{-k^2u} = c_\delta\,u^{-1/2-\delta}\sum_{n\in\ZZ}\hat\chi(n)\,n^\delta
  e^{-\pi^2n^2/(q^2u)},
\]
with $c_\delta = (-i)^\delta\pi^{1/2+\delta}q^{-1-\delta}$ and the Gauss sums
\[
  \hat\chi(n) = \sum_{a=1}^{q}\chi(a)e^{2\pi ian/q};
\]
the terms $\pm n$ are equal. Exactly as in the proof of
Theorem~\ref{thm:split},
\[
  \sum_{k=k_0}^{\infty}\chi(k)F(k)k^{-s} = \sum_{k=k_0}^{\infty}w(k)I_k(s+\delta)
  + c_\delta\sum_{n=1}^{\infty}\hat\chi(n)\,n^\delta J^\chi_n(s) + M^\chi(s),
\]
where $J^\chi_n(s) = Z_{1/2+\delta}[h_{s+\delta}](\pi^2n^2/q^2)$ and
\[
  M^\chi(s) = \int_0^{u_0}\Bigl([\delta = 0]\bigl(\tfrac12c_0\hat\chi(0)u^{-1/2} - \tfrac12\chi(0)\bigr)
  - \sum_{k=1}^{k_0-1}w(k)e^{-k^2u}\Bigr)h_{s+\delta}(u)\dd u,
\]
first for $\operatorname{Re}s$ large, then by continuation. For $\chi(k) = e^{2\pi i\alpha k}$ with real $\alpha$
(not periodic), the even and odd parts are $\cos2\pi\alpha k$ and $i\sin2\pi\alpha k$, Poisson summation gives
$\sum_{k\in\ZZ}\cos(2\pi\alpha k)e^{-k^2u} = \sqrt{\pi/u}\sum_{n\in\ZZ}e^{-\pi^2(n-\alpha)^2/u}$ and its derivative in
$\alpha$, and the dual terms are the same moments at the points $c = \pi^2(n-\alpha)^2$.
\end{remark}

\section{Damped functions: polylogarithms and damped zeta functions}\label{sec:beyond}

The proof of Theorem~\ref{thm:split} uses two facts: that the summand is a superposition of Gaussians,
\[
  F(x)x^{-s} = \int_0^\infty e^{-x^2u}h_s(u)\dd u \qquad (x \ge k_0),
\]
and the theta transformation. The algorithm needs a third: that the truncated Laplace transform $\mathcal L[h_s](\kappa)$
and the dual moment $Z_{1/2}[h_s](c)$ (Section~\ref{sec:twosides}) are holonomic in $\kappa$ and $c$, with series at
rational points whose terms are P-recursive.

This section extends the splitting, and with it the algorithm, to the damped functions $\Li_F$, $\Phi_F$ and
$\Lambda_F$ of \eqref{eq:functions}, whose summands are a regular function times an exponential factor. There are two
ways to add such a factor. A geometric factor $z^k$ can be kept on the theta side, as a twist with a complex
characteristic, at the price of complex dual points (Section~\ref{sec:geometric}); this generalizes Crandall's
splitting of the Lerch transcendent. A stretched exponential $e^{-\tau k^r}$ has no such interpretation, but it can be
moved into the kernel, whose expansion is then at $u = \infty$ instead of $u = 0$ (Sections~\ref{sec:stretched}
and~\ref{sec:products}).

Examples with $F \ne 1$ are the logarithm of Somos's quadratic recurrence constant and the generalized
Euler-constant function of Sondow and Hadjicostas (Section~\ref{sec:somos}),
\[
  \log\sigma = \sum_{k=1}^{\infty}\frac{1}{2^k}\log\Bigl(1+\frac1k\Bigr), \qquad
  \gamma(z) = \sum_{k=1}^{\infty}z^{k-1}\Bigl(\frac1k - \log\Bigl(1+\frac1k\Bigr)\Bigr).
\]
Derivatives in $s$ give logarithmic weights. For instance, Fisher's logarithmic series distribution
$P(k) = z^k/(kL)$, $L = -\log(1-z)$, $0 < z < 1$, a classical model of species abundance \cite{FCW43}, has the
entropy
\[
  H(z) = -\sum_{k=1}^{\infty}P(k)\log P(k) = \log L - \frac{z\log z}{(1-z)L} - \frac1L\,\partial_s\Li_s(z)\big|_{s=1},
\]
where $-\partial_s\Li_s(z)|_{s=1} = \sum_kz^k\log(k)/k$.

Damped zeta functions occur, for example, in partition functions over energy levels that grow like a fractional power
of the quantum number, such as the Landau levels of graphene, and in connection with
differential equations and probability distributions with a fractional parameter; see Section~\ref{sec:expsqrt}.

\subsection{The polylogarithm: geometric factors}\label{sec:geometric}
Let $F$ satisfy \eqref{eq:f} and let $0 < |z| < 1$. Write $z = e^{2\pi i\alpha}$ with $\alpha = \alpha_1 + i\alpha_2$,
$0 \le \alpha_1 < 1$, $\alpha_2 = \log(1/|z|)/(2\pi) > 0$, and put
\[
  c_n = \pi^2(n-\alpha)^2 \quad (n \in \ZZ), \qquad C_k = \frac{z^k + z^{-k}}2, \qquad S_k = \frac{z^k - z^{-k}}2.
\]
For $w \in \CC$ let $h_w$ be the kernel \eqref{eq:kernel} of $F(x)x^{-w}$, and
\begin{gather*}
  I_k(w) = \mathcal L[h_w](k^2), \qquad J^{(\delta)}_w(c) = Z_{1/2+\delta}[h_w](c), \\
  L_k(w) = \int_0^{u_0}e^{-k^2u}h_w\dd u, \qquad H(w) = \int_0^{u_0}h_w\dd u,
\end{gather*}
for $\delta \in \{0,1\}$ and $\operatorname{Re}c > 0$. The integrals $L_k(w)$ and $H(w)$ converge only for
$\operatorname{Re}w > -n_0$; for all $w$ they denote, as for $M$ in Theorem~\ref{thm:split}, the termwise
continuations
\[
  L_k(w) = \sum_{n=0}^{\infty}a_n\,k^{-w-n}P\Bigl(\frac{w+n}2, k^2u_0\Bigr), \qquad
  H(w) = \sum_{n=0}^{\infty}a_n\frac{u_0^{(w+n)/2}}{\Gamma(\frac{w+n}2+1)},
\]
which are entire functions of $w$. By \eqref{eq:startvals},
\[
  J^{(\delta)}_w(c) = \sum_{n=0}^{\infty}\frac{a_n}{\Gamma(\frac{w+n}2)}D_{(w+n-1)/2-\delta}(c), \qquad
  D_a(c) = c^a\Gamma\Bigl(-a, \frac c{u_0}\Bigr).
\]
Each $D_a$ is analytic on $\CC\setminus(-\infty,0]$: for $a \notin \ZZ_{\ge0}$ it is $c^a\Gamma(-a)$ minus $u_0^a$ times an entire function of $c$, and for $a \in \ZZ_{\ge0}$ it is $-\frac{(-c)^a}{a!}\log\frac c{u_0}$ plus an entire function. The series converges locally uniformly there, since its terms are
$O(R^n(|c|+u_0)^{n/2}e^{|c|/u_0}/\Gamma(\frac n2))$. Hence $J^{(\delta)}_w$ continues analytically to
$\CC\setminus(-\infty,0]$. All $c_n$ lie in this domain except $c_0 = -\pi^2\alpha_2^2$ when $\alpha_1 = 0$, i.e.\
$0 < z < 1$; there $J^{(\delta)}_w(c_0)$ denotes the boundary value from $\operatorname{Im}c > 0$.

\begin{theorem}\label{thm:geometric}
With the notation above, the following hold.
\begin{enumerate}
\item For all $s \in \CC$,
\begin{multline*}
  \Li_F^{\ge k_0}(s,z) = \sum_{k=k_0}^{\infty}\bigl(C_kI_k(s) + kS_kI_k(s+1)\bigr) \\
  + \frac{\sqrt\pi}2\sum_{n\in\ZZ}\bigl(J^{(0)}_s(c_n) - i\pi(n-\alpha)J^{(1)}_{s+1}(c_n)\bigr) \\
  - \frac12H(s) - \sum_{k=1}^{k_0-1}\bigl(C_kL_k(s) + kS_kL_k(s+1)\bigr),
\end{multline*}
where the $k$-terms are $O(|z|^{-k}e^{-(k^2-R^2)u_0})$ and the dual terms are $O(e^{-\operatorname{Re}c_n/u_0})$,
with $\operatorname{Re}c_n = \pi^2((n-\alpha_1)^2 - \alpha_2^2)$.
\item If $F \in \RegQ$, $\chi$ is periodic with
algebraic values, $z$ and $s_0$ are algebraic, and $j \ge 0$ is fixed, then the $j$-th Taylor coefficient at
$s_0$ of $\sum_{k=1}^{\infty}\chi(k)z^kF(k)k^{-s}$, for $\chi = 1$ of $\Li_F(s,z)$, can be computed to $p$ bits in
$O(p^{1/2}\Mul(p)\log^2p)$ bit operations.
\end{enumerate}
\end{theorem}

\begin{proof}
(1) Let first $\alpha$ be real with $0 < \alpha < 1$, i.e.\ $|z| = 1$, $z \ne 1$, and $\operatorname{Re}s$ large. Proceed
as in the proof of Theorem~\ref{thm:split}, with $\Theta$ replaced by $\sum_{k\ge k_0}e^{2\pi i\alpha k}e^{-k^2u}$,
split into even and odd parts as in Remark~\ref{rem:twist}: writing the odd part of the summand as $k\cdot F(k)k^{-s-1}$
and using
\begin{gather*}
  \sum_{k\in\ZZ}\cos(2\pi\alpha k)e^{-k^2u} = \sqrt{\frac\pi u}\sum_{n\in\ZZ}e^{-c_n/u}, \\
  \sum_{k\in\ZZ}k\sin(2\pi\alpha k)e^{-k^2u} = -\frac{\pi^{3/2}}{u^{3/2}}\sum_{n\in\ZZ}(n-\alpha)e^{-c_n/u}
\end{gather*}
gives the identity. Now fix $s$ with $\operatorname{Re}s$ large. The right-hand side is analytic in $\alpha$ on the
strip $0 < \operatorname{Re}\alpha < 1$, where no $c_n$ meets $(-\infty,0]$; the left-hand side is analytic for
$\operatorname{Im}\alpha > 0$ and continuous up to $\operatorname{Im}\alpha = 0$. Their difference vanishes on the
segment $0 < \alpha < 1$, hence, by the reflection principle, on the half-strip. For $\alpha_1 = 0$ let
$\alpha_1 \to 0^+$; then $\operatorname{Im}c_0 = 2\pi^2\alpha_1\alpha_2 > 0$, so $c_0$ tends to the cut from above, and the
other terms are continuous. Both sides are entire in $s$. The bounds follow from Theorem~\ref{thm:split} and from
$|D_a(c)| \le |e^{-c/u_0}|u_0^{\operatorname{Re}a}/\operatorname{Re}a$ for $\operatorname{Re}c > 0$,
$\operatorname{Re}a > 0$.
(2) See Appendix~\ref{app:geometric}.
\end{proof}

For $F = 1$ and $k_0 = 1$ the pieces are incomplete gamma functions:
\begin{gather*}
  I_k(s) = k^{-s}Q\Bigl(\frac s2,k^2u_0\Bigr), \qquad kI_k(s+1) = k^{-s}Q\Bigl(\frac{s+1}2,k^2u_0\Bigr), \\
  J^{(0)}_s(c) = \frac{c^{(s-1)/2}}{\Gamma(\frac s2)}\Gamma\Bigl(\frac{1-s}2,\frac c{u_0}\Bigr), \qquad
  H(s) = \frac{2u_0^{s/2}}{s\Gamma(\frac s2)}, \\
  J^{(1)}_{s+1}(c) = \frac{c^{s/2-1}}{\Gamma(\frac{s+1}2)}\Gamma\Bigl(1-\frac s2,\frac c{u_0}\Bigr).
\end{gather*}
Then part~(1) is Crandall's formula for $\Li_s(z)$ \cite[(34)]{Cra12} with $\lambda = u_0$. With a shift of the lattice
(Section~\ref{sec:twistshift}) it is his formula for the Lerch transcendent \cite[(14)]{Cra12}, which at $z = 1$ is
his series for the Hurwitz zeta function \cite[(30)]{Cra12}; see also \cite{BB15}. Crandall states it for
$\operatorname{Re}z \in [0,1)$, as an analytic continuation. What the theorem adds is the general regular factor $F$,
for which the incomplete gamma functions become the holonomic functions $I_k$ and $J^{(\delta)}_w$, and the bound
in~(2).

\subsection{The damped zeta function of $F = 1$}\label{sec:stretched}
For $\tau > 0$ and $0 < r < 2$ let
\[
  \Lambda(s;\tau,r) = \sum_{k=1}^{\infty}e^{-\tau k^r}k^{-s},
\]
an entire function of $s$. For $r = 1$ this is the polylogarithm $\Li_s(z)$ at $z = e^{-\tau}$, inside the unit disk;
for $r < 1$ the series converges only like $e^{-\tau k^r}$, and direct summation needs $(p\log2/\tau)^{1/r}$
terms. Expanding $e^{-\tau x^r}x^{-s}$ at $x = 0$ and transforming termwise gives the kernel
\[
  h_s(u) = \sum_{j=0}^{\infty}c_ju^{a_j-1}, \qquad a_j = \frac{s - rj}{2}, \qquad c_j = \frac{(-\tau)^j}{j!\,\Gamma(a_j)},
\]
which is $u^{s/2-1}$ times an entire function of $u^{-r/2}$, since $1/|\Gamma(a_j)| \ll \Gamma(rj/2)$ and $r < 2$. Let
$D_a(c) = \int_0^{u_0}u^{a-1}e^{-c/u}\dd u$ as in Section~\ref{sec:twosides}, and
\[
  M_\tau(z) = \int_0^\infty x^{z-1}e^{-\tau x^r}\dd x = \frac{\Gamma(z/r)\,\tau^{-z/r}}{r}.
\]

\begin{proposition}\label{prop:expsplit}
For $u_0 > 0$ and $s \notin \{rj, 1+rj : j \ge 0\}$ (where single terms have poles that cancel),
\[
  \Lambda(s;\tau,r) = \sum_{k=1}^{\infty}I_k + \sqrt\pi\sum_{m=1}^{\infty}J_m + M,
\]
where
\begin{gather*}
  I_k = \sum_{j=0}^{\infty}\frac{(-\tau)^j}{j!}\,k^{rj-s}\,Q(a_j, k^2u_0), \qquad
  J_m = \int_0^{u_0}u^{-1/2}e^{-\pi^2m^2/u}h_s(u)\dd u, \\
  M = M_\tau(1-s) - \frac{\sqrt\pi}2T_{1/2} + \frac12T_1, \qquad
  T_w = -\sum_{j=0}^{\infty}c_j\frac{u_0^{a_j+w-1}}{a_j+w-1}.
\end{gather*}
If $r < 1$, or $r = 1$ and $\tau < 2\pi m$, then $J_m = \sum_{j=0}^{\infty}c_j\,D_{a_j-1/2}(\pi^2m^2)$.
\end{proposition}

\begin{proof}
For $\operatorname{Re}s$ sufficiently negative, $\int_0^\infty x^{z-1}e^{-x^2u}\dd x = \frac12\Gamma(\frac z2)u^{-z/2}$
gives $\int_0^\infty u^{-z/2}h_s(u)\dd u = 2M_\tau(z-s)/\Gamma(\frac z2)$ in a strip, so $h_s$ is the inverse Laplace
transform in $x^2$ of $e^{-\tau x^r}x^{-s}$ (for $r \le 2$, $y \mapsto e^{-\tau y^{r/2}}$ is completely monotone),
and it is exponentially small as $u \to 0$. Now proceed as in the proof of Theorem~\ref{thm:split} with $k_0 = 1$:
the $k$-terms are integrated termwise, which converges since $u$ stays away from $0$. The $j$-th term of the
series for $J_m$ is $O\bigl((\tau(\pi m)^{-r})^j\,\Gamma(\frac{rj}2+O(1))^2/j!\bigr)$ (Appendix~\ref{app:stretched}),
so for $r < 1$, and for $r = 1$ with $\tau < 2\pi m$, the series of absolute values of the integrals converges and
$J_m$ may also be integrated termwise. The main term is
$M = \int_0^{u_0}(\frac12\sqrt{\pi/u} - \frac12)h_s\dd u$, and the integrals over $(0, u_0]$ are complete integrals,
$\int_0^\infty u^{-1/2}h_s\dd u = 2M_\tau(1-s)/\sqrt\pi$ and $\int_0^\infty h_s\dd u = 0$ ($z = 1$ and $z = 0$), minus
the tails $T_{1/2}$ and $T_1$, integrated termwise. Both sides are analytic in $s$.
\end{proof}

The series for $J_m$ converges geometrically with ratio about $\tau/(2\pi m)$ when $r = 1$, and faster when $r < 1$.
For $r = 1$ and $2\pi m \le \tau$, and for every $m$ when $1 < r < 2$, it diverges: by Stirling's formula its terms
grow like $(j!)^{r-1}$ times a geometric factor. The identity itself holds for all $0 < r < 2$, with $J_m$ defined by
the integral.

\subsection{The damped zeta function of a regular $F$}\label{sec:products}
Now let $F$ satisfy \eqref{eq:f}, let $r = 1/q$ with an integer $q \ge 1$, and consider the damped zeta function
$\Lambda_F(s;\tau,r)$ of \eqref{eq:functions}, or rather its tail (the first $k_0 - 1$ terms are entire in $s$),
\[
  \Lambda_F^{\ge k_0}(s;\tau,r) = \sum_{k=k_0}^{\infty}e^{-\tau k^r}F(k)\,k^{-s}.
\]
Expanding $F(x) = \sum_na_nx^{-n}$ and using the kernels of $e^{-\tau x^r}x^{-s-n}$, the kernel of the summand is
\[
  h_s(u) = \sum_{\ell\in\ZZ}b_\ell\,\frac{u^{w_\ell-1}}{\Gamma(w_\ell)}, \qquad w_\ell = \frac{s+\ell/q}{2}, \qquad
  b_\ell = \sum_{n \ge \max(0,\ell/q)}a_n\frac{(-\tau)^{qn-\ell}}{(qn-\ell)!},
\]
where $b_\ell$ is the coefficient of $t^{\ell/q}$ in the Laurent expansion of $f(t)e^{-\tau t^{-1/q}}$ on
$0 < t < 1/R$. The series converges for every $u > 0$, infinitely in both directions: for $\ell \to \infty$ like the
kernel of Section~\ref{sec:splitting}, and for $\ell \to -\infty$ like that of Proposition~\ref{prop:expsplit}.

Write $h_s = h_+ + h_-$, where $h_+$ collects the terms with $\ell \ge 0$. Then $h_-$ is the kernel of $x^{-s}V(x)$,
where
\[
  V(x) = \sum_{\ell<0}b_\ell x^{-\ell/q} = \sum_{n=0}^{\infty}a_nx^{-n}\Bigl(e^{-\tau x^r} - \sum_{j=0}^{qn}\frac{(-\tau x^r)^j}{j!}\Bigr)
\]
is an entire function of $x^r$. By the Cauchy--Saalschütz representation of $\Gamma$ \cite[(5.9.5)]{DLMF},
\begin{multline*}
  M_V(z) = \int_0^\infty x^{z-s-1}V(x)\dd x = \sum_{n=0}^{\infty}a_n\,q\,\Gamma\bigl(q(z-s-n)\bigr)\,\tau^{-q(z-s-n)} \\
  \Bigl(-\frac1q < \operatorname{Re}(z-s) < 0\Bigr),
\end{multline*}
continued meromorphically by the same series.

\begin{proposition}\label{prop:prodsplit}
For $u_0 > 0$, $k_0 > R$ and $s$ outside a discrete set,
\[
  \Lambda_F^{\ge k_0}(s;\tau,r) = \sum_{k=k_0}^{\infty}I_k + \sqrt\pi\sum_{m=1}^{\infty}J_m + M_+ + M_-,
\]
where, with $I^-_k$ the part of $I_k$ with $\ell < 0$, and the restriction of $h_-$ to $u \ge u_0$ integrated termwise,
\begin{gather*}
  I_k = \sum_{\ell\in\ZZ}b_\ell\,k^{-2w_\ell}\,Q(w_\ell, k^2u_0), \qquad
  J_m = \sum_{\ell\in\ZZ}\frac{b_\ell}{\Gamma(w_\ell)}\,D_{w_\ell-1/2}(\pi^2m^2), \\
  M_+ = \sum_{\ell\ge0}b_\ell\Bigl(\frac{\sqrt\pi\,u_0^{w_\ell-1/2}}{(2w_\ell-1)\Gamma(w_\ell)} - \frac{u_0^{w_\ell}}{2\Gamma(w_\ell+1)}
  - \sum_{k=1}^{k_0-1}\frac{P(w_\ell, k^2u_0)}{k^{2w_\ell}}\Bigr), \\
  M_- = M_V(1) - \int_{u_0}^\infty\Bigl(\frac{\sqrt\pi}{2\sqrt u} - \frac12\Bigr)h_-(u)\dd u
  - \sum_{k=1}^{k_0-1}\bigl(k^{-s}V(k) - I^-_k\bigr).
\end{gather*}
\end{proposition}

As in Proposition~\ref{prop:expsplit}, the series for $J_m$ converges for $q \ge 2$, but for $q = 1$ only when
$\tau < 2\pi m$; in general $J_m$ is the integral $\int_0^{u_0}u^{-1/2}e^{-\pi^2m^2/u}h_s(u)\dd u$. The series for
$I_k$ needs $k > R$. The termwise integrals behind $M_+$ and $M_-$ converge only for $\operatorname{Re}w_\ell > \frac12$
and $\operatorname{Re}w_\ell < 0$ respectively, so the terms of $M_\pm$ are their continuations in $s$; their poles
lie in the discrete exceptional set.

\begin{proof}
For $x > R$ the summand is $\int_0^\infty e^{-x^2u}h_s(u)\dd u$, and the proof of Theorem~\ref{thm:split} applies to
the $k$-terms, the dual terms and $h_+$ unchanged. For $h_-$ the integrals over $(0,u_0]$ in the main term are complete
integrals minus tails, as in Proposition~\ref{prop:expsplit}: $\int_0^\infty u^{-z/2}h_-\dd u = 2M_V(z)/\Gamma(\frac z2)$,
which is $2M_V(1)/\sqrt\pi$ at $z = 1$ and $0$ at $z = 0$, and $\int_0^\infty e^{-k^2u}h_-\dd u = k^{-s}V(k)$ for every
$k > 0$, since $V$ is entire.
\end{proof}

\subsection{Twists, shifts and the Lerch transcendent}\label{sec:twistshift}
A periodic weight $\chi(k)$ acts only on the theta side (Remark~\ref{rem:twist}) and combines with both mechanisms:
the kernels are unchanged, the odd part uses the kernel of the summand divided by $x$, and the dual points are shifted
as in Remark~\ref{rem:twist}. In the kernel route, a complex $z$ can also be handled by putting $|z|^k$ into the kernel
and treating $(z/|z|)^k = e^{2\pi i\alpha_1k}$ as an additive twist, with the real dual points $\pi^2(n-\alpha_1)^2$.
The power may also be shifted:
\[
  z^k(k+a)^{-s}F(k) = z^{-a}\psi(k+a), \qquad \psi(y) = z^yF(y-a)y^{-s},
\]
and $F(y-a)$ is again regular at infinity. Poisson summation over the shifted lattice $n + a$ (a theta function with
characteristic), applied to the even and the odd part of $\psi$ (the latter written as $y$ times the kernel of
$\psi(y)/y$), gives dual terms with the factors $\cos 2\pi ma$ and $\sin 2\pi ma$. The lattice points with
$|n+a| < k_0$ are treated like the terms $k < k_0$ in Theorem~\ref{thm:split}. This is Crandall's sign trick
\cite[\S2.4]{Cra12}. For $F = 1$ it gives the classical Lerch transcendent $\Phi(z,s,a) = \sum_{k=0}^{\infty}z^k(k+a)^{-s}$.

\subsection{Limits of the method}\label{sec:limits}
The splitting gains nothing when the kernel is itself a theta series: then each $k$-term is a rapidly convergent sum
and the method sees a two-dimensional lattice. For instance, by its partial fraction expansion,
$\operatorname{sech}(\pi x)$ has the kernel
\[
  \frac1\pi\sum_{n=0}^{\infty}(-1)^n(2n+1)e^{-(n+1/2)^2u},
\]
and the Lambert series $\sum_kk^{-s}/(e^{\tau k}-1) = \sum_{j=1}^{\infty}\Lambda(s;j\tau,1)$ needs $p/\tau$
polylogarithms. Then about $p$ terms are needed, and nothing is gained over direct summation.
The same happens for $L$-functions of degree $d \ge 2$, whose dual kernels decay like $e^{-c(n/\sqrt N)^{2/d}}$, and
for lattice sums $\sum_{k\in\ZZ^d}F(Q(k))$ with a positive definite quadratic form $Q$: replacing $e^{-k^2u}$ by
$e^{-Q(k)u}$ and using Ewald's theta transformation gives a splitting, but with $O(p^{d/2})$ terms.

Functions with AGM-type algorithms (theta, elliptic and modular functions) enter the splitting only through truncated
Laplace transforms and Gaussian moments of the kernel, not through point values. Complete Laplace transforms of theta
functions are elementary (Mittag-Leffler), but truncated ones lead back to the two-dimensional sums. We know no class
of summands, beyond periodic twists and the exponential factors above (and their products with regular functions),
for which both sides of the splitting are cheap; Section~\ref{sec:open} lists some candidates.

\section{Algorithms and complexity bounds}\label{sec:complexity}

In this section we study the computational complexity and practical costs
of theta splitting alongside several classical algorithms
for $\zeta_F$ and its relatives.
Table~\ref{tab:methods} summarizes the methods.

\begin{table}[ht]
\centering\small
\caption{Methods for $\zeta_F(s)$ to $p$ bits at a fixed point, for algebraic data ($s \in \Qbar$ and
$F \in \RegQ$) and for generic data ($s \in \CC$ or $F \in \Reg$ given by $p$-bit approximations).}
\label{tab:methods}
\begin{tabular}{llll}
\toprule
method & algebraic data & generic data & continuation \\
\midrule
direct summation & $2^{\Theta(p)}$ & $2^{\Theta(p)}$ & no \\
zeta-value expansion & $p^{2+o(1)}$ & $p^{2+o(1)}$ & yes \\
Euler--Maclaurin & $p^{2+o(1)}$ & $p^{2+o(1)}$ & in half-planes \\
alternating series$^\dagger$ & $p^{2+o(1)}$ & $p^{2+o(1)}$ & no \\
theta splitting (Theorem~\ref{thm:split}) & $p^{3/2+o(1)}$ & $p^{2+o(1)}$ & yes \\
partial fractions$^\ddagger$ & $p^{1+o(1)}$ & $p^{3/2+o(1)}$ & no \\
\bottomrule
\multicolumn{4}{l}{\footnotesize ${}^\dagger$Twisted series only. ${}^\ddagger$Rational $F$ at integer $s$ only
(Section~\ref{sec:rational}).}
\end{tabular}
\end{table}

Costs are bit complexities on a multitape Turing machine; $\Mul(p)$ is a bound for the cost of multiplying
$p$-bit integers satisfying the usual regularity conditions ($\Mul(p)/p$ nondecreasing and
$\Mul(p\log p) = O(\Mul(p)\log p)$); one can take $\Mul(p) = p\log p$ \cite{HvdH21} or even
$p(\log p)^{1-\kappa}$ for an explicit $\kappa > 0$ \cite{OAI26}.

\emph{Generic data.} When $s$ or the data of $F$ are arbitrary complex numbers, given by $p$-bit approximations, binary
splitting no longer applies, and a single value of a holonomic function to $p$ bits costs $p^{3/2+o(1)}$ by the
baby-step giant-step method \cite{Bor87,CC90,BGS07}, against $p^{2+o(1)}$ for summing its series
term by term. The theta splitting needs $O(\sqrt p)$ such values, hence $p^{2+o(1)}$. The classical methods need
$\Theta(p)$ values $F(k)$, which would cost $p^{5/2+o(1)}$ if computed one at a time, but which can be computed
together: the coefficients $a_0, \dots, a_M$ of the expansion at infinity, $M = O(p)$, cost $O(p)$ operations by their
recurrence, and the truncated expansion $\sum_{n\le M}a_nx^n$ is then evaluated at all the points $x = 1/k$,
$2R \le k < N$, by fast multipoint evaluation, which needs only $O(p)$ guard bits for points in a disk where the
series converges geometrically \cite{KZ08,Mor21}. This costs $\widetilde O(p^2)$, the $O(1)$ values with $k < 2R$
cost $p^{3/2+o(1)}$ each, and the powers $k^{-s}$ cost $\widetilde O(p)$ each, so the classical methods stay at
$p^{2+o(1)}$ (for the zeta-value expansion, see Section~\ref{sec:zetavalue}).

\subsection{Rational summands and the gamma function}\label{sec:rational}
The gamma function is the model for both kinds of data. For $\operatorname{Re}a > 0$ and $N \approx p\log2$,
\[
  \Gamma(a) = \gamma(a,N) + \Gamma(a,N), \qquad \gamma(a,N) = N^ae^{-N}\sum_{i=0}^{\infty}\frac{N^i}{a(a+1)\cdots(a+i)},
\]
where $\Gamma(a,N) = O(N^{\operatorname{Re}a}e^{-N})$ is negligible and $O(p)$ terms of the series suffice. For
algebraic $a$ the terms are P-recursive with coefficients of $O(\log p)$ bits, so binary splitting gives $\Gamma(a)$ in
$O(\Mul(p)\log^2p)$ \cite[Chapter~10]{BB87}, \cite{BB88,Bor87,BZ10,Joh23}, and the same series over truncated power series in $a$ gives $\log\Gamma$, $\psi$ and
the polygamma functions. For $a$ given only to $p$ bits the coefficients have $p$ bits, binary splitting no longer
applies, and the baby-step giant-step evaluation of the product of the term ratios gives $p^{3/2+o(1)}$
\cite[Chapter~10]{BB87}, \cite{BB88,Bor87,CC90,Joh23}. Theorem~\ref{thm:complexity} uses the same series for its incomplete gamma values.

For rational $F$, partial fractions reduce $\zeta_F$ at integer points to such values. Write $F(k)$ as a polynomial plus
a combination of terms $(k+b)^{-m}$. At an integer $s$, $F(k)k^{-s}$ is again of this form, and
$\sum_{k\ge1}(k+b)^{-m} = \zeta(m, 1+b) = (-1)^m\psi^{(m-1)}(1+b)/(m-1)!$ for $m \ge 2$, while the terms with $m = 1$
combine to $-\sum_bc_b\psi(1+b)$ (their coefficients $c_b$ sum to zero when the sum converges) and the polynomial part
gives values $\zeta(-j)$ in the regularized case. In the same way, $\sum_k\log R(k)$ for rational $R$ is a combination
of values $\log\Gamma(1+b)$ at the zeros and poles of $R$, by the Weierstrass product. These values therefore cost
$\widetilde O(p)$ when $F$ has algebraic coefficients, and $p^{3/2+o(1)}$ when its coefficients are arbitrary complex
numbers, the zeros and poles being computed to $p$ bits in $\widetilde O(p)$. No such reduction exists at
non-integer $s$, or for derivatives in $s$: the constants $\mathcal A$ and $\mathcal C$ of Table~\ref{tab:constants} are
derivatives at $s = 0$ of $\zeta_F$ for rational $F$, and for them the theta splitting, whose pieces are then incomplete
gamma values (Remark~\ref{rem:algebraic}), is the fastest method we know.

\subsection{Theta splitting}
\begin{theorem}\label{thm:complexity}
Let $F \in \RegQ$, let $\chi$ be periodic with algebraic values, let $s_0 \in \Qbar$, and let
$j \ge 0$ be fixed. Then the $j$-th coefficient of the Laurent expansion of $L_F(s,\chi)$ at $s_0$ can be computed to
$p$ bits in $O(p^{1/2}\Mul(p)\log^2 p)$ bit operations.
\end{theorem}

The proof is a cost analysis of each piece of Theorem~\ref{thm:split}, given in Appendix~\ref{app:cost} and
summarized in Table~\ref{tab:pieces}. After a reduction to $f$ analytic at $0$ and $\chi$ even
(Lemma~\ref{lem:reduction}, Remark~\ref{rem:twist}), each of the $O(\sqrt p)$ pieces is a series with P-recursive
terms and $O(\log p)$-bit coefficients, or the value of a fixed holonomic system at a point with $O(\log p)$ bits, and
so costs $O(\Mul(p)\log^2p)$; the derivatives in $s$ are obtained by the same computations over truncated power series
in $s - s_0$.

\begin{table}[ht]
\centering\footnotesize
\caption{The pieces of Theorem~\ref{thm:split} in the proof of Theorem~\ref{thm:complexity}; each costs
$O(\Mul(p)\log^2p)$ bit operations.}
\label{tab:pieces}
\begin{tabular}{@{}llll@{}}
\toprule
piece & number & form & tool \\
\midrule
$k$-terms $I_k$ & $O(\sqrt p)$ & double series \eqref{eq:Ik} & binary splitting \\
incomplete gamma & $O(\sqrt p)$ & hypergeometric series & binary splitting \\
dual terms $J_m$ & $O(\sqrt p)$ & value of $cY' = (B_0 + cB_1)Y$ at $\pi^2m^2$ & bit-burst \\
starting values $Y(c_0)$ & $O(\sqrt p)$ & series \eqref{eq:startvals} & binary splitting \\
main term $M$ & $1$ & series in $n$ & binary splitting \\
\bottomrule
\end{tabular}
\end{table}

The splitting is applied to the tail $L_F^{\ge k_0}$ with $k_0 > R$. If $F(k) = G(k)$ $(k \ge k_1)$ with $k_1 < k_0$, the finitely many
values $G(k)$, $k_1 \le k < k_0$, are obtained from the expansion of $G$ at $x = k_0$ by analytic continuation along
the real axis, in quasilinear time \cite{CC90,vdH99}, so they do not affect the bound.

\emph{Balancing.} The numbers of $k$-terms and dual terms needed for $p$ bits, $k_{\max}$ and $m_{\max}$ in
\eqref{eq:counts}, satisfy $k_{\max}m_{\max} \approx qp\log2/\pi$ ($q$ the period of a twist), independently of $u_0$. Hardy's uncertainty principle (a function and its Fourier transform cannot both decay
faster than Gaussians) suggests that no other smoothing weight does essentially better than the Gaussian; we do not
make this precise (compare Remark~\ref{rem:whyk2}). If a $k$-term costs $C_I$ and a dual term $C_J$ bit
operations, the total is of order $\sqrt{p/u_0}\,C_I + \sqrt{pu_0}\,C_J \ge 2\sqrt{p\,C_IC_J}$, with equality at
$u_0 = C_I/C_J$. The method is therefore subquadratic as soon as $C_IC_J = O(p^{3-\delta})$: the two sides need not be of
the same kind, and one of them may cost up to $p^{2-\delta}$ per term if the other costs $\widetilde O(p)$.

\begin{remark}[Practical points]\label{rem:practical}
(1) The terms are independent, so the algorithm parallelizes trivially, with $O(p)$ memory per term. (2) The
$k$-terms need only relative precision $p - (k^2u_0)\log_2 e$ and the dual terms $p - (c/u_0)\log_2e$, which also
shortens the series. (3) For $u_0 = 4^{-a}$ the arguments $k\sqrt{u_0}$ and $c_0/u_0$ are short dyadic numbers, so the
incomplete gamma values are cheap binary splitting series. (4) A dual term is typically several dozen times as
expensive as a $k$-term, so by the balancing argument above the optimal $u_0$ is below $1$ (in most of our applications
$u_0 = \frac1{16}$; for Somos's constant, whose $k$-terms are single incomplete gamma values, $u_0 = 1$). (5) In
floating-point evaluation of $M$, the functions $P(a,x)$ for $k < k_0$ must be computed directly from the lower
series and not as $1 - Q(a,x)$, since they are multiplied by $a_nk^{-n}$, which may be exponentially large.
\end{remark}

\begin{remark}[Where the factor $\pi$ goes]\label{rem:wherepi}
The $k$-terms depend only on $k$ and $x = k^2u_0$, which are rational for rational $u_0$. The dual terms depend on
$c = \pi^2m^2$ and $u_0$: with $y = c/u_0$,
\[
  J_m(s) = \sum_{n=0}^{\infty}a_n\,u_0^{(s+n-1)/2}\,\frac{y^{(s+n-1)/2}\,\Gamma(\frac{1-s-n}2, y)}{\Gamma(\frac{s+n}2)},
\]
so $u_0^{1/2}$ sets the scale of the kernel. Since $yu_0 = \pi^2m^2$, the dual side contains $\pi$ for every choice
of $u_0$: in the point $c$ when $u_0$ is rational, as above, or in the coefficients when $u_0 \in \pi^2\QQ$. The latter
choice makes $y$ rational, but $J_m$ then combines one incomplete gamma value and $e^{-y}$ with coefficients that are
values at $\pi$ of entire holonomic functions with rational coefficients, which again need the bit-burst algorithm;
and the $k$-terms become series at the transcendental points $x = k^2u_0$. Rational $u_0$ is the only choice that
keeps one side free of $\pi$. For $f = 1$ the kernel $u^{s/2-1}$ has no scale and $J_m$ depends on $y$ alone; then
the factor $\pi$ can be put on either side, or split between them as in Riemann's choice $u_0 = \pi$, at the same
cost.
\end{remark}

\subsection{Uniformity}\label{sec:uniform}
In Theorem~\ref{thm:complexity} everything is fixed except $p$. Its proof only uses that there are $O(\sqrt p)$
pieces, that their series have $O(p)$ terms and recurrences with $O(\log p)$-bit coefficients, and that all
intermediate quantities are bounded by $2^{O(p)}$. These properties survive when $s_0$ and the coefficients of $F$
grow moderately with $p$. This is what is needed to compute values at $s_0 = 1 - n$ exactly, such as Bernoulli numbers.

We write $H(\alpha)$ for the absolute multiplicative height of an algebraic number $\alpha$ and
$h(\alpha) = \log H(\alpha)$; in a fixed number field, an $\alpha$ with $H(\alpha) \le B$ can be written with
$O(\log B)$ bits.

\begin{theorem}[uniformity]\label{thm:uniform}
Fix a number field $\mathbb K \subset \CC$, an operator $L$ with coefficients in $\mathbb K[x]$ for which $\infty$ is
an ordinary or regular singular point with exponents in $\mathbb K$, and a basis $y_1, \dots, y_r$ of solutions of
the form \eqref{eq:frob} whose power series have coefficients in $\mathbb K$ and which continue analytically to
$[k_1, \infty)$. Fix also integers $q \ge 1$ and $j \ge 0$ and a constant $c > 0$. For $p \ge 2$, let
$F(k) = \sum_{i=1}^r\lambda_iy_i(k)$ $(k \ge k_1)$, let $\chi\colon\ZZ \to \mathbb K$ have period $q$, and let
$s_0 \in \mathbb K$, where
\[
  h(\lambda_i) \le cp, \qquad h(\chi(k)) \le cp, \qquad H(s_0) \le p^c, \qquad |s_0| \le \frac{cp}{\log p}.
\]
Then the $j$-th Laurent coefficient of $L_F(s,\chi)$ at $s_0$ can be computed with absolute error at most $2^{-p}$
in $O(p^{1/2}\Mul(p)\log^2p)$ bit operations and $O(p)$ bits of memory, with implied constants depending only on the
fixed data. If only $|s_0| \le cp$, the bounds become $O(p^{1/2}\Mul(p)\log^{7/2}p)$ bit operations and $O(p\log p)$
bits. The same holds for $\sum_{k=1}^{\infty}\chi(k)z^kF(k)k^{-s}$, in particular for $\Li_F(s,z)$, with a fixed
$z \in \mathbb K$, $0 < |z| < 1$.
\end{theorem}

The proof (Appendix~\ref{app:uniform}) checks the properties above piece by piece. The bound on $H(s_0)$ says that
$s_0$ has $O(\log p)$ bits, as binary splitting requires; an $s_0$ of $p$ bits is in effect generic data. The bound on
$|s_0|$ keeps the intermediate quantities below $2^{O(p)}$, or $2^{O(p\log p)}$ in the second case. The $\lambda_i$
and the values of $\chi$ enter linearly, which is why heights $e^{O(p)}$ are allowed for them. The theorem is not
uniform in the operator $L$ (Section~\ref{sec:openuniform}).

Let $B_n(x)$ and $E_n(x)$ be the Bernoulli and Euler polynomials, $B_n = B_n(0)$ and $E_n = 2^nE_n(\frac12)$, and, for a
function $\chi$ of period $f$, let $B_{n,\chi} = f^{n-1}\sum_{a=1}^f\chi(a)B_n(a/f)$ be the generalized Bernoulli
numbers.

\begin{corollary}\label{cor:bernoulli}
Fix a rational number $x$ and an integer $f \ge 1$. Then $B_n(x)$ and $E_n(x)$ can be computed exactly in
$O(n^{1/2}\Mul(n)\log^{7/2}n) \subset n^{3/2+o(1)}$ bit operations and $O(n\log n)$ bits of memory. In particular this
holds for $B_n$, for $E_n$, for the tangent and Genocchi numbers, and for $B_{n,\chi}$ for every $\chi$ of period $f$
with values in a fixed number field.
\end{corollary}

For $x = b/f'$ in $(0, 1]$ we have $B_n(x) = -nf'^{\,1-n}L_1(1-n, \mathbf 1_{b+f'\ZZ})$, an instance of
Theorem~\ref{thm:uniform} with $F = 1$ and $s_0 = 1 - n$. Since $B_n(x)$ has a denominator $e^{O(n)}$ and absolute value
at most $4e\cdot n!$, $p = n\log_2n + O(n)$ bits determine it (Appendix~\ref{app:sequences}). Equivalently,
$L(1-n,\chi) = -B_{n,\chi}/n$ for every $\chi$ of period $f$, and $\beta(-n) = L(-n, \chi_{-4}) = E_n/2$ for the
Dirichlet beta function, so the corollary computes the values of all Dirichlet $L$-functions of bounded conductor at
the negative integers, exactly.

This proves \cite[Theorem~2]{Joh21} for $B_n$ and $E_n$, extends it to $B_n(x)$ and $B_{n,\chi}$, and makes the bounds
explicit. The classical algorithms for $B_n$ (via the Euler product for $\zeta(n)$, or multimodular \cite{Har10}) take
$n^{2+o(1)}$ bit operations. Harvey's subquadratic modular algorithm \cite{Har14} has
a time--space tradeoff: with a parameter $\frac13 \le \alpha \le \frac12$ it uses $n^{1+\alpha}\log^{4-\alpha+o(1)}n$
bit operations and $O(n^{2-2\alpha}\log^{1+2\alpha}n)$ bits of memory.
At $\alpha = \frac13$ it is faster than Corollary~\ref{cor:bernoulli} but needs $n^{4/3+o(1)}$ bits of memory;
at $\alpha = \frac12$ its memory is $O(n\log^2n)$ and its time $n^{3/2}\log^{7/2+o(1)}n$, comparable to the $O(n^{3/2}\log^{9/2}n)$ of
Corollary~\ref{cor:bernoulli} with $\Mul(n) = O(n\log n)$.
Harvey's algorithm has only been stated for $B_n$, although
it may well extend to $E_n$ and $B_{n,\chi}$; we know of no other published subquadratic algorithm for $E_n$,
$B_n(x)$ or $B_{n,\chi}$.

These results are of theoretical interest. For $B_n$ the case $F = 1$ was implemented in \cite{Joh21}: the
subquadratic scaling is visible, barely, over the range tested, but the method was about $100$ times slower than the
quadratic algorithms there. For $F = 1$ the point $s_0 = 1 - n$ is equivalent to $s_0 = n$ by the functional
equation, and Riemann's formula, which is Theorem~\ref{thm:split} with $F = 1$ and $u_0 = \pi$
(Example~\ref{ex:riemann}), is symmetric under $s \mapsto 1-s$, so the theta splitting does the same work in both
cases. What makes this regime unfavourable is the competition: at $s = n$, with $p \approx n\log_2n$, the Euler
product for $\zeta(n)$ converges after about $n/\log n$ primes, so the classical algorithms are quadratic with very
small constants. With $F \ne 1$ the theorem gives, in the same way, the constant terms
of $\zeta_F$ at $s = 1-n$, the zeta-regularized values of the divergent sums $\sum_kk^{n-1}F(k)$, at uniform cost for
$n = O(p/\log p)$. These are convergent sums of the same type plus Bernoulli corrections \cite{Joh26}, and in general
transcendental.

The last statement of Theorem~\ref{thm:uniform} extends this from roots of unity to all algebraic poles of an
exponential generating function that is rational in $e^t$.

\begin{corollary}\label{cor:egf}
Let $\mathbb K$ be a number field and let $R \in \mathbb K(x)$ have no pole at $x = 1$ and no pole on the unit
circle other than roots of unity. Fix $\lambda \in \mathbb K^\times$ and an integer $m \ge 0$, and let
$a_n = n!\,[t^n]\,t^mR(e^{\lambda t})$. If the $a_n$ are rational, they can be computed exactly in
$O(n^{1/2}\Mul(n)\log^{7/2}n)$ bit operations and $O(n\log n)$ bits of memory.
\end{corollary}

By partial fractions in $e^t$, $a_n$ is a fixed combination of values $\Li_{-n-l}(z)$ at the points $z = \rho^{\pm1}$ in the closed
unit disk, $\rho$ running over the poles; these are values of $L$-functions with periodic weights when $|z| = 1$; again $n\log_2n + O(n)$ bits determine $a_n$
(Appendix~\ref{app:sequences}).

\begin{table}[t]
\centering\footnotesize
\renewcommand{\arraystretch}{1.25}\setlength{\tabcolsep}{4pt}
\caption{Sequences that can be computed exactly in $n^{3/2+o(1)}$ bit operations and $O(n\log n)$ bits of memory
(Corollaries~\ref{cor:bernoulli} and~\ref{cor:egf}), with their exponential generating functions and expressions as
special values. $x$, $z$ rational and $\chi$ of period $f$ are fixed; $\beta(s) = L(s,\chi_{-4})$,
$\eta(s) = (1-2^{1-s})\zeta(s)$; $E_n(x)$ are the Euler polynomials. The identities with $\Li_{-n}$ hold for $n \ge 1$.
Above the rule only $\zeta$ and Dirichlet $L$-functions occur; below it, polylogarithms inside the unit
disk.}
\label{tab:sequences}
\begin{tabular}{llll}
\toprule
sequence & OEIS & e.g.f. & value \\
\midrule
Bernoulli $B_n$ & A027641/2 & $\frac{t}{e^t-1}$ & $-n\zeta(1-n)$ \\
Bernoulli polynomials $B_n(x)$ & & $\frac{te^{xt}}{e^t-1}$ & $-n\zeta(1-n,x)$ \\
generalized Bernoulli $B_{n,\chi}$ & & $\sum_{a=1}^{f}\frac{\chi(a)te^{at}}{e^{ft}-1}$ & $-nL(1-n,\chi)$ \\
Euler (secant) $|E_{2n}|$ & A000364 & $\sec t$ & $2|\beta(-2n)|$ \\
tangent $T_{2n-1}$ & A000182 & $\tan t$ & $2^{2n}(2^{2n}-1)|\zeta(1-2n)|$ \\
Euler zigzag $A_n$ & A000111 & $\sec t + \tan t$ & $|E_n|$ or $T_n$ \\
Genocchi $G_{2n}$ & A110501 & $t\tan\frac t2$ & $4n|\eta(1-2n)|$ \\
Springer $S_n$ & A001586 & $\frac1{\cos t - \sin t}$ & $4^n|E_n(\frac14)|$ \\
Glaisher's $G$ & A002111 & $\frac{3t}{2+4\cos t}$ & $(-1)^{n+1}3^{2n+1}B_{2n+1}(\frac13)$ \\
Glaisher's $H'$ & A002114 & $\frac1{2(2\cos t - 1)}$ & $\frac12 3^{2n}|E_{2n}(\frac13)|$ \\
\midrule
Fubini (ordered Bell) & A000670 & $\frac1{2-e^t}$ & $\frac12\Li_{-n}(\frac12)$ \\
necklaces of partitions & A000629 & $\frac{e^t}{2-e^t}$ & $\Li_{-n}(\frac12)$ \\
generalized weak orders & A004123 & $\frac1{3-2e^t}$ & $\frac13\Li_{-n}(\frac23)$ \\
$\sum_k3^kk!\,S(n,k)$ & A032033 & $\frac1{4-3e^t}$ & $\frac14\Li_{-n}(\frac34)$ \\
ord.\ partitions, odd blocks & A006154 & $\frac1{1-\sinh t}$ & $\frac{\Li_{-n}(\sqrt2-1) + (-1)^n\Li_{-n}(1-\sqrt2)}{\sqrt2}$ \\
ord.\ partitions, even blocks & A094088 & $\frac1{2-\cosh t}$ & $\frac2{\sqrt3}\Li_{-2n}(2-\sqrt3)$ \\
Eulerian polynomials $A_n(z)$ & & $\frac{z-1}{z-e^{(z-1)t}}$ & $(1-z)^{n+1}z^{-1}\Li_{-n}(z)$ \\
\bottomrule
\end{tabular}
\end{table}

Table~\ref{tab:sequences} lists integer sequences covered by
Corollaries~\ref{cor:bernoulli} and~\ref{cor:egf}. Those above the rule have poles only at roots of unity. They need only the case $F = 1$ of Theorem~\ref{thm:uniform}, that is, Dirichlet $L$-functions and Hurwitz zeta
values at rational parameters. The others need polylogarithms $\Li_{-n}(z)$ at fixed algebraic $z$ inside the unit
disk. Some sequences are not of this form. The Bell numbers $\sum_kk^n/(e\,k!)$ have a summand $1/k!$ that is not
regular at infinity. The generalized Euler numbers with e.g.f. $1/\sum_kt^{qk}/(qk)!$, $q \ge 3$, are not covered
either. Their denominator is an exponential polynomial with frequencies $e^{2\pi ij/q}$, which are not all integer
multiples of a single $\lambda$.

\subsection{The zeta-value expansion}\label{sec:zetavalue}
The classical starting point is the expansion in Hurwitz zeta values \cite{FV96}.

\begin{proposition}\label{prop:cont}
Under \eqref{eq:f},
\begin{equation}\label{eq:zetavalue}
  \zeta_F^{\ge k_0}(s) = \sum_{n=0}^{\infty}a_n\,\zeta(s+n,k_0), \qquad \zeta(w,k_0) = \sum_{k=k_0}^{\infty}k^{-w},
\end{equation}
where the series converges locally uniformly on $\CC \setminus \{1, 0, -1, \dots\}$. Hence $\zeta_F$ is meromorphic
in $\CC$, with at most simple poles at $s = 1-n$, $n \ge n_0$, of residue $a_n$. For general $F \in \Reg$
\eqref{eq:frob}, the poles lie at $s = 1 - \sigma - n$ and have order at most $J+1$, with principal parts given
by Lemma~\ref{lem:reduction}.
\end{proposition}

\begin{proof}
For $\operatorname{Re}w = v > 1$, $|\zeta(w,k_0)| \le k_0^{-v}(1 + k_0/(v-1))$, so for $s$ in a compact set and
$n$ large the terms are $O((R/k_0)^n)$; for the finitely many remaining $n$, $\zeta(s+n,k_0)$ is meromorphic with a
single simple pole at $s = 1-n$ of residue $1$. For $\operatorname{Re}s > 1$ the double series converges absolutely,
which justifies the interchange.
\end{proof}

Its terms have size $(R/k_0)^n$, so $p$ bits need about $p/\log_2(k_0/R)$ Hurwitz zeta values. Increasing $k_0$ accelerates the convergence at the price of more
explicit terms $F(k)k^{-s}$, $k < k_0$; this technique is classical~\cite{FV96}.

\emph{Example.} The summand $F(k) = k\arcsin\frac1k - 1$ of Table~\ref{tab:constants} is holonomic but not algebraic.
Here $f(x) = x^{-1}\arcsin x - 1$, $R = 1$, and the coefficients
\[
  a_{2j} = \binom{2j}{j}\frac{1}{4^j(2j+1)}, \qquad a_{2j+1} = 0, \qquad \frac{a_{2j+2}}{a_{2j}} = \frac{(2j+1)^2}{(2j+2)(2j+3)},
\]
are hypergeometric. With $k_0 = 2$ the Hurwitz zeta values are $\zeta(2j,2) = \zeta(2j) - 1$, and
\begin{align*}
  \sum_{k=1}^{\infty}\Bigl(k\arcsin\frac1k - 1\Bigr) &= \frac\pi2 - 1 + \sum_{j=1}^{\infty}\binom{2j}{j}\frac{\zeta(2j) - 1}{4^j(2j+1)} \\
  &= 0.68538530275123375552\ldots,
\end{align*}
whose terms decrease like $4^{-j}j^{-3/2}$, so that $p$ bits need about $p/2$ even zeta values; with $k_0 = 1$ the
series would converge only like $j^{-3/2}$.

In general, the $N = \Theta(p)$ zeta values needed by zeta
expansion can be computed together in $\widetilde O(p^2)$ bit operations, for every $s$~ \cite{Joh14}. For
integer $s$ they are $\zeta(2), \dots, \zeta(N)$ minus finite power sums, and
\[
  \log\Gamma(1+x) = -\gamma x + \sum_{n=2}^{\infty}\frac{(-1)^n\zeta(n)}{n}x^n,
\]
so it suffices to compute $\Gamma(1+x)$ in $\RR[[x]]/(x^N)$ to $p$ bits and take the logarithm of the power series.
With $L \approx p\log 2$ and $T \approx eL$,
\[
  \Gamma(1+x) = L^{1+x}e^{-L}\sum_{k=0}^{T-1}\frac{L^k}{(1+x)(2+x)\cdots(k+1+x)} + O(2^{-p}),
\]
truncating the lower incomplete gamma series of $\Gamma(1+x) = \gamma(1+x,L) + \Gamma(1+x,L)$, and the sum is evaluated by
binary splitting with polynomials in $x$ truncated to degree $N$; the numbers at the top of the product tree have
$\widetilde O(Np)$ bits, so the cost is $\widetilde O(Np)$. (Equivalently one can apply Euler--Maclaurin summation to
$\log\Gamma(1+x)$ or $\psi(1+x)$ over $\RR[[x]]$, forming $\prod_{k<L}(k+x)$ or $\sum_{k<L}1/(k+x)$ by a product tree.) In
particular all $\zeta(n)$, $n \le N$, cost $\widetilde O(N^2)$ at precision $N$, which is quasi-optimal;
the classical alternatives (Bernoulli numbers for the even values \cite{Har10}, Borwein's algorithm \cite{Bor00} or
Euler products) are not. For non-integer~$s$, Euler--Maclaurin summation applied to the generating
function
\[
  \sum_{n=0}^{\infty}\zeta(s+n,k_0)\,y^n = \sum_{k=k_0}^{\infty}\frac{k^{-s}}{1-y/k}
\]
does the same: with a cutoff $M = \Theta(p)$, the direct terms $k < M$ are rational functions of $y$, combined by a
product tree; the integral is $\sum_ny^nM^{1-s-n}/(s+n-1)$; and the Bernoulli corrections for all $n$ at once,
\[
  \frac{M^{1-s-n}}{\Gamma(s+n)}\sum_{j=1}^{J}\frac{B_{2j}}{(2j)!\,M^{2j}}\,\Gamma(s+n+2j-1),
\]
form a single convolution with the sequence $\Gamma(s+m)$. This needs only $p$-bit approximations of $s$.

In summary, the zeta-value expansion costs $\widetilde O(p^2)$ bit operations at any $s$, provided that the $a_n$ are available
(for holonomic $F$ they are P-recursive, and $a_0, \dots, a_N$ to $p$ bits cost $\widetilde O(p^2)$ as well). This is optimal for the method since
the $\Theta(p)$ values to $p$ bits have total size $\Theta(p^2)$. It is the natural reference method for checking
other algorithms.

The zeta-value expansion and Theorem~\ref{thm:split} are closely related. Apply Theorem~\ref{thm:split} with
$f(x) = x^n$, i.e.\ Riemann's splitting of the tail $\zeta(s+n,k_0)$, with the same $u_0$ for every $n$, and sum over
$n$ with weights $a_n$: the result is Theorem~\ref{thm:split} for $f$. The only difference between the two methods is
the order of summation. The zeta-value expansion sums over $k$ and $m$ inside each $\zeta(s+n,k_0)$ and then over
$n$; Theorem~\ref{thm:split} sums over $n$ inside each $k$-term and each dual term, where the sum is a series with
P-recursive terms, computable by binary splitting, and then over the $O(\sqrt p)$ values of $k$ and $m$. Exchanging the
two summations is what turns $p^{2+o(1)}$ into $p^{3/2+o(1)}$.

\begin{figure}[ht]
\centering
\begin{tikzpicture}[x=0.52cm, y=0.62cm, font=\footnotesize]
  \def\P{4.159}
  \fill[gray!30] (2,0) rectangle (4,6.3);
  \fill[orange!30] (4,2.2) -- plot[domain=4:6.62, samples=40] (\x, {\P/ln(\x)}) -- cycle;
  \fill[blue!18] (4,0) rectangle (11.2,2.2);
  \draw[thick, domain=2:11.2, samples=80] plot (\x, {\P/ln(\x)});
  \draw[->] (2,0) -- (11.6,0) node[below left] {$k$};
  \draw[->] (2,0) -- (2,6.7) node[below left] {$n$};
  \draw (4,0) -- (4,-0.12) node[below] {$K_0$};
  \draw (2,2.2) -- (1.88,2.2) node[left] {$n_1$};
  \draw (2,0) node[below] {$k_0$};
  \node at (3,5.4) {rows};
  \node (dl) at (6.6,3.7) {direct};
  \draw[->] (dl) -- (4.7,2.65);
  \node at (7.6,1.1) {long columns $\zeta(s+n, K_0)$};
  \node[align=left] at (8.6,4.4) {negligible};
\end{tikzpicture}
\hfil
\begin{tikzpicture}[x=0.52cm, y=0.62cm, font=\footnotesize]
  \def\P{4.159}
  \def\a{0.0645}
  \fill[gray!30] (2,6.3) -- (2,{\a*4}) -- plot[domain=2:9.88, samples=40] (\x, {\a*\x*\x}) -- cycle;
  \fill[blue!18] (2,0) -- (2,{\a*4}) -- plot[domain=2:9.88, samples=40] (\x, {\a*\x*\x})
    -- (11.2,6.3) -- (11.2,0) -- cycle;
  \draw[dashed, domain=2:9.88, samples=40] plot (\x, {\a*\x*\x});
  \draw[thick, domain=2:11.2, samples=80] plot (\x, {\P/ln(\x)});
  \draw[->] (2,0) -- (11.6,0) node[below left] {$k$};
  \draw[->] (2,0) -- (2,6.7) node[below left] {$n$};
  \draw (2,0) node[below] {$k_0$};
  \node at (4.1,4.7) {$k$-terms};
  \node at (7.7,1.1) {dual terms and $M$};
  \node[above] at (9.4,6.3) {$n = 2u_0k^2$};
\end{tikzpicture}
\caption{The cells $(k, n)$ of $\sum_{k=k_0}^{\infty}\sum_{n=0}^{\infty}a_nk^{-s-n}$ (schematic); the cells above the
solid curve $|a_n|k^{-\sigma-n} = 2^{-p}$ are negligible. Left: zeta-value expansion with rows $k < K_0$, short columns
summed directly and long columns from zeta values. Right: Theorem~\ref{thm:split}, where row $k$ is weighted by
$Q(\frac{s+n}2, k^2u_0)$, which switches smoothly across the dashed parabola; the cells below it are replaced by
the dual terms.}
\label{fig:cells}
\end{figure}
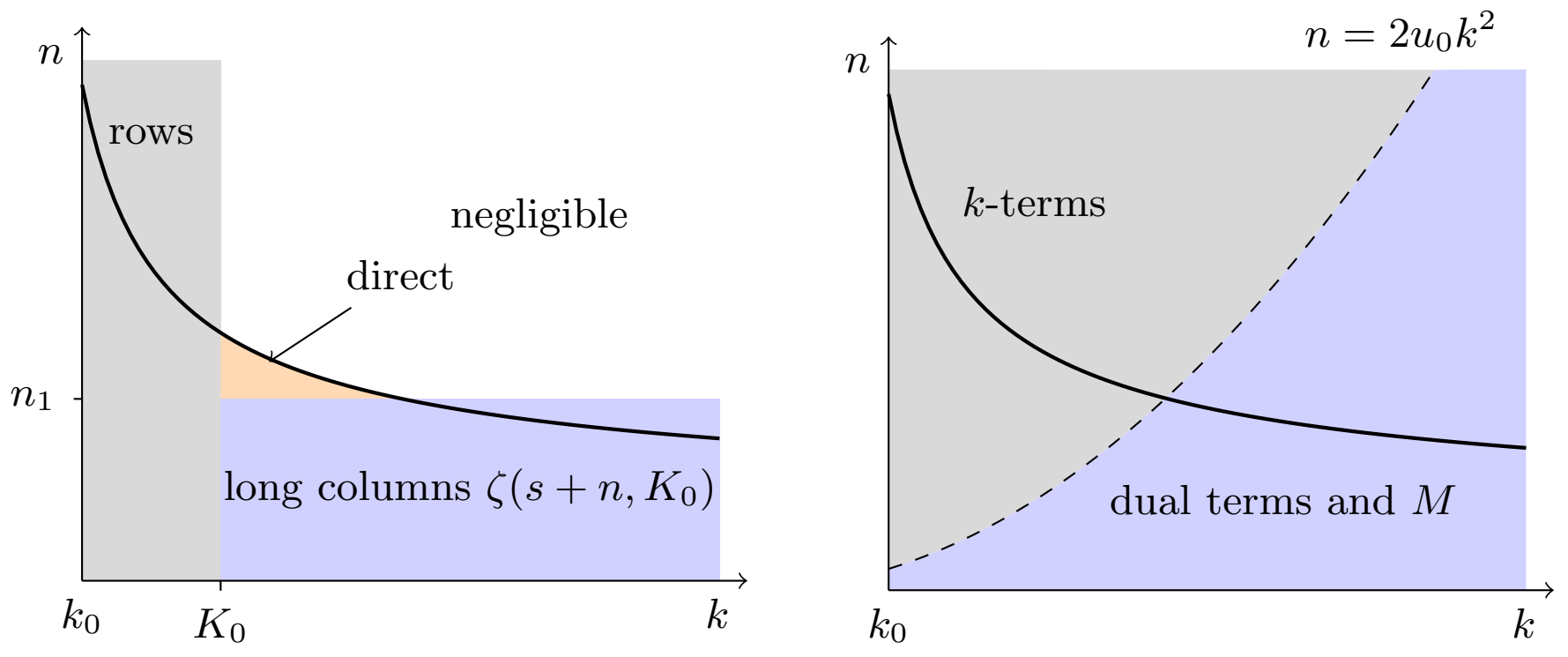

\emph{The cells of the double sum.} Both methods can be pictured as ways of covering the cells $(k, n)$,
$k \ge k_0$, $n \ge 0$, of the double series
\[
  \zeta_F^{\ge k_0}(s) = \sum_{k=k_0}^{\infty}\sum_{n=0}^{\infty}a_nk^{-s-n},
\]
of which only those with $|a_n|k^{-\sigma-n} \gtrsim 2^{-p}$ matter. A refined zeta-value expansion covers
them in three ways (Figure~\ref{fig:cells}, left):
\begin{itemize}
\item \emph{rows} $k_0 \le k < K_0$, each summed as $F(k)k^{-s}$. For holonomic $F$ a row is the series
$k^{-s}f(1/k)$, which converges like $(R/k)^n$ and costs $O(\Mul(p)\log^2p)$ by binary splitting, and less when $F$ is
elementary;
\item \emph{short columns} $n \ge n_1$, summed directly over $K_0 \le k \le K(n)$, where $K(n)$ is the largest $k$
with $|a_n|k^{-\sigma-n} \ge 2^{-p}$; the cell $(k, n)$ needs only about $p + \log_2|a_n| - (\sigma+n)\log_2k$ bits;
\item \emph{long columns} $n < n_1$, as $\zeta(s+n, K_0) = \zeta(s+n) - \sum_{k=1}^{K_0-1}k^{-s-n}$, where the
subtraction costs relative but not absolute precision.
\end{itemize}

No algorithm for the Hurwitz zeta function is needed, and only $n_1$ zeta values. In a column, by a dilation (Section~\ref{sec:twists}),
only $K_0 \le k < 2K_0$ and the odd $k \ge 2K_0$ need to be summed: if $X$ is that
partial sum, then $\zeta(s+n, K_0) = X/(1 - 2^{-s-n})$. The rows cost $O(K_0\Mul(p)\log^2p)$, while the
number $\Theta(p/\log(K_0/R))$ of columns decreases only logarithmically in $K_0$. Peeling off rows therefore saves at most a
factor $\log p$ (with $K_0 \approx p/\log p$), and the cost remains $p^{2+o(1)}$. In practice the gain is a constant factor.
An optimized implementation of the zeta expansion for Khinchin's constant in FLINT (Section~\ref{sec:khimpl}) uses $K_0$ between $6$ and $12$, so
about three times fewer columns than with $K_0 = 2$, and $K(n_1)$ between $2^3$ and $2^{17}$, so no zeta values for
$n \ge n_1$. Its rows are products of logarithms, all obtained from $\log q$ for the primes $q \le K_0$.

Theorem~\ref{thm:split} is a smooth version of the same picture (Figure~\ref{fig:cells}, right). By its series for
$I_k$, the $k$-term is row $k$ with the weight $Q(\frac{s+n}2, k^2u_0)$. This weight is close to $1$ above the parabola
$n = 2u_0k^2 - \sigma$ and close to $0$ below it, with a transition of width $O(k\sqrt{u_0})$. Once the parabola has
left the significant cells, the remaining weight of row $k$ is about $e^{-k^2u_0}$, so the $k$-terms end at
$k_{\max} \approx \sqrt{p\log2/u_0}$. The cells below the parabola are never summed individually. In column $n$ their
sum is $\sum_{k}k^{-s-n}P(\frac{s+n}2, k^2u_0)$, the part of Riemann's splitting of $\zeta(s+n, k_0)$ that Poisson
summation turns into dual terms, and the sum over $n$ of these columns is $\sqrt\pi\sum_mJ_m + M$. The zeta-value
expansion covers what the rows leave with $\Theta(p)$ columns of $p$ bits each, and the theta splitting covers it with
$O(\sqrt p)$ dual terms.

\subsection{Euler--Maclaurin summation}\label{sec:em}
For $N \ge k_0$ and $J \ge 1$, with $g(x) = F(x)x^{-s}$,
\[
  \zeta_F^{\ge k_0}(s) = \sum_{k=k_0}^{N-1} \frac{F(k)}{k^s} + \mathcal M_F(s;N) + \frac{g(N)}{2}
  - \sum_{j=1}^{J}\frac{B_{2j}}{(2j)!}g^{(2j-1)}(N) + R_J(s),
\]
with $|R_J(s)| \le 2\zeta(2J)(2\pi)^{-2J}\int_N^\infty|g^{(2J)}(x)|\dd x$. In our setting all the ingredients are
explicit: the Mellin part is $\mathcal M_F(s;N) = \int_N^\infty g(x)\dd x$, which equals
$\sum_na_nN^{1-s-n}/(s+n-1)$, and the
derivatives $g^{(i)}(N)$ are, up to factorials, the Taylor coefficients of $F(x)x^{-s}$ at $x = N$, which for
holonomic $F$ satisfy a linear recurrence, so that after the direct terms everything costs $O(J)$ arithmetic operations plus the Bernoulli
numbers.

Since $R_J$ is analytic for $\operatorname{Re}s > 1 - n_0 - 2J$, the formula gives the continuation in
growing half-planes, which is the classical proof that $\zeta_F^{\ge N}(s) - \mathcal M_F(s;N)$ is entire. For $p$ bits at fixed $s$ one
needs $N$ and $J$ proportional to $p$, so the cost is $p^{2+o(1)}$: $\Theta(p)$ values $F(k)$, each in
$O(\Mul(p)\log^2p)$ for $\smash{F \in \RegQ}$ and together in $\smash{\widetilde O(p^2)}$ for generic data (\emph{Generic data} at the beginning of
Section~\ref{sec:complexity}), and $\Theta(p)$ Bernoulli numbers.

\emph{Example.} For the summand $F(k) = k\arcsin\frac1k - 1$ of Section~\ref{sec:zetavalue} at $s = 0$, so that $g = F$, the
Mellin part is elementary, since $\frac{x^2}2\arcsin\frac1x + \frac12\sqrt{x^2-1}$ is an antiderivative of
$x\arcsin\frac1x$:
\[
  \mathcal M_F(0;N) = N - \frac{N^2}2\arcsin\frac1N - \frac12\sqrt{N^2-1}
  = \sum_{j=1}^{\infty}\frac{a_{2j}N^{1-2j}}{2j-1}.
\]
For the Taylor coefficients $F_i = F^{(i)}(N)/i!$, the second derivative $H = F'' = x^{-1}(x^2-1)^{-3/2}$ satisfies
$x(x^2-1)H' + (4x^2-1)H = 0$. Substituting $x = N + t$ gives a four-term recurrence for the Taylor coefficients $h_i$ of
$H$ at $N$,
\begin{multline*}
  N(N^2-1)(i+1)h_{i+1} \\ = -\bigl((3N^2-1)i + 4N^2 - 1\bigr)h_i - N(3i+5)h_{i-1} - (i+2)h_{i-2},
\end{multline*}
with $h_0 = N^{-1}(N^2-1)^{-3/2}$ and $F_{i+2} = h_i/((i+1)(i+2))$, together with
$F_0 = F(N)$ and $F_1 = \arcsin\frac1N - (N^2-1)^{-1/2}$. Hence
\begin{align*}
  \sum_{k=1}^{\infty}\Bigl(k\arcsin\frac1k - 1\Bigr) &= \sum_{k=1}^{N-1}F(k) + \mathcal M_F(0;N) + \frac12F(N) \\
  &\quad - \sum_{j=1}^{J}\frac{B_{2j}}{2j}F_{2j-1} + R_J(0),
\end{align*}
where, apart from the direct terms, the only transcendental inputs are $\arcsin\frac1N$ and $\sqrt{N^2-1}$. With
$N = 10$ and $J = 14$ the error is $1.6\cdot10^{-23}$. For fixed $N$ the corrections first decrease and then grow
like $(2j)!/(2\pi N)^{2j}$, which is why $N$ and $J$ must both grow like $p$.

\emph{Many derivatives.} Euler--Maclaurin summation gives many derivatives in $s$ at almost no extra cost, also for
$s$ given numerically and for $F \in \Reg \setminus \RegQ$. Replace $s$ by $s+t$ and work with power series in
$t$ truncated at order $n \le p$. The direct terms become $\sum_{k<N}w_ke^{-t\log k}$ with $w_k = F(k)k^{-s}$, that
is, the power sums $\sum_kw_k(\log k)^i$, $i < n$, which are the coefficients of the rational function
$\sum_kw_k/(1-t\log k)$ and can be computed by a product tree in $\widetilde O(N + n)$ operations on $p$-bit numbers.
The Mellin part $\sum_na_nN^{1-s-t-n}/(s+t+n-1)$ is a sum of the same kind. In the Bernoulli terms,
$g^{(m)}(N)/m! = N^{-s-t}\sum_{i\le m}F_{m-i}\binom{-s-t}{i}N^{-i}$ with the Taylor coefficients $F_i$ of $F$ at $N$;
exchanging the sums,
\[
  \sum_{j=1}^{J}\frac{B_{2j}}{(2j)!}\,g^{(2j-1)}(N) = N^{-s-t}\sum_{i}\gamma_i\binom{-s-t}{i}N^{-i},
\]
with $\gamma_i = \sum_jB_{2j}F_{2j-1-i}/(2j)$,
where the $\gamma_i$ are a correlation of two sequences (one multiplication of polynomials), and the sum over $i$, whose
terms satisfy a first-order recurrence with coefficients linear in $t$, is again a product tree. With $N, J = O(p)$
this costs $\widetilde O(p^2)$ bit operations for all $n \le p$ derivatives together, the same as for a single value,
provided the values $F(k)$, $k < N$, and the $F_i$ are available in that time (by binary splitting when
$F \in \RegQ$, and otherwise by multipoint evaluation of the expansion of $F$ at infinity as in \emph{Generic data} at the
beginning of Section~\ref{sec:complexity}, and by the recurrence for the $F_i$ from $O(1)$ derivatives at $N$). In particular $n$ derivatives
to $n$ bits cost $\widetilde O(n^2)$, as for the Hurwitz zeta function \cite{Joh15}. The theta splitting, in contrast,
gains nothing for many derivatives at a generic point.

Euler--Maclaurin summation is Poisson summation with a sharp cutoff: applied to $g\cdot\mathbf{1}_{[N,\infty)}$, the
dual terms are $\int_N^\infty g(x)e^{2\pi imx}\dd x$, and integrating by parts at the endpoint and summing over $m \ne 0$ with
$\sum_{m\ne0}(2\pi m)^{-2j} = 2\zeta(2j)/(2\pi)^{2j}$ gives the Bernoulli terms. Theorem~\ref{thm:split} instead smooths
the cutoff with the theta function and computes the dual terms exactly; with a sharp cutoff the dual terms decay only
polynomially in $m$, and their asymptotic expansion is divergent, which is why Euler--Maclaurin needs $N \asymp p$.

\subsection{Alternating series}
For the Riemann zeta function, $(1-2^{1-s})\zeta(s) = \sum_{k=1}^{\infty}(-1)^{k+1}k^{-s}$, and acceleration of the alternating
series gives Borwein's algorithm \cite{Bor00} and the method of Cohen, Rodriguez Villegas and Zagier \cite{CRVZ00}, with
$O(p)$ terms. For $\zeta_F$ the corresponding identity is
\[
  \sum_{k=1}^{\infty}(-1)^{k+1}F(k)k^{-s} = \zeta_F(s) - 2^{1-s}\zeta_{F(2\,\cdot)}(s),
\]
which relates $\zeta_F$ to the different function $\zeta_{F(2\,\cdot)}$ unless $F$ is a power. These methods therefore apply
to twisted functions such as $L_G(s,\chi)$ in \eqref{eq:K3}, provided the terms are moments of a positive measure, but
not to $\zeta_F$ itself; in either case they need $\Theta(p)$ values $F(k)$ to $p$ bits, hence $p^{2+o(1)}$, also
for generic data.

\subsection{Large height}\label{sec:rs}
At large height and moderate precision the zeta-value expansion combines with the Riemann--Siegel formula.

\begin{proposition}\label{prop:rs}
Let $F$ satisfy \eqref{eq:f}, and let $\sigma$ and $\varepsilon > 0$ be fixed. Then $\zeta_F(\sigma+it)$ can be computed
with absolute error at most $\varepsilon$ using $O(t^{1/2})$ arithmetic operations on numbers of $O(\log t)$ bits, with
implied constants depending on $F$, $\sigma$ and $\varepsilon$.
\end{proposition}

\begin{proof}
Write $\zeta_F(s) = \sum_{k<k_0}F(k)k^{-s} + \sum_{n<N}a_n\zeta(s+n,k_0) + R_N(s)$ with
$R_N(s) = \sum_{k=k_0}^{\infty}k^{-s}r_N(k)$ and $|r_N(k)| \le C'(R/k)^N$. Choose $N$ and then $k_2$, independently of $t$, so that
the terms of $R_N$ with $k > k_2$ contribute less than $\varepsilon/2$, and evaluate the others directly. Each of the
finitely many $\zeta(s+n,k_0)$ is $\zeta(s+n)$ minus a finite sum, and $\zeta(s+n)$ is computed by the Riemann--Siegel
formula in $O(t^{1/2})$ operations \cite{AdR11}.
\end{proof}

\begin{remark}\label{rem:complexu0}
Theorem~\ref{thm:split} with real $u_0$ is not efficient for large $t$: $\Gamma(\frac s2)^{-1}$ is of size
$e^{\pi t/4}$ and the terms cancel to that extent. For $L$-functions this is overcome by rotating the
integration path, i.e.\ taking $u_0$ complex \cite{Rub05}, which gives Riemann--Siegel-like formulas with
$O(\sqrt t)$ terms. All pieces of Theorem~\ref{thm:split} are holonomic in $u_0$ and analytic for
$\operatorname{Re}u_0 > 0$, so the same idea should apply here and give an exact, smoothed version of the
Hardy--Littlewood-type approximate functional equation for $\zeta_F$ proved in \cite{Joh26}; we have not worked this out (Section~\ref{sec:open}).
\end{remark}

\subsection{Other generating functions}\label{sec:ogf}
For $F$ analytic at infinity the ordinary generating function is, by the same expansion as \eqref{eq:zetavalue},
\[
  G_F(z) = \sum_{k=k_0}^{\infty}F(k)z^k = \sum_{n=0}^{\infty}a_n\Bigl(\Li_n(z) - \sum_{k=1}^{k_0-1}\frac{z^k}{k^n}\Bigr),
\]
an expansion in polylogarithms. The function $G_F(e^{-t})$ has Mellin transform $\Gamma(s)\zeta_F^{\ge k_0}(s)$. Hence the
expansion of $G_F$ at $z = e^{-t} = 1$ has the regular part
\[
  \sum_{m=0}^{\infty}\zeta_F^{\ge k_0}(-m)\frac{(-t)^m}{m!}
\]
(Abel summation of the divergent sums $\sum_kk^mF(k)$), plus logarithmic terms from the poles. Across $[1,\infty)$ each $\Li_n$ jumps by
$2\pi i(\log x)^{n-1}/(n-1)!$, so $G_F$ jumps by $2\pi i\,B(\log x)$, with the Borel transform $B$ of
Remark~\ref{rem:whyk2}: the generating function is the Stieltjes transform of $B\circ\log$, holonomic only up to this
logarithmic change of variable. On the unit circle the values $G_F(e^{2\pi i\alpha})$ are the twisted values
$L_F(0,e^{2\pi i\alpha k})$ of Remark~\ref{rem:twist}, so the theta splitting applies; inside the disk see
Section~\ref{sec:dampedcost}. The
Lambert series $\sum_kF(k)z^k/(1-z^k)$ is the generating function of $\sum_{d\mid n}F(d)$, with Dirichlet series
$\zeta(s)\zeta_F(s)$, and its expansion at $z = 1$ starts with $\zeta_F(1)/t$.

\subsection{Damped functions}\label{sec:dampedcost}
For the polylogarithm, Theorem~\ref{thm:geometric}(2) gives the same bound as Theorem~\ref{thm:complexity}. At
$s = 0$ it computes the generating function $\Li_F(0,z)$ of a sequence $F \in \RegQ$ at any algebraic $z$ inside the unit
disk. When $F(k)$ is itself P-recursive in $k$ (for example rational), the generating function is holonomic in $z$ and
costs $\widetilde O(p)$ directly; the gain is for summands such as $k^{-s}$ with $s \notin \ZZ$, $\log k$ or
$\log(1+\frac1k)$, where direct summation and the expansion in zeta values cost $p^{2+o(1)}$. The transcendental number $\log z$ enters only through the dual points
$c_n$, like $\pi^2m^2$ in Theorem~\ref{thm:split}, while $z^{\pm k}$ are algebraic weights. The $k$-terms with $z^{-k}$
and the dual terms with $\operatorname{Re}c_n < 0$ have size up to $e^{\pi^2\alpha_2^2/u_0} = e^{(\log|z|)^2/(4u_0)}$
and cancel to $O(1)$; this costs $(\log|z|)^2/(4u_0\log2)$ guard bits, independent of $p$.

For the Lerch transcendent, the shift of the power (Section~\ref{sec:twistshift}) can be removed when computing
Taylor coefficients at a fixed point $s_0$. For $k \ge 1$,
$(k+a)^{-s} = k^{-s}(1+\frac ak)^{-s}$, and $G_i(k) = F(k)\,(-\log(1+\frac ak))^i(1+\frac ak)^{-s_0}/i!$ is in
$\Reg$, and in $\RegQ$ when $F$ is and $a$, $s_0$ are algebraic. Multiplying the expansions in $s - s_0$,
the $j$-th Taylor coefficient of $\Phi_F(z,s,a)$ at $s_0$ is $F(0)$ times that of $a^{-s}$ plus
$\sum_{i\le j}$ of the $(j-i)$-th Taylor coefficient of $\Li_{G_i}(s,z)$ at $s_0$. With
Theorem~\ref{thm:geometric}(2) this proves Theorem~B(2) for $\Phi_F$, and with Theorem~\ref{thm:complexity} the same
reduction applies to the Hurwitz-type sums $\sum_{k\ge1}F(k)(k+a)^{-s}$.

For the damped zeta functions, Appendix~\ref{app:stretched} sketches why the analysis of Appendix~\ref{app:cost}
carries over (Claim~C). For $F = 1$ and rational $r$ the exponents $a_j$ of the kernel fall into finitely many classes
modulo $\ZZ$, within which all series of Proposition~\ref{prop:expsplit} have P-recursive terms, and the kernel is
holonomic, so that the dual moments are again values of a holonomic system. For $\Lambda_F$
(Proposition~\ref{prop:prodsplit}) the pieces are double series over $n$ and $j$ whose inner sums are P-recursive in
$n$. For $q = 1$ this is $\Li_F(s,e^{-\tau})$, which Theorem~\ref{thm:geometric} already covers; the new cases are
$q \ge 2$. We have not written out the orders of the recurrences, the guard bits or the starting values in full,
but we have implemented the splitting for $r = \frac12$ and $F = 1$, and computed $\sum_ke^{-\sqrt k}$ with it to
$10^6$ digits (Sections~\ref{sec:expsqrt} and~\ref{sec:appcomp}).
The classical alternatives are direct summation, with $p^{1+1/r+o(1)}$ bit operations, and the expansion
\[
\Lambda(s;\tau,r) = M_\tau(1-s) + \sum_{n=0}^{\infty}\zeta(s-rn)(-\tau)^n/n!
\]
in zeta values (from a
Mellin--Barnes integral; see \cite{Joh26}), which converges for $r < 1$, and for $r = 1$ when $\tau < 2\pi$, and costs
$p^{2+o(1)}$ like \eqref{eq:zetavalue}. (For $\sum_ke^{-\sqrt k}$ we used Euler--Maclaurin summation, up to $10^5$
digits, as the reference; we have not implemented this expansion.) For $r = 1$ Proposition~\ref{prop:expsplit} is an alternative to
Theorem~\ref{thm:geometric} with real dual points, but $\tau = \log(1/z)$ then enters the kernel, so that for algebraic
$z$ Theorem~\ref{thm:geometric} is the better choice.

Table~\ref{tab:expclass} summarizes the costs for the damped functions.

\begin{table}[ht]
\centering\small
\caption{Sums with exponential factors, to $p$ bits at algebraic parameters ($F \in \RegQ$ not
P-recursive, $s \notin \ZZ$). The last column says whether the bound $p^{3/2+o(1)}$ is proved or whether the identity
is proved and the cost analysis sketched.}
\label{tab:expclass}
\begin{tabular}{lllll}
\toprule
sum & direct & zeta values & splitting & bound \\
\midrule
$\sum_k\chi(k)z^kF(k)k^{-s}$, $|z| < 1$ & $p^{2}$ & $p^{2}$ & $p^{3/2}$ & Thm.~\ref{thm:geometric} \\
\quad e.g.\ $\Li_s(z)$, $\log\sigma$, $\gamma(z)$ & & & & \\
$\Phi(z,s,a) = \sum_{k\ge0}z^k(k+a)^{-s}$ & $p^{2}$ & $p^{2}$ & $p^{3/2}$ & Thm.~\ref{thm:geometric} \\
$\sum_ke^{-\tau k^r}k^{-s}$, $r \in \QQ$, $0 < r \le 1$ & $p^{1+1/r}$ & $p^{2}$ ($r < 1$) & $p^{3/2}$ & sketched \\
$\sum_ke^{-\tau k^{1/q}}F(k)k^{-s}$ & $p^{1+q}$ & $p^{2}$ & $p^{3/2}$ & sketched \\
\bottomrule
\end{tabular}
\end{table}

\section{Khinchin's constant}\label{sec:khinchin}

All earlier high-precision computations of Khinchin's constant $K$ use expansions in zeta values
with cost $p^{2+o(1)}$ (Section~\ref{sec:zetavalue}): the 66 digits by Shanks and
Wrench \cite{SW59}, $7\,350$ digits by Bailey,
Borwein and Crandall \cite{BBC97}, $110\,000$ by Gourdon (1997), $10^6$ by Sim\'o (2016) and $2\cdot10^6$ by Ghazi
\cite{Gha26,MWKdigits}.

Our method gives three subquadratic algorithms, which we call K1, K2 and K3, from
three representations of $K$ as a special value of a function of the class \eqref{eq:functions}:
\begin{align}
  \log K &= -Z'(0), \tag{K1}\label{eq:K1}\\
  \log K\log 2 &= \zeta_F(0), \tag{K2}\label{eq:K2}\\
  \log K\log 2 &= \log^22 + \Li_2(-\tfrac12) + \tfrac12 L_G(0,\chi), \tag{K3}\label{eq:K3}
\end{align}
where $Z(s)$ is the Khinchin zeta function introduced below in Section~\ref{sec:khinchinzeta},
\[
  F(k) = -\log\Bigl(1-\frac1k\Bigr)\log\Bigl(1+\frac1k\Bigr) \quad (k \ge 2),
\]
so that \eqref{eq:K2} is the formula \eqref{eq:twoconstants} of the introduction, and
\[
  G(k) = \Li_2\Bigl(\frac{4}{k^2}\Bigr) \quad (k \ge 2),
\]
twisted by $\chi(k) = (-1)^k$. The sums for both K2 and K3 start at $k = 2$; at $k = 1$, $\Li_2(4)$ is not real.

The three formulas for $K$ are linked by the operations of Section~\ref{sec:twists}.
Summation by parts turns $-Z'(0)$ into the product of logarithms in \eqref{eq:K2}; this holds only at $s = 0$, so $Z$
and $\zeta_F$ are different functions that share one value. A dilation and a telescoping identity then turn
\eqref{eq:K2} into \eqref{eq:K3} (Lemma~\ref{lem:dilog}). Each representation gives an algorithm with the complexity
of Theorem~\ref{thm:complexity}, by a different route: \eqref{eq:K1} is a derivative at a regular point, so
Algorithm~K1 differentiates the splitting in $s$; \eqref{eq:K2} is a value of $\zeta_F$; and \eqref{eq:K3} is a value
of the twisted zeta function of the non-elementary $G$.

The constant-factor overheads of these algorithms differ widely. Table~\ref{tab:khstates} lists the sizes of what each algorithm evaluates: the
series of each $k$-term, the moment systems of the dual terms (Proposition~\ref{prop:dual}), and the incomplete gamma
values; Section~\ref{sec:khcompare} explains the differences. After the probabilistic background and a closer look at
$Z$ (Section~\ref{sec:khinchinzeta}), we work out the three algorithms (Sections~\ref{sec:Zprime},
\ref{sec:logform} and~\ref{sec:dilog}), and compare implementations that share all their machinery
(Section~\ref{sec:khcompare}); to what extent such rewritings generalize is discussed in Section~\ref{sec:generalize}.
Algorithm~K3 is also the basis of an optimized implementation in FLINT, with which we computed $K$ to $10^7$ digits
(Sections~\ref{sec:khimpl}--\ref{sec:khcf}).

\begin{table}[t]
\centering\footnotesize
\caption{The three algorithms for Khinchin's constant at $s = 0$: the coefficients of the kernel and its exponential
type, the series evaluated for each $k$-term (number of series $\times$ dimension of the state), the incomplete gamma
values $\Gamma(b, \cdot)$ and $\partial_b\Gamma(b, \cdot)$ needed for each $k$-term and each dual term, and the moment
systems of the dual terms (number of systems
$\times$ dimension of the state, with the number of forcing constants in parentheses).}
\label{tab:khstates}
\setlength{\tabcolsep}{4pt}
\begin{tabular}{llll}
\toprule
 & K1 & K2 & K3 \\
\midrule
kernel coefficients & $2$ hypergeometric & alt.\ harmonic & hypergeometric \\
exponential type & $4$ & $1$ & $4$ \\
$k$-side series & $4\times4$ & $1\times4$ & $1\times2$ \\
gamma values, $k$-term & $E_1$, $\Gamma(\frac12)$, $\partial_b\Gamma(\frac12)$ & -- & -- \\
dual systems & $2\times16$ ($10$ each) & $1\times12$ ($5$) & $1\times4$ ($3$) \\
gamma values, dual term & $\Gamma(\frac12)$, $E_1$; $\partial_b$ of both & $\Gamma(\frac12)$ & $\Gamma(\frac12)$ \\
\bottomrule
\end{tabular}
\end{table}

\subsection{The Khinchin zeta function}\label{sec:khinchinzeta}

Let $x = [a_0; a_1, a_2, \dots]$ be the continued fraction of a real number $x$. For almost every $x$, the partial
quotients follow the Gauss--Kuzmin law: the proportion of the $a_i$ equal to $k$ is
\[
\mu_k = \log_2\Bigl(1 + \frac1{k(k+2)}\Bigr),
\]
the probability that
$a_1 = k$ when $x$ is drawn from the Gauss measure $\dd x/((1+x)\log2)$ on $(0,1)$, which is invariant under the
continued fraction map \cite{Khi64,IK02}. By the ergodic theorem the geometric means $(a_1\cdots a_n)^{1/n}$ tend
to $K = \prod_{k=1}^{\infty}k^{\mu_k}$ \cite{Khi35}, and the H\"older means $(\frac1n\sum_{i\le n}a_i^q)^{1/q}$ tend to
the generalized Khinchin constants $K_q = (\sum_kk^q\mu_k)^{1/q}$ for $q < 1$, $q \ne 0$ \cite{BBC97}.

All of these are values of the
\emph{Khinchin zeta function}
\begin{equation}\label{eq:Z}
  Z(s) = \sum_{k=1}^{\infty} \mu_k\,k^{-s} = \mathbb{E}\bigl[a_1^{-s}\bigr],
\end{equation}
the zeta function $\zeta_F$ of $F(k) = \mu_k$.
Indeed $Z(0) = 1$ (the product
$\prod_k(k+1)^2/(k(k+2))$ telescopes to $2$),
\[
  K = \exp(-Z'(0)), \qquad K_q = Z(-q)^{1/q},
\]
and in particular the harmonic mean is $K_{-1} = 1/Z(1)$.

By Proposition~\ref{prop:cont}, $Z$ is meromorphic and analytic in $\operatorname{Re}s > -1$, with simple poles at
$s = -1, -2, -3, \dots$; the first three residues are $1/\log 2$, $-2/\log 2$ and $7/(2\log 2)$.

\begin{proposition}\label{prop:Z}
The H\"older means
satisfy
\[
\log 2\,(K_q)^q = (1-q)^{-1} - 1 + O(1-q), \qquad q \to 1^-.
\]
Equivalently,
\[
Z(s) = \dfrac{1}{(s+1)\log2} - \dfrac{1}{\log2} + O(s+1), \qquad s \to -1.
\]
\end{proposition}

\begin{proof}
By summation by parts
\begin{multline*}
\sum_{k=1}^{N}k\,(2\log(k+1) - \log k - \log(k+2)) \\ = \log(N+1) - N\log\frac{N+2}{N+1} = \log N - 1 + o(1),
\end{multline*}
so $\sum_{k=1}^{\infty}(k\mu_k - \frac{1}{k\log2})$ converges to $-(1+\gamma)/\log 2$.
Since $Z(s) - \zeta(s+2)/\log 2$ is analytic at
$s = -1$ with value $\sum_{k=1}^{\infty}(k\mu_k - \frac1{k\log 2})$, and $\zeta(s+2) = \tfrac{1}{s+1} + \gamma + O(s+1)$, the constant term
is $-1/\log 2$.
\end{proof}

Table~\ref{tab:Zvalues} collects special values of $Z$.
Apart from the value at $0$, the constant term at the first pole
and the residues, we know of no other elementary values.
Numerical values of the complex zeros of $Z$
are computed in the companion note \cite{Joh26}.

\begin{table}[t]
\centering\small
\caption{Special values of the Khinchin zeta function $Z(s) = \sum_k\mu_kk^{-s}$ and the corresponding constants.}
\label{tab:Zvalues}
\begin{tabular}{@{}lll@{}}
\toprule
$s$ & $Z$ at $s$ & constant \\
\midrule
$0$ & $Z(0) = 1$, \ $Z'(0) = -\log K$ & $K = 2.68545200106530644530\ldots$ \\
$-1$ & residue $\frac1{\log2}$, constant term $-\frac1{\log2}$ & (Proposition~\ref{prop:Z}) \\
$-n$ & residue $a_{n+1} = (-1)^{n+1}\frac{2^{n+1}-2}{(n+1)\log2}$ & ($n \ge 2$; Proposition~\ref{prop:cont}) \\
$1$ & $0.57293271217004775403\ldots$ & $K_{-1} = 1.74540566240734686349\ldots$ \\
$2$ & $0.47540106853698467921\ldots$ & $K_{-2} = 1.45034032849563040605\ldots$ \\
$3$ & $0.44126831790660682396\ldots$ & $K_{-3} = 1.31350707868798574683\ldots$ \\
$\frac12$ & $0.69760711700293137124\ldots$ & $K_{-1/2} = 2.05484086835714429321\ldots$ \\
$\frac13$ & $0.76749642806804968695\ldots$ & $K_{-1/3} = 2.21192761765114288157\ldots$ \\
$-\frac12$ & $2.12910664715927745243\ldots$ & $K_{1/2} = 4.53309511497781997442\ldots$ \\
$-\frac13$ & $1.53570106170602789125\ldots$ & $K_{1/3} = 3.62176321599942938889\ldots$ \\
\bottomrule
\end{tabular}
\end{table}

By Theorem~\ref{thm:complexity} all H\"older means
$K_q$ with rational $q < 1$, not only the $K_{-n}$ computed by Bailey, Borwein and Crandall \cite{BBC97}, can be
computed to $p$ bits in $p^{3/2+o(1)}$ bit operations.
The series expansion at infinity is
\[
  f(x) = \frac{2\log(1+x) - \log(1+2x)}{\log 2}, \qquad a_n = \frac{(-1)^n(2^n - 2)}{n\log 2},
\]
with $\rho = \frac12$, so $k_0 = 3$, and $a_0 = a_1 = 0$, $n_0 = 2$. Strictly, $\mu_k$ is not in $\RegQ$ because of
the factor $1/\log 2$; the algorithms compute $Z(s)\log 2$.

\emph{Divisibility.} The Dirichlet series $\zeta(s)Z(s)$ has the coefficients $\sum_{d\mid n}\mu_d$, the probability
that $a_1$ divides $n$, so the mean of this probability over $n \le N$ tends to $Z(1) = 1/K_{-1} = 0.5729\ldots$ as
$N \to \infty$. The converse probabilities have closed forms: since $1 + \frac1{k(k+2)} = (k+1)^2/(k(k+2))$, the
product over $k = mn$, $m \ge 1$, telescopes in terms of the gamma function, and
\[
  \Pr[\,n \mid a_1\,] = \sum_{m=1}^{\infty}\mu_{mn} = \log_2\frac{\Gamma(1+\frac2n)}{\Gamma(1+\frac1n)^2}
  = \frac{\pi^2}{6n^2\log2} + O(n^{-3}),
\]
for instance $\log_2(4/\pi) = 0.34850387052768120195\ldots$ for $n = 2$, the probability that a partial quotient is
even (OEIS A340533; see also \cite{Nol90}). Summing over $n$ gives the mean number of divisors of a partial quotient,
$\mathbb E[d(a_1)] = 2.09285796492039431020\ldots$; this sum is not of the form \eqref{eq:functions}, since
$\log\Gamma(1+x)$ is not holonomic (Section~\ref{sec:openother}).

\subsection{K1: the derivative of $Z$}\label{sec:Zprime}
For \eqref{eq:K1} let $L(s) = \log 2\sum_{k=3}^{\infty}\mu_kk^{-s}$, so that $\log K = \log\frac98 - L'(0)/\log 2$,
with $f(x) = 2\log(1+x) - \log(1+2x)$, $a_n = (-1)^n(2^n - 2)/n$ and $k_0 = 3$. At $s = 0$ both parts
\eqref{eq:kernelparts} of the kernel are elementary,
\[
  G_0(u) = u\,g_0(u) = \tfrac12e^{4u} - e^u + \tfrac12, \qquad G_1(u) = g_1(u) = -\bigl(\omega(4u) - \omega(u)\bigr),
\]
where $\omega(y) = \sum_{m\ge0}y^m/\Gamma(m+\frac32) = y^{-1/2}e^y\operatorname{erf}\sqrt y$. We need the derivative of
the splitting,
\[
  L'(0) = \sum_{k=3}^{\infty}I_k'(0) + \sqrt\pi\sum_{m=1}^{\infty}J_m'(0) + M'(0),
\]
and $s$ enters in two ways: through $k^{-s}$ and $u^{s/2}$, which produce weights $\log k$ and $\log u$, and through
$1/\Gamma(\frac{s+n}2)$, which produces digamma values $\psi(\frac n2)$, that is, harmonic numbers.

\emph{$k$-terms.} With $x = k^2u_0$,
\[
  I_k'(0) = \sum_na_nk^{-n}\Bigl(-\log k\,Q\bigl(\tfrac n2,x\bigr) + \frac12\partial_aQ\bigl(\tfrac n2,x\bigr)\Bigr),
\]
and $\partial_aQ(a+1,x) = \partial_aQ(a,x) + x^ae^{-x}(\log x - \psi(a+1))/\Gamma(a+1)$ reduces each parity class to
its first member. For even $n = 2m$, with $e_m(x) = \sum_{i<m}x^i/i!$ and $H_i$ the harmonic numbers,
\[
  \sum_{m\ge1}a_{2m}k^{-2m}\Bigl(\frac{\log u_0 + \gamma}2e^{-x}e_m(x) + \frac12E_1(x)
  - \frac12e^{-x}\sum_{i<m}H_i\frac{x^i}{i!}\Bigr),
\]
and the odd terms are similar, with $x^i/\Gamma(i+\frac32)$, the partial sums $O_i = \sum_{j<i}1/(2j+1)$, and the
values $\Gamma(\frac12,x)$ and $\partial_a\Gamma(a,x)|_{a=1/2}$. Since $a_{2m} = \frac12\cdot4^m/m - 1/m$ and
$a_{2m+1} = 2(1 - 4^m)/(2m+1)$ combine two exponential scales, each $k$-term costs four series with states of
dimension $4$ (the harmonic numbers double the states), and three incomplete gamma values.

\emph{Dual terms.} Writing $Z^{\log}_t[g] = \int_0^{u_0}u^{-t}\log u\,e^{-c/u}g(u)\dd u = -\partial_tZ_t[g]$,
\[
  J_m'(0) = \tfrac12Z^{\log}_{3/2}[G_0] + Z_{3/2}[\partial_sG_0] + \tfrac12Z^{\log}_1[G_1] + Z_1[\partial_sG_1]
  \quad (c = \pi^2m^2),
\]
with
\begin{gather*}
  \partial_sG_0 = -\tfrac14\bigl(\varphi(4u) - 2\varphi(u) - \gamma(e^{4u} - 2e^u + 1)\bigr), \\
  \partial_sG_1 = \tfrac12\bigl(\psi(\tfrac12)(\omega(4u) - \omega(u)) + 2(\varrho(4u) - \varrho(u))\bigr),
\end{gather*}
where $\varphi(y) = e^y\Ein(y) - \Eip(y)$ and $\varrho(y) = \sum_{m\ge0}O_my^m/\Gamma(m+\frac32)$.
All these functions satisfy first-order relations in $\vartheta$: $\vartheta(e^y\Ein y) = y\,e^y\Ein y + e^y - 1$,
$\vartheta\Eip y = e^y - 1$, $\vartheta\omega = (y - \frac12)\omega + 1/\sqrt\pi$, $\vartheta\sigma = \frac12\sigma +
\frac12y\,\omega$ and $\vartheta\varrho = (y - \frac12)\varrho + \sigma$, with
$\sigma(y) = \sum_{m\ge1}y^m/((2m-1)\Gamma(m+\frac12))$, and the log-weighted moments satisfy \eqref{eq:ibp} with one
more term, $c\,Z^{\log}_{t+1}[g] = E\,u_0^{1-t}\log u_0\,g(u_0) - (1-t)Z^{\log}_t[g] - Z_t[g] - Z^{\log}_t[\vartheta g]$.
Proposition~\ref{prop:dual} then gives two moment systems, at $t = \frac32$ and $t = 1$, of dimension $16$ each. Their
starting values are the series \eqref{eq:startvals} and their derivatives in the index $a$ of $D_a$, which need
$\Gamma(\frac12,y)$, $E_1(y)$ and $\partial_b\Gamma(b,y)$ at $b = \frac12$ and $b = 0$.

\emph{Main term.} $M'(0)$ is a sum over $n$ of elementary terms with $\psi(\frac n2)$, and, for $k = 1, 2$, of the
derivatives of $k^{-s-n}P(\frac{s+n}2,k^2u_0)$; reorganized by the lower series of $P$, these are rational series with
harmonic weights, summed by binary splitting once.

\subsection{K2: the logarithm series}\label{sec:logform}
For \eqref{eq:K2}, $k_0 = 2$ and
\[
  f(x) = -\log(1-x)\log(1+x) = \sum_{n=1}^{\infty}\frac{A_n}{n}x^{2n}, \qquad A_n = \sum_{j=1}^{2n-1}\frac{(-1)^{j+1}}j.
\]
Since $f$ is even, only $g_0$ in \eqref{eq:kernelparts} occurs, and at $s = 0$ the kernel is
$H(u) = \sum_{n=0}^{\infty}\frac{A_{n+1}}{(n+1)!}u^n$, entire of exponential type $1$, with
\[
  uH(u) = \log 2\,(e^u-1) + \int_0^1\frac{e^{\beta u}-1}{2\beta(1+\sqrt\beta)}\dd\beta
\]
(from $A_n = \log 2 + \int_0^1 t^{2n-1}/(1+t)\dd t$). Its coefficients $h_n$ satisfy
\[
  (n+1)^2(n+2)(2n+3)h_{n+1} = (n+1)(4n^2+6n+3)h_n - n(2n+1)h_{n-1},
\]
and $H$ satisfies the inhomogeneous equation of order $4$
\[
  \vartheta^2(\vartheta+1)(2\vartheta+1)H - u(\vartheta+1)(4\vartheta^2+6\vartheta+3)H
  + u^2(\vartheta+1)(2\vartheta+3)H = -\frac u2.
\]
Theorem~\ref{thm:split} at $s = 0$ gives, for every $u_0 > 0$,
\begin{multline*}
  \log K\log 2 = \sum_{k=2}^{\infty}I_k + \sqrt\pi\sum_{m=1}^{\infty}Z_{1/2}[H](\pi^2m^2) \\
  + \int_0^{u_0}\Bigl(\frac12\sqrt{\frac\pi u} - \frac12 - e^{-u}\Bigr)H(u)\dd u,
\end{multline*}
with the $k$-terms
\[
  I_k = e^{-x}\sum_{n=0}^{\infty}\frac{A_{n+1}}{n+1}\,\frac{e_{n+1}(x)}{\kappa^{n+1}}, \qquad \kappa = k^2,\ x = \kappa u_0,
\]
where $e_n(x) = \sum_{i=0}^{n-1}x^i/i!$. The terms satisfy a $4\times4$ triangular recurrence (for the truncated
exponential, the powers of $x$ and $\kappa$, and their products with $A_n$), summed by binary splitting. The moments
$Z_t[H](c)$ satisfy a scalar recurrence of order $6$ in $t$ with coefficients of degree up to $4$ in $c$ and five
forcing constants $(\vartheta^iH)(u_0)$, $i < 4$, and $1$. As a first-order system in $c$
(Proposition~\ref{prop:dual}), for the moments of $(H, \vartheta H, \vartheta^2H, \vartheta^3H)$ and their shifts by
$u$ and $u^2$, it is linear in $c$, of dimension $12$. The starting values at a rational $c_0$ are the series
$Z_t[H](c_0) = \sum_jh_jD_{j-t+1}(c_0)$ of \eqref{eq:startvals}, with $t = \frac12$, so that all $D_a$ are reduced to
$D_{-1/2} = \sqrt{\pi/c}\erfc\sqrt{c/u_0}$ and $E = e^{-c/u_0}$. Since $H$ has type $1$, the $k$-terms decay like
$e^{-(k^2-1)u_0}$ and the dual terms like $e^{-\pi^2m^2/u_0}$, and $O(\sqrt p)$ terms of each kind suffice for $p$ bits.

\subsection{K3: the dilogarithm series}\label{sec:dilog}

\begin{lemma}[{Bailey, Borwein and Crandall \cite[\S3]{BBC97}}]\label{lem:dilog}
\[
  \log K \log 2 = \log^2 2 + \Li_2(-\tfrac12) + \frac12\sum_{k=2}^{\infty}(-1)^k \Li_2\Bigl(\frac{4}{k^2}\Bigr).
\]
\end{lemma}

In \cite{BBC97} this is attributed to Zagier and derived from an integral representation of $\log K$ via the Euler
product for $\sinc z$. We give a proof from \eqref{eq:K2} that uses only the three kinds of operations of
Section~\ref{sec:twists}: a dilation, functional equations of the dilogarithm (the duplication formula and
Landen's identity), and telescoping.

\begin{proof}
\emph{Dilation.} With $G(k) = \Li_2(4/k^2)$, $\sum_{k=2}^{\infty}(-1)^kG(k) = -\sum_{k=2}^{\infty}G(k) + 2\sum_{j=1}^{\infty}G(2j)$ and
$G(2j) = \Li_2(1/j^2)$, so
\[
  \frac12\sum_{k=2}^{\infty}(-1)^k\Li_2\Bigl(\frac4{k^2}\Bigr)
  = \frac{\pi^2}{6} + \sum_{k=2}^{\infty}\Bigl(\Li_2\Bigl(\frac1{k^2}\Bigr) - \frac12\Li_2\Bigl(\frac4{k^2}\Bigr)\Bigr).
\]
It therefore suffices to show that $\sum_{k=2}^{\infty} g(k) = \log^22 + \Li_2(-\frac12) + \frac{\pi^2}6$, where
\[
  g(k) = f\Bigl(\frac1k\Bigr) - \Li_2\Bigl(\frac1{k^2}\Bigr) + \frac12\Li_2\Bigl(\frac4{k^2}\Bigr).
\]
\emph{Telescoping.} By $\Li_2(z) + \Li_2(-z) = \frac12\Li_2(z^2)$,
\[
  g(k) = f\Bigl(\frac1k\Bigr) - 2\Li_2\Bigl(\frac1k\Bigr) - 2\Li_2\Bigl(-\frac1k\Bigr) + \Li_2\Bigl(\frac2k\Bigr) + \Li_2\Bigl(-\frac2k\Bigr).
\]
For $k \ge 3$ apply Landen's identity $\Li_2(z) + \Li_2(\frac{z}{z-1}) = -\frac12\log^2(1-z)$ to $z = \frac1k$ and
$z = \frac2k$:
\[
  g(k) = -2\bigl[\Li_2(-\tfrac1k) - \Li_2(-\tfrac1{k-1})\bigr] + \bigl[\Li_2(-\tfrac2k) - \Li_2(-\tfrac2{k-2})\bigr]
  + \Lambda_k,
\]
with $d_k = \log\frac{k}{k-1}$ and
\[
  \Lambda_k = d_kd_{k+1} + d_k^2 - \tfrac12(d_k + d_{k-1})^2 = (d_kd_{k+1} - d_{k-1}d_k) + \tfrac12(d_k^2 - d_{k-1}^2).
\]
All three parts telescope; since $d_k \to 0$,
\[
  \sum_{k=3}^{\infty} g(k) = 2\Li_2(-\tfrac12) - \Li_2(-2) - \Li_2(-1) - \log2\log\tfrac32 - \tfrac12\log^22 .
\]
Adding $g(2) = \log2\log\frac32 - \Li_2(\frac14) + \frac{\pi^2}{12}$ and using the classical values
\begin{gather*}
  \Li_2(-1) = -\frac{\pi^2}{12}, \qquad \Li_2(\tfrac12) = \frac{\pi^2}{12} - \frac12\log^22, \\
  \Li_2(\tfrac14) = 2\Li_2(\tfrac12) + 2\Li_2(-\tfrac12), \qquad
  \Li_2(-2) = -\frac{\pi^2}6 - \frac12\log^22 - \Li_2(-\tfrac12)
\end{gather*}
gives the claim.
\end{proof}

For $k \ge 3$,
\[
  \Li_2\Bigl(\frac4{k^2}\Bigr) = \int_0^\infty e^{-k^2u} G(u)\dd u, \qquad
  G(u) = \frac{\Eip(4u)}{u} = \sum_{n=1}^{\infty}\frac{4^nu^{n-1}}{n\cdot n!},
\]
where $\Eip(y) = \sum_{n=1}^{\infty}y^n/(n\,n!)$, so that $\vartheta\Eip(4u) = e^{4u} - 1$: the kernel has hypergeometric
coefficients and exponential type $4$. The weight $(-1)^k$ makes the relevant theta function
\[
  \theta_4(u) = \sum_{k\in\ZZ}(-1)^ke^{-k^2u} = 2\sqrt{\frac{\pi}{u}}\sum_{m=0}^{\infty}e^{-\pi^2(m+\frac12)^2/u},
\]
whose transform has no constant term (Remark~\ref{rem:twist} with $\hat\chi(n) = 1 - (-1)^n$).

\begin{theorem}\label{thm:khinchin}
For every $u_0 > 0$,
\[
  \log K\log 2 = \log^2 2 + \Li_2(-\tfrac12) + \frac{\pi^2}{12} + \frac12\Bigl(M + \sum_{k=3}^{\infty}(-1)^kI_k + \sum_{m=0}^{\infty}J_m\Bigr),
\]
where, with $x_k = k^2u_0$, $c_m = \pi^2(m+\frac12)^2$ and $y_m = c_m/u_0$,
\begin{align*}
  I_k &= e^{-x_k}\sum_{n=1}^{\infty}\Bigl(\frac{4}{k^2}\Bigr)^n\frac{e_n(x_k)}{n^2}, \\
  J_m &= \frac{2\pi}{\sqrt{c_m}}\erfc(\sqrt{y_m})\bigl(\Ci(4\sqrt{c_m}) - \gamma - \log(4\sqrt{c_m})\bigr) \\
      &\qquad + \sqrt\pi\, e^{-y_m}\mathcal B(c_m), \\
  \mathcal B(c) &= \sum_{n=1}^{\infty}\frac{4^n}{n\cdot n!}\,\beta_n(c), \qquad
  \beta_n(c) = \sum_{i=1}^{n}(-c)^{n-i}u_0^{i-\frac12}\frac{\Gamma(i-\frac12)}{\Gamma(n+\frac12)}, \\
  M &= \sum_{n=1}^{\infty}\frac{4^n}{n^2}\Bigl(\pi_n(u_0) - 4^{-n}\pi_n(4u_0)\Bigr) - \frac12\sum_{n=1}^{\infty}\frac{(4u_0)^n}{n^2\,n!},
\end{align*}
with $\pi_n(y) = 1 - e^{-y}e_n(y)$.
\end{theorem}

\begin{proof}
As in Theorem~\ref{thm:split}, with
$\sum_{k=3}^{\infty}(-1)^ke^{-k^2u} = \frac12\theta_4(u) - \frac12 + e^{-u} - e^{-4u}$,
\[
  T_3 := \sum_{k=3}^{\infty}(-1)^k\Li_2(4/k^2) = \sum_{k=3}^{\infty}(-1)^k I_k + \sum_{m=0}^{\infty} J_m + M,
\]
where
\begin{gather*}
  I_k = \mathcal L[G](k^2), \qquad J_m = \sqrt\pi\,Z_{1/2}[G](c_m), \\
  M = \int_0^{u_0}\Bigl(e^{-u} - e^{-4u} - \frac12\Bigr)G\dd u,
\end{gather*}
and $\log K\log2 = \log^22 + \Li_2(-\frac12) + \frac12(\frac{\pi^2}6 + T_3)$ by Lemma~\ref{lem:dilog}. Termwise,
\begin{gather*}
  \int_{u_0}^\infty e^{-\kappa u}u^{n-1}\dd u = \frac{(n-1)!}{\kappa^n}\,e^{-x}e_n(x), \\
  \int_0^{u_0}e^{-au}u^{n-1}\dd u = \frac{(n-1)!}{a^n}\,\pi_n(au_0)
\end{gather*}
give $I_k$ and $M$. For the dual terms,
\[
  J_m = \sqrt\pi\sum_{n=1}^{\infty}\frac{4^n}{n\cdot n!}D_{n-1/2}(c_m),
\]
and by \eqref{eq:Drec} $D_{n-1/2} = \alpha_nD_{-1/2} + \beta_nE$ with $\alpha_n = (-c)^n/(\frac12)_n$ and the $\beta_n$
as stated; $D_{-1/2} = \sqrt{\pi/c}\erfc\sqrt{c/u_0}$, and
\[
  \sum_{n=1}^{\infty}\frac{4^n}{n\cdot n!}\alpha_n = \sum_{n=1}^{\infty}\frac{(-16c)^n}{n\,(2n)!}
  = 2\bigl(\Ci(4\sqrt c) - \gamma - \log(4\sqrt c)\bigr).
\]
\end{proof}

All three series converge for every $u_0 > 0$: the $I_k$ series like $(4/k^2)^n$, and those for $\mathcal B$ and $M$
like $(16c)^n/(2n)!$ and $(4u_0)^n/n!$. For the constants, $\Li_2(-\frac12) = -\Li_2(\frac13) - \frac12\log^2\frac32$
and $\Li_2(\frac13) = \frac16\Li_2(\frac19) + \frac{\pi^2}{18} - \frac16\log^23$.

\subsubsection{Evaluation}
On the $k$-side, let $P_n = (4/\kappa)^ne_n(x)$ and $Q_n = (4u_0)^n/n!$. Then
\[
  P_{n+1} = \frac4\kappa(P_n + Q_n), \quad Q_{n+1} = \frac{4u_0}{n+1}Q_n, \quad I_k = e^{-x}\sum_{n=1}^{\infty}\frac{P_n}{n^2},
\]
a $2\times2$ triangular recurrence. On the dual side, with $\nu_t = Z_t[\Eip(4u)]$, $Y_t = Z_t[e^{4u}]$,
$C_t = Z_t[1]$ and $E = e^{-c/u_0}$, integration by parts gives
\begin{align*}
  c\,\nu_{t+1} &= E\,\Eip(4u_0)\,u_0^{1-t} - (1-t)\nu_t - Y_t + C_t, \\
  c\,Y_{t+1} &= E\,e^{4u_0}u_0^{1-t} - (1-t)Y_t - 4Y_{t-1}, \\
  c\,C_{t+1} &= E\,u_0^{1-t} - (1-t)C_t,
\end{align*}
and $J_m = \sqrt\pi\,\nu_{3/2}(c_m)$: a system of order $4$ (state $\nu_t, Y_t, Y_{t-1}, C_t$; three forcing constants
$\Eip(4u_0)$, $e^{4u_0}$, $1$) with coefficients linear in $c$ and $t$. Starting values at a rational $c_0$ are
\begin{gather*}
  \nu_{3/2} = \sum_{n=1}^{\infty}\frac{4^n}{n\cdot n!}D_{n-1/2}, \qquad Y_{3/2} = \sum_{n=0}^{\infty}\frac{4^n}{n!}D_{n-1/2}, \\
  Y_{1/2} = \sum_{n=0}^{\infty}\frac{4^n}{n!}D_{n+1/2}, \qquad C_{3/2} = D_{-1/2},
\end{gather*}
summed by one binary splitting. The bit-burst algorithm then steps from $c_0$ to $c_m$, and the main term and the
constants are rational series.

\subsubsection{Error bounds}\label{sec:khbounds}
With $U = u_0$ and $E = e^{-c/U}$, the following bounds make the evaluation rigorous in ball arithmetic (interval arithmetic with midpoint--radius
intervals, as in FLINT \cite{FLINT,Joh17}).
\begin{itemize}
\item \emph{Omitted terms.} Since $G(u) \le e^{4u}/U$ for $u \ge U$,
\[
  0 < I_k \le \frac{e^{-(k^2-4)U}}{(k^2-4)U}, \qquad 0 < J_m \le 2\sqrt{\pi U}\,G(U)\,e^{-\pi^2(m+\frac12)^2/U},
\]
and both tails are at most twice their first term (for the $k$-tail from $k_1$ once $(2k_1+1)U \ge \log 2$), so $p$ bits need $k_{\max} \approx \sqrt{4 + p\log2/U}$ and
$m_{\max} \approx \sqrt{pU\log 2}/\pi$.
\item \emph{$k$-series.} The $n$-th term of $e^{-x}\sum_{n=1}^{\infty}P_n/n^2$ is
\[
  \Bigl(\frac4\kappa\Bigr)^n\frac{\Pr[\mathrm{Poisson}(x) \le n-1]}{n^2},
\]
bounded by $\tau_n = (\frac4\kappa)^n\min(1, \frac{n\,x^{n-1}e^{-x}}{(n-1)!})\frac{1}{n^2}$ for $n - 1 < x$ and by
$(4/\kappa)^n/n^2$ beyond; for $U \le \frac14$ the tail after $N$ terms is at most
$2\tau_N + \frac95(\frac4\kappa)^{n_1}n_1^{-2}$ with $n_1 = \lceil x\rceil + 1$.
\item \emph{Taylor stages.} With $\rho = |\delta|/c_0$, the remainder after $L \ge 2$ terms of the Taylor series of
every state component at $c_0$ is at most
\[
  \frac{2e^{4U}\rho^L}{c_0^{1/2}(1-\rho)},
\]
and, if moreover $|\delta| \le U$, using
\[
  C_t(c_0) = c_0^{1-t}\,\Gamma\Bigl(t-1, \frac{c_0}U\Bigr), \quad \Gamma(a, y) \le \frac{y^{a-1}e^{-y}}{1 - (a-1)/y},
\]
also at most
\[
  8e^{4U}E\,\frac{(|\delta|/U)^L}{L!}\Bigl(\frac{U^{1/2}}{c_0} + \frac{U^{3/2}}{c_0^2}\Bigr)
  + \frac{2e^{4U}\rho^{\lfloor c_0/(2U)\rfloor+1}}{c_0^{1/2}(1-\rho)},
\]
so the stages only need to resolve the $p - (c/U)\log_2 e$ significant bits of $J_m$.
\item \emph{Starting values.} With $b_n = 4^n/n!$, $|b_n\alpha_n| = (16c)^n/(2n)!$, and
$E\beta_n = D_{n-1/2} - \alpha_nD_{-1/2}$ with $0 < D_{n-1/2} \le EU^{n-1/2}/(n-\frac12)$. Once
$16c \le \frac12(2N+1)(2N+2)$ and $4U \le \frac12(N+1)$, every tail of the $\alpha$- and $\beta$-series is at most
twice its first term.
\item \emph{The point.} If the last stage evaluates at $c_\ell$ with $|c - c_\ell| \le \delta$, the error is at most
$\delta\max|J'(\xi)|$ with
$|J'(\xi)| \le 2\sqrt\pi\,\Eip(4U)\,e^{-\xi/U}/(U^{1/2}\xi)$ for $\xi \ge c - \delta$, so
$p - (c/U)\log_2e$ bits of $c$ suffice.
\item \emph{Main term and constants.} The terms of $M$ are at most $3(8U)^N/N!$, and the tail of the series for
$\Li_2(\frac19)$ after $N$ terms is at most $9^{-N}$.
\end{itemize}

\subsection{Comparison}\label{sec:khcompare}
Table~\ref{tab:khcompare} compares the three algorithms
when implemented using the generic backend of Section~\ref{sec:appcomp}.
They share all their machinery: the binary splitting, the bit-burst evaluation of the moment
systems, which are built from the $\vartheta$-relations of the kernel functions as in Proposition~\ref{prop:dual},
and the starting values, from one binary splitting of the kernel coefficients times the recurrence \eqref{eq:Drec}.
Only the representations differ, and each is run at its best split point.

\begin{table}[t]
\centering\small
\caption{Khinchin's constant by the three algorithms:
CPU time (summed over $8$ threads, on the machine of Section~\ref{sec:khresults}), at the best split point
$u_0$ of each, with the fraction spent on the $k$-side in parentheses.
Here all three algorithms were implemented using the same generic backend of Section~\ref{sec:appcomp}
for a fair comparison; a further optimized implementation of K3 is discussed below in Section~\ref{sec:khimpl}.}
\label{tab:khcompare}
\begin{tabular}{lrrr}
\toprule
digits & K1, $u_0 = \frac1{16}$ & K2, $u_0 = \frac1{64}$ & K3, $u_0 = \frac1{16}$ \\
\midrule
$10^4$ & $80.0$\,s ($0.46$) & $18.4$\,s ($0.51$) & $4.04$\,s ($0.45$) \\
$10^5$ & $4\,415$\,s ($0.50$) & $1\,320$\,s ($0.41$) & $234$\,s ($0.48$) \\
\bottomrule
\end{tabular}
\end{table}

The ordering follows Table~\ref{tab:khstates}. Algorithm~K2 has the simpler summand, but the
alternating harmonic numbers $A_n$ are not hypergeometric (they are a first derivative with respect to a parameter,
$A_n = \log 2 + \frac12(\psi(n+\frac12)-\psi(n))$), so its kernel has order $4$ instead of $1$, and its dual system has
dimension $12$ instead of $4$. Algorithm~K3 pays for its hypergeometric coefficients with
exponential type $4$ instead of $1$ (slightly longer $k$-series, and $k = 2$ excluded), which costs far less.
Algorithm~K3 is $4.6$ and $5.6$ times faster than K2 at $10^4$ and $10^5$
digits. (The optimized implementation of Section~\ref{sec:khimpl} takes $88$ CPU seconds at $10^5$ digits on the
same machine, against $234$.)

Algorithm~K1 has the simplest kernel of all, but it is the slowest, by a factor of about
$19$ against K3. $\log K$ is a
derivative in $s$, and differentiating $1/\Gamma(\frac{s+n}2)$ introduces $\psi(\frac n2)$, that is, harmonic numbers
again, while the weight $\log u$ doubles the dual states. In addition the coefficients
$(-1)^n(2^n-2)/n$ combine two exponential scales, which doubles the states once more, and each term needs extra
incomplete gamma values. In short, the logarithm $\log k$, whether as a weight in \eqref{eq:K1} or after summation by parts
as a product of logarithms in \eqref{eq:K2}, shows up as harmonic numbers in the coefficients; the dilogarithm
identity removes it. The second identity of
\cite[\S3]{BBC97},
\[
  \log K\log2 = \frac{\pi^2}{6} - \frac12\log^22 + \sum_{k=2}^{\infty}\Li_2\Bigl(-\frac1{k^2-1}\Bigr),
\]
from Abel's dilogarithm identity, gives yet another representation, with a kernel $-e^u\Ein(u)/u$ of type $1$ but a
five-dimensional dual state.

\subsection{Implementation in FLINT}\label{sec:khimpl}
Our implementation is part of FLINT \cite{FLINT} version 3.7.0-dev.
The function \texttt{arb\_const\_khinchin} uses the new theta splitting algorithm above
$5\cdot10^6$ bits (about $1.5\cdot10^6$ digits) and an optimized zeta-value expansion \eqref{eq:zetavalue} below.
We only sketch it; Appendix~\ref{sec:khpseudo} lists a short annotated reference implementation of the same
algorithm.
\begin{itemize}
\item All linear recurrences (the $k$-terms, the Taylor stages and starting values of the dual terms,
$\Gamma(\frac12, y)$) are evaluated by exact binary splitting of products of small integer matrices over chunks of
terms whose products reach about half the working precision. Within a chunk, the two halves are merged by one
augmented product with FLINT's FFT-based matrix multiplication (\texttt{fmpz\_mat\_mul\_fft\_small}), which
transforms each entry once and exploits the sparsity structure.
Each chunk product is applied to the state, kept as integers scaled by the denominators,
by one truncated transformed matrix--vector product with per-entry exponents, so that all sizes stay $O(p)$.
\item The $k$-terms use the triangular $2\times2$ recurrence of Section~\ref{sec:dilog}; the lowest levels of their trees are done with
fixed-size limb arrays.
\item $\Gamma(\frac12, y)$ at the dyadic points $y = c_0/u_0$ is computed by the upper (asymptotic) series when it is
accurate enough, otherwise by binary splitting of the Legendre continued fraction (whose consecutive convergents
enclose the value) or of the lower series, whichever is shorter.
\item The bit-burst uses growth factor $4$ per stage (the stage cost is about $(g+1)/(g-1)$ times a plain stage,
because the coefficients are linear in the previous point); short final stages are summed directly.
\item Each term is computed with only the relative precision it needs (Remark~\ref{rem:practical}), and every omitted
term and Taylor remainder is bounded rigorously (Section~\ref{sec:khbounds}).
\item The $k$-terms are taken in chunks of consecutive $k$, which share powers of $e^{-u_0}$; the dual terms are taken
largest first, with at most half of the threads on dual terms while $k$-chunks remain. The finished terms are added
exactly in fixed point, so that no term is kept and the result does not depend on the scheduling. The memory is linear
in the precision times the number of threads, mostly for FFT transforms.
\end{itemize}

\subsection{Timings and a record computation}\label{sec:khresults}
Table~\ref{tab:khinchin} compares the theta splitting (with $u_0 = \frac1{16}$) with the zeta-value expansion in FLINT,
both run with $8$ threads on the same machine, with an AMD Ryzen 7 PRO 5850U processor and $32$ GB of memory. The
comparison is with a strong baseline: the zeta-value implementation in FLINT has itself
been optimized meticulously (compared to the program of Ghazi \cite{Gha26}, used for the previous record,
it is $3$ times faster at $10^5$ digits and $5$ times faster at $3\cdot10^5$ digits).
\begin{table}[ht]
\centering\footnotesize
\setlength{\tabcolsep}{3.5pt}
\caption{Khinchin's constant with $8$ threads: CPU time (summed over the threads), wall time and peak memory.}
\label{tab:khinchin}
\begin{tabular}{rrrrrrrr}
\toprule
& \multicolumn{3}{c}{zeta values} & \multicolumn{4}{c}{theta splitting} \\
\cmidrule(lr){2-4}\cmidrule(l){5-8}
digits & CPU & wall & memory & $k$/dual terms & CPU & wall & memory \\
\midrule
$10^5$ & $18.7$\,s & $2.43$\,s & $78$\,MB & $1\,919$/$38$ & $88.3$\,s & $11.2$\,s & $223$\,MB \\
$10^6$ & $4\,239$\,s & $544$\,s & $1.27$\,GB & $6\,069$/$121$ & $5\,416$\,s & $679$\,s & $1.08$\,GB \\
$2\cdot10^6$ & $23\,485$\,s & $3\,372$\,s & $1.96$\,GB & $8\,583$/$171$ & $18\,815$\,s & $2\,355$\,s & $1.78$\,GB \\
$10^7$ & $1\,852\,090$\,s & $235\,475$\,s & $15.3$\,GB & $19\,194$/$382$ & $327\,183$\,s & $40\,921$\,s & $10.5$\,GB \\
\bottomrule
\end{tabular}
\end{table}

Per decade the theta splitting grows by a factor $61$ from $10^5$ to $10^6$ digits and $60$ from $10^6$ to $10^7$ (about
$p^{1.78}$, consistent with $p^{3/2}$ times logarithmic factors), the zeta-value expansion by $227$ and $437$ (about
$p^{2.36}$ and $p^{2.64}$). The zeta-value code is not the $\widetilde O(p^2)$ variant of Section~\ref{sec:zetavalue}:
it computes the zeta values from power sums, which is roughly cubic in $p$ but has much smaller constants, and the
quadratic variant would only become competitive at unreasonably high precision. Part of the speedup at $10^7$ digits
therefore reflects the growth of the baseline rather than the factor $\sqrt p$. This is itself a point in favour of
the theta splitting: it is visibly subquadratic at the precisions actually reached, where the practical alternative is
not even quadratic. The crossover
lies between $10^6$ and $2\cdot10^6$ digits, as reflected in the cutoff of \texttt{arb\_const\_khinchin}. At $10^7$
digits the theta splitting is $5.7$ times faster ($11.4$ hours of wall time against $65.4$) and needs less memory
($10.5$ GB against $15.3$ GB). Both methods parallelize almost perfectly: the ratio of CPU to wall time is $7.9$--$8.0$ on
$8$ threads. The $10^7$-digit result of the theta splitting has error radius $1.8\cdot10^{-10\,000\,044}$, and the
two results agree. Since the theta splitting shares nothing with the zeta-value method (no Bernoulli numbers or power
sums), it is a genuinely independent check. The $10^7$ digits after the decimal point are available
in the repository.\footnote{\url{https://github.com/fredrik-johansson/holonomic-sums/blob/main/results/digits/khinchin_10000000.txt}}

\subsection{The continued fraction of $K$}\label{sec:khcf}
With $D = 10^7$ and $d = \lfloor 10^DK\rfloor$, we have $K \in I = [d, d+1] \cdot 10^{-D}$. We computed the
continued fraction expansions of both endpoints of $I$. They agree in the first $N + 1$ partial quotients, $N = 9\,699\,141$,
\[
  K = [2; 1, 2, 5, 1, 1, 2, 1, 1, 3, 10, 2, 1, 3, 2, 24, 1, 3, 2, 3, \ldots, a_N, \ldots],
\]
(OEIS A002211 \cite{OEIS}) and differ in $a_{N+1}$, which is $9$ for one endpoint and $26$ for the other. Since the set of
real numbers whose expansion begins with $a_0, \ldots, a_N$ is an interval, these are the first $N + 1$ partial quotients of
$K$. Let $p_n/q_n$ be the convergents. Then $q_N = 2.484246\ldots\cdot10^{4\,999\,999}$, and the largest of these
partial quotients is $a_{7\,384\,031} = 34\,770\,305$.

\begin{figure}[ht]
\centering
\begin{tikzpicture}
\begin{axis}[
  name=gm, width=0.46\textwidth, height=0.42\textwidth,
  xmode=log, xmin=10, xmax=1.2e7, ymin=1.9, ymax=3.5,
  xlabel={$n$}, ylabel={$G_n$},
  ytick={2.0,2.5,3.0,3.5}, yticklabel style={/pgf/number format/fixed, /pgf/number format/precision=1},
  tick label style={font=\footnotesize}, label style={font=\small},
  legend style={font=\footnotesize, at={(0.97,0.97)}, anchor=north east, draw=none, fill=none}]
\addplot[gray!60, thin, domain=10:1.2e7, samples=200, forget plot] {2.685452001*exp(2*1.10/sqrt(x))};
\addplot[gray!60, thin, domain=10:1.2e7, samples=200, forget plot] {2.685452001*exp(-2*1.10/sqrt(x))};
\addplot[black, dashed, domain=10:1.2e7, samples=2] {2.685452001};
\addlegendentry{$K$}
\addplot[blue!70!black, thick] table[x=n, y=gm] {\khcfdata};
\addlegendentry{$G_n$}
\end{axis}
\begin{axis}[
  at={(gm.east)}, anchor=west, xshift=0.8cm, width=0.46\textwidth, yticklabel pos=right, ylabel near ticks, height=0.42\textwidth,
  xmode=log, xmin=10, xmax=1.2e7, ymin=-4.2, ymax=4.2,
  xlabel={$n$}, ylabel={$z_n$},
  ytick={-4,-2,0,2,4},
  tick label style={font=\footnotesize}, label style={font=\small}]
\addplot[gray!60, thin, domain=10:1.2e7, samples=2] {2*1.10};
\addplot[gray!60, thin, domain=10:1.2e7, samples=2] {-2*1.10};
\addplot[black, dashed, domain=10:1.2e7, samples=2] {0};
\addplot[blue!70!black] table[x=n, y=z] {\khcfdata};
\end{axis}
\end{tikzpicture}
\caption{The geometric mean $G_n$ of the first $n$ partial quotients of $K$ (left) and the normalized error
$z_n = \sqrt n\,(\log G_n - \log K)$ (right), for $10 \le n \le N = 9\,699\,141$, at about $1\,100$ logarithmically spaced $n$. The grey curves show
$K e^{\pm2\sigma/\sqrt n}$ (left) and $\pm2\sigma$ (right), with $\sigma = 1.10$ estimated from the partial quotients.}
\label{fig:khcf}
\end{figure}
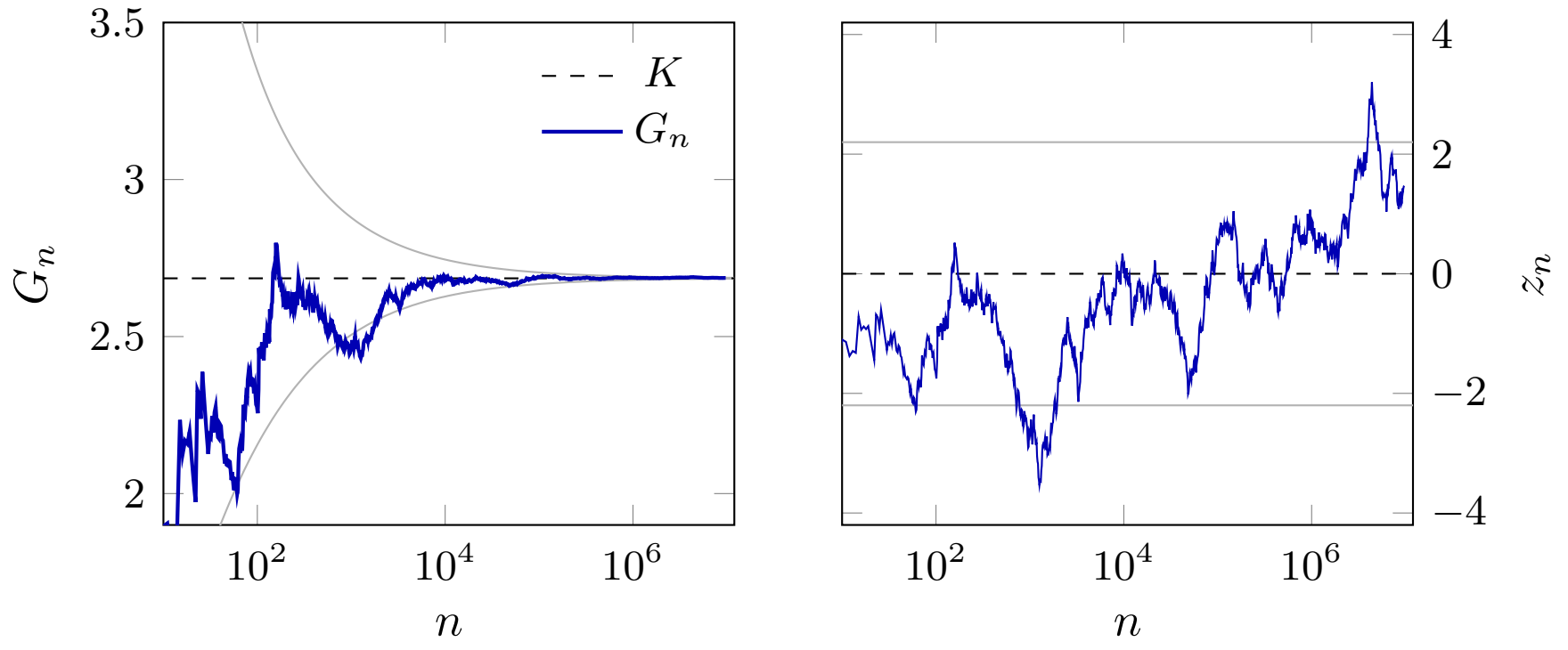

These partial quotients look like those of a typical real number. We have $N/D = 0.969914$ and
$\frac1N\log q_N = 1.187005$, against the constants $6\log2\log10/\pi^2 = 0.970270\ldots$ of Lochs' theorem
\cite{Loc64} and $\pi^2/(12\log2) = 1.186569\ldots$ of L\'evy's theorem \cite{Lev36}, and the proportions of the
partial quotients $1$, $2$ and $3$ are $0.414921$, $0.169996$ and $0.093136$, against the Gauss--Kuzmin probabilities
$0.415037$, $0.169925$ and $0.093109$.

\begin{proposition}\label{thm:khrat}
If $K = p/q$ with
integers $p$ and $q > 0$, then $q \ge 10q_N + q_{N-1} > 2.587\cdot10^{5\,000\,000}$, and $p > 6.94\cdot10^{5\,000\,000}$.
\end{proposition}

\begin{proof}
Let
\[
  x_t = \frac{tp_N + p_{N-1}}{tq_N + q_{N-1}} = [a_0; a_1, \ldots, a_N, t].
\]
Its denominator is $q(x_t) = tq_N + q_{N-1}$ (in lowest terms), and consecutive $x_t$ are Farey neighbours, since
\begin{multline*}
  \bigl((t+1)p_N + p_{N-1}\bigr)(tq_N + q_{N-1}) - (tp_N + p_{N-1})\bigl((t+1)q_N + q_{N-1}\bigr) \\
  = p_Nq_{N-1} - p_{N-1}q_N = \pm1.
\end{multline*}
Exact rational arithmetic shows that $I$ lies
between $x_9$ and $x_{27}$ and contains $x_{10}, \ldots, x_{26}$. A rational number strictly between two Farey
neighbours $a/b$ and $c/e$ has denominator at least $b + e$. Hence every rational number in $I$ is either one of
$x_{10}, \ldots, x_{26}$ or has denominator at least $q(x_9) + q(x_{10})$; in both cases its denominator is at least
$q(x_{10}) = 10q_N + q_{N-1}$. The bound for $p$ follows from $K > 2.685$.
\end{proof}

The proposition rests on the rigorous error bounds of Section~\ref{sec:khbounds} and on the agreement of two
independent computations.

By Khinchin's theorem the geometric means $G_n = (a_1\cdots a_n)^{1/n}$ of almost every $x$ tend to $K$; whether
$K$ itself has this property is not known. Its first $N$ partial quotients give $G_N = 2.686720\ldots$, and the
normalized error $z_n = \sqrt n\,(\log G_n - \log K)$ behaves as the central limit theorem for $\sum_k\log a_k$
\cite{IK02} predicts (Figure~\ref{fig:khcf}): batch means over blocks of $10^4$ partial quotients give a standard
deviation $\sigma \approx 1.10$ (less than $\operatorname{Var}(\log a_1)^{1/2} = 1.18$, since consecutive partial quotients
are weakly correlated), $|z_n| < 3.6$ for $n \ge 10^3$, and $z_N = 1.47 \approx 1.34\,\sigma$.

\section{Other constants}\label{sec:other}

This section treats several other constants of Table~\ref{tab:constants}. For the constants of Euler, Stieltjes, Stirling and
Glaisher (Section~\ref{sec:stieltjes}) we only compare the splitting with the classical formulas. For
$\mathcal A$, $\mathcal C$, $K_{-1}$, the constant of Theodorus and the Hardy--Littlewood function
(Sections~\ref{sec:AG}--\ref{sec:HL}), Somos's constant, a value of a damped function with a geometric factor
(Section~\ref{sec:somos}), and three constants from physics and probability, the vortex sum, $\operatorname{Tr}\Delta^{-3/2}$
on $S^2$ and the stretched-exponential sum $\sum_ke^{-\sqrt k}$ (Sections~\ref{sec:vortex}--\ref{sec:expsqrt}), we give the
kernels and the recurrences of both sides. Sections~\ref{sec:physics} and~\ref{sec:expsqrt} also survey where such sums
arise in spectral theory, statistical mechanics and probability.
Section~\ref{sec:appcomp} reports computations of all of them to $10^6$ digits with a generic program.

\subsection{The constants of Euler, Stieltjes, Stirling and Glaisher}\label{sec:stieltjes}
Several classical constants are the constant terms in the asymptotic expansions of partial sums.
Euler's constant is that of the harmonic numbers,
$\sum_{k\le n}\frac1k = \log n + \gamma + o(1)$, and the Stieltjes constants are, by
Stieltjes's definition, those of the sums of $(\log k)^m/k$,
\[
  \sum_{k=1}^{n}\frac{(\log k)^m}k = \frac{(\log n)^{m+1}}{m+1} + \gamma_m + o(1),
\]
with $\gamma_0 = \gamma$. Stirling's formula is the analogous statement for $\log n! = \sum_{k\le n}\log k$, with the
constant $\log\sqrt{2\pi}$, and Glaisher's constant $A = 1.2824271291\ldots$
plays the same role for the hyperfactorials $1 \cdot 2^2 \cdots n^n$:
\[
  \sum_{k=1}^{n}k\log k = \Bigl(\frac{n^2}2 + \frac n2 + \frac1{12}\Bigr)\log n - \frac{n^2}4 + \log A + o(1).
\]

In general, for $F \in \Reg$ the Euler--Maclaurin formula expands $\sum_{k\le n}F(k)$ in powers of $n$ and
$\log n$, and, with the antiderivative of $F$ normalized to have no constant term, the constant term of this
expansion is the regularized value $\zeta_F(0)$ (the constant term at $s = 0$ if $0$ is a pole) plus the constant terms
of $\frac12F(n)$ and of the corrections $B_{2j}F^{(2j-1)}(n)/(2j)!$, which are elementary (compare \cite{Zag06}).

For the examples above,
$\gamma_m = (-1)^m\CT_{s=1}\zeta^{(m)}(s)$, $\log\sqrt{2\pi} = -\zeta'(0)$ and $\log A = \frac1{12} - \zeta'(-1)$, and the generalized Glaisher constants of the products
$\prod_{k\le n}k^{k^m}$ involve $\zeta'(-m)$.

Infinite products are covered in the same way: Wallis's product
\[
  \frac\pi2 = \prod_{k=1}^{\infty}\frac{4k^2}{4k^2-1}, \qquad \log\frac\pi2 = \zeta_G(0), \quad
  G(k) = -\log\Bigl(1-\frac1{4k^2}\Bigr),
\]
is a value of the class, although computing $\pi$ this way would be circular, as $\pi$ enters the splitting through
the dual points $\pi^2m^2$. Constants of this kind with $F \ne 1$ include Hlawka's constant of Table~\ref{tab:constants},
the constant term of $\sum_{k<n}\arctan(1/\sqrt k) = 2\sqrt n + C_{\mathrm H} + o(1)$ (Section~\ref{sec:theodorus}).

For $F = 1$ all of these are values of $\zeta$ and its derivatives: $\gamma$ can be computed in quasilinear time
\cite{BM80}, and the $\gamma_m$ and $\zeta'(-m)$ in time $p^{3/2+o(1)}$ by Theorem~\ref{thm:complexity} with $F = 1$
(see also \cite{Joh21}). The splitting gives two further kinds of formulas for them. Expanding Riemann's formula
(Example~\ref{ex:riemann}) at $s = 1$ gives the classical series in which every term is a single incomplete gamma
function, for example
\[
  \gamma = 2\sum_{k=1}^{\infty}\Bigl(\frac{\erfc(k\sqrt\pi)}{k} + E_1(\pi k^2)\Bigr) - 2 + \log(4\pi)
\]
(Coffey \cite[Proposition~9]{Cof14}), and similar series for $\gamma_m$ \cite{Cof14}. Alternatively, the limits above
can be written as convergent sums of holonomic functions and hence as values at $s = 0$ of functions $\zeta_F$ and
their derivatives. Telescoping the main term of Stieltjes's definition gives
\[
  \gamma_n = \sum_{k=1}^{\infty}\Bigl(\frac{(\log k)^n}k - \frac{(\log(k+1))^{n+1} - (\log k)^{n+1}}{n+1}\Bigr),
\]
and expanding $\log(k+1) = \log k + \log(1+\frac1k)$,
\[
  \gamma_n = \sum_{j=0}^{n}(-1)^j\zeta_{\Phi_j}^{(j)}(0),
\]
where
\begin{gather*}
  \Phi_n(k) = \frac1k - \log\Bigl(1+\frac1k\Bigr), \\
  \Phi_j(k) = -\frac{1}{n+1}\binom{n+1}{j}\log^{n+1-j}\Bigl(1+\frac1k\Bigr) \quad (0 \le j < n).
\end{gather*}
For $n = 0$ this is $\gamma = \zeta_{\Phi_0}(0)$, the series \eqref{eq:twoconstants}. In the same way,
\[
  \sum_{k=1}^{\infty}\Bigl(\Bigl(k + \frac12\Bigr)\log\Bigl(1+\frac1k\Bigr) - 1\Bigr) = 1 - \log\sqrt{2\pi},
\]
\[
  \sum_{k=1}^{\infty}\Bigl(3k + \frac12 - k(3k+2)\log\Bigl(1+\frac1k\Bigr)\Bigr) = 6\log A - \frac12 - \log\sqrt{2\pi}.
\]
Theorem~\ref{thm:split} applied to these functions gives different series from the classical ones: for $\gamma$,
$f(x) = x - \log(1+x)$ has radius $1$, so $k_0 = 2$, and the kernel at $s = 0$ is
\[
  h_0(u) = \frac{e^u-1}{2u} - \sum_{m=1}^{\infty}\frac{u^{m-1/2}}{(2m+1)\Gamma(m+\frac12)},
\]
so that every $k$-term and dual term is a series rather than a single incomplete gamma function (the method is linear
in $f$, and the contributions of the terms $x^1$ of $x$ and $\log(1+x)$ cancel). The classical formulas are the
special case of the monomial kernel at the self-dual point $u_0 = \pi$; the new ones exist because the $\gamma_n$ are
also sums of holonomic functions. They are not meant to be competitive with the computation of the values of $\zeta$.

\begin{remark}[The Barnes $G$-function]
Various factorial-like products can be extended to analytic functions
via Weierstrass products similar to that for the gamma function.
An example is the Barnes $G$-function
\[
  \log G(1+x) = \frac x2\log2\pi - \frac{x + (1+\gamma)x^2}2 + \zeta_{F_x}(0)
\]
with
\[
F_x(k) = k\log\Bigl(1+\frac xk\Bigr) - x + \frac{x^2}{2k}
\]
which interpolates $G(n) = \prod_{i=0}^{n-2} i!$.
For algebraic $x$, the value $G(x)$ is thus computable in time $p^{3/2+o(1)}$;
unlike $\psi(x)$ and $\Gamma(x)$ no better method is known when $x \in \Qbar \setminus \ZZ$.
Glaisher's constant
is the particular value $A = (2^{1/24}e^{1/8}\pi^{-1/4}/G(\frac12))^{2/3}$.
\end{remark}

\subsection{The Alladi--Grinstead and L\"uroth constants}\label{sec:AG}
The constants $\mathcal A$ and $\mathcal C$ are the values $-\zeta_F'(0)$ for the rational functions
$F(k) = 1/(k^2-1)$ ($k \ge 2$) and $F(k) = 1/(k(k+1))$, whose zeta functions have poles at $s = -1, -3, -5, \dots$
and at $s = -1, -2, -3, \dots$. Partial fractions do not reduce these derivatives to values of $\zeta'$, since $\log k$
and the denominators $k \pm 1$ do not match. In this subsection and the next, $\chi = 1$ and $k_0 = 2$, so
\[
  \Theta(u) = \sum_{k=2}^{\infty}e^{-k^2u} = \frac12\sqrt{\frac\pi u} - \frac12 - e^{-u} + \sqrt{\frac\pi u}\sum_{m=1}^{\infty}e^{-\pi^2m^2/u}
\]
and a sum $S = \sum_{k=2}^{\infty}F(k)$ splits as
\[
  S = \sum_{k=2}^{\infty} I_k + \sqrt\pi\sum_{m=1}^{\infty}J_m + M, \qquad
  M = \int_0^{u_0}\bigl(\tfrac12\sqrt{\pi/u} - \tfrac12 - e^{-u}\bigr)h(u)\dd u.
\]
Here $h = h_0 + u^{-1/2}h_1$ is the kernel at $s = 0$, where we write $h_0 = g_0$ and $h_1 = g_1$ for the even and odd
kernels \eqref{eq:kernelparts}, and $J_m = Z_{1/2}[h_0](\pi^2m^2) + Z_1[h_1](\pi^2m^2)$.

The constant
\[
  \mathcal A = \sum_{k=2}^{\infty}\frac{\log k}{k^2-1} = 1.02313872642793929553\ldots
\]
(OEIS A340440; twice it is A244109, related to L\"uroth representations \cite{Kal88,Fin03}) equals
$-\sum_{i=1}^{\infty}\zeta'(2i)$ and also
\[
  \mathcal A = \tfrac12\log2 + \tfrac12 S_A, \qquad
  S_A = \sum_{k=2}^{\infty}\frac{2}{k}\arctanh\frac1k = \sum_{m=1}^{\infty}\frac{2(\zeta(2m)-1)}{2m-1}
\]
(the second form by summation by parts, Section~\ref{sec:twists}, or from the zeta-value expansion). The
splitting of $S_A$ is the example of Section~\ref{sec:workedexample}: the kernel is
$h_0 = \sqrt{\pi/u}\,\erfi\sqrt u$, the dual state $(\nu_t, Y_t, Y_{t-1})$ has dimension $3$ with two forcing
constants and $t_0 = \frac12$, and the $k$-terms are a $2\times2$ recurrence as for Khinchin's constant.

Next, let
\begin{align*}
  \mathcal C &= \sum_{k=2}^{\infty}\frac1k\log\frac{k}{k-1} = \sum_{k=1}^{\infty}\frac{\log k}{k(k+1)} = \sum_{n=1}^{\infty}\frac{\zeta(n+1)-1}{n} \\
  &= 0.78853056591150896106\ldots
\end{align*}
(A085361). Then
\begin{align*}
  e^{\mathcal C-1} &= 0.80939402054063913071\ldots, \\
  e^{\mathcal C} &= \prod_{k=1}^{\infty}k^{1/(k(k+1))} = 2.20016105809902655319\ldots
\end{align*}
are the Alladi--Grinstead constant \cite{AG77} (A085291) and Kalpazidou's analogue of Khinchin's constant for
L\"uroth expansions \cite{Kal88} (A245254). With
$f(x) = -x\log(1-x) = x\arctanh x - \frac x2\log(1-x^2)$ we have $\mathcal C = \frac12S_A + \mathcal O$, where
\[
  \mathcal O = -\sum_{k=2}^{\infty}\frac{1}{2k}\log\Bigl(1-\frac1{k^2}\Bigr) = \frac12\sum_{m=1}^{\infty}\frac{\zeta(2m+1)-1}{m},
\]
with $a_{2m+1} = \frac1{2m}$. The odd kernel and an auxiliary function are
\[
  h_1(u) = \sum_{m=1}^{\infty}\frac{u^m}{2m\,\Gamma(m+\frac12)}, \qquad
  \varphi(u) = \sum_{m=1}^{\infty}\frac{u^m}{\Gamma(m+\frac12)} = \sqrt u\,e^u\erf\sqrt u,
\]
with $\vartheta h_1 = \frac12\varphi$ and $\vartheta\varphi = (u+\frac12)\varphi + u/\sqrt\pi$. With $\nu_t = Z_t[h_1]$ and
$\Phi_t = Z_t[\varphi]$,
\begin{align*}
  c\,\nu_{t+1} &= Eu_0^{1-t}h_1(u_0) - (1-t)\nu_t - \tfrac12\Phi_t, \\
  c\,\Phi_{t+1} &= Eu_0^{1-t}\varphi(u_0) - (\tfrac32-t)\Phi_t - \Phi_{t-1} - C_{t-1}/\sqrt\pi,
\end{align*}
a state of dimension $4$ ($\nu_t, \Phi_t, \Phi_{t-1}, C_{t-1}$) with three forcing constants and $t_0 = 1$, so that the
starting values involve $E_1(c_0/u_0)$. The $k$-terms are, by \eqref{eq:Ik},
\[
  \frac1k\biggl[-\frac12\log\Bigl(1-\frac1{k^2}\Bigr)\erfc(k\sqrt{u_0})
  + \frac{2}{\sqrt\pi}e^{-x}\sqrt x\sum_{m=1}^{\infty} \frac{\kappa^{-m}}{2m}\sum_{j=0}^{m-1}\frac{(2x)^j}{3\cdots(2j+1)}\biggr].
\]
One run computes $S_A$ and $\mathcal O$ together.

\subsection{Khinchin's harmonic mean}\label{sec:KH}
Khinchin's harmonic mean $K_{-1} = 1/Z(1) = 1.7454056624\ldots$ is the limit of the harmonic means
$n/(a_1^{-1} + \dots + a_n^{-1})$ of the partial quotients of almost every $x$ (Section~\ref{sec:khinchin}). With
$j = k+1$,
\[
  K_{-1} = \frac{\log 2}{T}, \qquad T = \sum_{j=2}^{\infty}\frac{-\log(1-1/j^2)}{j-1},
\]
so that $f(x) = -\frac{x}{1-x}\log(1-x^2)$ and $a_n = H_{\lfloor(n-1)/2\rfloor}$.
Both kernels involve harmonic numbers:
\begin{align*}
  h_0(u) &= \sum_{j=0}^{\infty}\frac{H_j u^j}{j!} = e^u\Ein(u), & \vartheta h_0 &= uh_0 + e^u - 1, \\
  h_1(u) &= \sum_{m=1}^{\infty}\frac{H_mu^m}{\Gamma(m+\frac12)}, & \vartheta h_1 &= (u+\tfrac12)h_1 - \eta + \varphi,
\end{align*}
with $\eta$, $\varphi$ the functions $h_1$ and $\varphi$ of Section~\ref{sec:AG}. With $Y_t = Z_t[e^u]$, the even
and odd dual states are
\[
  (\nu_t, \nu_{t-1}, Y_t, Y_{t-1}, C_t), \qquad (\nu_t, \nu_{t-1}, \eta_t, \Phi_t, \Phi_{t-1}, C_{t-1}),
\]
and
the $k$-side recurrences have states of dimension $4$ (the harmonic numbers double the $2\times2$ systems). The
constant $\sum_{m=1}^{\infty} H_m\kappa^{-m} = -\log(1-\frac1\kappa)/(1-\frac1\kappa)$, with $\kappa = k^2$, multiplies $\erfc(k\sqrt{u_0})$.
The same applies to every $K_{-n}$, $n \ge 1$, with $f(x) = (x/(1-x))^n(-\log(1-x^2))$.

\subsection{The constant of Theodorus}\label{sec:theodorus}
The spiral of Theodorus, or \emph{Quadrat\-wurzel\-schnecke}, is the polygon formed by right triangles with legs $\sqrt k$
and $1$, placed one after the other around a common vertex, so that the $k$-th vertex lies at distance $\sqrt k$ from
it; it is named after Theodorus of Cyrene, who according to Plato proved that $\sqrt3, \dots, \sqrt{17}$ are
irrational. The angles of the triangles are $\arctan(1/\sqrt k)$, and Hlawka \cite{Hla80} showed that the angle swept
by the first $n-1$ triangles is $2\sqrt n + C_{\mathrm H} + o(1)$. Hlawka's constant $C_{\mathrm H} = -2.1577829966\ldots$ is
the regularized value $\zeta_G(0)$ of $G(k) = \arctan(1/\sqrt k)$ (Table~\ref{tab:constants}). Davis \cite{Dav93} interpolated the vertices by a smooth spiral, and its slope at the
point $(1,0)$ is given by the \emph{constant of Theodorus}
\[
  \mathcal T = \sum_{k=1}^{\infty}\frac{1}{(k+1)\sqrt k} = 1.8600250792\ldots
\]
(OEIS A226317), which has also been studied by Gautschi \cite{Gau10} and by Brink \cite{Bri12}. It is a classical
example of a sum of an algebraic function that is not known to reduce to other constants.

In our notation $\mathcal T = \zeta_F(\frac12)$ with $F(k) = 1/(k+1)$, so that $f(x) = x/(1+x)$ and
$a_n = (-1)^{n-1}$ for $n \ge 1$. Since $f$ has radius of convergence $1$, the expansion converges only from $k_0 = 2$
on, and Brink's formula $\mathcal T = \frac12 + \sum_{n=1}^{\infty}(-1)^{n-1}(\zeta(n+\frac12) - 1)$ is the
zeta-value expansion \eqref{eq:zetavalue} of the tail $\zeta_F^{\ge2}(\frac12)$. We use the splitting of this tail at
$s = \frac12$ (the term $k = 1$ enters the main term through $P(\frac{n+1/2}2, u_0)$).

Since $f$ has a single simple pole, the kernel parts \eqref{eq:kernelparts} are incomplete gamma functions:
\begin{align*}
  G_1(u) &= \sum_{j=0}^{\infty}\frac{u^j}{\Gamma(j+\frac34)} = u^{1/4}e^uP(-\tfrac14,u), &
  \vartheta\widehat G_1 &= (u + \tfrac14)\widehat G_1 - \tfrac14, \\
  G_0(u) &= -\sum_{j=0}^{\infty}\frac{u^{j+1}}{\Gamma(j+\frac54)} = -u^{3/4}e^uP(\tfrac14,u), &
  \vartheta\widehat G_0 &= (u + \tfrac34)\widehat G_0 + \tfrac14u,
\end{align*}
with $\widehat G_1 = \Gamma(\frac34)G_1$ and $\widehat G_0 = -\Gamma(\frac54)G_0$, which have rational Taylor
coefficients ($P$ and $Q$ continued analytically in the first argument). In all, at $s = \frac12$,
\[
  h_s(u) = e^u\bigl(Q(\tfrac14,u) - Q(-\tfrac14,u)\bigr).
\]

\emph{The $k$-terms.} On the $k$-side the series over $n$ closes. By
\[
  Q(b+m,x) = Q(b,x) + e^{-x}\sum_{i=0}^{m-1}\frac{x^{b+i}}{\Gamma(b+i+1)},
\]
the double sums are geometric in $1/k^2$ with ratio $x/k^2 = u_0$, and
\begin{gather*}
  I_k = \frac{1}{k^2-1}\Bigl(k^{1/2}Q(\tfrac34, k^2u_0) - k^{-1/2}Q(\tfrac14, k^2u_0) + C_*e^{-k^2u_0}\Bigr), \\
  C_* = \frac{u_0^{3/4}\Phi_{3/4}}{\Gamma(\frac74)} - \frac{u_0^{1/4}\Phi_{1/4}}{\Gamma(\frac54)},
\end{gather*}
with $\Phi_b = \sum_{i=0}^{\infty}u_0^i/(b+1)_i$. Each $k$-term thus costs two incomplete gamma functions and no series.

\emph{The dual terms.} They are $J_m = Z_{3/4}[G_1](c) + Z_{5/4}[G_0](c)$ at $c = \pi^2m^2$. With $C_t = Z_t[1]$,
the states are
\[
  \bigl(Z_t[\widehat G_1], Z_{t-1}[\widehat G_1], C_t\bigr), \qquad
  \bigl(Z_t[\widehat G_0], Z_{t-1}[\widehat G_0], C_t, C_{t-1}\bigr),
\]
of dimensions $3$ and $4$, at $t_0 = \frac34$ and $t_0 = \frac54$; the starting values need $\Gamma(\frac34,y)$ and
$\Gamma(\frac14,y)$. The main term and $C_*$ are eight hypergeometric series in $u_0$.

\begin{remark}[Algebraic summands]\label{rem:algebraic}
For algebraic $F$ the pieces of the theta splitting are not of a special type in general: the kernel parts satisfy the
operators of Lemma~\ref{lem:kernelhol}, whose order is that of the operator of $f$, as for any $F \in \Reg$. For
$f(x) = (1-bx)^{-\alpha}$, for instance, the even kernel part is $u^{-1}\,{}_2F_2(\frac\alpha2, \frac{\alpha+1}2;
\frac12, \frac s2; b^2u)/\Gamma(\frac s2)$, and products of such factors give kernels of higher order. What
simplifies is the case of poles ($\alpha = 1$, where the ${}_2F_2$ reduces to $e^{b^2u}$ times an incomplete gamma
function).

For rational $F$ the kernel parts are such functions, the dual states have small dimension and constant
forcing terms, and on the $k$-side the series over $n$ closes as above, with constants
$\sum_i(b^2u_0)^i/(\beta+1)_i$: each $k$-term is a fixed combination of incomplete gamma functions $Q(\beta, k^2u_0)$
and of $e^{-k^2u_0}$, with explicit coefficients (double poles add derivatives in $b$).

A different
simplification occurs for the Epstein--Hurwitz sums $\sum_k(k^2+a^2)^{-t}$ at $s = 2t$: then
$h_s(u) = u^{t-1}e^{-a^2u}/\Gamma(t)$ is elementary, each $k$-term is the single incomplete gamma function
$(k^2+a^2)^{-t}Q(t, (k^2+a^2)u_0)$, and each dual term is, up to the factor $1/\Gamma(t)$, an incomplete Bessel
function $\int_0^{u_0}u^{t-3/2}e^{-\pi^2m^2/u-a^2u}\dd u$; for $u_0 \to \infty$ this is the Chowla--Selberg formula with
Bessel functions $K_{t-1/2}$ \cite{Eli95,Kir02}, whose $\Theta(p)$ terms the finite split point reduces to
$\Theta(\sqrt p)$ on each side.
\end{remark}

\subsection{The Hardy--Littlewood function}\label{sec:HL}
The function
\[
  H(x) = \sum_{k=1}^{\infty}\frac1k\sin\frac xk,
\]
studied by Hardy and Littlewood, is a standard test case for the summation of slowly convergent series
\cite{Gau05,Mon10}. In our notation $H(x) = \zeta_F(1)$ with $F(k) = \sin(x/k)$, and by \eqref{eq:zetavalue}
\[
  \zeta_F(s) = \sum_{m=0}^{\infty}\frac{(-1)^mx^{2m+1}}{(2m+1)!}\,\zeta(s+2m+1),
\]
which is meromorphic with poles at $s = 0, -2, -4, \dots$, the pole at $s = -2m$ having residue
$(-1)^mx^{2m+1}/(2m+1)!$. Equivalently $H(x) = \zeta_G(0)$ with $G(k) = \frac1k\sin\frac xk$, and we split $\zeta_G$ at
$s = 0$: here $g(t) = t\sin(xt)$ is entire, so $k_0 = 1$ (the summand $G$ has an essential singularity at $k = 0$, which
does not matter). For integer $x$ the kernel is
\[
  h(u) = \sum_{j=0}^{\infty}\frac{(-1)^jx^{2j+1}u^j}{(2j+1)!\,j!} = x\,{}_0F_2\Bigl(;\frac32,1;-\frac{x^2u}4\Bigr),
\]
with $\vartheta^2(\vartheta+\frac12)h = -\frac{x^2}{4}u\,h$,
an entire function of order $\frac13$, so the $k$-side series converge for every $k \ge 1$. With
$P_t = Z_t[h]$, $Q_t = Z_t[\vartheta h]$ and $R_t = Z_t[\vartheta^2h]$, \eqref{eq:ibp} gives
\begin{align*}
  c\,P_{t+1} &= E\,u_0^{1-t}\,h(u_0) - (1-t)P_t - Q_t, \\
  c\,Q_{t+1} &= E\,u_0^{1-t}\,(\vartheta h)(u_0) - (1-t)Q_t - R_t, \\
  c\,R_{t+1} &= E\,u_0^{1-t}\,(\vartheta^2h)(u_0) - \bigl(\tfrac12 - t\bigr)R_t + \tfrac{x^2}{4}P_{t-1},
\end{align*}
and $J_m = Z_{1/2}[h](\pi^2m^2)$: a state of dimension $4$ ($P_t, Q_t, R_t, P_{t-1}$) with three forcing constants and
$t_0 = \frac12$, so that the starting values involve $\Gamma(\frac12, c_0/u_0)$. With $\kappa = k^2$ and
$x_k = \kappa u_0$, the $k$-terms are
\[
  I_k = e^{-x_k}\sum_{m=1}^{\infty}\frac{(-1)^{m-1}x^{2m-1}}{(2m-1)!}\kappa^{-m}e_m(x_k),
\]
a $2\times2$ recurrence as for Khinchin's constant.

\subsection{Somos's quadratic recurrence constant}\label{sec:somos}
Somos's quadratic recurrence constant is
\[
  \sigma = \sqrt{1\sqrt{2\sqrt{3\cdots}}} = \prod_{k=1}^{\infty}k^{1/2^k} = 1.6616879496\ldots,
\]
named after the recurrence $g_n = ng_{n-1}^2$, $g_0 = 1$, whose solutions grow like $\sigma^{2^n}/(n + 2 + O(1/n))$
\cite[\S6.10]{Fin03}, \cite[A112302]{OEIS}. Its logarithm is
\[
  \log\sigma = \sum_{k=1}^{\infty}2^{-k}\log k = -\partial_s\Li_s(\tfrac12)\big|_{s=0}
  = \sum_{k=1}^{\infty}2^{-k}\log\Bigl(1+\frac1k\Bigr).
\]
The same method applies to the generalizations $\sigma_t = \prod_kk^{1/t^k}$, $t > 1$, and to the generalized
Euler-constant function of Sondow and Hadjicostas \cite{SH07},
\[
  \gamma(z) = \sum_{k=1}^{\infty}z^{k-1}\Bigl(\frac1k - \log\Bigl(1+\frac1k\Bigr)\Bigr) \quad (|z| < 1),
  \qquad \gamma(\tfrac12) = 2\log\frac2\sigma.
\]

\emph{The splitting.} We compute $\log\sigma$ from Theorem~\ref{thm:geometric} with $F = 1$ and $k_0 = 1$ (Crandall's
formula), differentiated in $s$. Write $z = \frac12 = e^{2\pi i\alpha}$ with $\alpha = i\log2/(2\pi)$, and put
\begin{gather*}
  x_k = k^2u_0, \qquad y_n = \frac{\pi^2(n-\alpha)^2}{u_0}, \qquad C_k = \frac{2^{-k}+2^k}2, \qquad S_k = \frac{2^{-k}-2^k}2, \\
  L_0 = \log u_0 - \psi(\tfrac12), \qquad V_a(y) = e^yy^{-a}\Gamma(a,y).
\end{gather*}
Then
\[
  -\log\sigma = \sum_{k=1}^{\infty}\Bigl(C_k\frac{E_1(x_k)}2 + S_kB_k\Bigr) + \frac{\sqrt\pi}2\operatorname{Re}T_0
  + \sqrt\pi\sum_{n=1}^{\infty}\operatorname{Re}T_n - \frac{\log u_0 + \gamma}4,
\]
where
\begin{gather*}
  B_k = \frac12Q\Bigl(\frac12,x_k\Bigr)\Bigl(\partial_a\log\Gamma(a,x_k)\big|_{a=1/2} - \psi(\tfrac12) - 2\log k\Bigr), \\
  T_n = \frac{e^{-y_n}}{2\sqrt{u_0}}\Bigl(V_{1/2}(y_n) - i\sqrt{\frac{u_0}\pi}\,\frac{L_0 - V_0(y_n)}{\pi(n-\alpha)}\Bigr).
\end{gather*}
The point $c_0 = -\frac14\log^22$ lies on the cut, but only $\operatorname{Re}T_0$ is needed, which is the same from both
sides and is elementary in $\erfi(\log2/(2\sqrt{u_0}))$ and $\operatorname{Ei}(\log^22/(4u_0))$. The pieces of size $2^k$
in the $k$-terms cancel to $O(2^ke^{-x_k})$, which costs $k$ guard bits.

\emph{Evaluation.} For large $x_k$ the $k$-terms use Legendre's continued fraction for $\Gamma(0,x_k)$ and
$\Gamma(a,x_k)$, differentiated in $a$ by forward differentiation of the matrix product; for small $x_k$ they use the
lower series, with $\log k$ assembled from logarithms of primes. In the dual terms, $V_a$ satisfies
$yV_a' = (y-a)V_a - 1$. Its Taylor coefficients at a Gaussian dyadic point satisfy a two-term recurrence, which we sum
by binary splitting with Gaussian integers, and the bit-burst goes from the real point $\operatorname{Re}y_n$, where
$V_a$ is an incomplete gamma value at a dyadic argument, to $y_n$. We use $u_0 = 1$.

\emph{Reference.} The reference is the sum over prime powers
\[
  \log\sigma = \sum_p\log p\sum_{e\ge1}\frac1{2^{p^e}-1},
\]
with the logarithms of all primes up to $p$ (the precision) computed from
$\log q = \operatorname{arctanh}\frac1{2q^2-1} + \frac12\log\frac{q^2-1}{4} + \log 2$.

\subsection{Periodic arrays of vortex rings}\label{sec:vortex}

Consider an infinite row of identical coaxial thin-cored vortex rings of radius $r$ and circulation $\Gamma$, spaced
$\lambda r$ apart along their common axis. Each ring is carried along by the flow that the others induce at its core
(by the Biot--Savart law, the rings ahead and behind push it forward alike), so the row translates rigidly, faster
than an isolated ring by $U(\lambda)\,\Gamma/r$, where
\begin{equation}\label{eq:vortex}
  U(\lambda) = \frac{1}{2\pi}\sum_{n=1}^{\infty}\int_0^{2\pi}\frac{(1-\cos\varphi)\dd\varphi}{\bigl(2(1-\cos\varphi)+n^2\lambda^2\bigr)^{3/2}},
\end{equation}
a formula found independently by Vasilev (1916) and by Levy and Forsdyke (1928), see \cite{MS22}. It is a sum of the
form \eqref{eq:f}: the summand is a complete elliptic integral, and \cite{MS22} sums it by truncating after $500$ terms.
A few numerical values for different spacings are
\begin{align*}
  U(1)        &= 0.26882772504004215799420784633078761581\ldots, \\
  U(2)        &= 0.07770257288078689394748426153469590197\ldots, \\
  U(\tfrac23) &= 0.48599721401691389437533080303387632179\ldots, \\
  U(\tfrac12) &= 0.71291908153312931408325016665401004450\ldots;
\end{align*}
we compute $U(1)$ to high precision below.

\subsubsection{The series}
Substituting $v = \sin^2(\varphi/2)$, expanding the integrand binomially and using Wallis's integrals gives, for $a > 2$,
\[
  \frac1{2\pi}\int_0^{2\pi}\frac{(1-\cos\varphi)\dd\varphi}{(2(1-\cos\varphi)+a^2)^{3/2}}
  = a^{-3}\,{}_2F_1\Bigl(\frac32,\frac32;2;-\frac4{a^2}\Bigr),
\]
and the identity extends to all $a > 0$ by analytic continuation. By contiguity, ${}_2F_1(\frac32,\frac32;2;z)$ is a
combination of the complete elliptic integrals $K$ and $E$ with coefficients rational in $z$. Hence, with
$\beta = 4/\lambda^2$,
\[
  U(\lambda) = \lambda^{-3}\zeta_F(3), \qquad F(k) = {}_2F_1\Bigl(\frac32,\frac32;2;-\frac{\beta}{k^2}\Bigr),
\]
where $F(k) = f(1/k)$ with $f(x) = {}_2F_1(\frac32,\frac32;2;-\beta x^2)$, a holonomic function with radius of convergence
$\rho = \lambda/2$. Theorem~\ref{thm:complexity} applies with $k_0 > 2/\lambda$, the terms $n \le 2/\lambda$ being added
separately, so $U(\lambda)$ can be computed to $p$ bits in $p^{3/2+o(1)}$ bit operations for every rational $\lambda$ (and,
with algebraic $\beta$, for every algebraic $\lambda$). The leading term gives $U(\lambda) \sim \zeta(3)\lambda^{-3}$ as
$\lambda \to \infty$ \cite{MS22}. The zeta-value expansion of the terms $n \ge k_0$,
\[
  \lambda^{-3}\sum_{j=0}^{\infty}\frac{(3/2)_j^2}{(2)_j\,j!}(-\beta)^j\Bigl(\zeta(2j+3) - \sum_{n=1}^{k_0-1}n^{-2j-3}\Bigr),
\]
converges like $(\beta/k_0^2)^j$.

\subsubsection{A magnetic analogue}
The same elliptic integrals give the mutual inductance of two coaxial circular loops of radius $r$ (Maxwell's
formula \cite{RG12}); at distance $n\lambda r$, by Pfaff's transformation and with $\beta = 4/\lambda^2$,
\[
  M(n\lambda r) = \frac{\pi\mu_0r}{2\lambda^3n^3}\,{}_2F_1\Bigl(\frac32,\frac32;3;-\frac{\beta}{n^2}\Bigr).
\]
A chain of identical ring resonators spaced $\lambda r$ apart carries magnetoinductive waves \cite{SKRS02}. With coupling
between all rings, the dispersion relation of a current wave $I_n \propto e^{in\theta}$ in the quasistatic circuit
model involves the coupling sum
\[
  S_\lambda(\theta) = \frac1{\mu_0r}\sum_{n=1}^{\infty}M(n\lambda r)\cos n\theta
  = \frac{\pi}{2\lambda^3}\,L_F(3, \cos k\theta),
\]
with $F(k) = {}_2F_1(\frac32,\frac32;3;-\beta/k^2)$, which differs from \eqref{eq:vortex} only in the third parameter and in the twist. For rational $\theta/2\pi$ the
weight is periodic and Theorem~\ref{thm:complexity} applies. For rings spaced one diameter apart ($\lambda = 2$), at
the band edges,
\begin{align*}
  S_2(0) &= 0.14792841117146353358\ldots, \\
  S_2(\pi) &= -0.09695771912060612665\ldots
\end{align*}

\subsubsection{The splitting}
The kernel has a closed form. Let
\[
  y(u) = {}_1F_1\Bigl(\frac32; 2; -\beta u\Bigr) = e^{-\beta u/2}\Bigl(I_0\bigl(\tfrac{\beta u}{2}\bigr) - I_1\bigl(\tfrac{\beta u}2\bigr)\Bigr).
\]
Termwise integration gives
\[
  n^{-3}F(n) = \frac{2}{\sqrt\pi}\int_0^\infty u^{1/2}y(u)e^{-n^2u}\dd u.
\]
Since
$y(u) \sim (\beta u)^{-3/2}/\sqrt\pi$ as $u \to +\infty$, the kernel $u^{1/2}y(u)$ is bounded on $(0,\infty)$, and this
representation holds for every $n \ge 1$, not only for $n > 2/\lambda$. Splitting at $u_0$ as in
Theorem~\ref{thm:split}, with
\[
  \sum_{n=1}^{\infty}e^{-n^2u} = \frac12\Bigl(\sqrt{\frac\pi u} - 1\Bigr) + \sqrt{\frac\pi u}\sum_{m=1}^{\infty}e^{-\pi^2m^2/u},
\]
gives
\[
  U(\lambda) = \frac{2}{\sqrt\pi\,\lambda^3}\Bigl(\sum_{n=k_0}^{\infty}I_n + \sum_{n=1}^{k_0-1}(T_n - L_n) + M + \sqrt\pi\sum_{m=1}^{\infty}J_m\Bigr),
\]
where, with $\kappa = n^2$ and $x = \kappa u_0$,
\begin{align*}
  I_n &= \mathcal L[u^{1/2}y](\kappa)
       = \frac{1}{2\kappa^{3/2}}\Bigl(\Gamma(\tfrac12, x)\,F(n)
         + 2e^{-x}\sqrt x\sum_{j=0}^{\infty}A_jR_j\Bigr), \\
  &\qquad A_j = \frac{(3/2)_j^2}{(2)_j\,j!}\Bigl(-\frac\beta \kappa\Bigr)^j, \quad R_j = \sum_{i=0}^{j}\frac{\Gamma(\frac32)\,x^i}{\Gamma(i+\frac32)}, \\
  T_n &= \int_0^\infty u^{1/2}y\,e^{-\kappa u}\dd u = \frac{\sqrt\pi}{2}(\kappa+\beta)^{-3/2}\,{}_2F_1\Bigl(\frac32,\frac12;2;\frac{\beta}{\beta+\kappa}\Bigr), \\
  L_n &= \int_0^{u_0}u^{1/2}y\,e^{-\kappa u}\dd u, \\
  M &= \frac{\sqrt\pi}2\int_0^{u_0}y\dd u - \frac12\int_0^{u_0}u^{1/2}y\dd u, \\
  J_m &= Z_0[y](\pi^2m^2) .
\end{align*}
The $k$-side series converge like $(\beta/\kappa)^j$, so we take $k_0$ minimal with $k_0^2 > 2\beta$; the finitely many terms
$n < k_0$ are the full integral $T_n$ (by Pfaff's transformation) minus $L_n$, whose integrand $u^{1/2}y(u)e^{-\kappa u}$ has an
entire power series factor with coefficients satisfying a three-term recurrence. $F(n) = \sum_{j=0}^{\infty}A_j$ is summed in the
same binary splitting as $\sum_{j=0}^{\infty}A_jR_j$, with the state $(A_jR_j, A_jx^j/\Gamma(j+\frac32), A_j)$.

\subsubsection{Dual side} With $\vartheta y = w$, the equation of $y$ is $\vartheta w = -w - \beta u(w + \frac32 y)$. For
$P_t = Z_t[y]$ and $Q_t = Z_t[w]$ (Section~\ref{sec:splitting}), \eqref{eq:ibp} gives
\begin{align*}
  c\,P_{t+1} &= E\,u_0^{1-t}\,y(u_0) - (1-t)P_t - Q_t, \\
  c\,Q_{t+1} &= E\,u_0^{1-t}\,w(u_0) + t\,Q_t + \beta Q_{t-1} + \tfrac{3\beta}{2}P_{t-1},
\end{align*}
and $J_m = P_0(\pi^2m^2)$: a system of order $4$ (state $P_t, Q_t, P_{t-1}, Q_{t-1}$) with coefficients linear in $c$ and
two forcing constants. The moment indices are integers, so the starting values at $c_0$ are combinations of
$E_1(c_0/u_0)$ and $e^{-c_0/u_0}$: with $y = \sum_{j=0}^{\infty}y_ju^j$ and $D_a = \alpha_aE_1(c_0/u_0) + \beta_aE$ as in
Section~\ref{sec:splitting},
\begin{align*}
  P_0 &= \sum_{j=0}^{\infty}y_jD_{j+1}, & Q_0 &= \sum_{j=0}^{\infty} jy_jD_{j+1}, \\
  P_{-1} &= \sum_{j=0}^{\infty}y_jD_{j+2}, & Q_{-1} &= \sum_{j=0}^{\infty}jy_jD_{j+2},
\end{align*}
all summed by one binary splitting with a three-term state.

For small $\lambda$ the sum samples the function $a \mapsto a^{-3}\,{}_2F_1(\frac32,\frac32;2;-4/a^2)$ at the fine grid $a = n\lambda$, and the method
becomes more expensive: there are about $2.8/\lambda$ terms $n < k_0$, and the power series of $y$ at $u_0$, which
enter the main term and the starting values, have terms of size up to $e^{\beta u_0}$, so that about $5.8\,u_0/\lambda^2$ guard
bits are needed. For $\lambda \ge p^{-1/2}$ these costs stay within the bound of Theorem~\ref{thm:complexity}.

\subsection{Spectral sums and partition functions}\label{sec:physics}

Many quantities in mathematical physics are traces $\sum_kd_k\,\phi(\lambda_k)$ over a discrete spectrum whose
eigenvalues $\lambda_k$ and multiplicities $d_k$ are algebraic functions of a quantum number $k$. When
$d_k\phi(\lambda_k) = k^{-s}F(k)$ with $F \in \Reg$, the trace is a value of $\zeta_F$, and weights $\log\lambda_k$
come from derivatives in $s$. Sums of this kind include:
\begin{itemize}
\item spectral zeta functions $\zeta_A(s) = \operatorname{Tr}A^{-s} = \sum_kd_k\lambda_k^{-s}$ of Laplace-type operators $A$, with
their values at special points: one-loop effective actions, Casimir energies $\frac12\zeta_A(-\frac12)$ and
determinants $\exp(-\zeta_A'(0))$ \cite{Eli95,Kir02};
\item Kaluza--Klein towers, where a field in an extra dimension compactified on a circle of radius $R$ gives modes of mass
$m_k^2 = m^2 + k^2/R^2$, and a one-loop amplitude becomes a sum over $k$ of a loop function of $m_k$, such as the KK top
tower of Table~\ref{tab:constants} \cite{Pet02} and the Epstein--Hurwitz sums $\sum_k(k^2+a^2)^{-t}$;
\item partition functions $\operatorname{Tr}e^{-H/(k_BT)}$ and other thermal sums over bound states, regularized where
they diverge (energy levels that grow like a fractional power of the quantum number lead to stretched-exponential sums
instead, Section~\ref{sec:expsqrt});
\item integrals against the Bose--Einstein and Fermi--Dirac weights, which become sums over the Boltzmann factors
$e^{-kt}$.
\end{itemize}
The splitting itself has a familiar reading here. For $u > u_0$ the $k$-terms are a sum over the spectrum. For $u < u_0$
Poisson summation turns the spectrum into periodic orbits: the main term is the Weyl term (the orbit $m = 0$), and the
dual terms are the orbits of winding number $m$. Theorem~\ref{thm:split} is in this sense the simplest trace formula.

\subsubsection{The sphere} The Laplacian $\Delta$ on the unit sphere $S^2$ has the eigenvalues $\ell(\ell+1)$ with
multiplicities $2\ell+1$, so
\[
  \zeta_{S^2}(s) = \sum_{\ell=1}^{\infty}\frac{2\ell+1}{(\ell(\ell+1))^s} = \zeta_F(2s-1), \qquad
  F(k) = \Bigl(2+\frac1k\Bigr)\Bigl(1+\frac1k\Bigr)^{-s},
\]
for $\operatorname{Re}s > 1$. Here $F$ depends on $s$, so the continuation in $s$ is not that of $\zeta_F$ at fixed $F$:
where a pole of $\zeta(2s-1+j)$ meets a zero of the coefficient $c_j(s)$ of $k^{-j}$ in $F$, a rational correction appears.
For integer $s$ the summand is rational, and $\zeta_{S^2}(s)$ reduces to zeta values; for instance
$(2\ell+1)/(\ell(\ell+1))^2 = \ell^{-2} - (\ell+1)^{-2}$ telescopes to $\zeta_{S^2}(2) = 1$. Half-integer $s$ give algebraic
summands:
\begin{align*}
  \zeta_{S^2}(\tfrac32) &= \operatorname{Tr}\Delta^{-3/2} = 2.07054447632160590509\ldots, \\
  \zeta_{S^2}(\tfrac12) &= -1.75429230348530643626\ldots, \\
  \zeta_{S^2}(-\tfrac12) &= \zeta_F(-2) + \tfrac1{24} = -0.26509554911885852545\ldots
\end{align*}
The correction $\frac1{24}$ comes from the term $c_3(s)\zeta(2s+2)$ of the expansion, where
$c_3(s) = -s(s+1)(2s+1)/6$ vanishes at the pole $s = -\frac12$ of $\zeta(2s+2)$:
\[
  \lim_{s\to-1/2}c_3(s)\zeta(2s+2) = \frac{c_3'(-\frac12)}{2} = \frac1{24}.
\]
The last value gives the Casimir energy of a massless, minimally coupled scalar field on $\mathbb R\times S^2$, whose
modes have the frequencies $\omega_\ell = \sqrt{\ell(\ell+1)}$: the zeta-regularized value of
$\frac12\sum_\ell(2\ell+1)\,\omega_\ell$ is
\[
  E_{S^2} = \frac12\zeta_{S^2}(-\tfrac12) = -0.13254777455942926272\ldots
\]
For the conformally coupled field, with $\omega_\ell = \ell+\frac12$, the energy is $\zeta(-2,\frac12) = 0$ instead.
The determinant is a closed form \cite{Kir02},
\[
  \det{}'\Delta = \exp(-\zeta_{S^2}'(0)) = e^{1/2-4\zeta'(-1)} = 3.19531148605918608395\ldots,
\]
which gives a check: in our setting
\begin{gather*}
  \zeta_{S^2}'(0) = 2\zeta_{F_0}'(-1) - \zeta_{F_1}(-1) + \frac12, \\
  F_0(k) = 2+\frac1k, \qquad F_1(k) = \Bigl(2+\frac1k\Bigr)\log\Bigl(1+\frac1k\Bigr),
\end{gather*}
where the $\frac12$ comes from the term $c_2(s)\zeta(2s+1)$ with $c_2(s) = s^2$.

\subsubsection{Partition functions and Bose--Einstein integrals} The internal partition function
$\sum_n2n^2e^{\xi/n^2}$ of the hydrogen atom, $\xi = I/(k_BT)$ with $I$ the ionization energy, diverges. The
Planck--Larkin partition function removes the first two terms of each exponential \cite{EKR11},
\[
  \sigma(T) = \sum_{n=1}^{\infty}2n^2\Bigl(e^{\xi/n^2} - 1 - \frac{\xi}{n^2}\Bigr) = 2\zeta_F(-2), \qquad
  F(k) = e^{\xi/k^2} - 1 - \frac{\xi}{k^2}.
\]
Since $F$ is entire in $1/k$, $\sigma(T) = 2\sum_{j=2}^{\infty}\zeta(2j-2)\xi^j/j!$, and, as for the Hardy--Littlewood
function, the zeta-value expansion needs only even zeta values with factorially decaying coefficients. For example,
\[
  \sigma\big|_{\xi=1} = 2.1104548762585976160\ldots, \qquad \sigma\big|_{\xi=10} = 44148.268546579674\ldots,
\]
where $\xi = 1$ corresponds to about $1.58\cdot10^5$\,K.

Integrals against the Bose--Einstein weight lead to sums of the same kind. For an entire holonomic function $G$ with
$|G(t)| \le Ce^{ct}$, $c < 1$, on the positive axis,
\[
  \int_0^\infty\frac{G(t)}{e^t-1}\dd t = \sum_{k=1}^{\infty}F(k), \qquad F(k) = \int_0^\infty G(t)e^{-kt}\dd t,
\]
where the Laplace transform $F$ is again holonomic. The Fermi--Dirac weight $(e^t+1)^{-1}$ gives the corresponding
alternating sum. For instance, $G(t) = t\,J_0(at)$ gives $F(k) = k/(k^2+a^2)^{3/2}$, and
\[
  \int_0^\infty\frac{t\,J_0(t)}{e^t-1}\dd t = \sum_{k=1}^{\infty}\frac{k}{(k^2+1)^{3/2}} = 0.90052473534812592430\ldots
\]

\subsubsection{\texorpdfstring{The splitting for $\operatorname{Tr}\Delta^{-3/2}$}{The splitting for the trace}} We use the half-integer quantum number $y = \ell+\frac12$,
for which $\lambda_\ell = y^2-\frac14$ and
\[
  \operatorname{Tr}\Delta^{-3/2} = \sum_{\ell=1}^{\infty}G(\ell+\tfrac12), \qquad
  G(y) = \frac{2y}{(y^2-\frac14)^{3/2}} = 2\sum_{n=0}^{\infty}\frac{c_n}{y^{2n+2}},
\]
with $c_n = (\frac32)_n/(n!\,4^n)$, a shifted sum in the sense of Section~\ref{sec:twistshift}. Termwise, $G(y) = \int_0^\infty e^{-y^2u}h(u)\dd u$ with the
kernel
\[
  h(u) = 2\sum_{n=0}^{\infty}\frac{c_n}{n!}u^n = 2\,{}_1F_1\Bigl(\frac32;1;\frac u4\Bigr), \qquad
  \vartheta^2h = \frac u4\Bigl(\vartheta+\frac32\Bigr)h,
\]
and the theta function of the half-integer lattice is
\[
  \sum_{y\in\ZZ+\frac12}e^{-y^2u} = \sqrt{\frac\pi u}\sum_{m\in\ZZ}(-1)^me^{-\pi^2m^2/u}.
\]
Splitting at $u_0$ gives
\[
  \operatorname{Tr}\Delta^{-3/2} = \sum_{y\ge3/2}I_y + M + \sqrt\pi\sum_{m=1}^{\infty}(-1)^mJ_m,
\]
where, with $x = y^2u_0$ and $e_n(x) = \sum_{i=0}^{n}x^i/i!$,
\begin{align*}
  I_y &= \int_{u_0}^\infty e^{-y^2u}h(u)\dd u = 2e^{-x}\sum_{n=0}^{\infty}c_ny^{-2n-2}e_n(x), \\
  M &= \int_0^{u_0}\Bigl(\frac12\sqrt{\frac\pi u} - e^{-u/4}\Bigr)h(u)\dd u \\
    &= \sqrt\pi\sum_{n=0}^{\infty}\frac{c_n}{n!}\frac{u_0^{n+1/2}}{n+\frac12}
       - 2u_0\sum_{n=0}^{\infty}\frac{(-\frac12)_n}{n!^2(n+1)}\Bigl(-\frac{u_0}4\Bigr)^n, \\
  J_m &= Z_{1/2}[h](\pi^2m^2).
\end{align*}
The term $e^{-u/4}$ in $M$ removes $y = \frac12$, the zero eigenvalue, and the second series in $M$ comes from Kummer's
transformation $e^{-u/4}h(u) = 2\,{}_1F_1(-\frac12;1;-\frac u4)$. The half-integer lattice makes the dual terms
alternate. The moments on the $k$-side have integer indices, so the $k$-terms need no incomplete gamma functions: each
is a double series converging like $(4y^2)^{-n} \le 9^{-n}$, summed by binary splitting with the
state $c_ny^{-2n-2}\cdot(e_n(x), x^{n+1}/(n+1)!)$. The dual terms use the moment system of $(h, \vartheta h)$ of
order $4$, with starting values from $\Gamma(\frac12, c_0/u_0)$. The references are Euler--Maclaurin summation, with the
Taylor coefficients of $F(x) = (2x+1)(x(x+1))^{-3/2}$ at the cutoff $N$ from $2x(x+1)(2x+1)F' = -(8x^2+8x+3)F$ and
$\int_N^\infty F = 2(N(N+1))^{-1/2}$, and the zeta-value expansion
\[
  \operatorname{Tr}\Delta^{-3/2} = 2\sum_{n=0}^{\infty}c_n\zeta(2n+2,\tfrac32),
\]
where $\zeta(2n+2,\frac32) = (2^{2n+2}-1)\zeta(2n+2) - 2^{2n+2}$, which needs only even zeta values and converges like
$9^{-n}$.

\subsection{Stretched-exponential sums}\label{sec:expsqrt}
The summands of Section~\ref{sec:physics} are regular at infinity. A Boltzmann factor $e^{-E_n/(k_BT)}$ over energy
levels that grow like a fractional power of the quantum number is not, and leads instead to the damped zeta functions
of Section~\ref{sec:beyond}; so does a heat kernel whose generator has a fractional power in its spectrum. We give one
example of each; at the simplest parameters both lead to the same constant.

\emph{Landau levels of graphene.} In graphene in a perpendicular magnetic field $B$, the charge carriers behave as
massless Dirac fermions, and their Landau levels are not equidistant, as for a nonrelativistic electron gas, but
\[
  E_{\pm n} = \pm E_1\sqrt n, \qquad E_1 = v_F\sqrt{2e\hbar B} \approx 36\,\mathrm{meV}\cdot(B/1\,\mathrm T)^{1/2},
\]
with $v_F \approx 10^6$\,m/s the Fermi velocity \cite{CGPNG09}. The Boltzmann sums over the conduction-band levels
$n \ge 1$, which give the partition function in the nondegenerate limit and its moments, are values of
$\Lambda(s;\tau,\frac12)$ with $\tau = E_1/(k_BT)$:
\[
  \sum_{n=1}^{\infty}n^{-s}e^{-\tau\sqrt n} = \Lambda(s;\tau,\tfrac12).
\]
For instance, the mean energy over these levels is $E_1\Lambda(-\frac12;\tau,\frac12)/\Lambda(0;\tau,\frac12)$. For
$\tau = 1$, which is $B = 10$\,T at about $1\,330$\,K,
\begin{gather*}
  \Lambda(-\tfrac12;1,\tfrac12) = \sum_{n=1}^{\infty}\sqrt n\,e^{-\sqrt n} = 3.86299603659638723326\ldots, \\
  \frac{\Lambda(-\frac12;1,\frac12)}{\Lambda(0;1,\frac12)} = 2.31260791984760101746\ldots
\end{gather*}

\emph{Fractional diffusion.} The wrapped symmetric $\alpha$-stable distribution on the circle \cite{MJ00}, which is
also the heat kernel of the fractional Laplacian $(-\Delta)^{\alpha/2}$ there, has the density
\[
  p_t(x) = \frac1{2\pi}\Bigl(1 + 2\sum_{k=1}^{\infty}e^{-tk^\alpha}\cos kx\Bigr) \qquad (0 < \alpha < 2).
\]
At $x = 0$ this is $(1 + 2\Lambda(0;t,\alpha))/(2\pi)$, and when $x/2\pi$ is rational it is a value of $\Lambda$ with
a periodic weight (Section~\ref{sec:twistshift}).

\emph{The constant.} For $\tau = 1$ in the first example and for $\alpha = \frac12$, $t = 1$ in the second, both lead to
\[
  S = \sum_{k=1}^{\infty}e^{-\sqrt k} = \Lambda(0;1,\tfrac12), \qquad p_1(0) = \frac{1+2S}{2\pi} = 0.69086194719938\ldots
\]
To compute $S$ we use Proposition~\ref{prop:expsplit} with $r = \frac12$, $\tau = 1$ and $s = 0$. Then $a_j = -j/4$, and the class $j \equiv 0 \pmod 4$ drops out at $s = 0$.

\emph{The kernel.} For $j = 4m+\varrho$, $\varrho \in \{1,2,3\}$, the kernel is $\omega_\varrho w^{1+\varrho/4}H_\varrho(w)$
with $w = 1/u$ and $\omega_\varrho = 1/\Gamma(-\varrho/4)$, where
\begin{gather*}
  H_\varrho(w) = \sum_{m=0}^{\infty}g_mw^m, \qquad g_0 = \frac{(-1)^\varrho}{\varrho!}, \\
  \frac{g_{m+1}}{g_m} = -\frac{m+1+\varrho/4}{(4m+\varrho+1)(4m+\varrho+2)(4m+\varrho+3)(4m+\varrho+4)},
\end{gather*}
and
\[
  \prod_{i=0}^{3}(4\vartheta+\varrho-i)\,H_\varrho = -w\Bigl(\vartheta+1+\frac\varrho4\Bigr)H_\varrho.
\]

\emph{The $k$-terms.} With $x = k^2u_0$, the relation $Q(a,x) = Q(a+1,x) - x^ae^{-x}/\Gamma(a+1)$ gives
\[
  I_k = \sum_{\varrho=1}^{3}(-1)^\varrho\Bigl(k^{\varrho/2}Q\bigl(1-\tfrac\varrho4,x\bigr)\sum_{m\ge0}\frac{k^{2m}}{(4m+\varrho)!}
  - \frac{e^{-x}u_0^{-\varrho/4}}{\Gamma(1-\frac\varrho4)}\sum_{m\ge0}\frac{k^{2m}G_m}{(4m+\varrho)!}\Bigr),
\]
where
\[
  G_m = \sum_{i=0}^{m}\prod_{l=1}^{i}\frac{1-\frac\varrho4-l}{x}:
\]
three incomplete gamma values and two rational series per class.

\emph{The dual terms.} The dual term is $\sum_\varrho\omega_\varrho Z_\varrho(\pi^2m^2)$ with
$Z_\varrho(c) = \sum_mg_mD_{-1/2-\varrho/4-m}(c)$. Let $t = \frac32 + \frac\varrho4$ and
\[
  Z = \bigl(Z_t[\vartheta^iH_\varrho]\bigr)_{i<4}, \qquad
  \Phi = e^{-c/u_0}u_0^{1-t}\bigl((\vartheta^iH_\varrho)(1/u_0)\bigr)_{i<4}.
\]
Then
\[
  c^2Z' = -(A_1A_0' + cA_0')Z - (A_1 + c)\Phi, \qquad \Phi' = -\Phi/u_0,
\]
where $A_0$ is the companion matrix of $\prod_{i=0}^{3}(\vartheta + \frac{\varrho-i}4)$, $A_0' = A_0 + (t-1)I$, and
$A_1$ has the single nonzero row $(-(1+\frac\varrho4)/256, -\frac1{256}, 0, 0)$. Unlike in \eqref{eq:fuchsys}, the
singular point $c = 0$ is irregular, since the kernel is now an expansion at $u = \infty$. The starting values at a
dyadic $c_0$ with $\min(\max(64, p/256), 1024)$ bits come from the series for $Z_\varrho$ via \eqref{eq:Drec}, with one
incomplete gamma value $\Gamma(\frac12+\frac\varrho4, c_0/u_0)$.

For the Taylor steps we do not use the system above, whose recurrence has three terms, but the moment shift
$-\partial_cZ_t = Z_{t+1}$. By \eqref{eq:ibp} and the equation of $H_\varrho$,
\[
  (cI - A_1)Z_{t+1} = (A_0 + (t-1)I)Z_t + \Phi,
\]
and $A_1^2 = 0$, so that $(cI - A_1)^{-1} = (I + A_1/c)/c$. The Taylor coefficients then satisfy a first-order
recurrence with a state of dimension $8$ (as in Proposition~\ref{prop:dual}), which we sum by binary splitting.

\emph{Main term and reference.} The main term is $M = 2 - \frac{\sqrt\pi}2T_{1/2} - \frac12(1 - T_1)$, and we use
$u_0 = \frac1{16}$. The reference is Euler--Maclaurin summation, with the Taylor coefficients of $F(x) = e^{-\sqrt x}$
at the cutoff from $4xF'' + 2F' - F = 0$, and $\int_M^\infty e^{-\sqrt x}\dd x = 2(\sqrt M+1)e^{-\sqrt M}$.

\subsection{Computations}\label{sec:appcomp}
We computed the constants of Sections~\ref{sec:AG}--\ref{sec:expsqrt} with a generic implementation of
Theorem~\ref{thm:complexity} at $s = 0$ (at $s = \frac12$ for $\mathcal T$), the program \texttt{holsum}
in the repository.\footnote{\url{https://github.com/fredrik-johansson/holonomic-sums}} It is much less optimized than the implementation of
Section~\ref{sec:khimpl}: ball arithmetic
throughout, FLINT's default matrix products, heuristic (but conservative) truncation of the
series, the same growth factor $4$, and split point $u_0 = \frac1{16}$ for the constants of
Sections~\ref{sec:AG}--\ref{sec:theodorus} and for $\operatorname{Tr}\Delta^{-3/2}$, $u_0 = \frac14$ for the Hardy--Littlewood function and the vortex sum, and
the split points of Sections~\ref{sec:somos} and~\ref{sec:expsqrt} for the damped constants. (The program adds a margin of two terms
to the counts \eqref{eq:counts}.)

Tables~\ref{tab:apps} and~\ref{tab:appsref} show its timings with $8$ threads (on
the machine of Section~\ref{sec:khresults}) for $\mathcal A$, $\mathcal C$, $T$ (for $K_{-1}$), $\mathcal T$, $H(1)$, $\sigma$,
the vortex sum at $\lambda = 1$ ($\beta = 4$, $k_0 = 3$), $\operatorname{Tr}\Delta^{-3/2}$ on $S^2$ and $S = \sum_ke^{-\sqrt k}$. Besides the theta splitting, the
program implements the reference methods of Sections~\ref{sec:somos} and~\ref{sec:expsqrt} for $\sigma$ and $S$, Euler--Maclaurin
summation for $\mathcal T$, and two reference methods for the others. The
zeta-value expansion \eqref{eq:zetavalue} is applied after taking out the terms $k \le 31$, so that it converges
like $32^{-n}$, except for
\[
  H(x) = \sum_{m=0}^{\infty}(-1)^m\zeta(2m+2)\frac{x^{2m+1}}{(2m+1)!},
\]
whose coefficients decay factorially, and for $\operatorname{Tr}\Delta^{-3/2}$, where only the term $y = \frac12$ is
taken out and the expansion converges like $9^{-n}$ (Section~\ref{sec:physics}); the zeta values come from FLINT, in blocks of consecutive arguments. Euler--Maclaurin summation uses the
cutoff $M \approx p/4$ for $p$ bits; the Taylor coefficients of the summand at $M$ come from a linear recurrence (from
its differential equation), the Bernoulli numbers from FLINT's reverse iterator. All results agree to all digits
where they overlap, and the $10^6$-digit values agree with the $10^5$-digit ones. At $10^5$ digits, the theta
splitting also agrees with runs at a second split point, which change every $k$-term, dual term and the main term.

\begin{table}[tp]
\centering\small
\caption{Timings of the theta splitting in \texttt{holsum} with $8$ threads: CPU time summed over the threads, with
the fraction spent on the $k$-side in parentheses, wall time and peak memory.}
\label{tab:apps}
\begin{tabular}{llrrrr}
\toprule
constant & digits & $k$/dual terms & CPU ($k$-side) & wall & memory \\
\midrule
$\mathcal A$ & $10^4$ & $606$/$12$ & $2.35$\,s ($0.52$) & $0.39$\,s & $58$\,MB \\
 & $10^5$ & $1\,918$/$38$ & $137$\,s ($0.58$) & $17.9$\,s & $247$\,MB \\
 & $10^6$ & $6\,068$/$120$ & $8\,435$\,s ($0.55$) & $1\,063$\,s & $1.43$\,GB \\
\midrule
$\mathcal C$ & $10^4$ & $606$/$12$ & $7.00$\,s ($0.66$) & $1.01$\,s & $64$\,MB \\
 & $10^5$ & $1\,918$/$38$ & $469$\,s ($0.68$) & $59.7$\,s & $368$\,MB \\
 & $10^6$ & $6\,068$/$120$ & $29\,201$\,s ($0.66$) & $3\,662$\,s & $2.11$\,GB \\
\midrule
$T$ ($K_{-1}$) & $10^4$ & $606$/$12$ & $14.7$\,s ($0.58$) & $2.05$\,s & $88$\,MB \\
 & $10^5$ & $1\,918$/$38$ & $997$\,s ($0.59$) & $127$\,s & $560$\,MB \\
 & $10^6$ & $6\,068$/$120$ & $51\,551$\,s ($0.55$) & $6\,458$\,s & $4.01$\,GB \\
\midrule
$\mathcal T$ & $10^4$ & $606$/$12$ & $7.17$\,s ($0.60$) & $1.09$\,s & $64$\,MB \\
 & $10^5$ & $1\,918$/$38$ & $476$\,s ($0.65$) & $60.4$\,s & $277$\,MB \\
 & $10^6$ & $6\,068$/$120$ & $27\,804$\,s ($0.62$) & $3\,482$\,s & $1.65$\,GB \\
\midrule
$H(1)$ & $10^4$ & $303$/$24$ & $6.15$\,s ($0.35$) & $0.86$\,s & $73$\,MB \\
 & $10^5$ & $959$/$76$ & $419$\,s ($0.41$) & $52.8$\,s & $419$\,MB \\
 & $10^6$ & $3\,034$/$241$ & $28\,465$\,s ($0.36$) & $3\,564$\,s & $2.44$\,GB \\
\midrule
$\sigma$ & $10^4$ & $152$/$48$ & $10.4$\,s ($0.30$) & $1.41$\,s & $72$\,MB \\
 & $10^5$ & $480$/$152$ & $652$\,s ($0.29$) & $82.8$\,s & $148$\,MB \\
 & $10^6$ & $1\,517$/$483$ & $41\,423$\,s ($0.27$) & $5\,268$\,s & $948$\,MB \\
\midrule
$U(1)$ & $10^4$ & $303$/$24$ & $9.85$\,s ($0.52$) & $1.34$\,s & $73$\,MB \\
 & $10^5$ & $959$/$76$ & $677$\,s ($0.56$) & $85.6$\,s & $305$\,MB \\
 & $10^6$ & $3\,034$/$241$ & $37\,795$\,s ($0.50$) & $4\,736$\,s & $2.19$\,GB \\
\midrule
$\operatorname{Tr}\Delta^{-3/2}$ & $10^4$ & $607$/$12$ & $4.12$\,s ($0.29$) & $0.65$\,s & $55$\,MB \\
 & $10^5$ & $1\,919$/$38$ & $255$\,s ($0.31$) & $32.7$\,s & $310$\,MB \\
 & $10^6$ & $6\,069$/$120$ & $16\,144$\,s ($0.29$) & $2\,021$\,s & $2.10$\,GB \\
\midrule
$\sum_ke^{-\sqrt k}$ & $10^4$ & $608$/$12$ & $15.9$\,s ($0.57$) & $2.11$\,s & $73$\,MB \\
 & $10^5$ & $1\,920$/$38$ & $1\,025$\,s ($0.56$) & $129$\,s & $400$\,MB \\
 & $10^6$ & $6\,069$/$120$ & $66\,975$\,s ($0.50$) & $8\,379$\,s & $3.13$\,GB \\
\bottomrule
\end{tabular}
\end{table}

\begin{table}[tp]
\centering\small
\caption{Timings of the reference methods in \texttt{holsum} with $8$ threads (CPU time summed over the threads, and
wall time), and the speedup of the theta splitting: the time of the faster reference method divided by that of the
theta splitting (Table~\ref{tab:apps}), for CPU and for wall time separately. ${}^\dagger$For $\sigma$, direct
summation over prime powers instead of Euler--Maclaurin summation.}
\label{tab:appsref}
\begin{tabular}{llrrrrrr}
\toprule
& & \multicolumn{2}{c}{zeta values} & \multicolumn{2}{c}{Euler--Maclaurin} & \multicolumn{2}{c}{speedup} \\
\cmidrule(lr){3-4}\cmidrule(lr){5-6}\cmidrule(l){7-8}
constant & digits & CPU & wall & CPU & wall & CPU & wall \\
\midrule
$\mathcal A$ & $10^4$ & $4.26$\,s & $0.81$\,s & $2.51$\,s & $0.82$\,s & $1.1$ & $2.1$ \\
 & $10^5$ & $1\,059$\,s & $134$\,s & $1\,695$\,s & $353$\,s & $7.7$ & $7.5$ \\
\midrule
$\mathcal C$ & $10^4$ & $29.8$\,s & $3.89$\,s & $2.49$\,s & $0.79$\,s & $0.36$ & $0.78$ \\
 & $10^5$ & $27\,817$\,s & $3\,481$\,s & $1\,888$\,s & $389$\,s & $4.0$ & $6.5$ \\
\midrule
$T$ ($K_{-1}$) & $10^4$ & $31.9$\,s & $4.12$\,s & $3.36$\,s & $1.16$\,s & $0.23$ & $0.57$ \\
 & $10^5$ & $30\,050$\,s & $3\,762$\,s & $2\,287$\,s & $569$\,s & $2.3$ & $4.5$ \\
\midrule
$\mathcal T$ & $10^4$ & -- & -- & $0.98$\,s & $0.50$\,s & $0.14$ & $0.46$ \\
 & $10^5$ & -- & -- & $181$\,s & $115$\,s & $0.38$ & $1.9$ \\
\midrule
$H(1)$ & $10^4$ & $1.01$\,s & $0.15$\,s & $4.34$\,s & $0.93$\,s & $0.16$ & $0.17$ \\
 & $10^5$ & $304$\,s & $39.1$\,s & $1\,925$\,s & $345$\,s & $0.73$ & $0.74$ \\
\midrule
$\sigma^\dagger$ & $10^4$ & -- & -- & $1.07$\,s & $0.18$\,s & $0.10$ & $0.13$ \\
 & $10^5$ & -- & -- & $142$\,s & $19.1$\,s & $0.22$ & $0.23$ \\
\midrule
$U(1)$ & $10^4$ & $53.5$\,s & $8.58$\,s & $61.3$\,s & $8.16$\,s & $5.4$ & $6.1$ \\
 & $10^5$ & $53\,348$\,s & $6\,905$\,s & -- & -- & $79$ & $81$ \\
\midrule
$\operatorname{Tr}\Delta^{-3/2}$ & $10^4$ & $2.38$\,s & $2.38$\,s & $1.31$\,s & $0.54$\,s & $0.32$ & $0.83$ \\
 & $10^5$ & $663$\,s & $663$\,s & $233$\,s & $124$\,s & $0.91$ & $3.8$ \\
\midrule
$\sum_ke^{-\sqrt k}$ & $10^4$ & -- & -- & $2.92$\,s & $0.69$\,s & $0.18$ & $0.33$ \\
 & $10^5$ & -- & -- & $1\,703$\,s & $302$\,s & $1.7$ & $2.3$ \\
\bottomrule
\end{tabular}
\end{table}

The theta splitting grows by a factor of $52$ to $69$ per decade, about $p^{1.75}$ to $p^{1.85}$, like the optimized
program for Khinchin's constant, and parallelizes as well (the ratio of CPU to wall time is $7.7$--$8.0$ at $10^5$ and
$10^6$ digits). The generic code is not far behind the optimized one: at $10^6$ digits, $\mathcal A$, whose states have
dimensions close to those for Khinchin's constant, takes $8\,435$ CPU seconds against $5\,416$. As expected from the
dimensions of the states, the cost grows from $\mathcal A$ (states of dimension $2$ and $3$) to $\mathcal C$ (odd
parts, $\Gamma(0,y)$ starting values) and to $T$ (harmonic numbers on both sides). For the vortex sum and for $H$ the
dual systems have order $4$ but few forcing constants, so that the dual terms are relatively cheap and the larger
split point $u_0 = \frac14$ is better; at $10^5$ digits, $u_0 = \frac1{16}$ is about $1.3$ times slower for the vortex
sum.

The zeta-value expansion grows by factors of $250$ to $1\,000$ per decade. At $10^5$ digits it is $8$ times slower than
the theta splitting for $\mathcal A$, which needs only even zeta values (Bernoulli numbers), and $30$ to $80$ times
slower for $\mathcal C$, $T$ and the vortex sum, where the odd zeta values dominate (for the vortex sum they are
needed to about $1.25p$ bits, since the coefficients grow like $\beta^j$). The exception is $H(1)$: its expansion needs
only even zeta values and, as its coefficients decay factorially, only about $p/\log p$ of them. It is faster than the
theta splitting at $10^5$ digits ($304$ against $419$ CPU seconds), but it grows by a factor of $300$ per decade
against $68$, so the crossover should be near $1.6\cdot10^5$ digits; it remains quasi-quadratic, since the Bernoulli
numbers involved have total size of order $p^2/\log p$. Euler--Maclaurin summation is the fastest method for the
logarithmic sums at $10^4$ digits, but at $10^5$ digits it needs $2$ to $12$ times more CPU time than the theta
splitting and $4.5$ to $20$ times more wall time, since its Bernoulli corrections are sequential. For the vortex sum its
direct part requires a ${}_2F_1$ value per term, and we did not run it beyond $10^4$ digits.

For $\mathcal T$, whose $k$-terms are closed forms, the theta splitting costs about as much as for $\mathcal C$; the
$k$-side is dominated by the two incomplete gamma functions per term, and the growth factor from $10^5$ to $10^6$
digits is $58$. Euler--Maclaurin summation is unusually cheap here, since the summand costs only a square root and a
division: at $10^5$ digits it needs $2.6$ times less CPU time than the theta splitting, though $1.9$ times more wall
time. It grew by a factor of $184$ from $10^4$ to $10^5$ digits; extrapolating, it would need about $3\cdot10^4$ CPU
seconds at $10^6$ digits, about as much as the theta splitting, and some six times more wall time. For
$\mathcal T$ the crossover in CPU time should be near $10^6$ digits.

For $\operatorname{Tr}\Delta^{-3/2}$ the theta splitting is cheaper than for $\mathcal T$: its $k$-terms need no
incomplete gamma functions, and the $k$-side takes $0.29$ of the time against $0.62$. It grows by a factor of $63$ from
$10^5$ to $10^6$ digits, and $u_0 = \frac1{64}$ is about as fast as $u_0 = \frac1{16}$. As for $\mathcal T$, the summand
costs only a square root, and Euler--Maclaurin summation is cheap: at $10^5$ digits it needs slightly less CPU time than the
theta splitting ($233$ against $255$ seconds), but $3.8$ times more wall time. The zeta-value expansion needs only even
zeta values, but converges only like $9^{-n}$; our implementation computes them in a single thread, with peak memory
$4.1$\,GB at $10^5$ digits.

For the damped constants the theta splitting grows by factors of $64$ ($\sigma$) and $65$ ($S$) from $10^5$ to $10^6$
digits, like the others. Their runs also agree with runs at $u_0 = \frac14$ and $u_0 = \frac1{16}$ ($\sigma$) and
$u_0 = 1$ and $u_0 = \frac14$ ($S$) at $10^3$ and $10^4$ digits. For $S$ Euler--Maclaurin summation is slower already
at $10^5$ digits. For $\sigma$ the reference is not: the special structure of $\sum_k2^{-k}\log k$ makes the sum over
prime powers a very cheap quasi-quadratic method (about $p/\log p$ logarithms of primes, each a short binary splitting
series), which grows by a factor of about $130$ per decade but is $4.6$ times faster than the theta splitting at
$10^5$ digits, so that the crossover should be near $10^7$ digits. For the generating function $\sum_kF(k)2^{-k}$ of a
general $F \in \Reg$ there is no such shortcut.

The values are, to $50$ digits (the $10^6$-digit files are in the repository\footnote{\url{https://github.com/fredrik-johansson/holonomic-sums/tree/main/results/digits}}),
\begin{align*}
  \mathcal A &= 1.02313872642793929553508807697521809749214527936608\ldots, \\
  \mathcal C &= 0.78853056591150896106027632345455466647274966822328\ldots, \\
  e^{\mathcal C-1} &= 0.80939402054063913071793188059409131721595399242500\ldots, \\
  K_{-1} &= 1.74540566240734686349459630968366106729493661877798\ldots, \\
  \mathcal T &= 1.86002507922119030718069591571714332466652412152345\ldots, \\
  H(1) &= 1.47282823195618529629494738382314582532386592787930\ldots, \\
  \sigma &= 1.66168794963359412129581892274995074996441863502506\ldots, \\
  U(1) &= 0.26882772504004215799420784633078761581126503327691\ldots, \\
  \operatorname{Tr}\Delta^{-3/2} &= 2.07054447632160590509314982808257430800874099821595\ldots, \\
  S &= 1.67040681796633972123962780986372002147147238406408\ldots
\end{align*}
To our knowledge $\mathcal A$, $\mathcal C$ and $K_{-1}$ had previously been published to about $10^2$--$10^3$ digits
(OEIS b-files), except $K_{-1}$, which Bailey, Borwein and Crandall computed to $7\,350$ digits with the zeta-value
expansion of their Corollary~7 \cite{BBC97}; the OEIS lists $\mathcal T$ and $\sigma$ to about $1\,000$ digits
\cite[A226317, A112302]{OEIS}.

The cross-checks are not a formality. A first version of the vortex program truncated the series for the starting
values too early and the computations at different $u_0$ disagreed
beyond $75\,000$ digits; a first version of \texttt{holsum} lost about $\sqrt{\beta c}$ bits to cancellation in the
dual recurrences of the vortex sum, which the ball arithmetic revealed at $10^4$ digits.

\section{Open problems}\label{sec:open}

\subsection{General implementation}

In all the examples given in the paper, explicit formulas,
recurrences and error bounds for the terms
were derived and implemented case by case.
It should be possible to develop a general implementation of $\zeta_F$ and its relatives
that takes any $F$ as input (for example in the form of an
annihilating differential operator and generalized series coordinates at infinity)
and automatically performs all applicable optimizations
including series rewriting (see also Section~\ref{sec:generalize} below),
parameter tuning, and dispatch to different algorithms
(zeta-value expansion, Euler--Maclaurin summation, theta splitting)
depending on the regime.
In principle an implementation should be able to match the speed of
our hand-optimized FLINT implementation of Khinchin's constant
while being completely general.

\subsection{The exponent}\label{sec:openexp}
Can $\zeta_F(s)$ be computed in quasilinear time for $F \in \RegQ$ and algebraic $s$? This is open even for
$\zeta(s)$ at algebraic $s \notin \ZZ$. The exponent $\frac32$ comes from the balance $k_{\max}m_{\max} \asymp p$
between the two sides, which the uncertainty principle suggests cannot be improved by another smoothing weight
(Section~\ref{sec:complexity}); lowering it would require evaluating a sum of many terms at once, or an entirely
different method.

\subsection{Irregular singularities at infinity}\label{sec:openirreg}
The summands in \eqref{eq:functions} are a function regular at infinity times a periodic weight, a geometric factor or a
stretched exponential. Can sums of holonomic functions with other irregular singularities at infinity be computed in
subquadratic time?

At an irregular singularity at infinity, the formal solutions are
\[
  e^{Q(k^{1/q})}k^{\varrho}\phi(k^{-1/q}),
\]
with $Q$ a polynomial and $\phi$ a power series (possibly with logarithms) that is in general divergent but multisummable
\cite{Bal00}. The exponential factors are rarely the problem. If the sum converges, the leading term of $Q$ is either
damped, which gives a stretched exponential when its degree $r$ is less than $2$ (up to routine extensions, such as
several terms) and is harmless when $r \ge 2$, or purely oscillatory. Oscillatory factors with $r < 1$, such as
$e^{i\sqrt k}$, are stretched exponentials with imaginary $\tau$, and Proposition~\ref{prop:expsplit} continues
analytically in $\tau$: with $\tau = -i$, $r = \frac12$ and $s = 2.3$ it reproduces $\sum_{k=1}^{\infty}e^{i\sqrt k}k^{-s}$
to $25$ digits, with the same value for $u_0 = \frac14$ and $u_0 = 1$. For $r = 1$ an oscillatory factor is a twist.
Only oscillatory factors with $r > 1$ are excluded (for $r = 2$, apart from the periodic case); for these, as for
$k^{-it}$ at large height, the Poisson dual of a chirp is again a chirp. By stationary phase, the Fourier transform of
$e^{i\tau x^r}$ is, up to a slowly varying amplitude, $e^{i\tau'|\xi|^{r'}}$ with $r' = r/(r-1) > 1$, which does not
decay either, so that smoothing one side lengthens the other and no splitting shortens both. For $r < 1$ the exponent
$r'$ is negative, and the dual side decays.

The problem is the divergence of $\phi$, and the threshold is set by the Gaussian kernel. If $\phi$ has Gevrey order
$1/\rho$ (coefficients of size $(n!)^{1/\rho}$ up to geometric factors), as it does when another solution of the
operator has an exponential factor of degree $\rho$, the coefficients of $E_s$ in \eqref{eq:kernel} grow like
$\Gamma(n/\rho)/\Gamma(n/2)$. For $\rho > 2$ the kernel is still entire. For $\rho = 2$ it has a finite radius of
convergence, as in the first of
\begin{gather*}
  e^{k^2}\erfc(k) = \frac1\pi\int_0^\infty e^{-k^2u}\frac{u^{-1/2}}{1+u}\dd u, \\
  e^{-qk} = \frac{q}{2\sqrt\pi}\int_0^\infty e^{-k^2u}u^{-3/2}e^{-q^2/(4u)}\dd u.
\end{gather*}
For $\rho < 2$ its series diverges, and $h_s$ has an irregular singular point at $u = 0$, where the other exponentials
reappear: by the second formula, a factor $e^{-qk}$ becomes $e^{-q^2/(4u)}$. The $k$-side needs $h_s$ only on $[u_0,\infty)$, but the dual side integrates it over $(0,u_0]$. The
starting values \eqref{eq:startvals} integrate the Taylor series of the kernel at $0$ termwise, which now diverges for
every $u_0$, and condition~(H) is what makes the kernel's equation regular at $u = 0$. For $\Lambda_F$ the
kernel also has an essential singularity at $u = 0$, but it is a convergent series in $u^{-r/2}$, because the
exponential is an explicit factor. The other exponentials of an irregular $F$ are present on the positive axis only
through the divergence of $\phi$ (in the language of resurgence, as the singularities of its Borel transform), and
cannot be factored out in this way.

We do not think this obstruction is fundamental. When $F$ is Borel summable in the direction of the positive real
axis, writing $F(k) = \int_0^\infty e^{-kt}\hat\phi(t)\dd t$ as a Borel--Laplace integral, with $\hat\phi$ holonomic,
gives
$\zeta_F(s) = \int_0^\infty\hat\phi(t)\Li_s(e^{-t})\dd t$, a superposition of the splittings of
Theorem~\ref{thm:geometric}; alternatively, the starting values of the dual side could be obtained by effective
resummation at $u = 0$ \cite{vdH07}. Whether the cost then remains $p^{3/2+o(1)}$ is open. The case $\rho = 2$ also
suggests that (H) is stronger than necessary: what the proof of Lemma~\ref{lem:kernelhol} uses is that the operator
rescaled by $1/\Gamma$ is regular at $u = 0$, which should allow other solutions with exponential factors of degree at
least $2$.

Examples from the literature are the Schl\"omilch series $\sum_ka_kJ_\nu(kx)$ \cite{Wat44}, such as
$\sum_kJ_0(kx)k^{-s}$, and $\sum_ke^{-kz}J_0(kr)$, which solves Laplace's equation in cylindrical coordinates. By Hankel's expansion,
$J_\nu(kx)$ is a combination of $e^{\pm ikx}k^{-1/2}$ times divergent series in $1/k$, so it is not a twist
(Remark~\ref{rem:twist}) of a regular function.

Some summands with irregular singularities nevertheless have explicit kernels. In the relativistic Boltzmann sums
$\sum_ke^{-\beta\sqrt{k^2+m^2}}$, the summand is $e^{-\beta k}F(k)$ with $F(x) = e^{\beta(x-\sqrt{x^2+m^2})}$, which is
analytic at infinity but not regular there: as for the example in Section~\ref{sec:defs}, every operator annihilating
$F$ also annihilates the conjugate $e^{\beta(x+\sqrt{x^2+m^2})}$. But the summand is a function of $k^2$ with an
elementary kernel,
\[
  e^{-\beta\sqrt{k^2+m^2}} = \frac{\beta}{2\sqrt\pi}\int_0^\infty e^{-k^2u}\,u^{-3/2}e^{-\beta^2/(4u)}e^{-m^2u}\dd u.
\]
The factor $e^{-\beta^2/(4u)}$ merges with the theta exponent, so the dual terms are moments of $e^{-m^2u}$ at the
shifted points $\pi^2\ell^2 + \beta^2/4$, to which Proposition~\ref{prop:dual} applies, as for the Epstein--Hurwitz sums
(Remark~\ref{rem:algebraic}). The same holds for sums $\sum_kK_0(kx)$, with the kernel $\frac12u^{-1}e^{-x^2/(4u)}$;
each dual term is then a single value of $\erfc$. With $\beta = m = x = 1$ both splittings reproduce the sums to
$26$ digits. We expect the cost $p^{3/2+o(1)}$ at algebraic parameters, but have not worked out the details. Such sums
of Bessel functions $K_\nu$ give the finite-size and finite-temperature corrections for massive fields, and with
divisor weights they appear in the Chowla--Selberg formula for Epstein zeta functions \cite{Eli95,Kir02}. With a
factor $k^{-s}$ the kernel becomes a convolution and is no longer explicit; that case remains open, like the
Schl\"omilch series.

\subsection{Other summands}\label{sec:openother}
Is there any other natural class of summands, not holonomic or not regular at infinity in any of the senses above,
whose sums can be computed in subquadratic time, by a splitting (compare Section~\ref{sec:limits}) or otherwise? We
list two candidates.

\subsubsection{P-recursive summands} Some summands are P-recursive in $k$ but not holonomic as functions of a
continuous variable, such as $k^n/k!$ in Dobinski's formula $B_n = e^{-1}\sum_kk^n/k!$ for the Bell numbers, which
Corollary~\ref{cor:egf} does not cover. Others give Dirichlet series with P-recursive coefficients, such as the
harmonic zeta function $\sum_kH_kk^{-s}$ \cite{AV84,CC20}. P-recursive sequences of polynomial growth typically have
divergent asymptotic expansions ($H_k = \psi(k+1) + \gamma$, or $\binom{2k}{k}4^{-k}$), so they are not values of a
function regular at infinity; conversely $\log k$ and $\sqrt k$ are not P-recursive \cite{FGS05}. By the following
proposition, P-recursive sequences and holonomic functions analytic at infinity overlap only in the rational
functions.

\begin{proposition}\label{prop:precursive}
If $F$ is holonomic and analytic at infinity and the sequence $F(k)$ is P-recursive, then $F(k)$ agrees for large $k$
with a rational function of $k$.
\end{proposition}

\begin{proof}
Suppose $\sum_{i=0}^dc_i(k)F(k+i) = 0$ for large integers $k$, with polynomials $c_i$. The left side, with $k$
replaced by $x$, is a function of $x$ meromorphic at $x = \infty$. It vanishes at the points $x = k$, which accumulate
at $\infty$, so it vanishes identically, and $F(x)$ satisfies the difference equation. It also satisfies a linear differential equation, and it lies in
$\CC[[1/x]]$. By a theorem of Sch\"afke and Singer \cite[Corollary~5]{SS19}, a series in $\CC[[1/x]][x]$ that
satisfies both a linear differential equation and a linear difference equation $x \mapsto x+1$ over $\CC(x)$ is
rational.
\end{proof}

\subsubsection{Non-holonomic summands and divisor-weighted sums} Some summands are analytic at infinity but not
holonomic, although their zeta-value expansions converge geometrically. Examples are the logarithm of the
Kepler--Bouwkamp constant and two constants from the OEIS (A342680, A336405),
\[
  \sum_{k=3}^{\infty}\log\cos\frac\pi k, \qquad \sum_{n=1}^{\infty}\Bigl(\tan\frac1n - \sin\frac1n\Bigr), \qquad
  \sum_{n=1}^{\infty}\log\Bigl(n\sin\frac1n\Bigr)
\]
(the derivative of $\log\cos$ is $-\tan$, which has infinitely many poles), and sums involving
$(1 - \frac2{n+1})^{2n}$, such as the oscillator strengths of the hydrogen atom. Sums of Clausen functions
$\operatorname{Cl}_2(2\pi/n)$ are similar, but their expansion at infinity contains a term $(\log n)/n$.

Several of these are double sums of holonomic functions. By the product formulas for the cosine and the sine,
\begin{gather*}
  \log\cos\frac\pi k = \sum_{m=1}^{\infty}\log\Bigl(1 - \frac{4}{(2m-1)^2k^2}\Bigr), \\
  \log\Bigl(n\sin\frac1n\Bigr) = \sum_{m=1}^{\infty}\log\Bigl(1 - \frac{1}{\pi^2m^2n^2}\Bigr),
\end{gather*}
so the first and third constants are sums of a function of the product of the two indices, weighted by the number of
(odd) divisors; in the third the function contains the transcendental parameter $\pi$. The simplest sums of this kind
are $\sum_{k=1}^{\infty}d(k)F(k) = \sum_{n=1}^{\infty}\sum_{m=1}^{\infty}F(mn)$ with $F \in \RegQ$. An example is the
mean number of divisors of a partial quotient (Section~\ref{sec:khinchinzeta}),
\[
  \mathbb E[d(a_1)] = \sum_{k=1}^{\infty}d(k)\mu_k
  = \sum_{n=1}^{\infty}\log_2\frac{\Gamma(1+\frac2n)}{\Gamma(1+\frac1n)^2} = 2.09285796492039431020\ldots,
\]
for which we found no closed form. Since $\sum_{k=1}^{\infty}d(k)k^{-j} = \zeta(j)^2$,
\[
  \mathbb E[d(a_1)]\log2 = \log\frac43 + 2\log\frac98
  + \sum_{j=2}^{\infty}(-1)^j\,\frac{2^j-2}{j}\bigl(\zeta(j)^2 - 1 - 2^{1-j}\bigr),
\]
which gives $\widetilde O(p^2)$. Malmsten's formula and a Mellin transform,
\[
  \log\frac{\Gamma(1+2x)}{\Gamma(1+x)^2} = \int_0^\infty\frac{(1-e^{-xt})^2}{t(e^t-1)}\dd t, \qquad
  \int_0^\infty\log\Bigl(1+\frac ax\Bigr)x^{s-1}\dd x = \frac{\pi a^s}{s\sin\pi s},
\]
give the integral representations
\begin{align*}
  \mathbb E[d(a_1)]\log2 &= \int_0^\infty\sum_{n=1}^{\infty}\bigl(1-e^{-t/n}\bigr)^2\frac{\dd t}{t(e^t-1)} \\
  &= \frac1{2\pi i}\int_{c-i\infty}^{c+i\infty}\frac{\pi(2-2^s)\,\zeta(s)^2}{s\sin\pi s}\dd s
\end{align*}
with $1 < c < 2$. In the double sum $\sum_{n=1}^{\infty}\sum_{m=1}^{\infty}F(mn)$, the inner sum is, for each fixed
$n$, a sum of a holonomic function of $m$, which theta splitting handles; for $F(k) = \mu_k$ it is
$\Pr[n \mid a_1]$, given in closed form above. As a function of $n$, however, the inner sum is not holonomic: its
derivative in $1/n$ involves the digamma function, which has infinitely many poles. The factor $\zeta(s)^2$ shows the
degree-$2$ structure. Voronoi summation, the analogue of theta splitting for the coefficients $d(k)$, needs about $p$
terms, as for other $L$-functions of degree $2$ (Section~\ref{sec:limits}), so the question is whether these sums can
be computed in subquadratic time by any method.

\subsection{Damped zeta functions}
For $\Lambda_F$ with general $F \in \Reg$ (Appendix~\ref{app:stretched}), make the cost analysis explicit, with the orders
and degrees of the recurrences, and treat irrational $r$ and $r$ not of the form $1/q$. For $F = 1$ and $1 < r < 2$
the identity of Proposition~\ref{prop:expsplit} holds, but the series for the dual terms and for the starting values
of the dual moments diverge; find a quasi-optimal way to evaluate $J_m$ there.

\subsection{Large height}
Give an exact, smoothed Riemann--Siegel formula for $\zeta_F$ with $O(\sqrt t)$ terms at height $t$
(Remark~\ref{rem:complexu0}), and an algorithm uniform in $\operatorname{Im}s$.

\subsection{Uniformity in the operator}\label{sec:openuniform}
Theorem~\ref{thm:uniform} is uniform in $s_0$ and in the coordinates of $F$, but the operator $L$ is fixed. Make the
bound explicit in the order, degree and height of $L$ and in the degree of the number field. Besides the sizes of the
recurrences, this involves the radius $R$, which can be exponential in the height of $L$, since a polynomial of height
$h$ can have roots of size $e^h$, and with it the number $k_0 - k_1$ of terms added directly, the analytic
continuation to them and the growth $e^{R^2u}$ of the kernel. A version uniform in $L$ would need, among other
things, a bound such as $R = O(\sqrt p)$, which keeps the $O(R)$ direct terms within the bound. Families of interest where $R$ grows with a parameter include the Epstein--Hurwitz sums with
large $a$ ($R = a$) and the vortex sums with small spacing $\lambda$ ($R = 2/\lambda$, Section~\ref{sec:vortex}).

\subsection{Good representatives}\label{sec:generalize}
Given a constant $\sum_{k=1}^{\infty}F(k)$, find a representative that minimizes the cost of
Theorem~\ref{thm:complexity}. The operations of Section~\ref{sec:twists} (dilations, twists and shifts, telescoping,
and functional equations) generate an equivalence relation on summands that preserves the value at $s = 0$. For
instance, the first step of the proof of Lemma~\ref{lem:dilog} is a dilation,
$\sum_{k=2}^{\infty}(-1)^kG(k) = -\sum_{k=2}^{\infty}G(k) + 2\sum_{k=1}^{\infty}G(2k)$, and summation by parts turns
\eqref{eq:K1} into \eqref{eq:K2}.
The cost of Theorem~\ref{thm:complexity}, on the other hand, depends on the representative: on the order and degree of
the recurrence of the coefficients $a_n$ (which determine the $k$-side recurrences), on the dimension and degree of
the dual system, and on the exponential type of the kernel. For Khinchin's constant the best known representative has
hypergeometric coefficients. Other examples point the same way:
\begin{itemize}
\item For $\mathcal A$ and $\mathcal C$, summation by parts gives the representations of Section~\ref{sec:AG}, with
hypergeometric coefficients $1/(2m-1)$ and $1/n$ and elementary or $\operatorname{erfi}$-type kernels
(Section~\ref{sec:workedexample}).
\item For the harmonic mean, $1/K_{-1} = Z(1)$, and the choice of the variable matters: in $1/k$ the summand
$\mu_k/k = x\log_2((1+x)^2/(1+2x))$ has coefficients that are a sum of two hypergeometric terms, while after the shift
$j = k+1$, as in $\log 2/K_{-1} = \sum_{j=2}^{\infty}-\log(1-j^{-2})/(j-1)$, the coefficients are harmonic numbers
$H_{\lfloor (n-1)/2\rfloor}$. Section~\ref{sec:KH} uses the second form. The first form trades the harmonic
numbers for two exponential scales: it splits into the pieces $x\log(1+x)$ and $x\log(1+2x)$, with kernels of
types $1$ and $4$, each with dual states like those of $\mathcal A$ and $\mathcal C$. The total dimension of the dual
states is about the same as for the second form, so it is not clear which is faster in practice.
\end{itemize}
This suggests the question whether every constant $\sum_{k=k_0}^{\infty}\log k\cdot\log R(k)$ (``weight two'') with
$R(k) = S(k+1)/S(k)$ for a rational function $S$, so that summation by parts applies as for Khinchin's constant, can be
brought by these operations to a combination of sums $\sum_{k=k_0}^{\infty}\chi(k)\Li_2(c/k^d)$ with hypergeometric
coefficients plus classical constants. For other rational $R$ the antidifference of $\log R(k)$ involves log-gamma
functions, such as $\log\Gamma(k \pm i)$ for $R(k) = 1 + k^{-2}$, which are outside the class. More generally, one can
ask whether sums of products of $w$ logarithms of rational functions can be
reduced to single polylogarithms $\Li_w$, as functional equations of polylogarithms would suggest. A systematic
search would amount to creative telescoping for functions regular at infinity, modulo dilations.

\subsection{Closed forms and algebraic values}
When does $\zeta_F$ have closed-form values? For rational $F$ at integer $s$, partial fractions reduce $\zeta_F(s)$
to polygamma and log-gamma values at algebraic points (Section~\ref{sec:rational}). For other summands we know of no
such reductions: most constants of Table~\ref{tab:constants} are not known to be related to
other periods or special values. Are there identities that relate them?

A sharper version of the question concerns the negative integers. For $F = 1$ the values $\zeta(-m) = -B_{m+1}/(m+1)$
are rational, and every sequence in Table~\ref{tab:sequences} comes from $F = 1$, with a periodic weight or a geometric
factor. Is there an integer sequence of this kind for a nontrivial $F$? To make this precise, let $F \in \RegQ$ be
analytic at infinity, let $z \in \Qbar$ with $0 < |z| < 1$ or $z = 1$, and consider the regularized moments
\[
  M_m = \sum_{k=1}^{\infty}F(k)\,k^mz^k \qquad (m \ge 0),
\]
that is, $M_m = \Li_F(-m, z)$ for $|z| < 1$ and $M_m = \CT_{s=-m}\zeta_F(s)$ for $z = 1$. Up to the factors
$(-1)^m/m!$ they are the Taylor coefficients of the regular part of $\sum_{k=1}^{\infty}F(k)e^{-kt}$ at $t = -\log z$
(Section~\ref{sec:ogf}), except that for $z = 1$ at a pole $s = -m$ of $\zeta_F$ the coefficient is
$M_m + a_{m+1}\psi(m+1)$. Let $V_F(z)$ be the $\Qbar$-vector space spanned by $1$ and the $M_m$. For $F = 1$ the sums
$\sum_kk^mz^k$ are rational functions of $z$, so $V_1(z) = \Qbar$.

\begin{proposition}\label{prop:rationalF}
If $F(k)$ agrees for large $k$ with a rational function of $k$, then $\dim V_F(z) < \infty$.
\end{proposition}

\begin{proof}
The finitely many terms with $k < k_0$ are algebraic. For $k \ge k_0$ write $F(k) = c + \sum_{a,i}c_{a,i}(k+a)^{-i}$
with algebraic $a$ and $c_{a,i}$. Division with remainder gives $k^m(k+a)^{-i} = Q_m(k) + \sum_{l=1}^{i}d_{m,l}(k+a)^{-l}$,
with a polynomial $Q_m$ and numbers $d_{m,l}$, all algebraic. The polynomial parts contribute values $\Li_{-j}(z)$,
which are rational functions of $z$, or $\zeta(-j) \in \QQ$ for $z = 1$. The other parts contribute the finitely many
constants $\sum_{k\ge k_0}z^k(k+a)^{-l}$, $l \le \max i$; for $z = 1$ and $l = 1$ the constant is
$\CT_{s=0}\sum_{k\ge k_0}k^{-s}(k+a)^{-1}$ for all $m$.
\end{proof}

For rational $F$, then, the moments are combinations of finitely many Hurwitz--Lerch values at algebraic shifts, such as
$\psi(a)$ and $\log(1-z)$. Another natural source of such finiteness would be a linear recurrence with polynomial
coefficients for the values $F(k)$, since multiplying it by $k^mz^k$ and summing relates moments of different orders.
But among the summands considered here, which are holonomic and analytic at infinity, the only ones with
P-recursive values are the rational ones (Proposition~\ref{prop:precursive}). For other $F$ we know of no mechanism that would confine the moments to a
finite-dimensional space, and we expect none to exist.

\begin{conjecture}\label{conj:algebraic}
Let $F \in \RegQ$ be analytic at infinity, and let $z \in \Qbar$ with $0 < |z| < 1$ or $z = 1$. If $F(k)$ does not
agree for large $k$ with a rational function of $k$, then $V_F(z)$ is infinite-dimensional. In particular,
infinitely many of the moments $M_m$ are transcendental.
\end{conjecture}

The conjecture does not settle the question about integer sequences: moments $M_m \in \omega^m\Qbar$ for a
transcendental $\omega$, as for $\zeta(2n) \in \pi^{2n}\QQ$, would be consistent with it.

The conjecture is out of reach even in simple cases. For $F(k) = \log(1+\frac1k)$ and $z = 1$, shifting the index in
$\sum_{k=1}^{\infty}\log(k+1)k^{-s} = \sum_{j=2}^{\infty}\log j\,(j-1)^{-s}$ and expanding $(j-1)^{-s}$ in powers of
$1/j$ shows that $M_m$ is a rational linear combination of $1$, $\log 2\pi$ and $\zeta'(-j)$, $1 \le j < m$; for
instance $M_0 = 0$ and $M_1 = \frac12(1 - \log2\pi)$. Since $\zeta'(-2n) = (-1)^n(2n)!\,\zeta(2n+1)/(2(2\pi)^{2n})$,
the space $V_F(1)$ contains combinations of $\zeta(3)/\pi^2$, $\zeta(5)/\pi^4$, and so on. The conjecture for this $F$
would follow from Grothendieck's period conjecture, but unconditionally it does not even seem to be known that
$\dim V_F(1) \ge 2$, which would require the irrationality of one of these combinations, for instance of
$\log 2\pi$. For summands such as $e^{1/k}$ the
moments are not obviously periods at all.

Values that are rational up to a period do occur for $L$-functions of degree~$2$. The Eisenstein series of the
Gaussian lattice are
\[
  \sum_{\lambda\in\ZZ[i]\setminus\{0\}}\lambda^{-4n} = \frac{(2\varpi)^{4n}}{(4n)!}H_{4n}, \qquad
  \varpi = \frac{\Gamma(\frac14)^2}{2\sqrt{2\pi}},
\]
with rational Hurwitz numbers $H_{4n}$ \cite{Hur99}, \cite[Chapter~12]{AIK14}, in analogy with
$\zeta(2n) \in \pi^{2n}\QQ$. But these are two-dimensional lattice sums, outside the subquadratic class
(Section~\ref{sec:limits}).

\section*{Acknowledgements}

This research was done in collaboration with Claude Opus 5.5.
The author thanks Drew Sutherland for the extremely generous donation of a
Claude subscription which enabled this work, and Vincent Neiger for running an additional
verification computation of Khinchin's constant.

\appendix

\section{Formalization and proofs of the complexity bounds}\label{app:cost}

This appendix proves Theorems~\ref{thm:complexity} and~\ref{thm:uniform} (Sections~\ref{app:tools}--\ref{app:uniform}),
Corollaries~\ref{cor:bernoulli} and~\ref{cor:egf} (Section~\ref{app:sequences}), Theorem~\ref{thm:geometric}(2)
(Section~\ref{app:geometric}), and sketches Claim~C (Section~\ref{app:stretched}). Much of the mathematics of the paper,
including every estimate these proofs use, has also been formalized, and we first describe that formalization.

\emph{The formalization.} The directory \texttt{lean/} of the repository\footnote{\url{https://github.com/fredrik-johansson/holonomic-sums/tree/main/lean}}
of this paper contains a formalization in
Lean~4 with its mathematics library Mathlib (Lean v4.37.0-rc1, Mathlib of November 2026; about $20\,000$ lines). It
contains no unproved statements (\texttt{sorry}) and introduces no axioms; the main theorems depend only on the three
standard axioms of Lean's logic. Like the proof text below, the formalization was produced by an LLM-based agent, not
by the author. Its proofs are checked by the Lean kernel, but whether a formal statement says what the paper says must
be checked by reading it: the file \texttt{lean/README.md} lists, for each result of the paper, the Lean names of the
corresponding statements and any differences in hypotheses, and \texttt{ZetaF/Open.lean} records the statements that
remain open.

\emph{What is formalized.} In Section~\ref{sec:splitting}: the abstract theta splitting and Theorem~\ref{thm:split} for
every $s \in \CC$ (the $k$-terms and dual terms are entire, the main term is meromorphic with poles only at
$s = 1 - n$, the sums converge, the identity holds, and the right-hand side does not depend on $u_0$), hence, with
Lemma~\ref{lem:reduction}, the meromorphic continuation of $\zeta_F$ for $F \in \Reg$; the recurrence part of
Lemma~\ref{lem:kernelhol}; Proposition~\ref{prop:dual} for real $c > 0$; and the twisted splittings of
Remark~\ref{rem:twist}. In Section~\ref{sec:beyond}: Theorem~\ref{thm:geometric}(1) on the punctured disk and on
$|z| = 1$, $z \ne 1$, for every $s$, and Propositions~\ref{prop:expsplit} and~\ref{prop:prodsplit} in termwise form. In Section~\ref{sec:complexity}:
Proposition~\ref{prop:cont} and the identity behind Corollary~\ref{cor:bernoulli}. In Section~\ref{sec:khinchin}:
Proposition~\ref{prop:Z}, the formulas \eqref{eq:K1}--\eqref{eq:K3} with Lemma~\ref{lem:dilog}, Theorem~\ref{thm:khinchin}
with the closed forms of its pieces, Algorithms~K1 and~K2 as instances of Theorem~\ref{thm:split},
Proposition~\ref{thm:khrat}, and the error bounds of Section~\ref{sec:khbounds}. In this appendix: every estimate of
Section~\ref{app:estimates}, among them the Taylor-step lemma that controls the bit-burst algorithm.

\emph{What is not formalized.} Bit complexity itself. Mathlib's interface for Turing machines has no primitives for
integer arithmetic with time and space accounting, nor for the standard algorithms built on it (fast multiplication,
binary splitting, rounding), so expressing the algorithms at that level would have meant building such a framework
first. We formalized instead the bounds on which the cost analysis rests (Section~\ref{app:estimates}); the cost model
of Section~\ref{app:tools} is assumed, and the account of the algorithm in Sections~\ref{app:pieces}
and~\ref{app:uniform}, which combines the formal bounds with that model, is informal. Three mathematical steps enter
the formalization as explicit hypotheses or are left out. The first is that the even and odd parts of a function
satisfying (H) again satisfy (H) (the first step of the proof of Lemma~\ref{lem:kernelhol}). The proof uses Fuchs's
criterion, that a linear differential operator is regular singular at a point if and only if all its solutions have
moderate growth there, together with the fact that the solutions of the least common left multiple of two operators
are the sums of their solutions. Mathlib has neither the analytic theory of linear differential equations (solution
spaces near a singular point, Frobenius bases) nor least common left multiples of differential operators; the
formalization works with recurrences on coefficient sequences, and the statement is kept as the hypothesis that
sequences satisfying (H) are closed under addition. The second is the representation of the damped kernels of
Propositions~\ref{prop:expsplit} and~\ref{prop:prodsplit} as inverse Laplace transforms, which needs Hankel or
Mellin--Barnes contour integrals, and the third is the monodromy argument in the reduction of
Section~\ref{app:uniform}. The identity $\Ci(x) = \gamma + \log x - \int_0^x(1-\cos t)\,\mathrm dt/t$ used in
Theorem~\ref{thm:khinchin} is classical and is not formalized, and neither are the applications of
Section~\ref{sec:other} and the computations.

\emph{What the formalization changed.} Formalizing the arguments exposed a handful of minor errors (unstated
conditions, a misstated constant or bound, a definition by an integral that diverges in part of the range), which have
been corrected without changing the overall argument. The exception is a genuine gap in Propositions~\ref{prop:expsplit}
and~\ref{prop:prodsplit}. Their proofs integrate the damped kernel $h_s(u) = \sum_jc_ju^{a_j-1}$ termwise, on the
grounds that $e^{-c/u}$ suppresses every power of $u$, which gives the dual terms as series
$J_m = \sum_jc_jD_{a_j-1/2}(\pi^2m^2)$. The $j$-th term of this series has size about
$(\tau(\pi m)^{-r})^j\Gamma(\frac{rj}2)^2/j!$, that is $(j!)^{r-1}$ times a geometric factor, so the series converges
for $r < 1$, for $r = 1$ only when $\tau < 2\pi m$, and diverges for every $m$ when $1 < r < 2$; for $r = \frac32$,
$\tau = 1$ and $m = 1$ its terms exceed $10^{232}$ at $j = 800$. The splitting identity itself is unaffected when
$J_m$ is defined by its integral, and this is how it is now stated. But the evaluation of the dual terms behind Claim~C
needs the series, so Claim~C is now restricted to $0 < r \le 1$, and the case $1 < r < 2$ is listed as an open problem
(Section~\ref{sec:open}).

The following proof text is LLM-generated and unedited by the author.

\begin{small}

\subsection{Cost model}\label{app:tools}
Costs are bit operations on a multitape Turing machine. We use the following standard facts as assumptions.
\begin{itemize}
\item[(M)] $\Mul(n)/n$ is nondecreasing and $\Mul(n\log n) = O(\Mul(n)\log n)$. Then $\Mul(An) = O(\Mul(n))$ for fixed
$A$, and $(\log p)\,\Mul(Ap\log p) = O(\Mul(p)\log^2p)$ (\nolinkurl{regular_scale}, \nolinkurl{regular_cost}).
\item[(K)] Elements of a fixed number field $\mathbb K$ are stored as integer vectors over an integral basis with a
common denominator; an element of height $H$ takes $O(\log H)$ bits, and a product of elements of $b$ bits costs
$O(\Mul(b))$ and has $b + O(1)$ bits.
\item[(BS)] \emph{Binary splitting.} The ordered product of $N$ matrices of fixed size over $\mathcal O_{\mathbb K}$
with entries of $b$ bits, computed by a balanced product tree, costs $O(\Mul(Nb)\log N)$
(\nolinkurl{treeCostL_le}, \nolinkurl{matSize_mul_le}, \nolinkurl{treeCost_matrix_le}). A series whose terms satisfy
$X_{n+1} = q_n^{-1}A_nX_n$, $s_n = s_{n-1} + q_n^{-1}B_nX_n$ is summed by the product tree of the matrices
$\bigl[\begin{smallmatrix}A_n & 0\\ B_n & q_n\end{smallmatrix}\bigr]$ \cite{HP98}.
\item[(A)] Approximate arithmetic on $P$-bit fixed-point numbers costs $O(\Mul(P))$ per operation, with the usual
rounding errors.
\item[(E)] The constants $\pi$ and $\gamma$, and $\exp$ and $\log$ at algebraic points with $O(\log p)$ bits, can be
computed to $p$ bits in $O(\Mul(p)\log^2p)$ \cite{BM80,HP98,BZ10}.
\end{itemize}

\emph{Certified series.} By (M), (BS) and (A), a series is summed to absolute error $2^{-p}$ in
$O(\Mul(Nb)\log N + \Mul(p+G))$ bit operations if it comes with a \emph{certificate} $(N, b, G)$: its terms satisfy a
recurrence of fixed order over $\mathbb K$ whose coefficients have at most $b$ bits for $n \le N$, the tail after $N$
terms is at most $2^{-p}$, and all solutions of the recurrence, normalized as in the product tree, are at most $2^G$
for $n \le N$ (so that $G$ guard bits absorb the rounding errors). With $N = O(p)$, $b = O(\log p)$ and $G = O(p)$
this is $O(\Mul(p)\log^2p)$. Every piece below is a fixed number of certified series, or a bit-burst evaluation.

\emph{Bit-burst.} Let $Y(c) \in \CC^D$ satisfy $cY'(c) = (B_0 + cB_1)Y(c)$, with $B_0, B_1$ over $\mathbb K$, whose
only finite singular point is $c = 0$. Its Taylor coefficients at $c_0 \ne 0$ satisfy
$c_0(l+1)Y_{l+1} = (B_0 + c_0B_1 - l)Y_l + B_1Y_{l-1}$ (\nolinkurl{taylor_rec_iff}). The bit-burst algorithm
\cite{CC90,vdH99} steps from $c_0$ to $c$ through rational points $c_i$ with $2^ib_0$-bit expansions, summing the
Taylor series at $c_i$ in $h = c_{i+1} - c_i$ by binary splitting. The step is controlled by the \emph{Taylor-step
lemma} (\nolinkurl{taylor_step_bound}, \nolinkurl{taylor_step_tail}): if
\[
  \theta \ge |h|\,\frac{\|B_0 + c_iB_1\|}{|c_i|}, \qquad \theta \ge |h|\,\|B_1\|, \qquad \theta \ge \frac{|h|}{|c_i|},
\]
then every solution of the recurrence satisfies $\|h^lY_l\| \le (2\theta)^l\|Y_0\|$; for $\theta < \frac12$ the
step has norm at most $1/(1-2\theta)$ and the tail after $L$ terms is at most $(2\theta)^L\|Y_0\|/(1-2\theta)$.
Take $|c - c_0| \le \frac12|c_0|p^{-A}$, where $p^{A-1} \ge \|B_0\| + |c|\,\|B_1\| + 1$, and
$b_0 \ge 2A\log_2p$. Then $\theta_0 \le \frac12p^{-1}$ for the first step, which therefore has $O(p/\log p)$ terms
with $O(\log p)$-bit coefficients, and $\theta_i \le p^A2^{-2^ib_0} \le 2^{-2^{i-1}b_0}$ for the later ones, which have
$O(p/2^ib_0)$ terms with $O(2^ib_0)$-bit coefficients. Since $\sum_i\theta_i \le \frac14$, the steps amplify errors by at most $e$
(\nolinkurl{error_propagation_linear}), so no a priori bound on $Y$ is needed beyond $|Y(c_0)|$; the total cost is
$O(\Mul(p)\log^2p)$ (\nolinkurl{bitBurst_cost_le}). The convergent Taylor series at $c_i$ continues $Y$ along the
segment $[c_i, c_{i+1}]$.

\subsection{Estimates}\label{app:estimates}
Fix the data of Theorem~\ref{thm:uniform}, let $S \ge 2$ bound $|s_0|$ and $f(x) = \sum a_nx^n$ with
$|a_n| \le CR^n$. The following are proved in the formalization; $A$ denotes a constant depending only on the fixed
data.
\begin{itemize}
\item[(G1)] $|1/\Gamma(w)| \le e^{\pi|\operatorname{Im}w|/2}/\Gamma(\operatorname{Re}w)$ and
$|\Gamma(w)| \le \Gamma(\operatorname{Re}w)$ for $\operatorname{Re}w \ge \frac12$ (\nolinkurl{norm_inv_Gamma_le_G1},
\nolinkurl{norm_Gamma_le_Gamma_re}).
\item[(G2)] $|1/\Gamma(w)| \le e^{1+\pi T/2}(2T+2)^{\lceil T\rceil+1}$ for $|w| \le T$ (\nolinkurl{norm_inv_Gamma_le_G2}).
\item[(G3)] $|\Gamma(w)| \le 2^N\Gamma(\operatorname{Re}w + N)/\min(\delta, 1)$ if $\operatorname{Re}w + N \ge 1$ and
$|w + j| \ge \delta$ for $j < N$ (\nolinkurl{norm_Gamma_le_G3}).
\item[(Li)] \emph{Liouville.} $|\alpha|_v \ge H(\alpha)^{-1}$ for $\alpha \in \mathbb K^\times$ and every archimedean
place $v$, and $|\alpha + \beta|_v \ge (2^{[\mathbb K:\QQ]}H(\alpha)H(\beta))^{-1}$ if $\alpha + \beta \ne 0$
(\nolinkurl{inv_mulHeight}$_1$\nolinkurl{_le_abv}, \nolinkurl{small_divisor}). Hence the divisors $s_0 + \sigma + n$, $a_* + n$ and
the $\delta$ of (G3) are at least $p^{-A}$.
\item[(GS)] $\sum_jx^j|1/\Gamma(\sigma + j + 1)| \le e^{AS\log S}\max(1,x)^{-\operatorname{Re}\sigma}e^x
\le e^{AS\log S}(1+x)^Se^x$ for $|\sigma| \le S$, $x \ge 0$; the tail of
$P(a,y) = y^ae^{-y}\sum_iy^i/\Gamma(a+i+1)$ (\nolinkurl{hasSum_gammaPTerm}) after $N \ge 2y + 2S$ terms is at most
$e^{AS\log S}y^{\operatorname{Re}a}(1+y)^S4^y2^{-N}$ for $|a| \le S$ (the factor $4^y$ cannot be omitted); and
$|\Gamma(w,x)| \le 2x^{\operatorname{Re}w-1}e^{-x}$ for $\operatorname{Re}w \le \frac x2 + 1$
(\nolinkurl{sum_pow_mul_norm_inv_Gamma_le_sharp}, \nolinkurl{sum_pow_mul_norm_inv_Gamma_le},
\nolinkurl{norm_gammaPTerm_tail_uniform}, \nolinkurl{norm_upperGamma_le}).
\item[(KB)] \emph{Kernel.} For $|s| \le S$, $|h_s(u)| \le e^{AS\log S}e^{R^2u}$ for $u \ge u_0$ and
$|h_s(u)| \le e^{AS\log S}u^{-S/2-1}$ for $0 < u \le u_0$; hence $|I_k(s)| \le e^{AS\log S}e^{-(k^2-R^2)u_0}$ for
$k > R$ and $|J_m(s)| \le e^{AS\log S}e^{-\pi^2m^2/(2u_0)}$, for every $s$ with $|s| \le S$, also where the pieces
are defined by continuation (\nolinkurl{norm_kernel_le_far}, \nolinkurl{norm_kernel_le_near},
\nolinkurl{norm_kTerm_le_uniform}, \nolinkurl{norm_dualTerm_le_uniform}). For a twist of period $q$, $\pi^2$ becomes
$\pi^2/q^2$.
\item[(T)] \emph{Truncation.} For fixed $s$, by the bounds in Theorem~\ref{thm:split}, the tails beyond the first
neglected $k$-term and dual term are bounded by geometric series, so $p$ bits need
\begin{equation}\label{eq:counts}
  k \le k_{\max} \approx \sqrt{\frac{p\log 2}{u_0}}, \qquad m \le m_{\max} \approx \frac{q}{\pi}\sqrt{p\log 2\cdot u_0}
\end{equation}
(\nolinkurl{Bounds/Truncation.lean}); uniformly in $|s| \le S$, (KB) gives the same with $p$ replaced by
$p + AS\log S$ and $m_{\max}$ multiplied by $\sqrt2$. Either way there are $O(\sqrt p)$ terms on each side, and the
tails are at most twice their first terms (\nolinkurl{tail_le_two_mul}).
\item[(D)] \emph{Dual moments.} The moments $Z_t[g](c) = \int_0^{u_0}u^{-t}e^{-c/u}g(u)\dd u$ satisfy
\eqref{eq:ibp} and $-\partial_cZ_t = Z_{t+1}$, and $D_a = Z_{1-a}[1]$ satisfies \eqref{eq:Drec}
(\nolinkurl{dualMoment_ibp}, \nolinkurl{hasDerivAt_dualMoment}, \nolinkurl{momentD_rec}); the system of
Proposition~\ref{prop:dual} holds with explicit $B_0, B_1$ (\nolinkurl{dual_system}).
\item[(S)] \emph{Starting values.} $D_{a+n} = \alpha_nD_a + \beta_nE$ with $\alpha_n = (-c)^n/(a+1)_n$ for every
complex $a$ with $a + 1 + i \ne 0$ (\nolinkurl{momentD_add_nat_eq}), and $\sum_nx^n/(n!)^2 \le e^{2\sqrt x}$
(\nolinkurl{sum_pow_div_factorial_sq_le}).
\item[(B)] \emph{Bernoulli numbers.} We have $B_n(b/f) = -nf^{1-n}L(1-n, \mathbf 1_{b+f\ZZ})$ and
$f^n(n+1)\#\,B_n(b/f) \in \ZZ$, where $N\#$ is the product of the primes $\le N$; $|B_n(x)| \le 4e\cdot n!$ on $[0,1]$; and a real number within $1/(2D)$ of a rational number whose denominator divides
$D$ determines that rational number by rounding (\nolinkurl{bernoulli_eq_LFunction}, \nolinkurl{isInt_den_mul_bernoulli_eval},
\nolinkurl{abs_bernoulli_eval_le}, \nolinkurl{bernoulli_eval_eq_round}).
\end{itemize}

\subsection{The pieces}\label{app:pieces}
By Theorem~\ref{thm:split}, after the reduction of Section~\ref{app:uniform}, the value is a combination of the
$k$-terms $I_k(s)$, $k_0 \le k \le k_{\max}$, the dual terms $J_m(s)$, $1 \le m \le m_{\max}$, and the main term
$M(s)$, at $s = s_0 + \varepsilon$ over $\mathbb K[\varepsilon]/(\varepsilon^{j+1})$. Table~\ref{tab:certificates}
lists how each is computed; we comment on the entries. Throughout, $x = k^2u_0$, $c = \pi^2m^2/q^2$ and
$S = |s_0| + 2$; in Theorem~\ref{thm:complexity}, $S = O(1)$.

\begin{table}[ht]
\centering\footnotesize
\caption{The pieces and their certificates $(N, b, G)$ for $p$ bits (Section~\ref{app:tools}; for the dual terms, the stages of the bit-burst); each costs $O(\Mul(p)\log^2p)$. The guard bits $G$ come from (GS), (KB), (S), (G1)--(G3) and (Li).}
\label{tab:certificates}
\begin{tabular}{@{}lllll@{}}
\toprule
piece & number & $N$ & $b$ & $G$ \\
\midrule
$k$-terms $I_k$, double series \eqref{eq:Ik} & $O(\sqrt p)$ & $O(p)$ & $O(\log p)$ & $O(S\log S + \log p)$ \\
$P(a, x)$ and $1/\Gamma(w)$ & $O(\sqrt p)$, $O(1)$ & $O(p)$ & $O(\log p)$ & $O(p)$ \\
dual terms $J_m$, bit-burst for \eqref{eq:fuchsys} & $O(\sqrt p)$ & stages & stages & $O(p)$ \\
starting values $Y(c_0)$, series \eqref{eq:startvals} & $O(\sqrt p)$ & $O(p)$ & $O(\log p)$ & $O(\sqrt{c_0} + S\log S)$ \\
main term $M$, series in $n$ & $1$ & $O(p)$ & $O(\log p)$ & $O(S\log S + \log p)$ \\
\bottomrule
\end{tabular}
\end{table}

\emph{$k$-terms.}\label{app:kterms} Splitting the series of Theorem~\ref{thm:split} by the parity of $n$ and using
$Q(\sigma_0+m, x) = Q(\sigma_0, x) + e^{-x}\sum_{j<m}x^{\sigma_0+j}/\Gamma(\sigma_0+j+1)$ gives,
for $f(x) = A_0(x^2) + xA_1(x^2)$ and $\kappa = k^2$ (\nolinkurl{kTerm_double_series}),
\begin{equation}\label{eq:Ik}
\begin{split}
  I_k(s) &= k^{-s}\Bigl(Q\bigl(\tfrac s2, x\bigr)A_0\bigl(\tfrac1\kappa\bigr)
    + Q\bigl(\tfrac{s+1}2, x\bigr)\frac{A_1(1/\kappa)}{k}\Bigr) \\
  &\quad + e^{-x}u_0^{s/2}\sum_{m=1}^{\infty}\frac{a_{2m}}{\kappa^m}\sum_{j=0}^{m-1}\frac{x^j}{\Gamma(\frac s2+j+1)} \\
  &\quad + e^{-x}u_0^{(s+1)/2}\sum_{m=1}^{\infty}\frac{a_{2m+1}}{\kappa^m}\sum_{j=0}^{m-1}\frac{x^j}{\Gamma(\frac{s+1}2+j+1)}.
\end{split}
\end{equation}
Factor out $1/\Gamma(\sigma_0 + j_1 + 1)$, $\sigma_0 \in \{\frac s2, \frac{s+1}2\}$, with $j_1$ the least index for
which $\operatorname{Re}(\sigma_0 + j_1 + 1) \ge \frac12$; the $O(S)$ terms with $j < j_1$ are polynomials in $s$
times that factor, and the others follow a recurrence that divides only by units of
$\mathbb K[\varepsilon]/(\varepsilon^{j+1})$. The double sums are then certified series in $m$ over $\mathbb K$: their
terms are at most $|a_n|\kappa^{-n/2}$ times the sum in (GS), so $O(p/\log(k/R))$ terms and $O(S\log S + \log p)$
guard bits suffice. $A_0(1/\kappa)$ and $A_1(1/\kappa)$ are holonomic values inside the disk of convergence, and
$k^{-s}$ costs one logarithm per $k$.

\emph{Incomplete gamma functions.} For $a \in \ZZ_{\le0}$, $Q(a,y) = 0$. Otherwise $P(a,y)$ is the series of (GS),
with the factor $1/\Gamma(a + i_1 + 1)$ taken out as above; it has $O(p + S)$ terms, and by (GS) its tail is at most
$2^{-p}$ after $N = 2y + 2S + p + O(y + S\log S) = O(p)$ terms, since $y = O(p)$. The
$O(1)$ values $1/\Gamma(w)$ with $\operatorname{Re}w \ge \frac12$, $|w| \le 3S$ come from
$\Gamma(w) = \gamma(w, N) + \Gamma(w, N)$ with $N = O(p)$, where $|\Gamma(w,N)| \le 2N^{3S}e^{-N}$ by (GS) and
$|\Gamma(w)| \ge 0.88\,e^{-3\pi S/2}$ by (G1).

\emph{Dual terms.}\label{app:dual} By Lemma~\ref{lem:kernelhol} and Proposition~\ref{prop:dual},
$J_m(s) = Z_{(3-s)/2}[G_0](c) + Z_{1-s/2}[G_1](c)$ is a fixed linear form in the solution $Y(c)$ of a system
\eqref{eq:fuchsys} of fixed dimension. We multiply $G_0$, $G_1$ by $\Gamma(m_2 + \frac s2)$, $\Gamma(m_2 + \frac{s+1}2)$,
with $m_2$ the least integer making the real parts at least $\frac12$, so that their coefficients are rational
functions of $s$, and replace $E$ by $\widetilde E = Eu_0^{1-t}$. The inhomogeneity $b$ of \eqref{eq:Gsys} then has
entries $\Gamma(\sigma_0+m_2)/\Gamma(\sigma_0+m)$, $m < m_1 + r$, times polynomials of bounded degree, with the common
factor $\beta = (\sigma_0 + m_1 + r)_{m_2-m_1-r}$ of degree $O(S)$; we use $\beta C_{t-i}$ and $\beta\widetilde E$ as
components of $Y$ (the equations of this block are homogeneous in it). Then $B_0, B_1$ are polynomials in $s$ of bounded
degree with coefficients in $\mathbb K$, so their entries have $O(\log p)$ bits and norm $p^{O(1)}$, and the bit-burst
of Section~\ref{app:tools} from a rational $c_0 \le c$ with $c - c_0 \le \frac12cp^{-A}$ costs $O(\Mul(p)\log^2p)$;
$c_0$ has $O(\log p)$ bits. The factors $\beta$, $1/\Gamma$ and $u_0^{1-t}$ are computed once.

\emph{Starting values.} By \eqref{eq:startvals} the components of $Y(c_0)$ are series $\sum_jg_jD_{j-t+1}(c_0)$ whose
exponents lie in two classes modulo $\ZZ$. In each class choose a base exponent $a_*$: the least exponent that occurs,
or $a_* = 0$ in the integer class, with $D_0 = E_1(c_0/u_0) = \Ein(c_0/u_0) - \gamma - \log(c_0/u_0)$. By (S),
$D_a = \alpha_aD_{a_*} + \beta_aE$, where going up from $a_*$ divides by $a_* + n$, at least $p^{-A}$ by (Li), and
going down from $0$ divides by $c_0$. Since $|g_j| \le CR^{2j}/|\Gamma(j+\sigma_0)|$ and
$|\alpha_a| \le c_0^n/|(a_*+1)_n|$, (G1)--(G3) and (S) bound the terms of $\sum_jg_j\alpha_{j-t+1}$ by
$e^{AS\log S}e^{2R\sqrt{c_0}}c_0^{AS}$, and $E\beta_a = D_a - \alpha_aD_{a_*}$ gives the same for the $\beta$-series:
$O(\sqrt{c_0} + S\log S) = O(p)$ guard bits suffice. The base values $D_{a_*}(c_0) = c_0^{a_*}(\Gamma(-a_*) - \gamma(-a_*, c_0/u_0))$
need $\Gamma(-a_*) \le 2^{O(p)}$, by (G3), or $E_1$. For the jets, the divisions by $a + \frac\varepsilon2$ are by
units, and $\partial_a^iD_a$ reduce by \eqref{eq:Drec} to jets of the base values.

\emph{Main term.} $M(s)$ is a series in $n$ whose terms are $a_n$ times powers of $u_0^{1/2}$, $1/\Gamma$ at
$\frac{s+n}2$ or $\frac{s+n}2 + 1$, rational functions of $n$ and $s$ with small divisors $s + n - 1$, and, for $k < k_0$,
$k^{-s-n}P(\frac{s+n}2, k^2u_0) = e^{-x}u_0^{(s+n)/2}\sum_ix^i/\Gamma(\frac{s+n}2+i+1)$; in this form $k^{-n}$ has
cancelled and the double series has positive majorants. At a pole $s_0 = 1 - n_1$ one term is singular and is
expanded explicitly.

\emph{Derivatives in $s$.} All pieces are computed over $\mathbb K[\varepsilon]/(\varepsilon^{j+1})$. The recurrences
are polynomial in $s$ except for the factors $1/\Gamma(\sigma_0 + \varepsilon)$, $u_0^{\varepsilon/2}$, $k^{-\varepsilon}$
and $\beta$, computed once (or once per $k$). The jets of $G_0, G_1$ satisfy a block-triangular system of the form
\eqref{eq:Gsys}, and Proposition~\ref{prop:dual} applies with the moments
$Z^{(i)}_t[g] = \int_0^{u_0}u^{-t}(\log u)^ie^{-c/u}g\dd u$, which satisfy
$c\,Z^{(i)}_{t+1}[g] = Eu_0^{1-t}(\log u_0)^ig(u_0) - (1-t)Z^{(i)}_t[g] - Z^{(i)}_t[\vartheta g] - iZ^{(i-1)}_t[g]$
(\nolinkurl{dualMomentLog_ibp}).
This multiplies all costs by a constant depending on $j$.

\subsection{Proof of Theorem~\ref{thm:uniform}}\label{app:uniform}
\begin{proof}[Proof of Theorems~\ref{thm:complexity} and~\ref{thm:uniform}]
Theorem~\ref{thm:complexity} is the case of fixed data: $F \in \RegQ$ is then a fixed $\mathbb K$-linear combination of
a basis as in Theorem~\ref{thm:uniform}, for a number field $\mathbb K$ containing all data, and the hypotheses hold
for $p \ge p_0$. Let $S = |s_0| + 2$, so that $S\log S \le c'p$ when $|s_0| \le cp/\log p$.

\emph{Reduction.} By Lemma~\ref{lem:reduction}, Remark~\ref{rem:twist} and the decomposition of $\chi$ into the
indicators of the $q$ residue classes, the Laurent coefficient is a combination, with coefficients $\lambda_i\chi(a)$
times fixed algebraic numbers, of Laurent coefficients of bounded order of the functions
$\sum_{k\ge k_0,\,k\equiv a}f(1/k)k^{-s}$ at the points $s_0 + \sigma$, and of the terms $F(k)\chi(k)k^{-s}$ for
$k_1 \le k < k_0$. Here $f$ and $\sigma$ range over fixed finite sets, and each $f$ satisfies (H): each
$B_\sigma = x^{-\sigma}\sum_j(\log x)^jf_{\sigma,j}(1/x)$ is obtained from a solution by a polynomial in the
monodromy operator at $t = 1/x = 0$, so the functions $x^{-\sigma}\sum_j(\log x - 2\pi im)^jf_{\sigma,j}(1/x)$,
$0 \le m \le J$, are solutions, and solving for the powers of $m$ exhibits $f_{\sigma,j}$ as a polynomial in
$\log t$ with coefficients $t^{-\sigma}y(1/t)$ of moderate growth. The values $y_i(k)$, $k < k_0$, are computed once
by analytic continuation \cite{CC90,vdH99}. The coefficients have $O(p)$ bits and size $e^{O(p)}$, so each piece is
needed to absolute error $2^{-O(p)}$, and $|k^{-s_0}| \le k_0^S = 2^{O(p)}$. In the odd case of Remark~\ref{rem:twist}
the kernel is that of $h_{s+\delta}$ with the weights $k^\delta$.

\emph{Cost.} By (T) there are $O(\sqrt p)$ pieces, and by Section~\ref{app:pieces} each is a fixed number of certified
series with $N = O(p)$, $b = O(\log p)$, $G = O(p)$, or a bit-burst evaluation, so each costs $O(\Mul(p)\log^2p)$.

\emph{Memory.} The pieces are computed one after another and added to a running sum; the constants ($\pi$, the factors
$1/\Gamma$, $\beta$, the powers of $u_0$, the base values, and $\log k$ as needed) are $O(1)$ numbers of $O(p)$ bits.
A certified series with $Nb = O(p\log p)$ is split into blocks of $\lceil p/b\rceil$ terms; the product tree of a
block, computed depth-first, occupies $O(p)$ bits and costs $O(\Mul(p)\log p)$, and the $O(Nb/p)$ block matrices are
multiplied into a running product rounded to $O(p)$ bits. The normalized partial products are fundamental matrices
of the recurrence, so all its solutions must be at most $2^{O(p)}$: for the Taylor stages this is the Taylor-step
lemma, for the recurrences of $a_n$ it follows from (H), and the $\Gamma$-twists are of first order. Hence $O(p)$ bits
of memory suffice; without blocking the top of the tree has $O(p\log p)$ bits.

\emph{Large $|s_0|$.} If only $|s_0| \le cp$, run the algorithm at precision $P = \lceil p\log_2p\rceil$; for
$p \ge p_0$ the hypotheses hold with $(P, 2c)$, and the bounds become $O(P^{1/2}\Mul(P)\log^2P) =
O(p^{1/2}\Mul(p)\log^{7/2}p)$ and $O(p\log p)$ bits, by (M).

\emph{Geometric factors.} For the last statement, follow Section~\ref{app:geometric} with the modifications above: the
weights $C_k$, $S_k$ have $O(\sqrt p)$ bits and size $2^{O(\sqrt p)}$, the truncation bound gains the factor
$|z|^{-k}$, and for the $O(1)$ dual points with $\operatorname{Re}c_n < 1$ the values come from the series of
Section~\ref{sec:geometric} in the $D_a(c_n)$, bounded by $2^{O(p)}$ with (G1)--(G3) and (Li).
\end{proof}

\emph{Remarks.} The memory bounds assume multiplication in time $\Mul(p)$ and space $O(p)$, as with FFT-based
algorithms. Some bound on $|s_0|$ is needed with real $u_0$: for $|\operatorname{Im}s_0| \gg p$ the cancellation
$e^{\pi|\operatorname{Im}s_0|/4}$ exceeds $2^{O(p)}$ (Remark~\ref{rem:complexu0}), and for $s_0 \to -\infty$ the value
itself may have more than $p$ bits before the point. For rational $s_0$, and for $s_0$ in an imaginary quadratic
field, $|s_0| \le H(s_0)$; in a general number field a bound on $H(s_0)$ does not bound $|s_0|$. The constants depend
on the operator $L$ (Section~\ref{sec:openuniform}).

\subsection{Bernoulli numbers and exponential generating functions}\label{app:sequences}
\begin{proof}[Proof of Corollary~\ref{cor:bernoulli}]
By $B_n(x+1) = B_n(x) + nx^{n-1}$ we may assume $0 < x = b/f' \le 1$. By (B), $B_n(x) = -nf'^{\,1-n}L_1(1-n,
\mathbf 1_{b+f'\ZZ})$, $D_nB_n(x) \in \ZZ$ with $D_n = f'^{\,n}(n+1)\# = e^{O(n)}$, and $|B_n(x)| \le 4e\cdot n!$, so an
approximation of $L_1(1-n, \cdot)$ with absolute error $e^{-An}$ determines $B_n(x)$, and $n\log_2n + O(n)$ bits suffice.
Theorem~\ref{thm:uniform} applies with $F = 1$ ($L = \partial_x$), $\mathbb K = \QQ$, $j = 0$, $s_0 = 1 - n$ and
$p = \Theta(n\log n)$: $H(s_0) \le n \le p$ and $|s_0| = O(p/\log p)$. The cost is $O(p^{1/2}\Mul(p)\log^2p)$, and
$\Mul(n\log n) = O(\Mul(n)\log n)$. For the Euler polynomials,
$E_n(x) = \frac2{n+1}(B_{n+1}(x) - 2^{n+1}B_{n+1}(\frac x2))$; the tangent and Genocchi numbers are
$2^{2n}(2^{2n}-1)|B_{2n}|/(2n)$ and $2(1-2^n)B_n$, and $B_{n,\chi}$ is a fixed combination of the $f^{n-1}B_n(a/f)$.
\end{proof}

\begin{proof}[Proof of Corollary~\ref{cor:egf}]
Since $a_n = n(n-1)\cdots(n-m+1)\,b_{n-m}$ and $b_n = \lambda^nc_n$, with $c_n = n![t^n]R(e^t)$, it suffices to compute
$c_n$. Over a fixed extension of $\mathbb K$ containing the poles, $R(x)$ is a Laurent polynomial plus a combination of
terms $(1 - x/\rho)^{-i}$. A Laurent polynomial contributes $\sum_j\gamma_jj^n$. For $|\rho| > 1$,
$(1 - e^t/\rho)^{-i} = \sum_{k\ge0}\binom{k+i-1}{i-1}\rho^{-k}e^{kt}$ contributes a fixed combination of the
$\Li_{-n-l}(1/\rho)$, $0 \le l < i$; for $|\rho| < 1$, expand $(-\rho e^{-t})^i(1 - \rho e^{-t})^{-i}$ in the same way;
for a root of unity $\rho \ne 1$, $\Li_{-n-l}(1/\rho) = L_1(-n-l,\chi)$ with $\chi(k) = \rho^{-k}$, by continuity on
$|z| \le 1$, $z \ne 1$. Theorem~\ref{thm:uniform} computes each of these $O(1)$ values. Finally
$|a_n| \le C^nn!$, and with $R = P/Q$, $P, Q \in \mathcal O_{\mathbb K}[x, x^{-1}]$, $Q(1) \ne 0$, $d\lambda \in \mathcal O_{\mathbb K}$
and $N = |N_{\mathbb K/\QQ}(Q(1))|$, the expansion $1/Q(e^{\lambda t}) = \sum_i(Q(1) - Q(e^{\lambda t}))^i/Q(1)^{i+1}$ in
the ring of series $\sum e_nt^n/n!$ with $e_n \in d^{-n}\mathcal O_{\mathbb K}$ gives $d^nN^{n+1}a_n \in \ZZ$. Hence
$n\log_2n + O(n)$ bits determine $a_n$.
\end{proof}

\subsection{Geometric factors}\label{app:geometric}
\begin{proof}[Proof of Theorem~\ref{thm:geometric}(2)]
Expanding $\chi$ into additive characters replaces $z$ by the algebraic numbers $ze^{2\pi ib/q}$, so we may take
$\chi = 1$; by Lemma~\ref{lem:reduction} and the derivatives in $s$, we may take $j = 0$ and $f$ satisfying~(H), and
we use Theorem~\ref{thm:geometric}(1). Since $|z|^{-k}e^{-(k^2-R^2)u_0} \le 2^{-p}$ for
$k \ge \sqrt{p\log2/u_0} + O(1)$ and the dual terms are $O(e^{-\operatorname{Re}c_n/u_0})$, $O(\sqrt p)$ terms suffice;
the cancellation between them is at most $e^{(\log|z|)^2/(4u_0)}$. The $k$-terms at $s$ and $s+1$ are computed as in
Section~\ref{app:pieces}, $C_k$ and $S_k$ incrementally, and $L_k(w)$, $H(w)$ are series like the main term.

For the dual terms, Proposition~\ref{prop:dual} holds for complex $c$ by analytic continuation: $J^{(\delta)}_w(c)$ is a
component of the solution of $cY' = (B_0 + cB_1)Y$, continued to $\CC\setminus(-\infty, 0]$. Compute $c_n$ to $p$
bits and choose a Gaussian rational $c_n'$ with $O(\log p)$ bits, $|c_n - c_n'| \le \frac12|c_n|p^{-A}$, and
$\operatorname{Im}c_n' > 0$ if $c_n$ lies on the cut. The base values $D_a(c_n') = (c_n')^a(\Gamma(-a) - \gamma(-a, c_n'/u_0))$
need $\Gamma(-a)$ once and $\gamma(-a, y)$ at the Gaussian rational $y = c_n'/u_0$, an everywhere convergent series with
$O(p)$ P-recursive terms and cancellation $e^{O(|y|)} = 2^{O(p)}$; the series \eqref{eq:startvals} at $c_n'$ is of the
same kind. The bit-burst from $c_n'$ to $c_n$ is controlled by the Taylor-step lemma, which needs only $|Y(c_n')|$
and no bound on a disk; when $c_n$ lies on the cut, the Taylor series at $c_n'$ continue $Y$ along the segment, which
gives the boundary value from above. Each of the $O(\sqrt p)$ terms costs $O(\Mul(p)\log^2p)$.
\end{proof}

\subsection{Stretched exponentials}\label{app:stretched}
We sketch the cost analysis for $\Lambda(s;\tau,r)$ (Proposition~\ref{prop:expsplit}) with $r = r_1/r_2$ in lowest
terms, $0 < r \le 1$, and algebraic $s$ and $\tau > 0$. Write $a_j = (s-rj)/2$; then $a_j$ modulo $\ZZ$ depends only
on $j$ modulo $2r_2$, and within each class the coefficients $c_j = (-\tau)^j/(j!\Gamma(a_j))$ are hypergeometric in
$j$. We use $|\Gamma(a,x)| \le x^{\operatorname{Re}a-1}e^{-x}$ for $\operatorname{Re}a \le 1$, $x > 0$, and
$|1/\Gamma(a)| \le \Gamma(1-\operatorname{Re}a)\cosh(\pi\operatorname{Im}a)/\pi$ for $\operatorname{Re}a \le 0$.

\emph{$k$-terms.} The $j$-th term of $I_k$ is at most
$e^{-k^2u_0}\,O(1)\,(\tau u_0^{-r/2})^j\,\Gamma(1+\frac{rj}2-\frac{\operatorname{Re}s}2)/j!$, which decays like
$(j!)^{r/2-1}$; within each class the terms are P-recursive by $Q(a+1,x) = Q(a,x) + x^ae^{-x}/\Gamma(a+1)$.

\emph{Dual terms.} For $c > 0$ and $\operatorname{Re}a < 0$, $|D_a(c)| \le c^{\operatorname{Re}a}\Gamma(-\operatorname{Re}a)$,
so the $j$-th term of the series for $J_m$ at $c_0$ is at most $O(1)(\tau c_0^{-r/2})^j\Gamma(\frac{rj}2+O(1))^2/j!$.
This decays like $(j!)^{r-1}$ for $r < 1$, and geometrically with ratio $\tau/(2\sqrt{c_0})$ for $r = 1$, which needs
$c_0 > \tau^2$: for $m \le \tau/\pi$ one starts at such a $c_0$ and continues along the real axis. For $1 < r < 2$ the
terms grow like $(j!)^{r-1}$ for every $c_0$; we do not treat this case. With $v = u^{1/(2r_2)}$ the kernel is a power
of $u$ times an entire function $H(v^{-r_1})$ whose coefficients satisfy
$c_{j+2r_2}/c_j = \tau^{2r_2}(a_j-1)(a_j-2)\cdots(a_j-r_1)/(j+1)_{2r_2}$, so $H$ satisfies an equation with
property~(H). The integration by parts of Proposition~\ref{prop:dual}, with $t$ shifted in steps of $1/(2r_2)$, now
raises the moment index, and solving for the highest index gives a linear system in $c$ whose only finite singular
point is $c = 0$, possibly of higher Poincar\'e rank; the bit-burst applies at $c \gg 0$ unchanged.

\emph{Main term.} $M_\tau(1-s) = \Gamma(\frac{1-s}r)\tau^{(s-1)/r}/r$ needs one value of $\Gamma$ at an algebraic point;
the tails $T_w$ are convergent series with P-recursive terms.

Altogether $\Lambda(s;\tau,r)$ costs $p^{3/2+o(1)}$, the extra factors coming only from the $O(\log p)$ variation in the
lengths of the series. For the products of Proposition~\ref{prop:prodsplit} every piece is a double sum of terms
holonomic in $(n, j)$; by closure under definite summation \cite{Zei90} the inner sums are P-recursive in $n$, and
since unwanted solutions grow at most like $(n!)^{O(1)}C^n$, $O(p\log p)$ guard bits suffice, which again gives
$p^{3/2+o(1)}$.

\end{small}

\section{Reference implementation for Khinchin's constant}\label{sec:khpseudo}

\lstdefinestyle{ref}{language=Python, basicstyle=\ttfamily\scriptsize, columns=fixed, basewidth=0.5em, keepspaces=true,
  numbers=left, numberstyle=\tiny, numbersep=6pt, xleftmargin=1.5em, showstringspaces=false,
  keywordstyle=\ttfamily, commentstyle=\itshape, aboveskip=0.6\baselineskip, belowskip=0.6\baselineskip}

The Python program \texttt{khinchin\_ref.py} in the repository evaluates Theorem~\ref{thm:khinchin} with the
bounds of Section~\ref{sec:khbounds}. We list it in full, with annotations; the line numbers are those of the
file. It uses python-flint: \texttt{fmpq} and \texttt{fmpq\_mat} are exact rationals and rational matrices, and
\texttt{arb} is a real ball. Quantities computed from $u_0$, $k$, $n$, $j$, $c_0$, $c$ and $\delta$ alone are exact
rationals, and so are all matrices. Every other quantity is a ball at working precision $p + 30$ bits, and
\texttt{pm(x, r)} adds $r$ to the radius of $x$. The program is written for clarity, not speed; at $p = 5000$ it
takes a few seconds and returns a ball of radius $6\cdot10^{-1507}$ containing $K$.

\emph{Parameters and helpers.} The split point $u_0 = \frac1{16}$, the number $r_0 = 16$ of bits of the first
bit-burst point and the growth factor $g = 4$ are global. A line \texttt{N = least(n0, cond)} sets $N$ to the least
$N \ge n_0$ for which an upper bound of a tail bound $R(N)$ is at most $\varepsilon = 2^{-p-10}$. The black boxes are
$\exp$, $\sqrt{\cdot}$, $\erfc$, $\pi$, $\log 2$, $\Li_2(-\frac12)$ and $\Eip(y) = \operatorname{Ei}(y) - \gamma - \log y$.
\begin{lstlisting}[style=ref, firstnumber=12]
import sys
from flint import arb, fmpq, fmpq_mat, ctx

u0 = fmpq(1, 16)         # splitting point (rational)
r0 = 16                  # bits of the first bit-burst point
g = 4                    # bit-burst growth factor
half = fmpq(1, 2)

def least(n, cond):                     # min{n' >= n : cond(n')}
    while not cond(n):
        n += 1
    return n

def mat(rows):                          # rational matrix from a list of rows
    return fmpq_mat(len(rows), len(rows[0]),
                    [fmpq(x) for r in rows for x in r])

def matvec(S, v):                       # exact matrix times vector of balls
    return [sum((arb(S[i, j]) * v[j] for j in range(S.ncols())
                 if S[i, j] != 0), arb(0)) for i in range(S.nrows())]

def pm(x, r):                           # x +/- r
    return x + arb(0, arb(r).upper())

def Eip(y):                             # sum_{n>=1} y^n/(n n!)
    return y.ei() - arb.const_euler() - y.log()
\end{lstlisting}

\emph{Binary splitting.} All series are summed by one primitive. If $v_{n+1} = T_nv_n$ with
$T_n \in \QQ^{d\times d}$ and $W_n \in \QQ^{e\times d}$, then $v_b = Pv_a$ and $\sum_{n=a}^{b-1}W_nv_n = Sv_a$. An
efficient implementation writes $T_n = \tilde T_n/q_n$ with integer $\tilde T_n$ and $q_n$, uses the triangular or
sparse structure of the $T_n$, and applies partial products to the state instead of forming $S$ exactly
(Section~\ref{sec:khimpl}).
\begin{lstlisting}[style=ref, firstnumber=39]
# Binary splitting
def BinSplit(T, W, a, b):
    """(P, S): P = T(b-1)...T(a), S = sum_{a<=n<b} W(n) T(n-1)...T(a)."""
    if b == a + 1:
        return T(a), W(a)
    m = (a + b) // 2
    P1, S1 = BinSplit(T, W, a, m)
    P2, S2 = BinSplit(T, W, m, b)
    return P2 * P1, S1 + S2 * P1
\end{lstlisting}

\emph{The $k$-terms.} $I_k = e^{-x}\sum_{n=1}^{\infty}P_n/n^2$ is summed with the state
$(P_n, Q_n) = ((4/\kappa)^ne_n(x), (4u_0)^n/n!)$, starting from $(P_1, Q_1) = (q, 4u_0)$, where $\kappa = k^2$ (\texttt{K} in the code), $x = \kappa u_0$ and
$q = 4/\kappa$; $R(N)$ is the tail bound of Section~\ref{sec:khbounds}.
\begin{lstlisting}[style=ref, firstnumber=49]
# The k-terms I_k
def KTerm(k, eps):
    K = fmpq(k * k); x = K * u0; q = 4 / K; n1 = int(x.ceil()) + 1
    X = arb(x)

    def tau(n):
        p = n * X ** (n - 1) * (-X).exp() / arb(n - 1).fac()
        return arb(q) ** n * min(arb(1), p) / n ** 2

    def R(N):
        if N < n1:
            return 2 * tau(N) + arb(9) / 5 * arb(q) ** n1 / n1 ** 2
        return arb(9) / 5 * arb(q) ** (N + 1) / (N + 1) ** 2

    N = least(1, lambda N: R(N) <= eps)
    T = lambda n: mat([[q, q], [0, 4 * u0 / (n + 1)]])
    W = lambda n: mat([[fmpq(1, n * n), 0]])
    _, S = BinSplit(T, W, 1, N + 1)
    s = matvec(S, [arb(q), arb(4 * u0)])[0]
    return pm((-X).exp() * s, R(N))
\end{lstlisting}

\emph{The main term.} Since $\pi_n(y) = e^{-y}\sum_{i=n}^{\infty}y^i/i!$, exchanging the order of summation gives
$M = e^{-u_0}A - e^{-4u_0}B - \frac12C$ with $A = \sum_{i=1}^{\infty}w_i$, $B = \sum_{i=1}^{\infty}\hat w_i$ and
$C = \sum_{i=1}^{\infty}v_i/i^2$, where
\[
  v_i = \frac{(4u_0)^i}{i!}, \qquad w_i = \frac{u_0^i}{i!}\sum_{n=1}^{i}\frac{4^n}{n^2}, \qquad
  \hat w_i = v_i\sum_{n=1}^{i}\frac1{n^2},
\]
and $v_1 = w_1 = \hat w_1 = 4u_0$; the state is $(v_i, w_i, \hat w_i)$.
\begin{lstlisting}[style=ref, firstnumber=70]
# The main term M
def MainTerm(eps):
    U = arb(u0)
    R = lambda N: 6 * (8 * U) ** N / arb(N).fac()
    N = least(1, lambda N: R(N) <= eps)

    def T(i):
        a = 4 * u0 / (i + 1); b = u0 / (i + 1); r = 4 * u0 / fmpq(i + 1) ** 3
        return mat([[a, 0, 0], [r, b, 0], [r, 0, a]])
    W = lambda i: mat([[0, 1, 0], [0, 0, 1], [fmpq(1, i * i), 0, 0]])
    _, S = BinSplit(T, W, 1, N + 1)
    A, B, C = matvec(S, [arb(4 * u0)] * 3)
    return pm((-U).exp() * A - (-4 * U).exp() * B - C / 2, R(N))
\end{lstlisting}

\emph{Starting values.} The dual state at $t = \frac32$ is
\[
  V = \bigl(\nu_{3/2},\ Y_{3/2},\ Y_{1/2},\ C_{3/2},\ \operatorname{Forcing}(E)\bigr),
\]
where the last three components are $E\,u_0^{-1/2}(\Eip(4u_0), e^{4u_0}, 1)$. At a rational point $c_0$, with $D = D_{-1/2}(c_0)$ and
$\hat E = e^{-c_0/u_0}u_0^{-1/2}$, the recurrence \eqref{eq:Drec} gives $D_{n-1/2} = \alpha_nD + \tilde\beta_n\hat E$
with $\alpha_0 = 1$, $\tilde\beta_0 = 0$ and
\[
  \alpha_{n+1} = -\frac{c_0\alpha_n}{n+\frac12}, \qquad \tilde\beta_{n+1} = \frac{u_0^{n+1} - c_0\tilde\beta_n}{n+\frac12}.
\]
With $b_n = 4^n/n!$ the binary splitting state is $(b_n\alpha_n, b_n\tilde\beta_n, b_nu_0^n)$, and the six rows of
$W_n$ are the $n$-th terms of the $D$- and $\hat E$-parts of $\nu_{3/2} = \sum_{n=1}^{\infty}\frac{b_n}nD_{n-1/2}$,
$Y_{3/2} = \sum_{n=0}^{\infty}b_nD_{n-1/2}$ and $Y_{1/2} = \sum_{n=0}^{\infty}b_nD_{n+1/2}$; the terms $n = 0$ are added
separately (line~116). The two parts have size about $e^{4\sqrt{c_0} - c_0/u_0}$ and their sum about $e^{-c_0/u_0}$;
the cancellation costs no absolute precision, since both parts are smaller than $1$.
\begin{lstlisting}[style=ref, firstnumber=84]
# Forcing components of the dual state
def Forcing(E):
    U = arb(u0)
    Eh = E / U.sqrt()
    return [Eh * Eip(4 * U), Eh * (4 * U).exp(), Eh]

# Starting values of the dual state at a rational point c0
def StartValues(c0, eps):
    U, C0 = arb(u0), arb(c0)
    D = (arb.pi() / C0).sqrt() * (C0 / U).sqrt().erfc()
    E = (-C0 / U).exp(); Eh = E / U.sqrt()
    alpha = lambda n: C0 ** n / arb.rising(arb(half), n)
    b = lambda n: arb(4) ** n / arb(n).fac()

    def R(N):
        return 2 * b(N) * max(2 * alpha(N + s) * D
                              + E * U ** (N + s - half) / (N + s - half)
                              for s in (0, 1))
    N = least(1, lambda N: 32 * c0 <= (2 * N + 1) * (2 * N + 2)
              and 8 * u0 <= N + 1 and R(N) <= eps)

    def T(n):
        h = n + half; d = 4 / ((n + 1) * h)
        return mat([[-c0 * d, 0, 0], [0, -c0 * d, u0 * d],
                    [0, 0, 4 * u0 / (n + 1)]])

    def W(n):
        h = n + half
        return mat([[fmpq(1, n), 0, 0], [0, fmpq(1, n), 0], [1, 0, 0],
                    [0, 1, 0], [-c0 / h, 0, 0], [0, -c0 / h, u0 / h]])
    _, S = BinSplit(T, W, 1, N)
    s = matvec(S, [arb(-8 * c0), arb(8 * u0), arb(4 * u0)])
    s = [si + arb(ti) for si, ti in zip(s, [0, 0, 1, 0, -2 * c0, 2 * u0])]
    nu = pm(D * s[0] + Eh * s[1], R(N))
    Y = pm(D * s[2] + Eh * s[3], R(N))
    Ym = pm(D * s[4] + Eh * s[5], R(N))
    return [nu, Y, Ym, D] + Forcing(E)
\end{lstlisting}

\emph{Taylor stages.} At a fixed point $c$, let $V_j$ be the state at $t = \frac32 + j$,
\[
  V_j = \bigl(\nu_t,\ Y_t,\ Y_{t-1},\ C_t,\ \Eip(4u_0)Eu_0^{1-t},\ e^{4u_0}Eu_0^{1-t},\ Eu_0^{1-t}\bigr).
\]
The integration-by-parts recurrences of Section~\ref{sec:dilog} read $V_{j+1} = M_jV_j$ with
\[
  M_j = \frac1c\begin{pmatrix}
  \eta_j & -1 & 0 & 1 & 1 & 0 & 0 \\
  0 & \eta_j & -4 & 0 & 0 & 1 & 0 \\
  0 & c & 0 & 0 & 0 & 0 & 0 \\
  0 & 0 & 0 & \eta_j & 0 & 0 & 1 \\
  0 & 0 & 0 & 0 & c/u_0 & 0 & 0 \\
  0 & 0 & 0 & 0 & 0 & c/u_0 & 0 \\
  0 & 0 & 0 & 0 & 0 & 0 & c/u_0
  \end{pmatrix}, \qquad \eta_j = j + \tfrac12.
\]
Since $\partial_cZ_t = -Z_{t+1}$ and $E(c+\delta) = E(c)e^{-\delta/u_0}$, every component satisfies
$V(c+\delta) = \sum_{j=0}^{\infty}\frac{(-\delta)^j}{j!}V_j(c)$, so $T_j = -\frac{\delta}{j+1}M_j$. The first four
components are summed this way and the last three are recomputed.
\begin{lstlisting}[style=ref, firstnumber=122]
# One bit-burst stage: from V(c) to V(c + delta)
def TaylorStage(c, delta, V, eps):
    U = arb(u0)
    rho = arb(abs(delta) / c)
    R = lambda L: 2 * (4 * U).exp() * rho ** L / (arb(c).sqrt() * (1 - rho))
    L = least(2, lambda L: R(L) <= eps)

    def T(j):
        h = -delta / (j + 1); e = j + half
        M = [[e / c, -1 / c, 0, 1 / c, 1 / c, 0, 0],
             [0, e / c, -4 / c, 0, 0, 1 / c, 0],
             [0, 1, 0, 0, 0, 0, 0],
             [0, 0, 0, e / c, 0, 0, 1 / c],
             [0, 0, 0, 0, 1 / u0, 0, 0],
             [0, 0, 0, 0, 0, 1 / u0, 0],
             [0, 0, 0, 0, 0, 0, 1 / u0]]
        return mat([[h * x for x in row] for row in M])
    W = lambda j: mat([[int(i == l) for l in range(7)] for i in range(4)])
    _, S = BinSplit(T, W, 0, L)
    E = (-arb(c + delta) / U).exp()
    return [pm(si, R(L)) for si in matvec(S, V)] + Forcing(E)
\end{lstlisting}

\emph{Dual terms.} This is the bit-burst algorithm. The point $c_m = \pi^2(m+\frac12)^2$ is truncated to
$r_0, gr_0, g^2r_0, \dots$ fractional bits, up to the number $r^*$ of bits required by the bound for $|J'|$ in
Section~\ref{sec:khbounds}, and $J_m = \sqrt\pi\,\nu_{3/2}(c_m)$.
\begin{lstlisting}[style=ref, firstnumber=144]
# The dual terms J_m
def DualTerm(m, eps):
    U = arb(u0)
    c = arb.pi() ** 2 * (m + arb(half)) ** 2
    trunc = lambda r: fmpq(int((c.mid() * 2 ** r).floor().unique_fmpz()),
                           2 ** r)
    cm = c.lower() - arb(2) ** (-r0)
    K0 = 2 * arb.pi().sqrt() * Eip(4 * U) * (-cm / U).exp() / (U.sqrt() * cm)
    x = ((K0 / (eps - K0 * c.rad())).log() / arb.const_log2()).upper()
    rstar = max(r0, int(x.ceil().unique_fmpz()))
    r = r0; cc = trunc(r)
    V = StartValues(cc, eps)
    while r < rstar:
        r = min(g * r, rstar)
        V = TaylorStage(cc, trunc(r) - cc, V, eps)
        cc = trunc(r)
    return pm(arb.pi().sqrt() * V[0], K0 * (arb(2) ** (-rstar) + c.rad()))
\end{lstlisting}

\emph{Khinchin's constant.} The functions \texttt{tk} and \texttt{tm} are the bounds for the omitted $k$-terms and
dual terms of Section~\ref{sec:khbounds}, and the constant is assembled as in Theorem~\ref{thm:khinchin}. For
example, $p = 5000$ gives $k_{\max} = 235$ and $m_{\max} = 4$, and the dual term $m = 0$ uses the points with
$r = 16, 64, 256, 1024, 4096, 4951$ fractional bits, and $m = 4$ those with $r = 16, 64, 256, 391$.
\begin{lstlisting}[style=ref, firstnumber=162]
# Khinchin's constant
def Khinchin(p):
    ctx.prec = p + 30
    eps = arb(2) ** (-p - 10)
    U, pi = arb(u0), arb.pi()
    G = Eip(4 * U) / U
    tk = lambda k: (2 * (-((k + 1) ** 2 - 4) * U).exp()
                    / (((k + 1) ** 2 - 4) * U))
    tm = lambda m: (4 * (pi * U).sqrt() * G
                    * (-pi ** 2 * (m + 1 + arb(half)) ** 2 / U).exp())
    kmax = least(3, lambda k: tk(k) <= eps)
    mmax = least(0, lambda m: tm(m) <= eps)
    S = MainTerm(eps)
    for k in range(3, kmax + 1):
        S += (-1) ** k * KTerm(k, eps)
    for m in range(0, mmax + 1):
        S += DualTerm(m, eps)
    S = pm(S, tk(kmax) + tm(mmax))
    log2 = arb.const_log2()
    L0 = log2 ** 2 + arb(-half).polylog(2) + pi ** 2 / 12
    return ((L0 + S / 2) / log2).exp(), kmax, mmax
\end{lstlisting}


\end{document}